\documentclass[twocolumn,prd,superscriptaddress,nofootinbib,preprintnumbers]{revtex4-1}

\usepackage{placeins}
\usepackage{mathtools,slashed}
\usepackage{booktabs,multirow}
\usepackage{graphicx}
\usepackage{amsmath,bm,amssymb,amsfonts,dsfont,xspace}
\usepackage{textgreek}
\usepackage{derivative}
\usepackage[usenames,dvipsnames]{xcolor}
\usepackage[normalem]{ulem}
\usepackage{url}
\usepackage{footmisc}
\usepackage[colorlinks  = true,
            linkcolor   = NavyBlue,
            urlcolor    = NavyBlue,
            citecolor   = NavyBlue,
            anchorcolor = NavyBlue, unicode]{hyperref}

\newcommand{\lsim}{\mathrel{\mathop{\kern 0pt \rlap
  {\raise.2ex\hbox{$<$}}}
  \lower.9ex\hbox{\kern-.190em $\approx$}}}
\newcommand{\gsim}{\mathrel{\mathop{\kern 0pt \rlap
  {\raise.2ex\hbox{$>$}}}
  \lower.9ex\hbox{\kern-.190em $\approx$}}}

\usepackage{stfloats}
\usepackage{lineno}

\usepackage[compat=1.1.0]{tikz-feynman} % TikZ-Feynman (TikZ by Till Tantau)
\tikzfeynmanset{every vertex/.style={/tikz/font=\small}}

\usepackage[normalem]{ulem} %temporary

\usepackage{pifont}

\newcommand{\diag}{\mathrm{diag}}

\newcommand{\dd}{\mathrm{d}}
\newcommand{\ii}{\mathrm{i}}

\newcommand{\MKK}{M_{\mathrm{KK}}}

\begin{document}

\raggedbottom

%\title{A sequestered radiative neutrino portal for secluded dark matter}
\title{A geometric origin for the radiative neutrino portal to secluded dark matter}

\author{Mattia Di Mauro}\email{dimauro.mattia@gmail.com}
\affiliation{Istituto Nazionale di Fisica Nucleare, Sezione di Torino, Via P. Giuria 1, 10125 Torino, Italy}

\date{\today}

%==========================================================
% Abstract
%==========================================================
\begin{abstract}

We study a secluded dark-matter scenario in which the smallness of laboratory signals is linked to the origin of neutrino masses. In the four-dimensional theory, the dark sector contains a hidden scalar, a fermionic dark-matter particle, and heavy Majorana neutrinos. The same heavy-neutrino sector that generates light neutrino masses via the seesaw mechanism also induces the Higgs–dark-scalar portal at one loop. This portal is tied to the small observed neutrino masses, naturally leading to very small Higgs–singlet mixing and suppressed signals in direct-detection and collider experiments. 
We then provide a geometric origin for the boundary condition of the tree level portal coupling being zero $\kappa(\Lambda_{\rm UV})=0$ via a five-dimensional sequestered setup. In this construction, the Standard Model fields and the hidden sector are localized on different branes, with sterile neutrinos propagating in the bulk. Five-dimensional locality forbids a fundamental local tree-level Higgs--dark-scalar contact interaction, while heavy-neutrino loops induce a small residual portal coupling.
The main phenomenological parameter controlling the viability of the model is the heavy-neutrino mass scale: direct detection bounds it from above, while Big Bang nucleosynthesis bounds it from below through the requirement that the hidden scalar decays sufficiently early. Depending on the amount of five-dimensional sequestering, the viable region can span heavy-neutrino masses from the multi-TeV scale to the PeV scale, or even higher.
\end{abstract}

\maketitle
%\tableofcontents

%==========================================================

\section{Introduction}
\label{sec:intro}

Dark matter (DM) is required by a broad set of gravitational observations, from galactic rotation curves and cluster dynamics to gravitational lensing and the cosmic microwave background anisotropies~\cite{Zwicky:1933gu,Rubin:1970zza,Clowe:2006eq,Planck:2018vyg}. These observations indicate the presence of a non-luminous and non-baryonic matter component that cannot be accounted for by the known particles of the Standard Model (SM). A viable particle candidate must be sufficiently stable, electrically
neutral, and non-relativistic during structure formation, while its nongravitational interactions must satisfy laboratory and astrophysical constraints~\cite{Jungman:1995df,Bertone:2004pz,Cirelli:2024ssz}.

The search for DM proceeds in complementary directions. Direct-detection experiments look for the scattering of Galactic-halo DM particles in detector nuclei or electrons \cite{Schumann:2019eaa}. Collider searches attempt to produce invisible states in high-energy \(pp\) or \(e^+e^-\) collisions \cite{Boveia:2018yeb}, while indirect-detection experiments search for cosmic photons, neutrinos, or charged particles produced by DM annihilation or decay \cite{Gaskins:2016cha}. The null results of these searches place strong pressure on many standard weakly interacting massive particle (WIMP) scenarios. 
In particular, XENONnT and LZ constrain the spin-independent DM--nucleon cross
section to the $10^{-48}$--$10^{-47}\,{\rm cm}^2$ range for weak-scale DM masses ($m_\chi$)~\cite{Aprile:2023XENONnT,Aalbers:2024LZ}. This has made WIMP models in which the same interaction controls both thermal freeze-out and present-day scattering highly constrained, with large portions of the viable thermal parameter space now excluded~\cite{DiMauro:2023tho,Arcadi:2024ukq,Kong:2025ccv,DiMauro:2025jia,Koechler:2025ryv}.

One possible way to reconcile a thermal relic with small direct-detection rates is to exploit resonant annihilation. If the mediator mass satisfies $m_\phi\simeq 2m_\chi$, the annihilation cross section is enhanced in the early Universe and the observed abundance can be obtained with small couplings~\cite{Griest:1990kh,Arcadi:2017kky,Arcadi:2019lka,Arcadi:2024ukq,DiMauro:2025jia}. This mechanism provides a region of the parameter space compatible with direct detection constraints and relic density observations, but it typically relies on a special relation between the DM and mediator masses, which relies on a special mass relation that is not generic and calls for an explanation in the underlying theory~\cite{DiMauro:2025jia}.

A different possibility is that the interaction setting the relic abundance is not the same one that controls visible-sector searches. This is the basic idea of secluded DM, originally proposed in Refs.~\cite{Pospelov:2007mp,Pospelov:2008jd}. In this class of models, the DM particle $\chi$ annihilates dominantly into lighter hidden-sector mediators $\phi$,
\begin{equation}
\chi\bar\chi\to\phi\phi, \qquad m_\chi>m_\phi,
\end{equation}
while the mediator communicates with the SM only through a portal parametrized by a coupling $\epsilon$. Therefore, the thermal abundance is controlled mainly by a dark-sector coupling $g_X$, whereas direct detection, collider production and visible indirect signals are controlled by the coupling $\epsilon$ that can take very small values. In particular, values of $g_X=\mathcal{O}(0.1\text{--}1)$ can give a standard thermal cross section through annihilations within the hidden sector, while the connection to the SM remains very weak for $\epsilon\ll1$.

The phenomenology of secluded scenarios has recently been comprehensively revisited in concrete BSM realizations in Ref.~\cite{DiMauro:2025jsb}. A central point of that analysis was to go beyond the simple requirement of obtaining the correct relic density and to check the consistency of the full thermal history. In particular, Ref.~\cite{DiMauro:2025jsb} derived the conditions under which the mediator remains in thermal contact with the SM bath in the early Universe, identifying the range of portal couplings for which the visible and hidden sectors are efficiently equilibrated before freeze-out. At the same time, it determined the minimum portal strength required for the mediator to decay before Big Bang nucleosynthesis (BBN), so that late electromagnetic or hadronic energy injection does not spoil the observed light-element abundances~\cite{Cooke:2018qzw,Kurichin:2021,Sbordone:2010}.

These results show that a viable secluded thermal history can be maintained even for very small visible portals. For portal couplings well below the percent level, and in particular for $\epsilon\ll10^{-3}$, the relic abundance is controlled mainly by the dark-sector annihilation process $\chi\chi\to\phi\phi$, rather than by annihilation directly into SM particles. The corresponding spin-independent scattering rate is then strongly suppressed and can lie below current bounds, and even below the projected reach of experiments limited by coherent neutrino-scattering backgrounds~\cite{DARWIN:2016hyl}. 
The price of this phenomenological success is that the smallness of the portal coupling calls for a theoretical explanation.

This theoretical question is especially sharp in secluded DM models with scalar mediators. In general, the renormalizable Higgs–singlet operator $\kappa R^2 \Phi^\dagger \Phi$, where $\Phi$ is the SM Higgs doublet and $R$ is a hidden-sector scalar, is allowed by SM gauge invariance.  In a generic effective theory formulated in four spacetime dimensions (4D), one would therefore expect $\kappa \neq0$ at tree level, unless some symmetry or UV mechanism forces $\kappa=0$.
For a generic portal coupling, $\kappa =\mathcal{O}(1)$, the induced Higgs--singlet mixing would lead to detectable direct-detection rates, possible visible indirect-detection signatures through mediator decays, modifications of Higgs properties, and exotic Higgs decay channels \cite{Kong:2025ccv}. Choosing $\kappa$ to be very small can of course make the model compatible with laboratory and astrophysical constraints at the EFT level. However, without an underlying symmetry or UV mechanism, this choice is not theoretically well motivated: it does not explain why this renormalizable interaction should be so strongly suppressed relative to the other allowed couplings of the theory.

A possible way to address this problem is to make the Higgs--dark portal a radiative quantity rather than an independent tree-level parameter. In Ref.~\cite{DiMauro:2025nmsdm}, it was proposed that the leading portal can be neutrino-aligned: the same heavy-neutrino interactions that generate light neutrino masses through a type-I seesaw also induce the scalar operator $R^2\Phi^\dagger\Phi$ at one loop~\cite{Minkowski:1977sc,Yanagida:1979as,GellMann:1979,Glashow:1979vf,Mohapatra:1979ia,Casas:2001sr}. In this framework the smallness of the portal is not imposed as an unrelated tuning, but is linked to the smallness of the light-neutrino masses and to the loop suppression of the heavy-neutrino sector.
The corresponding EFT assumption is that the tree-level portal vanishes at the UV scale,
\begin{equation}
\kappa(\Lambda_{\mathrm{UV}})=0.
\label{eq:kUV0}
\end{equation}
This is a nontrivial assumption. In the low-energy EFT, when the theory scale is below the lightest heavy-neutrino mass, the renormalization-group evolution of the Higgs--dark-scalar portal is multiplicative: if $\kappa_{\rm tree}$ is exactly zero at a given scale, ordinary running does not generate it at lower energies. A nonzero portal therefore requires an additional physical source, such as the finite threshold correction induced by the heavy-neutrino sector. The central question addressed in this work is therefore how the boundary condition $\kappa(\Lambda_{\mathrm{UV}})=0$ can theoretically arise, rather than being imposed by hand.

The goal of this work is to give this boundary condition a meaningful physical interpretation through five-dimensional (5D) sequestering. The use of an extra compact dimension follows the long tradition of Kaluza--Klein (KK) theories~\cite{Kaluza:1921tu,Klein:1926tv}, brane-world constructions~\cite{ArkaniHamed:1998rs,Randall:1999ee,Randall:1999vf}, and extra-dimensional sequestering mechanisms~\cite{Randall:1998uk,Kaplan:1999ac}. The basic idea is simple: the visible sector and the hidden sector are localized on different branes of the 5D higher-dimensional space, the only nongravitational messenger fields assumed to connect the two branes are the bulk sterile neutrinos.
Similar bulk-neutrino constructions have been used to address the smallness of neutrino masses in extra-dimensional models~\cite{Dienes:1998sb,Grossman:1999ra,Neubert:2000zb}. In our case, the geometrical separation of the two sectors forbids a direct Higgs--dark contact interaction at tree level. The only fields that can communicate between the two branes are the bulk neutrino messengers. The Higgs--dark portal is therefore generated radiatively and inherits both the loop suppression and the neutrino-sector structure.

The phenomenological consequence is that the model provides a physical reason why direct-detection and collider signals are highly suppressed, without abandoning thermal secluded DM. The relic abundance can still be set by thermalization and annihilations inside the hidden sector, while the smallness of the SM neutrino masses is explained through the same heavy-neutrino sector via the seesaw mechanism. At the same time, laboratory probes are not absent in principle. 
In the aligned benchmark with fixed light neutrino mass $m_\nu$ and Yukawa coupling $y_N$, the loop-induced mixing grows with the heavy-neutrino mass ($M_N$). Therefore, direct detection and Higgs-mixing observables can become relevant for sufficiently large $M_N$, typically around or above the PeV scale for the benchmarks considered below. For smaller, multi-TeV heavy-neutrino masses, the most relevant constraints instead come from the lifetime and decay channels of the hidden mediator $R$.

The paper is organized as follows. In Sec.~\ref{sec:4dmodel}, we
introduce the four-dimensional model and discuss its main
phenomenological implications. In Sec.~\ref{sec:5D}, we present its
five-dimensional sequestered realization with bulk sterile neutrinos.
We summarize our conclusions in Sec.~\ref{sec:conclusion}.

%==========================================================
\section{Four-dimensional model}
\label{sec:4dmodel}
%==========================================================

In this section, we present the ingredients of the model in
four-dimensional spacetime. We first define the low-energy field
content and the assumptions concerning the tree-level Higgs--dark
scalar portal. We then present the one-loop matching calculation,
discuss the corresponding additive contribution to the beta function,
and explain how the smallness of the Standard Model neutrino masses is
accounted for through the seesaw mechanism. Finally, we examine the
consequences for scalar mixing and for laboratory and astroparticle
searches.

\subsection{Field content, assumptions and scalar potential}
\label{sec:field_content_4D}

The four-dimensional low-energy theory we consider is the same of Ref.~\cite{DiMauro:2025nmsdm} to which we refer for the full details. 
The model contains the SM fields,
including the Higgs doublet $\Phi$ and the lepton doublets
$L_\alpha$, three gauge-singlet right-handed Weyl fermions
$N_{RI}$, a real scalar singlet $R$, and a Dirac dark-matter
fermion $\chi$. After the Majorana mass matrix of the sterile
fermions is generated and diagonalized, we denote the corresponding
four-component Majorana mass eigenstates by $N_I$. In the
five-dimensional realization introduced in
Sec.~\ref{sec:5D}, the fields $N_{RI}$ arise from the orbifold-even
right-handed components of five-dimensional Dirac fermions. The
massive KK levels are vector-like before the hidden-brane Majorana
interaction is included, whereas the physical four-dimensional
eigenstates obtained after diagonalizing the complete mass matrix are
Majorana or pseudo-Dirac states.

We embed the dark fermion in a minimal renormalizable
$U(1)_X$ sector with gauge boson $Z'_\mu$ and gauge coupling $g_X$.
The fermion $\chi$ is a singlet under the SM gauge group and carries
a vector-like $U(1)_X$ charge, $q_{\chi_L}=q_{\chi_R}\equiv q_X^\chi$.
The scalar $R$, the sterile neutrinos, and all SM fields are taken to
be neutral under $U(1)_X$. The relevant renormalizable interactions
may then be written as
\begin{widetext}

\begin{eqnarray}
\mathcal L_{\rm int} 
\supset 
-\frac{1}{4}X_{\mu\nu}X^{\mu\nu}
-\frac{\epsilon_X}{2}X_{\mu\nu}B^{\mu\nu}
+\overline{\chi}
\left(i\slashed D-m_{\chi,0}\right)\chi
+\partial_\mu R\,\partial^\mu R
-V(\Phi,R) + \nonumber \\ -
\left[
(Y_\nu)_{\alpha I}
\overline{L_\alpha}\widetilde\Phi N_{RI}
+
\frac{1}{2}(Y_N)_{IJ}
R\,\overline{N_{RI}^{\,c}}N_{RJ}
+\mathrm{h.c.}
\right]
-y_pR\,\overline{\chi}\chi ,
\label{eq:Yukawa4D}
\end{eqnarray}
\end{widetext}
where
\begin{equation}
\widetilde\Phi=i\sigma_2\Phi^\ast,
\qquad
D_\mu=\partial_\mu+i q_X^\chi g_X Z'_\mu,
\end{equation}
and $Y_N=Y_N^T$ is a complex symmetric matrix. We take $y_p$
to be real without loss of generality. With the convention used in
this work, the kinetic term of the real scalar is
$\partial_\mu R\,\partial^\mu R$, so that the fluctuation $\rho$
defined below is canonically normalized.

Because the $U(1)_X$ charge of $\chi$ is vector-like, the bare Dirac
mass $m_{\chi,0}$ is gauge invariant. The unbroken $U(1)_X$ charge
also guarantees the stability of $\chi$, provided that it is the
lightest particle carrying this charge. Since neither $R$ nor the
sterile neutrinos carry $U(1)_X$ charge, the gauge boson does not
participate in the neutrino portal. We assume that its mass is
generated by a Stueckelberg mechanism
\cite{Ruegg:2003yd,Kors:2004dx}. The kinetic mixing
$\epsilon_X$ with hypercharge is taken to be negligible
\cite{Holdom:1985ag,Essig:2013lka}. Since the low-energy spectrum
contains no field charged simultaneously under $U(1)_X$ and
hypercharge, vanishing kinetic mixing is not regenerated at one loop
within the effective theory.

The $Z'$ therefore plays a spectator role in the analysis below. In
particular, we restrict ourselves to the scalar-dominated secluded
regime by assuming either $m_{Z'}>m_\chi$
so that $\chi\overline\chi\to Z'Z'$ is kinematically forbidden, or
that $g_X$ is sufficiently small for this channel to be subdominant
to $\chi\overline\chi\to H_pH_p$. 
We additionally assume that the spectator gauge boson either decays promptly into dark-sector states, for example when $m_{Z'}>2m_\chi$, or is never appreciably populated. This avoids an
additional long-lived $Z'$ relic in the limit of negligible kinetic mixing.
Under these assumptions, the relic
density, direct-detection rate, and mediator lifetime considered in
this work depend only negligibly on the $U(1)_X$ gauge sector.

The part of the scalar potential containing operators even in $R$ is
\begin{equation}
V(\Phi,R)
=
\mu_H^2\Phi^\dagger\Phi
+\mu_R^2R^2
+\lambda_H(\Phi^\dagger\Phi)^2
+\lambda_RR^4
+\kappa R^2\Phi^\dagger\Phi .
\label{eq:V4D}
\end{equation}
The coefficient $\kappa$ controls the small mixing between $H_p$ and $H$ and hence the strength of visible-sector probes.
At tree level, boundedness of the scalar potential requires $\lambda_H>0$, $\lambda_R>0$, $\kappa>-2\sqrt{\lambda_H\lambda_R}$,
with the last coefficient understood in the normalization of Eq.~\eqref{eq:V4D}. We additionally require both CP-even mass eigenvalues to be positive and all couplings to remain perturbative up to the cutoff of the effective theory.
For a generic real singlet, the additional renormalizable operators $t_RR$, $\mu_3/3! R^3$, $\mu_{R\Phi}R\Phi^\dagger\Phi$ are also allowed. Two consistent treatments of these operators are discussed in Appendix~\ref{app:odd_scalar_operators}. In the general real-singlet theory, five-dimensional locality imposes the boundary
conditions
\begin{equation}
\kappa(\Lambda_{\rm UV})=0,
\qquad
\mu_{R\Phi}(\Lambda_{\rm UV})=0,
\label{eq:UV_portal_conditions_general}
\end{equation}
because both operators connect fields localized on opposite branes.
They may nevertheless be regenerated radiatively by the bulk
messenger sector. In this case, observables that depend only on the
physical scalar mixing constrain the combination
\begin{equation}
\kappa_{\rm eff}
\equiv
\kappa+\frac{\mu_{R\Phi}}{\sqrt{2}v_r},
\label{eq:kappa_eff_main}
\end{equation}
rather than the two coefficients separately.

Alternatively, one may impose a $Z_4$ symmetry that forbids all
operators containing an odd number of $R$ fields while preserving
the Yukawa interactions in Eq.~\eqref{eq:Yukawa4D}. The vector-like
mass $m_{\chi,0}$ then provides a soft breaking confined to the
$\chi$ sector. It generates hidden-sector tadpole and cubic
counterterms, but does not induce the linear Higgs portal at the
leading heavy-neutrino loop order considered here. In this
realization,
\begin{equation}
\kappa_{\rm eff}\simeq\kappa
\end{equation}
to the accuracy of the present calculation. The explicit loop
matching performed below corresponds to the quartic contribution
$\kappa$; all expressions involving scalar mixing may be extended to
the general real-singlet theory through the replacement
$\kappa\to\kappa_{\rm eff}$.
Unless stated otherwise, the numerical analysis adopts the softly broken $Z_4$ realization.

A bare Majorana mass for the sterile fermions is also allowed in the
absence of an additional symmetry,
\begin{equation}
\Delta\mathcal L_N
=
-\frac{1}{2}
\overline{N_{RI}^{\,c}}
(M_{\rm bare})_{IJ}
N_{RJ}
+\mathrm{h.c.}
\label{eq:bareMajorana}
\end{equation}
We denote this matrix by $M_{\rm bare}$ to distinguish it from the
mass $M_0$ of the lightest physical state used later in the
five-dimensional analysis. The most general heavy-neutrino mass
matrix after symmetry breaking is therefore
\begin{equation}
M_N
=
M_{\rm bare}
+\frac{v_r}{\sqrt{2}}Y_N .
\label{eq:general_heavy_mass}
\end{equation}
Our baseline scenario assumes $M_{\rm bare}=0$
so that the heavy-neutrino scale originates from the same
$RNN$ interaction that enters the radiative portal. This condition
is symmetry protected in the $Z_4$ realization described in
Appendix~\ref{app:odd_scalar_operators}; in the unrestricted
real-singlet theory it should instead be understood as a simplifying
UV boundary condition.

After electroweak and hidden-sector symmetry breaking, we write
\begin{equation}
\Phi
=
\begin{pmatrix}
0\\
(v_h+h)/\sqrt{2}
\end{pmatrix},
\qquad
R=\frac{v_r+\rho}{\sqrt{2}},
\label{eq:scalar_expansion_4D}
\end{equation}
where $v_h\simeq246\,{\rm GeV}$. 

In the baseline limit $M_{\rm bare}=0$, the heavy-neutrino masses are
generated dominantly by the Yukawa interaction proportional to $Y_N$ once $R$
acquires a vev
\begin{equation}
-\frac{1}{2}\,Y_N^{ij}\,R\,\overline{N_i^{\,c}}N_j+\text{h.c.}
\;\supset\;
-\frac{1}{2}\,\overline{N_i^{\,c}}\,(M_N)^{ij}\,N_j+\text{h.c.},
\label{eq:MNfromvr}
\end{equation}
where $(M_N)^{ij}\equiv v_r/\sqrt{2}\,Y_N^{ij}$.
Therefore, the effective heavy-neutrino and light-neutrino mass matrixes that enters the seesaw mechanism are 
\begin{equation}
M_N=\frac{v_r}{\sqrt{2}}Y_N,
\qquad
m_D=\frac{v_h}{\sqrt{2}}Y_\nu .
\label{eq:VEVs4D}
\end{equation}
The matrix $M_N$ in Eq.~\eqref{eq:VEVs4D} is the physical
four-dimensional Majorana mass matrix and should not be confused
with the five-dimensional Dirac bulk mass $M_5$ introduced in
Sec.~\ref{sec:5D}. The light-neutrino mass matrix is obtained through
the type-I seesaw relation
\begin{equation}
m_\nu
\simeq
-m_DM_N^{-1}m_D^T .
\label{eq:seesaw4D}
\end{equation}

The same symmetry breaking gives the physical dark-matter mass
\begin{equation}
m_\chi
=
m_{\chi,0}
+\frac{y_pv_r}{\sqrt{2}} .
\label{eq:physical_DM_mass}
\end{equation}
The independent vector-like mass therefore permits $m_\chi$ and
$y_p$ to be treated as separate renormalized inputs. We note,
however, that for
$y_pv_r/\sqrt{2}\gg m_\chi$ a weak-scale physical mass requires a
cancellation between the two contributions in
Eq.~\eqref{eq:physical_DM_mass}. This possible naturalness issue does
not alter the phenomenological parameterization adopted here and is
discussed further in
Appendix~\ref{app:odd_scalar_operators}.

In the even-potential realization, the CP-even scalar mass matrix in
the $(h,\rho)$ basis is
\begin{equation}
\mathcal M^2
=
\begin{pmatrix}
2\lambda_Hv_h^2 & \kappa v_hv_r\\
\kappa v_hv_r & 2\lambda_Rv_r^2
\end{pmatrix}.
\label{eq:massmatrix}
\end{equation}
The scalar interaction eigenstates are related to the mass
eigenstates by
\begin{equation}
\begin{pmatrix}
H\\
H_p
\end{pmatrix}
=
\begin{pmatrix}
\cos\alpha & \sin\alpha\\
-\sin\alpha & \cos\alpha
\end{pmatrix}
\begin{pmatrix}
h\\
\rho
\end{pmatrix},
\label{eq:scalar_rotation}
\end{equation}
where $H$ is mostly SM-like and $H_p$ is mostly singlet-like. In the
general real-singlet theory, the linear and quartic portals combine
in the off-diagonal mass term as
\begin{equation}
\left(\mathcal M^2\right)_{h\rho}
=
v_h
\left(
\kappa v_r+\frac{\mu_{R\Phi}}{\sqrt{2}}
\right)
=
\kappa_{\rm eff}v_hv_r .
\label{eq:general_offdiagonal_mass}
\end{equation}
The physical mixing angle therefore satisfies
\begin{equation}
\tan(2\alpha)
=
\frac{2\kappa_{\rm eff}v_hv_r}
     {m_{H_p}^2-m_H^2}.
\label{eq:eigenmix}
\end{equation}
For the softly broken $Z_4$ realization,
$\kappa_{\rm eff}\simeq\kappa$ at the order considered.
We will use simply $\kappa$ in the rest of the paper. In the
small-mixing regime relevant for this work,
\begin{equation}
H\simeq h,
\qquad
H_p\simeq\rho .
\end{equation}

For later reference, the leading couplings of the scalar mass
eigenstates to SM fermions and to DM are
\begin{align}
\mathcal L_{H,H_p-f\overline f}
&=
-\sum_f\frac{m_f}{v_h}
\left(
\cos\alpha\,H-\sin\alpha\,H_p
\right)
\overline f f ,
\label{eq:scalar_SMfermion_couplings}
\\
\mathcal L_{H,H_p-\chi\overline\chi}
&=
-\frac{y_p}{\sqrt{2}}
\left(
\sin\alpha\,H+\cos\alpha\,H_p
\right)
\overline\chi\chi .
\label{eq:scalar_DM_couplings}
\end{align}
Thus, for $|\alpha|\ll1$, SM particles couple predominantly to $H$,
whereas the dark fermion couples predominantly to $H_p$.

The DM particle is assumed to lie in the secluded regime, in which
the interaction controlling its thermal abundance is predominantly
a hidden-sector interaction rather than the small coupling to the
SM. We impose
\begin{itemize}
\item $m_\chi>m_{H_p}$, so that
      $\chi\overline\chi\to H_pH_p$ is kinematically open;
\item $y_p=\mathcal O(0.1\text{--}1)$, so that this process can
      efficiently determine the thermal relic abundance;
\item $|\tan(2\alpha)|\ll1$, so that the singlet-like scalar
      communicates only weakly with the SM;
\item either $m_{Z'}>m_\chi$ or a sufficiently small $g_X$, so that
      annihilation into gauge bosons does not dominate freeze-out.
\end{itemize}
In this region, freeze-out is controlled mainly by $y_p$, while the
small physical mixing angle governs direct detection, collider
production, and the visible decay rate of $H_p$. These observables
therefore constrain $\kappa_{\rm eff}$ in the general real-singlet
theory and reduce to constraints on the radiatively generated
quartic coupling $\kappa$ in the softly broken $Z_4$ realization.

\newpage

\subsection{Seesaw mechanisms and Casas--Ibarra parametrization}

It is useful to express the neutrino-sector dependence of the portal in terms of physical seesaw parameters. We work in the basis in which the heavy-neutrino mass matrix is diagonal,
\begin{equation}
\widehat M_N=\diag(M_1,M_2,M_3),
\end{equation}
and the light-neutrino mass matrix is diagonalized by the PMNS matrix $U_\nu$ defined as $U_\nu^T m_\nu U_\nu=\widehat m_\nu$, where $\widehat m_\nu=\diag(m_{\nu,1},m_{\nu,2},m_{\nu,3})$. The Casas--Ibarra parameterization~\cite{Casas:2001sr} expresses the Dirac mass matrix $m_D$ as
\begin{equation}
m_D =
\mathrm{i}\,U_\nu^*\sqrt{\widehat m_\nu}\, \mathcal O\sqrt{\widehat M_N},
\qquad
\mathcal O^T\mathcal O=\mathbf 1. \end{equation}
Here $\mathcal{O}$ is a complex orthogonal matrix. This parameterization is useful because it automatically reproduces the type-I seesaw relation in Eq.~\eqref{eq:seesaw4D} for any choice of $\mathcal{O}$. The matrix $\mathcal{O}$ contains the remaining freedom in the neutrino Yukawa sector that is not fixed by the measured light-neutrino masses and mixings.

Since $Y_\nu=\sqrt{2}m_D/v_h$, the quantity entering the loop-induced portal (see Sec.~\ref{sec:radiativeloop}) is
\begin{equation}
(Y_\nu^\dagger Y_\nu)_{II}=\frac{2M_I}{v_h^2}\sum_{i=1}^3m_{\nu,i}|\mathcal{O}_{iI}|^2.
\label{eq:YnuCI}
\end{equation}

A particularly transparent limit is the aligned case where $\mathcal{O}=\mathbf{1}_{3\times 3}$. In this limit each light-neutrino mass eigenstate is associated with one heavy-neutrino state. Equivalently, there is no additional rotation in the heavy-neutrino sector beyond the one already fixed by the diagonal mass basis. The off-diagonal contributions controlled by $\mathcal{O}_{iI}$ are then absent, and one obtains
\begin{equation}
(Y_\nu^\dagger Y_\nu)_{II}
=\sum_{\alpha=e,\mu,\tau}|y_{\alpha I}|^2
=\frac{2m_{\nu,I}M_I}{v_h^2}.
\label{eq:kaligned}
\end{equation}
This gives the effective seesaw mass parameter
\begin{equation}
\widetilde m_I\equiv\frac{(m_D^\dagger m_D)_{II}}{M_I}=m_{\nu,I}
=\frac{v_h^2}{2}\frac{|y_{\nu,I}|^2}{M_I},
\label{eq:mnuI}
\end{equation}
where $y_{\nu,I}\equiv (Y_\nu)_{II}$ in the aligned basis. This makes the physical scaling manifest: in the aligned limit the contribution of each heavy neutrino to the portal is directly proportional to the corresponding light-neutrino mass eigenvalue.

\subsection{Radiative generation of the portal}
\label{sec:radiativeloop}

\begin{figure}[!t]
\centering
\begin{tikzpicture}[baseline={(current bounding box.center)}]
\begin{feynman}
  \vertex (A) at (0,  1.2);
  \vertex (B) at (3.0, 1.2);
  \vertex (C) at (3.0,-1.2);
  \vertex (D) at (0, -1.2);
  \vertex[left=1.2cm of A] (PhiL) {$\rho$};
  \vertex[right=1.2cm of B] (PhiR) {$\rho$};
  \vertex[right=1.2cm of C] (HbotR) {$\Phi^\dagger$};
  \vertex[left=1.2cm of D] (HbotL) {$\Phi$};
  \diagram*{
    (A) -- [fermion, edge label'=$N$] (B)
        -- [fermion, edge label'=$N$] (C)
        -- [fermion, edge label=$L$]   (D)
        -- [fermion, edge label=$N$]   (A),

    (PhiL)  -- [scalar, edge label=$Y_N/\sqrt{2}$] (A),
    (PhiR)  -- [scalar, edge label'=$Y_N^\ast/\sqrt{2}$] (B),
    (HbotR) -- [scalar, edge label=$Y_\nu^\ast$] (C),
    (HbotL) -- [scalar, edge label'=$Y_\nu$] (D),
  };
\end{feynman}
\end{tikzpicture}
\caption{One-loop box diagram generating the radiative Higgs--dark
portal after hidden-sector symmetry breaking. The two external
singlet fluctuations $\rho$ couple to the heavy-neutrino line through
$Y_N/\sqrt{2}$, while the two external Higgs fields couple through
$Y_\nu$ and the lepton doublet. At momenta below the heavy-neutrino
masses, this four-point amplitude is local and matches onto the
$\rho^2\Phi^\dagger\Phi$ component of the gauge-invariant operator
$R^2\Phi^\dagger\Phi$. In the five-dimensional realization the same
topology is present, but the heavy-neutrino propagators are replaced
by brane-to-brane bulk propagators or, equivalently, by a KK-tower
sum.}
\label{fig:box}
\end{figure}
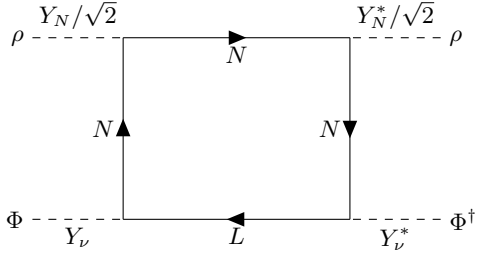

If the tree-level coefficient $\kappa$ is absent at the UV scale, the
leading visible--hidden interaction is generated by the one-loop box
diagram shown in Fig.~\ref{fig:box}. The diagram contains two
insertions of the visible-sector Yukawa coupling $Y_\nu$ and two
insertions of the hidden-sector Yukawa coupling $Y_N$. Since, in the
baseline scenario, the heavy-neutrino masses are generated by
$M_N=Y_Nv_r/\sqrt{2}$, the finite threshold matching is performed
after hidden-sector symmetry breaking but before electroweak symmetry
breaking. We therefore calculate the amplitude with two external
singlet fluctuations $\rho$, defined by
$R=(v_r+\rho)/\sqrt{2}$, and two external Higgs doublets. The full
one-loop matching calculation is given in
Appendix~\ref{app:loop_matching_complete}. In this section we
summarize the main steps and the resulting phenomenological scaling.

The physical meaning of the matching is simple. At external momenta
$p_i$ much smaller than the heavy-neutrino masses $M_N$, the loop
cannot resolve the short-distance propagation of the heavy fields.
The renormalized one-loop four-point amplitude with external fields
$\rho,\rho,\Phi,\Phi^\dagger$ is therefore local at leading order in
the external momenta and can be expanded as
\begin{equation}
\ii\mathcal M_{\rm loop}^{\rm full}
(\rho\rho\to\Phi\Phi^\dagger)
=
\ii\mathcal M_{\rm loop}^{\rm full}(0)
+
\mathcal O\left(\frac{p_i^2}{M_N^2}\right).
\label{eq:loop_low_momentum_expansion}
\end{equation}
The leading momentum-independent term has the same field structure as
the $\rho^2\Phi^\dagger\Phi$ component of the local operator
$R^2\Phi^\dagger\Phi$ and can therefore be absorbed into the
coefficient of the Higgs--dark-scalar portal.

We write the corresponding low-energy interaction as
\begin{equation}
\mathcal L_{\rm EFT}
\supset
-\kappa_{\rm EFT}(\mu)R^2\Phi^\dagger\Phi
\supset
-\frac{\kappa_{\rm EFT}(\mu)}{2}
\rho^2\Phi^\dagger\Phi,
\label{eq:EFT_portal_lagrangian}
\end{equation}
where the EFT portal coefficient receives tree-level and loop-induced contributions,
\begin{equation}
\kappa_{\rm EFT}(\mu)
=
\kappa(\mu)+\kappa_{\rm loop}(\mu).
\label{eq:ktot}
\end{equation}
In the radiative-portal scenario considered here, the UV boundary
condition is $\kappa(\Lambda_{\rm UV})=0$. 
The heavy-neutrino box diagram in Fig.~\ref{fig:box} is the leading radiative source of this operator.

Matching the loop-induced part of the full-theory amplitude gives
\begin{equation}
\ii\mathcal M_{\rm loop}^{\rm full}(0)
=
-\ii\kappa_{\rm loop}(\mu).
\label{eq:portal_matching_condition}
\end{equation}
The absence of a factor of two in this relation follows from the
normalization
$-(\kappa/2)\rho^2\Phi^\dagger\Phi$ after expanding
$R=(v_r+\rho)/\sqrt{2}$.

The matching is performed after hidden-sector symmetry breaking but
in the unbroken electroweak phase, at vanishing external momenta, and
treating the SM leptons as massless. 
Higher-order terms in
the expansion in $p_i^2/M_N^2$ generate derivative operators and are
not relevant for the local portal. The complete result in dimensional
regularization and in the $\overline{\rm MS}$ scheme can be written as
\begin{equation}
\ii\mathcal{M}_{\rm loop}^{\rm ren}(0)
=
\frac{\ii |Y_N|^2|Y_\nu|^2}{8\pi^2}
\left[
\ln\left(\frac{M_N^2}{\mu^2}\right)
+
2
\right],
\label{eq:Mloop_ren_new}
\end{equation}
where $\mu$ is the renormalization scale, which is conveniently chosen
close to the heavy-neutrino threshold. The finite constant depends on
the subtraction and matching convention, while the logarithmic term
and the Yukawa structure are fixed.

Using the matching relation in
Eq.~\eqref{eq:portal_matching_condition}, the loop-induced portal
coupling for one heavy neutrino is
\begin{equation}
\kappa_{\rm loop}(\mu)
= -\frac{|Y_N|^2|Y_\nu|^2}{8\pi^2} \left[ \ln\left(\frac{M_N^2}{\mu^2}\right) + 2 \right].
\label{eq:ksingle_new}
\end{equation}
Choosing the matching scale $\mu=M_N$ removes the logarithm and gives,
in the zero-momentum $\overline{\rm MS}$ matching convention used
here,
\begin{equation}
\kappa_{\rm loop}(M_N)
=
-\frac{|Y_N|^2|Y_\nu|^2}{4\pi^2}.
\label{eq:ksinglethreshold}
\end{equation}
In the aligned one-generation limit, Eq.~\eqref{eq:YnuCI} gives
\begin{equation}
\kappa_{\rm loop}
=
-\frac{y_N^2 M_N m_\nu}{2\pi^2 v_h^2},
\label{eq:ksinglethreshold_exact}
\end{equation}
which numerically becomes
\begin{equation}
\kappa_{\rm loop} \simeq
-4.2\times10^{-13} \left(\frac{y_N}{1}\right)^2 \left(\frac{M_N}{10\,{\rm TeV}}\right) \left(\frac{m_\nu}{0.05\,{\rm eV}}\right).
\label{eq:ksinglethreshold_new}
\end{equation}
The finite numerical coefficient in
Eq.~\eqref{eq:ksinglethreshold_new} corresponds to the matching
convention specified above. The scale dependence and the additive
beta-function term are controlled by the logarithmic part of
Eq.~\eqref{eq:ksingle_new}.

For three non-degenerate heavy neutrinos, the EFT matching should be
performed at the corresponding thresholds. In the basis in which the
heavy-neutrino mass matrix and the hidden Yukawa matrix are diagonal,
one obtains
\begin{equation}
\kappa_{\rm loop}^{\rm step}
=
-\sum_{I=1}^3
\frac{y_{N,I}^2\left(Y_\nu^\dagger Y_\nu\right)_{II}}{4\pi^2}.
\label{eq:kstep_general_new}
\end{equation}
Using the Casas--Ibarra parameterization,
\begin{equation}
\left(Y_\nu^\dagger Y_\nu\right)_{II}
=
\frac{2M_I}{v_h^2}
\sum_i m_{\nu,i}|\mathcal{O}_{iI}|^2 ,
\label{eq:YdagY_CI_new}
\end{equation}
and taking the aligned limit $\mathcal O=\mathbf{1}$, this reduces to
\begin{equation}
\kappa_{\rm loop}^{\rm step}
=
-\sum_{I=1}^3
\frac{y_{N,I}^2M_I m_{\nu,I}}{2\pi^2v_h^2}.
\label{eq:kstep_aligned_new}
\end{equation}
This structure is the minimal Yukawa combination that connects the
hidden scalar to the Higgs doublet. It is also the reason why the
portal is aligned with the neutrino sector: the same couplings that
participate in the seesaw mechanism determine the leading radiative
quartic interaction.

The following mixing estimate applies directly to the softly broken
$Z_4$ realization, for which
$\kappa_{\rm eff}\simeq\kappa$. In the general real-singlet theory,
the total mixing instead depends on $\kappa_{\rm eff}$ defined in
Eq.~\eqref{eq:kappa_eff_main}. Using the aligned single-generation
estimate and Eq.~\eqref{eq:eigenmix}, one obtains
\begin{eqnarray}
&&|\tan(2\alpha)|
\simeq
\frac{\sqrt{2}\,m_\nu y_N M_N^2}
{\pi^2v_h\left(m_{H_p}^2-m_H^2\right)} \simeq \\
&\simeq&
\frac{1.86\times10^{-10}}
{\left|1-(\frac{m_{H_p}}{125\,{\rm GeV}})^2\right|}
\left(\frac{y_N}{1}\right)
\left(\frac{|m_\nu|}{0.05\,{\rm eV}}\right)
\left(\frac{M_N}{10\,{\rm TeV}}\right)^2. \nonumber
\label{eq:tanaligned_new_signed}
\end{eqnarray}
This expression makes explicit the scaling
$|\tan(2\alpha)|\propto M_N^2$ in the aligned seesaw limit. 
Naively extrapolating the small-mixing expression suggests that order-one mixing would occur only for hundreds-of-PeV values for $M_N$, while phenomenologically relevant direct-detection effects can appear at lower masses depending on $m_{H_p}$, $y_p$, and the experimental sensitivity.

\subsection{Additive beta-function term and threshold interpretation}
\label{sec:beta_explained}

The purpose of this subsection is to clarify how the condition
$\kappa(\Lambda_{\rm UV})=0$ should be interpreted in the full and
low-energy effective theories. Here $\kappa(\mu)$ denotes the
renormalized coefficient of the local operator
$R^2\Phi^\dagger\Phi$ in the theory in which the heavy neutrinos are
active. It is therefore a local Lagrangian parameter. At tree level,
it coincides with the classical portal coupling. At the quantum
level, its counterterm cancels the ultraviolet divergence of the
four-point amplitude, and the renormalized coefficient evolves with
the renormalization scale. By contrast, $\kappa_{\rm loop}(\mu)$ is
not an independent Lagrangian coupling: it denotes the finite
contribution extracted from the explicit renormalized heavy-neutrino
loop amplitude at zero external momentum.

The portal is assumed to vanish at the UV scale because of a physical
mechanism, such as the five-dimensional sequestering construction
discussed in Sec.~\ref{sec:5D}. This does not mean that the local
portal coupling remains zero once the heavy neutrinos are included.
The reason is that the heavy neutrinos couple to both sectors: to the
SM through $Y_\nu$ and to the hidden scalar through $Y_N$. Therefore,
in the theory in which the heavy neutrinos are active fields, loops
containing $N_I$ can generate the operator
$R^2\Phi^\dagger\Phi$ even if its classical coefficient vanishes.

The same radiative effect appears through two complementary
ingredients: renormalization-group evolution above the heavy-neutrino
thresholds and threshold matching when the heavy neutrinos are
integrated out. We first consider the full-theory
renormalization-group description. For one heavy neutrino, the
renormalized loop contribution is given in
Eq.~\eqref{eq:ksingle_new}. Neglecting the running of the Yukawa
couplings within this one-loop expression, its explicit dependence on
the renormalization scale is
\begin{equation}
\frac{\mathrm{d}\kappa_{\rm loop}}{\mathrm{d}\ln\mu}
=
\frac{|Y_N|^2|Y_\nu|^2}{4\pi^2}.
\label{eq:dkloop_single_beta_explained}
\end{equation}

With the normalization adopted for the portal operator, the
renormalized zero-momentum four-point amplitude contains the
combination
\begin{equation}
\ii\mathcal{M}_{\rm full}^{\rm ren}(0,\mu)
=
-\ii\left[\kappa(\mu)+\kappa_{\rm loop}(\mu)\right].
\label{eq:ktot}
\end{equation}
At the order considered, a physical amplitude cannot depend on the
arbitrary renormalization scale. Isolating the contribution
proportional to $|Y_N|^2|Y_\nu|^2$ and independent of $\kappa$, scale
independence therefore requires
\begin{equation}
\frac{\mathrm{d}}{\mathrm{d}\ln\mu}
\left[
\kappa(\mu)+\kappa_{\rm loop}(\mu)
\right]_{\rm add}
=
0.
\label{eq:portal_scale_independence}
\end{equation}
It follows that the additive running of the local coupling cancels
the explicit scale dependence of the loop contribution:
\begin{equation}
\beta_\kappa^{\rm add}
\equiv
\left.\frac{\mathrm{d}\kappa}{\mathrm{d}\ln\mu}\right|_{\rm add}
=
-\frac{\mathrm{d}\kappa_{\rm loop}}{\mathrm{d}\ln\mu}
=
-\frac{|Y_N|^2|Y_\nu|^2}{4\pi^2}.
\label{eq:betasingle}
\end{equation}

The important point is that this contribution is additive: it remains
nonzero even when $\kappa=0$. Above the heavy-neutrino threshold, the
full beta function therefore has the schematic form
\begin{equation}
\beta_\kappa^{\rm full}
\equiv
\frac{\mathrm{d}\kappa}{\mathrm{d}\ln\mu}
=
\gamma_\kappa^{\rm full}\kappa
+
\beta_\kappa^{\rm add}.
\label{eq:beta_full_additive}
\end{equation}
Here $\gamma_\kappa^{\rm full}$ denotes the anomalous-dimension
coefficient of the portal operator in the full theory, including the
contributions of all fields active above the heavy-neutrino thresholds;
it governs the multiplicative running proportional to
$\kappa(\mu)$.
Consequently, in the theory containing active heavy neutrinos, the
condition $\kappa=0$ is not preserved by renormalization-group
evolution. If the boundary condition
$\kappa(\Lambda_{\rm UV})=0$ is imposed at a scale above the
heavy-neutrino masses, running toward the heavy-neutrino thresholds
generates a nonzero local portal coupling.

For three generations, while the heavy neutrinos are active, the
additive term has the spurion structure
\begin{equation}
16\pi^2\beta_\kappa^{\rm add}
=
-4\operatorname{Tr}\left[
Y_NY_N^\dagger Y_\nu^\dagger Y_\nu
\right].
\label{eq:ck}
\end{equation}
In the aligned basis used below,
$Y_N=\operatorname{diag}(y_{N,1},y_{N,2},y_{N,3})$. Equation~\eqref{eq:ck}
is the renormalization-group expression of the same physical
statement: the heavy-neutrino loop connects the visible and hidden
sectors and therefore radiatively generates the Higgs--dark-scalar
portal.

The second ingredient is threshold matching. When the renormalization
scale crosses a heavy-neutrino mass $M_I$, the field $N_I$ is removed
from the low-energy EFT. Its effect does not disappear; it is encoded
in the Wilson coefficient of the portal operator immediately below
the threshold. Choosing the matching scale close to the
heavy-neutrino mass avoids large logarithms. At $\mu=M_I$, the
logarithm in the threshold correction vanishes and the finite matching
contribution is
\begin{equation}
\Delta\kappa_I(M_I)
\equiv
\kappa_{{\rm loop},I}(M_I)
=
-\frac{y_{N,I}^2\left(Y_\nu^\dagger Y_\nu\right)_{II}}{4\pi^2}.
\label{eq:kloop_match_I}
\end{equation}
For a single threshold, the matching relation is
\begin{equation}
\kappa_{\rm EFT}(M_I^-)
=
\kappa(M_I^+)
+
\Delta\kappa_I(M_I),
\label{eq:kappa_single_threshold_matching}
\end{equation}
where $M_I^+$ and $M_I^-$ denote scales immediately above and below
the threshold, respectively.

For non-degenerate heavy neutrinos, the matching is performed
sequentially. Below the lightest heavy-neutrino threshold, all $N_I$
have been integrated out. We denote the portal coefficient in this
low-energy theory by $\kappa_{\rm EFT}(\mu)$. It is the Wilson
coefficient of $R^2\Phi^\dagger\Phi$ at a generic scale
$\mu<M_{\rm min}$, where $M_{\rm min}$ is the lightest heavy-neutrino
mass. Summing the finite threshold contributions gives
\begin{equation}
\kappa_{\rm loop}^{\rm step}
=
-\sum_I
\frac{y_{N,I}^2\left(Y_\nu^\dagger Y_\nu\right)_{II}}{4\pi^2}.
\label{eq:kstep_general_beta_explained}
\end{equation}
Neglecting the running between the thresholds for this estimate, the
boundary value below the lightest threshold is
\begin{equation}
\kappa_{\rm EFT}(M_{\rm min}^-)
=
\kappa(M_{\rm min}^+)
+
\kappa_{\rm loop}^{\rm step}.
\label{eq:kappa_EFT_boundary}
\end{equation}
Here $\kappa(M_{\rm min}^+)$ denotes the renormalized local
coefficient in the theory immediately above the lightest threshold.
If this local contribution is neglected, the matching condition
reduces to
\begin{equation}
\kappa_{\rm EFT}(M_{\rm min}^-)
\simeq
\kappa_{\rm loop}^{\rm step}.
\label{eq:kappa_EFT_boundary_approx}
\end{equation}
The symbol $M_{\rm min}^-$ indicates that the coefficient is evaluated
just below the lightest heavy-neutrino threshold. This equation fixes
the initial value of the low-energy coupling; it does not mean that
the renormalization scale is fixed at that value.

For scales below the heavy-neutrino thresholds, the additive
contribution $\beta_\kappa^{\rm add}$ is absent. The low-energy EFT no
longer contains the fields $N_I$ and therefore no longer contains the
loop involving both $Y_\nu$ and $Y_N$ that can generate
$R^2\Phi^\dagger\Phi$ from zero. The effect of that loop has already
been included through the matching condition in
Eq.~\eqref{eq:kappa_EFT_boundary}. In the absence of any remaining
light field that couples to both sectors, loops of the light fields
renormalize the portal only if the portal operator is already present.
The low-energy beta function is therefore multiplicative,
\begin{equation}
\beta_{\kappa_{\rm EFT}}^{\rm low}(\mu)
\equiv
\frac{\mathrm{d}\kappa_{\rm EFT}(\mu)}{\mathrm{d}\ln\mu}
=
\gamma_\kappa(\mu)\,\kappa_{\rm EFT}(\mu).
\label{eq:beta_low_multiplicative}
\end{equation}

Here $\gamma_\kappa$ is the multiplicative running coefficient of the
portal operator in the low-energy EFT. It describes the fractional
running of an already existing portal coupling. For the minimal low-energy theory containing the SM fields, the real scalar $R$, and
the hidden fermion $\chi$, and using the scalar-potential normalization in Eq.~\eqref{eq:V4D}
one may write schematically
\begin{equation}
\gamma_\kappa(\mu)
=
\frac{12\lambda_H+6\lambda_R+8\kappa_{\rm EFT}+6y_t^2+c_\chi y_p^2-\frac{9}{2}g_2^2-\frac{3}{2}g_1^2}{16\pi^2}.
\label{eq:gamma_kappa_explicit}
\end{equation}
All couplings in Eq.~\eqref{eq:gamma_kappa_explicit} are evaluated at
the scale $\mu$. Here $g_1$ and $g_2$ are the SM hypercharge and weak
gauge couplings, $y_t$ is the top-quark Yukawa coupling, $\lambda_H$
and $\lambda_R$ are the Higgs and singlet quartic couplings, and $y_p$
is the hidden-sector Yukawa coupling between $R$ and $\chi$. The
coefficient $c_\chi$ depends on the normalization of the hidden
Yukawa interaction and on whether $\chi$ is Dirac or Majorana. For
example, $c_\chi=4$ for a Dirac fermion with interaction
$-y_pR\overline{\chi}\chi$, while the Majorana case differs by a
convention-dependent factor. Its precise value is not important for
the present discussion. What matters here is that all these terms
multiply $\kappa_{\rm EFT}(\mu)$. There is no additive term
proportional to $Y_N^2Y_\nu^2$ below the heavy-neutrino thresholds,
because the heavy neutrinos have been integrated out of the
low-energy EFT.

The solution of Eq.~\eqref{eq:beta_low_multiplicative} is
\begin{equation}
\kappa_{\rm EFT}(\mu)
=
\kappa_{\rm EFT}(M_{\rm min}^-)
\exp\left[
\int_{\ln M_{\rm min}}^{\ln\mu}
\gamma_\kappa(\mu')\,\mathrm{d}\ln\mu'
\right].
\label{eq:k_low_running_solution}
\end{equation}
Thus, the heavy-neutrino thresholds determine the boundary value of
the portal, while the subsequent low-energy running only rescales
this value. In the phenomenological estimates considered in this
work, this running is a mild multiplicative correction. The reason is
that $\gamma_\kappa$ is loop suppressed and the light-sector
couplings run only logarithmically as long as they remain
perturbative. Moreover, the term proportional to
$8\kappa_{\rm EFT}$ in Eq.~\eqref{eq:gamma_kappa_explicit} is
numerically negligible in the radiative-portal regime because
$\kappa_{\rm EFT}$ is already very small. Therefore, unless stated
otherwise, we approximate
\begin{equation}
\kappa_{\rm EFT}(\mu_{\rm low})
\simeq
\kappa_{\rm EFT}(M_{\rm min}^-)
\simeq
\kappa_{\rm loop}^{\rm step}.
\label{eq:kappa_low_approx}
\end{equation}

This is the EFT interpretation used in the phenomenological analysis.
Above the heavy-neutrino masses, the local coefficient $\kappa$
receives an additive beta-function contribution because the heavy
neutrinos are active messengers between the visible and hidden
sectors. At the heavy-neutrino thresholds, the same loop produces
finite matching corrections. Below the thresholds, the heavy
neutrinos are absent and the Wilson coefficient
$\kappa_{\rm EFT}$ runs only multiplicatively. The role of the
five-dimensional construction is therefore not to remove the
loop-generated portal, but to explain why the boundary condition
$\kappa(\Lambda_{\rm UV})=0$ is meaningful in the first place.

\begin{figure}
\centering
\includegraphics[width=0.99\columnwidth]{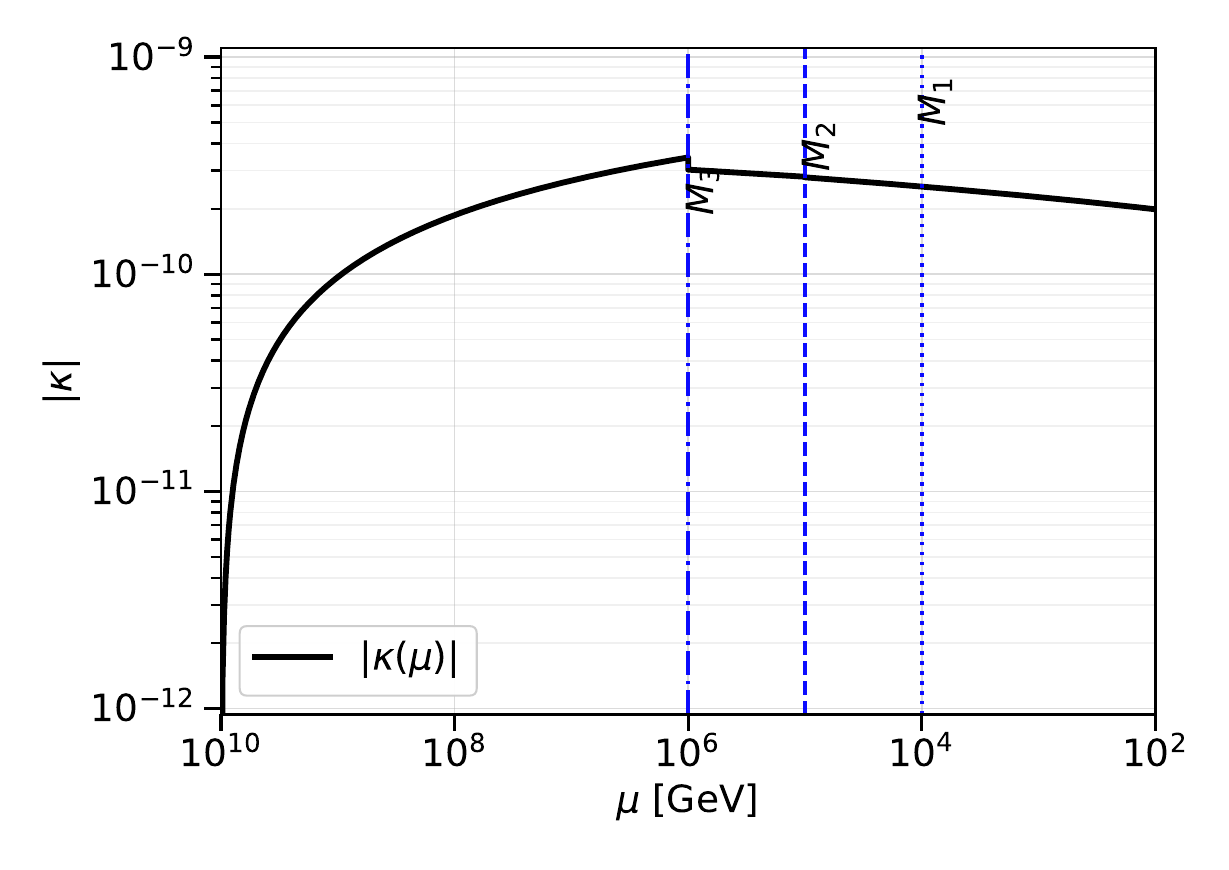}
\caption{Running and threshold matching of the radiatively generated
Higgs--dark-scalar portal. The plot shows the magnitude of the running
coefficient of the local operator $R^2\Phi^\dagger\Phi$ as a function
of the renormalization scale $\mu$, starting from the UV boundary
condition $\kappa(\Lambda_{\rm UV})=0$. In each energy interval,
$\kappa(\mu)$ denotes the renormalized portal coefficient in the theory
containing the corresponding active heavy neutrinos; below the lightest
threshold, it is identified with the Wilson coefficient
$\kappa_{\rm EFT}(\mu)$ of the low-energy effective theory. The vertical
lines mark the heavy-neutrino thresholds
$M_1=10^4\,{\rm GeV}$, $M_2=10^5\,{\rm GeV}$, and
$M_3=10^6\,{\rm GeV}$. We take
$\Lambda_{\rm UV}=10^{10}\,{\rm GeV}$, $y_{N,I}=1$, and use the aligned
seesaw relation with
$m_{\nu,I}=(0.01,0.02,0.05)\,{\rm eV}$. The Standard Model couplings
entering the multiplicative contribution to the portal beta function
are evolved at one loop, while the hidden-sector couplings are kept
fixed in this illustrative benchmark.}
\label{fig:kappa_running_thresholds}
\end{figure}

Figure~\ref{fig:kappa_running_thresholds} illustrates the EFT
interpretation of the radiatively generated portal. We impose the
boundary condition $\kappa(\Lambda_{\rm UV})=0$ at the UV scale and
evolve the local portal coefficient toward lower energies. For the
illustrative benchmark shown in the figure, we take
$\Lambda_{\rm UV}=10^{10}\,{\rm GeV}$, $M_1=10^4\,{\rm GeV}$,
$M_2=10^5\,{\rm GeV}$, $M_3=10^6\,{\rm GeV}$, $y_{N,I}=1$, and use
the aligned seesaw with
$m_{\nu,I}=(0.01,0.02,0.05)\,{\rm eV}$. The SM couplings entering
$\gamma_\kappa$ are evolved at one loop, while the hidden-sector
parameters $\lambda_R$ and $y_p$ are kept fixed because their running
depends on the detailed hidden-sector completion.

Above the heavy-neutrino thresholds, the fields $N_I$ are active and
generate an additive contribution to $\beta_\kappa$. When the scale
crosses each heavy-neutrino mass, the corresponding field is
integrated out and its finite threshold correction is included in the
Wilson coefficient of the portal operator below the threshold. The
vertical lines indicate the three heavy-neutrino masses. The apparent
jumps at the thresholds are not physical discontinuities in
observables, but finite matching corrections relating the theory above
the threshold to the EFT below it. Below the lightest threshold, the
additive heavy-neutrino source is absent and
$\kappa_{\rm EFT}$ runs only multiplicatively. In the benchmark shown,
the low-energy running gives only a mild rescaling of the
threshold-generated value, so
$\kappa_{\rm EFT}(\mu_{\rm low})\simeq
\kappa_{\rm loop}^{\rm step}$ is a good approximation for the
phenomenological estimates.

In the phenomenological estimates used in the rest of the paper, we
adopt the finite threshold result $\kappa_{\rm loop}^{\rm step}$ in
Eq.~\eqref{eq:kstep_general_beta_explained}, obtained by matching at
$\mu\simeq M_I$. The additional logarithmic contribution generated by
running above and between the heavy-neutrino thresholds depends on the
scale at which the boundary condition
$\kappa(\Lambda_{\rm UV})=0$ is imposed and is not included in the
benchmark estimates unless explicitly stated. Thus, these estimates
retain the finite threshold contribution while neglecting the
UV-to-threshold logarithmic evolution.

\medskip 

\subsection{Thermalization of the hidden sector}
\label{sec:hidden_thermalization_short}

The relic-density calculation in this work assumes that the hidden sector has a thermal origin. A detailed discussion of the thermal history of this class of secluded models is given in Ref.~\cite{DiMauro:2025nmsdm}; here we only summarize the ingredients that are relevant for the present construction.

At temperatures above the lightest heavy-neutrino mass, the visible and hidden sectors can be brought into thermal contact by the same neutrino-portal interactions that generate the radiative Higgs--dark-scalar portal. The relevant processes include decays and inverse decays of the heavy neutrinos, such as
\begin{equation}
N_I \leftrightarrow L_\alpha \Phi,
\end{equation}
as well as scatterings involving the hidden scalar,
\begin{equation}
N_I R \leftrightarrow L_\alpha \Phi.
\end{equation}
For seesaw-motivated parameters, these reactions are efficient at temperatures of order a few times the lightest heavy-neutrino mass. Thus the hidden scalar $R$, and through the Yukawa interaction $-y_pR\overline{\chi}\chi$ also the dark-matter particle $\chi$, can be thermally connected to the SM bath at early times.

When the temperature drops below the lightest heavy-neutrino mass, the abundance of the heavy neutrinos becomes Boltzmann suppressed and the neutrino-portal reactions rapidly decouple. The visible and hidden sectors then evolve approximately as two separate thermal baths. We denote the hidden-sector temperature by $T'$ and define
\begin{equation}
\zeta\equiv \frac{T'}{T},
\end{equation}
where $T$ is the visible-sector temperature. The value of $\zeta$ is fixed by entropy conservation in the two sectors after decoupling and is typically of order unity, up to factors depending on the particle content and on the decoupling temperature.

In the secluded regime considered here, which is valid for $m_\chi>m_{H_p}$ and $|\alpha|\ll 1$, the relic abundance is controlled mainly by annihilation inside the hidden sector $\chi\bar\chi\to H_pH_p$ rather than by annihilation directly into SM particles. For a scalar coupling $-y_pR\overline{\chi}\chi$, this process is p-wave suppressed, but it can still efficiently set the relic abundance at freeze-out, where the DM velocity is not negligible. The small Higgs--singlet mixing generated by the radiative portal is therefore not required to determine the relic density. Its main role is instead to allow the mediator $H_p$ to decay into SM particles before BBN.

For the scalar coupling $-y_p R\overline{\chi}\chi$, the annihilation cross section for $\chi\bar\chi\to H_pH_p$ is p-wave suppressed, but it can still reproduce the observed relic abundance for moderately large hidden-sector Yukawa couplings. Denoting by $T'$ the hidden-sector temperature and defining $x'_f=m_\chi/T'_f$, the value of $y_p$ required to obtain $\Omega_\chi h^2\simeq0.12$ scales approximately as
\begin{widetext}
\begin{equation}
y_p
\simeq
0.43
\left(\frac{x'_f}{25}\right)^{1/2}
\left(\frac{m_\chi}{100\,{\rm GeV}}\right)^{1/2}
\left(\frac{\zeta}{1}\right)^{1/4}
\left(\frac{g_{\ast S}^{\rm vis}}{86.25}\right)^{-1/4}
\left(\frac{g_\ast^H}{100}\right)^{1/8},
\label{eq:yp_relic_scaling_short}
\end{equation}
\end{widetext}
$g_{\ast S}^{\rm vis}$ is the visible-sector entropy number of degrees of freedom at freeze-out, and $g_\ast^H$ is the effective number of degrees of freedom entering the Hubble rate. For $\zeta=\mathcal{O}(1)$ and $x'_f\simeq20$--$30$, this gives $y_p=\mathcal{O}(0.1\text{--}1)$ over the GeV--TeV mass range. Thus the relic abundance can be fixed by an ordinary hidden-sector coupling, while the much smaller loop-induced Higgs--singlet mixing controls only the residual connection to the Standard Model and the lifetime of $H_p$.

In the phenomenological analysis below, we therefore assume that the hidden sector was thermalized at early times through the heavy-neutrino bridge and that dark-matter freeze-out occurred in the hidden bath. The detailed Boltzmann treatment, including the visible--hidden decoupling temperature, the temperature ratio $\zeta=T'/T$, and the p-wave freeze-out calculation, follows Ref.~\cite{DiMauro:2025nmsdm}. The main consistency requirement for the present paper is that the hidden scalar decays sufficiently early, while the relic abundance itself is controlled by the hidden-sector coupling $y_p$.

\subsection{Expected signals in laboratory and astroparticle searches}
\label{sec:expected_signals}

We now summarize the main experimental consequences of the loop-induced portal. The relevant low-energy parameter controlling the visible connection of the hidden scalar is the Higgs--singlet mixing angle $\alpha$. Therefore direct-detection rates, collider production of the mostly hidden scalar $H_p$, Higgs-signal-strength deviations and the lifetime of $H_p$ are all controlled by powers of $\sin\alpha$ or equivalently $\tan 2\alpha$.
%In the five-dimensional realization this mixing is further suppressed by the brane-to-brane factor,
%\begin{equation}
%\sin\alpha_{\rm 5D}\simeq F(M_5L)\sin\alpha_{\rm 4D}.
%\label{eq:alpha5D_expected_signals}
%\end{equation}

For fermionic DM coupled as $-y_p R\overline{\chi}\chi$, spin-independent scattering on nucleons proceeds through $t$-channel exchange of the two CP-even scalars $H$ and $H_p$. The corresponding cross section is
\begin{equation}
\sigma_{\rm SI}\simeq
\frac{\mu_{\chi N}^2m_N^2f_N^2y_p^2}{2\pi v_h^2}
\sin^2\alpha\cos^2\alpha
\left(\frac{1}{m_H^2}-\frac{1}{m_{H_p}^2}\right)^2 ,
\label{eq:sigmaSI}
\end{equation}
where $\mu_{\chi N}=m_\chi m_N/(m_\chi+m_N)$ is the DM--nucleon reduced mass and $f_N\simeq0.30$ is the scalar nucleon form factor.
%In the five-dimensional realization,
%\begin{equation}
%\sigma_{\rm SI}^{\rm 5D}\simeq F^2(M_5L)\,\sigma_{\rm SI}^{\rm 4D}.
%\label{eq:sigmaSI_5D_expected}
%\end{equation}
For $m_{H_p}\ll m_H$ and very small mixing, the $H_p$ propagator dominates and the SI cross section scales as
\begin{equation}
\sigma_{\rm SI}\propto y_p^2\sin^2\alpha\,m_{H_p}^{-4}.
\label{eq:sigmaSI_scaling_basic}
\end{equation}
Substituting the loop-induced, neutrino-aligned expression for the mixing angle gives the useful numerical estimate~\cite{DiMauro:2025nmsdm}\footnote{Note that in the original version of Ref.~\cite{DiMauro:2025nmsdm} there was an error in the prefactor of Eq.~\eqref{eq:sigmaSI_MN_numeric} that we have corrected here.}
\begin{widetext}
\begin{eqnarray}
\sigma_{\rm SI}
\sim
1.0\times10^{-47}\,{\rm cm}^2
\left(\frac{y_p}{0.4}\right)^2
\left(\frac{y_N}{1}\right)^2
\left(\frac{m_\nu}{0.05\,{\rm eV}}\right)^2
\left(\frac{M_N}{10\,{\rm PeV}}\right)^4
\left(\frac{10\,{\rm GeV}}{m_{H_p}}\right)^4 .
\label{eq:sigmaSI_MN_numeric}
\end{eqnarray}
\end{widetext}
Thus, for multi-TeV heavy neutrinos the prediction is far below current LZ and XENONnT limits, which are of order $10^{-48}$--$10^{-47}\,{\rm cm}^2$ for WIMP masses around the electroweak scale~\cite{LZ:2024zvo,Aprile:2023XENONnT}. Instead, for very light mediators and tens-of-PeV-scale heavy neutrinos, the strong $M_N^4$ scaling can bring the cross section closer to the reach of present or future direct-detection experiments.

Collider probes are suppressed by the same Higgs--singlet mixing angle. The couplings of the mostly SM-like Higgs boson $H$ to SM particles are rescaled by its SM-Higgs component,
\begin{equation}
g_{HXX}\simeq \cos\alpha\,g_{hXX}^{\rm SM},
\label{eq:H_coupling_rescaling}
\end{equation}
where $X$ denotes a SM fermion or gauge boson. As a result, production rates and partial decay widths into SM states scale approximately as $\xi_H=\cos^2\alpha$. If no additional exotic Higgs decay mode is relevant, the Higgs signal deviates from the SM one as $1-\xi_H\simeq \sin^2\alpha$.
The mostly hidden scalar $H_p$ couples to SM states through its small Higgs component,
\begin{equation}
g_{H_pXX}\simeq -\sin\alpha\,g_{hXX}^{\rm SM}\big|_{m_h=m_{H_p}} .
\label{eq:Hp_coupling_rescaling}
\end{equation}
Therefore its production rate at colliders is approximately
\begin{equation}
\sigma(pp\to H_p)
\simeq
\sin^2\alpha\,
\sigma_{\rm SM}(pp\to h_{\rm SM})\big|_{m_h=m_{H_p}},
\label{eq:Hp_collider_scaling}
\end{equation}
up to the model-dependent branching fractions of $H_p$~\cite{Cheung:2015dta}. For the loop-induced values of $\alpha$ obtained in the multi-TeV seesaw regime, these rates are far below present collider sensitivities. Collider constraints can become relevant only in less sequestered regions, for much larger heavy-neutrino masses, or if additional interactions enhance the production or visible decay branching ratios of $H_p$. If $m_{H_p}<m_H/2$, exotic Higgs decays such as $H\to H_pH_p$ may also be possible, but their rate depends on the scalar potential and on the trilinear scalar coupling, and is therefore more model dependent than the mixing-induced production rate.
Collider constraints for hidden sector models mediated by Higgs or Higgs-like scalars are typically not as strong as direct and indirect detection constraints (see e.g., \cite{DiMauro:2023tho}).

Heavy-neutrino production through active--sterile mixing is small because the seesaw estimate gives
\begin{equation}
\theta_{\nu N}\sim \sqrt{\frac{m_\nu}{M_N}}
\simeq 7\times10^{-8}
\left(\frac{m_\nu}{0.05\,{\rm eV}}\right)^{1/2}
\left(\frac{10\,{\rm TeV}}{M_N}\right)^{1/2}.
\label{eq:active_sterile_scaling}
\end{equation}
Therefore, for the multi-TeV heavy-neutrino masses relevant here, prompt collider signatures associated with active--sterile mixing are expected to be very suppressed.

Indirect detection deserves a separate discussion. In secluded DM models, indirect searches can be important because the dominant annihilation may proceed into light mediators that subsequently decay into SM particles~\cite{Pospelov:2007mp,Arina:2023eic}. In the present notation the relevant cascade is
\begin{equation}
\chi\bar\chi\to H_pH_p\to 4\,{\rm SM}.
\label{eq:cascade_ID}
\end{equation}
If the two mediators are produced on shell, the inclusive cascade rate factorizes as
\begin{equation}
\sigma(\chi\bar\chi\to H_pH_p\to 4\,{\rm SM})
\simeq
\sigma(\chi\bar\chi\to H_pH_p)\,{\rm Br}^2(H_p\to{\rm SM}).
\label{eq:cascade_factorization}
\end{equation}
The decay width of $H_p$ into SM particles is mixing suppressed,
\begin{equation}
\Gamma(H_p\to{\rm SM})\simeq \sin^2\alpha\,\Gamma_H^{\rm SM}(m_{H_p}),
\label{eq:Hp_width}
\end{equation}
where $\Gamma_H^{\rm SM}(m_{H_p})$ is the width of a SM Higgs boson evaluated at mass $m_{H_p}$. If no additional hidden decay channel is open, the total width is controlled by the same mixing-suppressed SM modes, and therefore ${\rm Br}(H_p\to{\rm SM})\simeq1$. In that case the small mixing does not suppress the on-shell cascade branching ratio; it suppresses the decay rate and can make $H_p$ long lived.

For the specific scalar-mediator realization considered here, the more important point is the velocity dependence of the annihilation. For a Dirac fermion with a scalar coupling $-y_p R\overline{\chi}\chi$, the secluded annihilation into two scalars is p-wave suppressed. Thus the same channel can set the thermal relic abundance at freeze-out, where $v_{\rm rel}^2=\mathcal{O}(0.1)$, but it is strongly suppressed today in the Galactic halo, where $v_{\rm rel}^2\sim10^{-6}$. Therefore, although secluded cascade signals can be important in more general models, present-day indirect-detection limits are strongly suppressed in the minimal scalar-mediator scenario studied here. An unsuppressed indirect signal would require a different dark-sector structure, for example a pseudoscalar coupling or another interaction generating an s-wave annihilation rate today.

The most relevant constraint associated with $H_p$ is instead its lifetime. For decays into fermions through Higgs mixing,
\begin{equation}
\Gamma(H_p\to f\bar f)
=
\sum_f
\frac{N_c^f m_{H_p}}{8\pi}
\frac{m_f^2}{v_h^2}
\sin^2\alpha
\left(1-\frac{4m_f^2}{m_{H_p}^2}\right)^{3/2},
\label{eq:Hp_ff_width}
\end{equation}
where the sum runs over kinematically open fermion channels. Neglecting threshold effects and summing over the dominant open fermions, a useful scaling is
\begin{eqnarray}
\tau_{H_p}
&\sim&
0.5\,{\rm s}
\left(\frac{1}{y_N}\right)^2
\left(\frac{20\,{\rm GeV}}{m_{H_p}}\right)
\left(\frac{10\,{\rm TeV}}{M_N}\right)^4
\nonumber\\
&\times&
\left(\frac{0.05\,{\rm eV}}{m_\nu}\right)^2
\left[1-\left(\frac{m_{H_p}}{125\,{\rm GeV}}\right)^2\right]^2 .
\label{eq:Hp_lifetime_MN_scaling}
\end{eqnarray}
The numerical result reported in Eq.~\eqref{eq:Hp_lifetime_MN_scaling} does not take into account threshold effects, QCD corrections, and the running fermion masses entering $\Gamma_H^{\rm SM}(m_{H_p})$.
This expression shows that, away from the resonant region $m_{H_p}\sim m_H$, the mediator lifetime is strongly sensitive to the heavy-neutrino scale: $\tau_{H_p}\propto M_N^{-4}$. In particular, for the benchmark in Eq.~\eqref{eq:Hp_lifetime_MN_scaling}, values of $M_N$ in the tens-of-TeV range are typically required for $H_p$ to decay before BBN.
%In the five-dimensional realization the physical mixing is reduced by $F(M_5L)$, and therefore
%\begin{equation}
%\tau_{H_p}^{\rm 5D}\simeq \frac{\tau_{H_p}^{\rm 4D}}{F^2(M_5L)}.
%\label{eq:Hp_lifetime_5D}
%\end{equation}
This is an important consistency condition. The same geometrical suppression that makes direct detection and collider probes even weaker can make $H_p$ too long lived. If $H_p$ decays after BBN, late electromagnetic or hadronic energy injection may spoil the successful predictions for the light-element abundances.

Therefore, direct detection and BBN constraints typically provide, respectively, an upper and a lower bound on the heavy-neutrino scale. For the benchmark choices considered here, the viable interval is roughly
\begin{equation}
10\,{\rm TeV}\lesssim M_N\lesssim10\,{\rm PeV},
\label{eq:MN_viable_rough}
\end{equation}
although the precise range depends on $m_{H_p}$, $y_N$, $m_\nu$, the direct-detection limit at the relevant DM mass, and possible additional hidden-sector decay channels. We will discuss in the next section how this constraint is modified in the five-dimensional sequestered model.
%Therefore, in the sequestered realization, BBN, CMB energy injection, displaced decays or additional hidden-sector decay modes can become more important than standard prompt Higgs-portal searches. In particular, if $F(M_5L)$ is very small, viability requires either a sufficiently large heavy-neutrino scale, a larger underlying loop-induced mixing, a suppressed abundance of $H_p$, or additional decay channels that allow $H_p$ to decay before BBN.

\subsection{Radiative stability and naturalness}
\label{sec:naturalness}

The hierarchy between the singlet-scalar and heavy-neutrino masses is
not radiatively protected when $M_N\gg m_{H_p}$ and
$y_N=\mathcal O(1)$. Heavy-neutrino loops generate threshold
corrections to the singlet-scalar mass of order
\begin{equation}
\left|\delta m_{H_p}^2\right|
\sim
\frac{y_N^2M_N^2}{16\pi^2},
\label{eq:singlet_mass_naturalness}
\end{equation}
up to logarithmic and scheme-dependent terms. Consequently, a light
singlet scalar in the presence of multi-TeV--PeV heavy neutrinos
requires a cancellation among the renormalized parameters of the
scalar potential. In the phenomenological analysis, we treat
$m_{H_p}$ as a low-energy input and do not attempt to address this
singlet-sector naturalness problem.

At the upper end of the heavy-neutrino mass range, the model also
inherits the familiar electroweak-naturalness issue of the type-I
seesaw~\cite{Vissani:1997ys,Clarke:2015hta}. Integrating out the
heavy Majorana neutrinos generates a threshold correction to the
quadratic Higgs parameter of order
\begin{equation}
\left|\delta\mu_H^2\right|
\sim
\frac{1}{8\pi^2}
\sum_I M_I^2
\left(Y_\nu^\dagger Y_\nu\right)_{II}.
\label{eq:higgs_mass_naturalness}
\end{equation}
In the aligned limit, this becomes
\begin{equation}
\left|\delta\mu_H^2\right|
\sim
\sum_I
\frac{m_{\nu,I}M_I^3}{4\pi^2v_h^2}.
\label{eq:higgs_mass_naturalness_aligned}
\end{equation}
For $m_\nu\simeq0.05\,\mathrm{eV}$, this correction remains below $|\mu_H^2|$ for masses around the PeV scale, but becomes
comparable to or larger than $|\mu_H^2|$ for
$M_N\gtrsim\mathcal O(10^7)\,\mathrm{GeV}$. We do not impose
electroweak naturalness as a hard constraint in the phenomenological
analysis. The highest-mass region should therefore be understood as
requiring a cancellation in the Higgs quadratic parameter unless an
additional protective mechanism is present.

%==========================================================
\section{Five-dimensional sequestering}
\label{sec:5D}
%==========================================================

We now provide a geometrical realization of the boundary condition
$\kappa(\Lambda_{\rm UV})=0$ discussed in the previous section by
embedding the model in a five-dimensional sequestered construction.
The SM and the hidden sector are localized on opposite
boundaries of a compact fifth dimension, while sterile fermions
propagate in the bulk and provide the only non-gravitational
communication between the two sectors. Five-dimensional locality
forbids a fundamental local tree-level operator
$R^2\Phi^\dagger\Phi$ at the cutoff scale, because $R$ and $\Phi$ are
confined to spatially separated boundaries. Nevertheless, nonlocal
brane-to-brane propagation of the bulk fermions generates the portal
radiatively. The condition $\kappa(\Lambda_{\rm UV})=0$ is therefore
not imposed arbitrarily in the four-dimensional theory at a UV scale, but follows
from the geometrical separation of the visible and hidden sectors.
We summarize the construction here and give the detailed KK reduction in Appendix~\ref{app:5D_Dirac_KK}.

%----------------------------------------------------------
\subsection{Geometry and field localization}
%----------------------------------------------------------

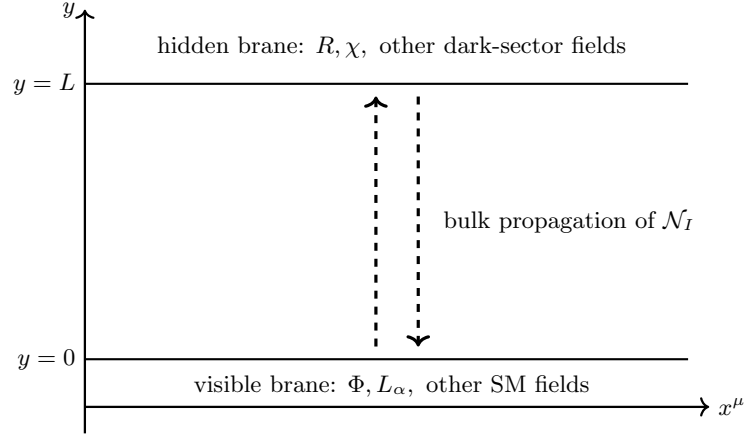
\begin{figure*}
\centering
\begin{tikzpicture}[scale=1.4,line width=0.85pt]
  \draw[->] (0,0) -- (0,4) node[left] {$y$};
  \draw[->] (0,0.25) -- (5.9,0.25) node[right] {$x^\mu$};
  \draw[thick] (0,0.7) -- (5.7,0.7);
  \draw[thick] (0,3.3) -- (5.7,3.3);
  \node[left] at (0,0.7) {$y=0$};
  \node[left] at (0,3.3) {$y=L$};
  \node at (2.9,0.45)
  {visible brane: $\Phi,L_\alpha,\text{ other SM fields}$};
  \node at (2.9,3.65)
  {hidden brane: $R,\chi,\text{ other dark-sector fields}$};
  \draw[very thick,dashed,->] (2.75,0.82) -- (2.75,3.18);
  \draw[very thick,dashed,->] (3.15,3.18) -- (3.15,0.82);
  \node[right] at (3.30,2.0)
  {bulk propagation of $\mathcal{N}_I$};
\end{tikzpicture}
\caption{Schematic representation of the five-dimensional sequestered
setup. The visible sector is localized on the brane at $y=0$, while
the hidden sector is localized on the brane at $y=L$. The two branes
share the same four-dimensional spacetime coordinates $x^\mu$ but are
separated along the compact fifth dimension $y$. The sterile fermions
$\mathcal{N}_I$ propagate in the bulk and, by assumption, provide the
only non-gravitational messengers between the two sectors.
Five-dimensional locality forbids a fundamental local
Higgs--dark-scalar contact operator involving fields confined to
opposite branes, while bulk propagation generates the corresponding
interaction radiatively in the low-energy four-dimensional theory.}
\label{fig:brane}
\end{figure*}

We consider a flat five-dimensional spacetime with coordinates
\begin{equation}
X^A=(x^\mu,y),
\qquad
A=0,1,2,3,5,
\end{equation}
and metric
\begin{equation}
\eta_{AB}
=
{\rm diag}(+,-,-,-,-).
\end{equation}
The fifth coordinate spans the physical interval
\begin{equation}
0\leq y\leq L,
\end{equation}
which may be regarded as the fundamental domain of the orbifold
$S^1/\mathbb Z_2$, as commonly done in compactified
five-dimensional field theories~\cite{Appelquist:2000nn,Csaki:2004ay}.
The visible boundary is located at $y=0$ and contains the SM fields,
including the Higgs doublet $\Phi$ and the lepton doublets
$L_\alpha$. The hidden boundary is located at $y=L$ and contains the
real scalar $R$, the DM field $\chi$, and the remaining hidden-sector
degrees of freedom. The localization of matter sectors on
four-dimensional boundaries, with selected fields propagating in the
higher-dimensional bulk, is a standard ingredient of brane-world
constructions~\cite{ArkaniHamed:1998rs,Rubakov:2001kp}. In the present
model, the sterile fermions $\mathcal N_I(x,y)$ propagate in the bulk,
as in higher-dimensional neutrino
models~\cite{Dienes:1998sb,Grossman:1999ra,ArkaniHamed:1999dc}.

The two boundaries share the same four-dimensional coordinates
$x^\mu$ and are separated only along the compact coordinate $y$.
Thus, the visible and hidden sectors are not separated in ordinary
three-dimensional space, but are localized at different positions in
the fifth dimension. 
Nongravitational communication between the two boundaries requires a bulk messenger or an explicitly nonlocal operator; in the minimal
local EFT considered here, the sterile fermions provide the only such messengers.

The setup is illustrated in Fig.~\ref{fig:brane}. The horizontal
direction schematically represents the four-dimensional spacetime
coordinates $x^\mu$ common to both branes, whereas the vertical
direction represents the compact coordinate $y$. Five-dimensional
locality forbids a fundamental local contact interaction involving
fields confined to the two different boundaries. Nevertheless, the
bulk sterile fermions $\mathcal N_I$ can connect the visible and
hidden sectors through propagation across the interval and thereby
generate the Higgs--dark-scalar portal radiatively.

%----------------------------------------------------------
\subsection{Five-dimensional action and orbifold projection}
%----------------------------------------------------------

The total action is written as
\begin{equation}
S
=
S_{\rm bulk}
+
S_{\rm vis}
+
S_{\rm hid}.
\label{eq:Stotal5D}
\end{equation}
The field $\mathcal N_I$ is a five-dimensional Dirac fermion. From
the four-dimensional point of view, it decomposes into two
four-dimensional chiral components of opposite chirality,
\begin{equation}
\mathcal N_I
=
\mathcal N_{IL}
+
\mathcal N_{IR},
\,\,\,
\mathcal N_{IL}=P_L\mathcal N_I,
\,\,\,
\mathcal N_{IR}=P_R\mathcal N_I,
\end{equation}
with
\begin{equation}
P_{L,R}
=
\frac{1\mp\gamma^5}{2}.
\end{equation}
% At each massive KK level, these two components combine into an
% ordinary four-dimensional Dirac fermion. The orbifold projection can
% instead remove one of the two zero modes and leave a chiral massless
% state.

The bulk action on the physical interval is
\begin{equation}
S_{\rm bulk}
=
\int\dd^4x
\int_0^L\dd y\,
\overline{\mathcal N}_I
\left(
\ii\Gamma^A\partial_A
-
M_{5,I}
\right)
\mathcal N_I,
\label{eq:Sbulk5D}
\end{equation}
where
\begin{equation}
\Gamma^A
=
(\gamma^\mu,\ii\gamma^5),
\qquad
\{\Gamma^A,\Gamma^B\}
=
2\eta^{AB}.
\end{equation}
The parameter $M_{5,I}$ is a five-dimensional Dirac bulk mass. It
couples the two four-dimensional chiral components in the bulk and
controls the localization of the chiral zero mode, the KK spectrum,
and propagation along the fifth dimension. It must not be identified
with the four-dimensional Majorana mass $M_I$ appearing in the seesaw
relation of Eq.~\eqref{eq:seesaw4D}. The latter arises only after
dimensional reduction, when the hidden-brane interaction proportional
to $\widehat Y_{5,N}$ is included and the scalar $R$ acquires a vacuum
expectation value; see Appendix~\ref{sec:5D_to_4D_matching}.

On the covering circle, with $y\sim y+2L$ and $y\sim-y$, we impose
\begin{equation}
\mathcal N_I(x,-y)
=
\gamma^5\mathcal N_I(x,y),
\label{eq:orbifold_full_parity_5D}
\end{equation}
which is equivalent to
\begin{align}
\mathcal N_{IR}(x,-y)
&=
+\mathcal N_{IR}(x,y),
\\
\mathcal N_{IL}(x,-y)
&=
-\mathcal N_{IL}(x,y).
\label{eq:orbifold_chiral_parity_5D}
\end{align}
The right-handed component is therefore even, while the left-handed
component is odd. Since $y=0$ and $y=L$ are orbifold fixed points,
\begin{equation}
\mathcal N_{IL}(x,0)
=
\mathcal N_{IL}(x,L)
=
0.
\label{eq:NL_boundary_5D}
\end{equation}
Before the hidden-brane Majorana interaction is included, this choice
retains a massless right-handed chiral zero mode and allows the brane
interactions written below. The left-handed component
$\mathcal N_{IL}$ is present in the bulk, but it is odd under the
orbifold symmetry and therefore vanishes at the branes. Such orbifold
projections are commonly used in higher-dimensional neutrino
constructions~\cite{Dienes:1998sb,Grossman:1999ra,
ArkaniHamed:1999dc}.

The fermion bilinear
$\overline{\mathcal N}_I\mathcal N_I$ is odd under the orbifold
reflection. A strict covering-space description therefore requires
the bulk mass to be odd,
\begin{equation}
M_{5,I}(y)
=
M_{5,I}\epsilon(y),
\qquad
\epsilon(-y)=-\epsilon(y),
\label{eq:orbifold_odd_mass_5D}
\end{equation}
so that the mass term
$M_{5,I}(y)\overline{\mathcal N}_I\mathcal N_I$ is invariant under the
orbifold reflection. Inside the physical interval,
$\epsilon(y)=+1$, so the same theory is described by the constant mass
appearing in Eq.~\eqref{eq:Sbulk5D}, together with the orbifold
boundary conditions. Thus, the covering-space odd mass and the
constant interval mass are not two independent interactions; they are
two descriptions of the same orbifold-equivalent theory.

The visible-boundary action is
\begin{widetext}
\begin{equation}
S_{\rm vis}
=
\int\dd^4x\int_0^L\dd y\,
\delta(y)
\Big[
\mathcal L_{\rm SM}(x)
-
\left(
\widehat Y_{5,\nu}
\right)_{\alpha I}
\overline{L_\alpha}(x)
\widetilde\Phi(x)\,
P_R\mathcal N_I(x,y)
+
{\rm h.c.}
\Big].
\label{eq:Svis5D_full_field}
\end{equation}
\end{widetext}
Writing $P_R\mathcal N_I=\mathcal N_{IR}$ makes explicit that the
visible interaction involves the brane-accessible chirality. The same
result would be obtained by omitting the projector, because
$\overline{L_\alpha}\widetilde\Phi\,\mathcal N_I$ automatically selects
the right-handed component. Nevertheless, the explicit projector is
useful for keeping the orbifold and chirality assignments manifest.

The hidden-boundary action is
\begin{widetext}
\begin{equation}
S_{\rm hid}
=
\int\dd^4x\int_0^L\dd y\,
\delta(y-L)
\Big[
\mathcal L_{\rm hid}(x)
-
\frac{1}{2}
\left(
\widehat Y_{5,N}
\right)_{IJ}
R(x)\,
\overline{\mathcal N_{IR}^{\,c}}(x,y)
\mathcal N_{JR}(x,y)
+
{\rm h.c.}
\Big].
\label{eq:Shid5D}
\end{equation}
\end{widetext}
Only the symmetric part of $\widehat Y_{5,N}$ contributes to the
Majorana bilinear, and we therefore take
$\widehat Y_{5,N}=\widehat Y_{5,N}^{\,T}$. The coefficients
$\widehat Y_{5,\nu}$ and $\widehat Y_{5,N}$ are five-dimensional
brane couplings whose four-dimensional counterparts are obtained
after inserting the boundary values of the normalized bulk profiles.
The matching between the five- and four-dimensional descriptions is
presented in Appendix~\ref{sec:5D_to_4D_matching}.

Here $\mathcal L_{\rm SM}(x)$ is the SM Lagrangian, while
$\mathcal L_{\rm hid}(x)$ is the hidden-sector Lagrangian containing
the singlet scalar $R$, the Dirac DM field $\chi$, the Yukawa
interaction $-y_pR\overline\chi\chi$, and the hidden-sector scalar
self-interactions. These are four-dimensional Lagrangian densities
and depend only on $x^\mu$; the delta functions localize them at the
two endpoints of the interval.

After integrating Eqs.~\eqref{eq:Svis5D_full_field} and
\eqref{eq:Shid5D} over $y$, the visible fields couple to
$\mathcal N_{IR}(x,0)$, whereas the hidden fields couple to
$\mathcal N_{IR}(x,L)$. Therefore, only the even component
$\mathcal N_{IR}$ appears in the boundary-localized visible- and
hidden-sector interactions, because the odd component vanishes at
both boundaries. The left-handed component remains part of the bulk
Dirac field and is essential for the massive KK levels, but it does
not enter a strictly localized boundary vertex.

%----------------------------------------------------------
\subsection{Locality and absence of a tree-level portal}
%----------------------------------------------------------

The essential feature of the construction is that the visible and
hidden fields are localized at different points in the compact
dimension. The Higgs doublet $\Phi(x)$ is confined to the visible
brane at $y=0$, while the hidden scalar $R(x)$ is confined to the
hidden brane at $y=L$. A local five-dimensional contact interaction
cannot contain both fields, because all fields entering a local
operator must be evaluated at the same point in the fifth dimension.
This statement assumes the usual local five-dimensional effective
field theory and excludes fundamental nonlocal operators stretching
between the two branes.

To display this geometrical statement explicitly, consider the
formal expression that one would obtain by trying to localize a
tree-level contact operator simultaneously on both boundaries:
\begin{equation}
S_{\mathrm{portal}}^{\mathrm{tree}}
=
-\int\dd^4x\int_0^L\dd y\,
\kappa_5
R^2(x)\Phi^\dagger(x)\Phi(x)
\delta(y)\delta(y-L).
\label{eq:portal5Dtree}
\end{equation}
Here $\kappa_5$ is introduced only as a formal coefficient of the
would-be five-dimensional local operator. It is not the coefficient
of an allowed interaction in the five-dimensional theory. The two
delta functions encode the fact that $\Phi$ lives at $y=0$, whereas
$R$ lives at $y=L$. After integrating over the fifth dimension, the
formal expression would correspond to the four-dimensional
tree-level portal
\begin{equation}
S_{\mathrm{portal}}^{\mathrm{tree}}
=
-\int\dd^4x\,
\kappa(\mu_{\rm match})\,
R^2(x)\Phi^\dagger(x)\Phi(x),
\label{eq:portal4Dtree_def}
\end{equation}
with
\begin{equation}
\kappa(\mu_{\rm match})
=
\kappa_5
\int_0^L\dd y\,\delta(y)\delta(y-L),
\label{eq:ktree_from_kappa5}
\end{equation}
where $\mu_{\rm match}$ is the matching scale between the 5D and the 4D theories.
For separated branes, $L\neq0$, the two boundary distributions have
disjoint support. Equivalently, this conclusion follows by regulating
the branes as nonoverlapping profiles of finite width and then taking
the zero-width limit. Therefore, in this regulated sense,
\begin{equation}
\int_0^L\dd y\,\delta(y)\delta(y-L)=0,
\qquad
L\neq0,
\label{eq:deltaproduct_zero}
\end{equation}
and hence
\begin{equation}
S_{\mathrm{portal}}^{\mathrm{tree}}=0,
\qquad
\kappa(\mu_{\rm match})=0.
\label{eq:ktree5Dzero}
\end{equation}
If the matching scale is identified with the UV scale used in the
four-dimensional discussion, this condition can equivalently be
written as
\begin{equation}
\kappa(\Lambda_{\rm UV})=0.
\end{equation}

This result is the distributional representation of the locality
argument. There is no point in the fifth dimension at which the
visible-brane field $\Phi$ and the hidden-brane field $R$ are
simultaneously local degrees of freedom. A direct Higgs--dark-scalar
contact term is therefore absent at tree level, not because its
coefficient has been tuned to zero, but because the corresponding
local five-dimensional operator has no common support. The two sectors
can communicate only through fields that propagate in the bulk. In
the present model, this role is played by the sterile-neutrino fields
$\mathcal N_I$, which generate the leading nonzero Higgs--dark-scalar
portal radiatively.

%----------------------------------------------------------
\subsection{Orbifold parity and KK decomposition}
%----------------------------------------------------------

In this subsection, we perform the KK decomposition of the two
four-dimensional chiral components of each bulk fermion
$\mathcal N_I$. We derive the profile equations and boundary
conditions implied by the orbifold projection, identify the chiral
zero mode and the massive free KK spectrum, and then show how the
hidden-boundary Majorana interaction mixes the free KK states.
Finally, we define the physical masses and profiles obtained after
diagonalizing the complete Dirac-plus-Majorana mass matrix. A detailed
derivation is provided in Appendix~\ref{sec:5Ddd}.

Before the hidden-boundary Majorana interaction is included, the KK
decomposition is
\begin{align}
\mathcal N_{IR}(x,y)
&=
N_{IR}^{(0)}(x)f_{R,I}^{(0)}(y)
+
\sum_{n=1}^{\infty}
N_{IR}^{(n)}(x)f_{R,I}^{(n)}(y),
\label{eq:KKR_5D}
\\
\mathcal N_{IL}(x,y)
&=
\sum_{n=1}^{\infty}
N_{IL}^{(n)}(x)f_{L,I}^{(n)}(y).
\label{eq:KKL_5D}
\end{align}
Only the right-handed component contains a zero mode as a
consequence of the chosen orbifold parity in Eq.~\eqref{eq:orbifold_chiral_parity_5D}. Because the right-handed
component is even, it may be nonzero at both fixed points and admits a
normalizable solution with four-dimensional mass $m_{I,0}=0$. The
left-handed component is odd and vanishes at the boundaries, so no
left-handed zero-mode partner survives. The resulting
four-dimensional zero mode is therefore chiral and appears only in
the right-handed KK expansion. Its vanishing four-dimensional mass
does not imply a constant profile in $y$; the bulk mass can localize
the massless mode exponentially along the interval.

The profiles obey
\begin{align}
\left(
\partial_y+M_{5,I}
\right)
f_{R,I}^{(n)}
&=
m_{I,n}f_{L,I}^{(n)},
\\
\left(
-\partial_y+M_{5,I}
\right)
f_{L,I}^{(n)}
&=
m_{I,n}f_{R,I}^{(n)}.
\label{eq:first_order_profile_equations_5D}
\end{align}
The orbifold conditions imply
\begin{equation}
f_{L,I}^{(n)}(0)
=
f_{L,I}^{(n)}(L)
=
0,
\end{equation}
and hence
\begin{equation}
\left.
\left(
\partial_y+M_{5,I}
\right)
f_{R,I}^{(n)}
\right|_{y=0,L}
=
0.
\label{eq:Robin_fR_5D}
\end{equation}
The right-handed profiles therefore satisfy Robin rather than ordinary
Neumann boundary conditions when $M_{5,I}\neq0$.

For the chiral zero mode,
\begin{equation}
\left(
\partial_y+M_{5,I}
\right)
f_{R,I}^{(0)}(y)
=
0,
\end{equation}
so that
\begin{equation}
f_{R,I}^{(0)}(y)
=
\left(
\frac{2M_{5,I}}{1-e^{-2M_{5,I}L}}
\right)^{1/2}
e^{-M_{5,I}y}.
\label{eq:zero_profile_5D}
\end{equation}
The normalization factor is positive for either sign of $M_{5,I}$.
For $M_{5,I}>0$, the zero mode is localized toward the visible
boundary at $y=0$, while for $M_{5,I}<0$ it is localized toward the
hidden boundary at $y=L$. The latter case is more directly related to
the sequestered DM realization considered here. Indeed, since the SM
fields are localized at $y=0$ and the DM sector at $y=L$, a negative
bulk mass suppresses the overlap of the sterile-fermion zero mode with
the visible brane relative to its overlap with the hidden brane:
\begin{equation}
\frac{f_{R,I}^{(0)}(0)}
     {f_{R,I}^{(0)}(L)}
=
e^{M_{5,I}L}
=
e^{-|M_{5,I}|L},
\,\,\,
M_{5,I}<0.
\label{eq:zero_mode_overlap_ratio_5D}
\end{equation}
Consequently, the effective coupling to the SM brane can be
exponentially suppressed, while the coupling to the hidden brane
remains comparatively large. This provides an additional geometrical
sequestering of the SM and DM sectors, although it is not required for
the absence of the fundamental tree-level portal. Similar
localization mechanisms are commonly used to generate hierarchical
four-dimensional couplings in extra-dimensional
models~\cite{Grossman:1999ra,ArkaniHamed:1999dc}.

In the approximately flat regime $|M_{5,I}|L\ll1$ the zero-mode profile becomes
\begin{equation}
f_{R,I}^{(0)}(y) = \frac{1}{\sqrt L}.
\label{eq:approximately_flat_zero_mode_5D}
\end{equation}

The Dirac KK modes free masses are defined as
\begin{equation}
m_{I,n}^2
=
M_{5,I}^2
+
\frac{n^2\pi^2}{L^2},
\qquad
n\geq1.
\label{eq:KKmasses_5D}
\end{equation}
Each massive level contains both chiralities and forms a
four-dimensional Dirac fermion. In the absence of brane-localized mass
terms, the characteristic KK scale is
\begin{equation}
\MKK\equiv\frac{\pi}{L}.
\label{eq:MKKdef}
\end{equation}

After the hidden scalar develops the vacuum expectation value
\begin{equation}
R
=
\frac{1}{\sqrt2}
(v_r+r),
\end{equation}
and the bulk fields are decomposed into KK modes, the right-handed KK
components acquire a boundary-induced Majorana mass term
\begin{equation}
\mathcal{L}_{\rm KK}^{\rm Majorana}
=
-
\frac{1}{2}
\sum_{I,J}
\sum_{m,n=0}^{\infty}
\left(
\mathcal{M}_{N}^{mn}
\right)_{IJ}
\overline{
\left(
N_{IR}^{(m)}
\right)^{c}
}
N_{JR}^{(n)}
+
{\rm h.c.},
\label{eq:KK_Majorana_Lagrangian}
\end{equation}
where the hidden-boundary interaction generates the Majorana matrix
\begin{equation}
\left(
\mathcal M_N^{mn}
\right)_{IJ}
=
\frac{v_r}{\sqrt2}
\left(
\widehat Y_{5,N}
\right)_{IJ}
f_{R,I}^{(m)}(L)
f_{R,J}^{(n)}(L).
\label{eq:brane_Majorana_KK_matrix_5D}
\end{equation}
The entry $\mathcal M_N^{00}$ gives the zero-mode Majorana mass in the
zero-mode truncation, the entries $\mathcal M_N^{0n}$ mix the zero
mode with the massive tower, and the entries $\mathcal M_N^{mn}$ with
$m,n\geq1$ generate Majorana masses and mixing among the massive KK
states.

For one sterile generation, the boundary matrix factorizes in KK
space,
\begin{equation}
\mathcal M_N^{mn}
=
\frac{v_r}{\sqrt2}
\widehat y_{5,N}
f_R^{(m)}(L)f_R^{(n)}(L),
\end{equation}
and is an outer product of the boundary values of the free profiles.
The complete mass matrix nevertheless contains the different Dirac KK
masses $m_n$ and generally shifts and mixes the full tower. The
physical masses $M_a$ are obtained through a Takagi diagonalization of
the complete complex symmetric Dirac-plus-Majorana mass matrix, written
in a common Weyl basis,
\begin{equation}
U_N^T\mathcal M_{\rm full}U_N
=
{\rm diag}(M_0,M_1,\ldots),
\qquad
M_a\geq0.
\end{equation}
The physical right-handed profiles are defined by expanding each
five-dimensional field in the physical Majorana basis,
\begin{equation}
\mathcal N_{IR}(x,y)
=
\sum_a
F_{R,Ia}(y)P_RN_a(x),
\end{equation}
with
\begin{equation}
F_{R,Ia}(y)
=
\sum_{n=0}^{\infty}
f_{R,I}^{(n)}(y)
\left(U_N^R\right)_{(I,n)a}.
\label{eq:physical_profiles_5D}
\end{equation}
Here $U_N^R$ denotes the block of the Takagi matrix that rotates the
right-handed free KK components into the physical Majorana mass
eigenstates. Thus, the lower-case profiles $f_{R,I}^{(n)}$ refer to
the free KK basis, whereas the capital profiles $F_{R,Ia}$ describe
the right-handed wave-function components of the physical Majorana
mass eigenstates. The index $I$ labels the original five-dimensional
sterile field, while the index $a$ labels a physical mass eigenstate.

%----------------------------------------------------------
\subsection{Effective four-dimensional couplings and one-generation case}
%----------------------------------------------------------

In this subsection, we express the effective four-dimensional Yukawa
couplings in terms of the boundary values of the bulk-fermion
profiles. We first give the couplings in the free KK basis and in the
physical mass basis. We then specialize to one sterile generation and
to the regime in which a light, mostly zero-mode state is well
separated from the massive KK tower. Finally, we show explicitly how
its visible and hidden Yukawa couplings depend on the localization
parameter $M_5L$.

Before diagonalization, using the Majorana matrix defined in
Eq.~\eqref{eq:brane_Majorana_KK_matrix_5D}, the effective couplings of
the free KK modes can be written as
\begin{align}
\left(
Y_\nu^{(n)}
\right)_{\alpha I}
&=
\left(
\widehat Y_{5,\nu}
\right)_{\alpha I}
f_{R,I}^{(n)}(0),
\label{eq:Ynu_KK_overlap_5D}
\\
\left(
Y_N^{mn}
\right)_{IJ}
&=
\left(
\widehat Y_{5,N}
\right)_{IJ}
f_{R,I}^{(m)}(L)
f_{R,J}^{(n)}(L).
\label{eq:YN_KK_overlap_5D}
\end{align}
After diagonalization, the corresponding physical couplings are
\begin{align}
\left(\lambda_\nu\right)_{\alpha a}
&=
\sum_I
\left(
\widehat Y_{5,\nu}
\right)_{\alpha I}
F_{R,Ia}(0),
\\
\left(g_N\right)_{ab}
&=
\sum_{I,J}
F_{R,Ia}(L)
\left(
\widehat Y_{5,N}
\right)_{IJ}
F_{R,Jb}(L).
\end{align}

To obtain a controlled low-energy description and quantify the
phenomenological consequences of the model, we are particularly
interested in the regime in which one physical state $N_0$ is much
lighter than the first massive KK level,
\begin{equation}
M_0\ll m_1,
\qquad
m_1
=
\sqrt{
M_5^2+\frac{\pi^2}{L^2}
},
\label{eq:light_hierarchy_5D}
\end{equation}
and the boundary-induced zero-mode--KK mixing is perturbative,
\begin{equation}
\epsilon_n
\equiv
\frac{
|\mathcal M_N^{0n}|
}{
m_n-M_0
}
\ll1,
\qquad
n\geq1.
\label{eq:small_mixing_5D}
\end{equation}
Since $M_0\ll m_n$, one may estimate
\begin{equation}
\epsilon_n
\simeq
\frac{
|\mathcal M_N^{0n}|
}{
m_n
}.
\end{equation}
In this regime, the physical light-state profile is related to the
free zero-mode profile by
\begin{equation}
F_{R,0}(y)
=
f_R^{(0)}(y)
+
\mathcal O(\epsilon),
\label{eq:F_approx_f_5D}
\end{equation}
where $\epsilon$ denotes the characteristic size of the
zero-mode--KK mixing. Therefore, for $\epsilon\ll1$, one can use
$F_{R,0}(y)\simeq f_R^{(0)}(y)$. For one sterile generation, the
Yukawa couplings are then
\begin{align}
Y_\nu^{(0)}
&=
\widehat y_{5,\nu}F_{R,0}(0)
\simeq
\widehat y_{5,\nu}f_R^{(0)}(0),
\label{eq:Ynu_zero_5D}
\\
Y_N^{(00)}
&=
\widehat y_{5,N}
\left[F_{R,0}(L)\right]^2
\simeq
\widehat y_{5,N}
\left[f_R^{(0)}(L)\right]^2,
\label{eq:YN_zero_5D}
\end{align}
where $\widehat y_{5,N}$ and $\widehat y_{5,\nu}$ are the
one-generation five-dimensional Yukawa couplings. As a consequence,
the physical Majorana mass is
\begin{equation}
M_0
\simeq
\frac{v_r}{\sqrt2}
Y_N^{(00)}.
\label{eq:M0_5D}
\end{equation}

For a hidden-localized state, we define
\begin{equation}
M_5=-\mu,
\qquad
\mu>0,
\qquad
x\equiv\mu L.
\end{equation}
The Yukawa couplings then become
\begin{align}
Y_\nu^{(0)}
&=
\widehat y_{5,\nu}
\left[
\frac{2\mu}{1-e^{-2\mu L}}
\right]^{1/2}
e^{-\mu L}
=
\frac{\widehat y_{5,\nu}}{\sqrt L}
\left[
\frac{2x}{e^{2x}-1}
\right]^{1/2},
\label{eq:Ynu_hidden_localized_5D}
\\
Y_N^{(00)}
&=
\widehat y_{5,N}
\frac{2\mu}{1-e^{-2\mu L}}
=
\frac{\widehat y_{5,N}}{L}
\frac{2x}{1-e^{-2x}}.
\label{eq:YN_hidden_localized_5D}
\end{align}
The choice $M_5<0$ is the one most naturally aligned with the
sequestered DM construction, because the light state is localized
toward the hidden brane and has a suppressed overlap with the SM
brane. For $x=\mu L\ll1$, the profile is approximately flat and the
couplings approach
\begin{equation}
Y_\nu^{(0)}
\simeq
\frac{\widehat y_{5,\nu}}{\sqrt L},
\qquad
Y_N^{(00)}
\simeq
\frac{\widehat y_{5,N}}{L}.
\end{equation}
In the strongly localized regime, $x\gg1$, the boundary values of the
light-state profile scale as
\begin{equation}
F_{R,0}(0)
\simeq
\sqrt{\frac{2x}{L}}\,e^{-x},
\,\,\,
F_{R,0}(L)
\simeq
\sqrt{\frac{2x}{L}},
\end{equation}
and therefore
\begin{equation}
Y_\nu^{(0)}
\simeq
\frac{\widehat y_{5,\nu}}{\sqrt L}
\sqrt{2x}\,e^{-x},
\,\,\,
Y_N^{(00)}
\simeq
\frac{2x}{L}\widehat y_{5,N},
\,\,\,
M_0
\simeq
\frac{\sqrt2\,v_r x}{L}\widehat y_{5,N}.
\label{eq:hidden_localized_scaling_5D}
\end{equation}
Thus, increasing $\mu L$ exponentially suppresses the visible Yukawa
coupling, whereas the hidden Yukawa coupling and the Majorana mass are
not exponentially suppressed and instead scale linearly with $\mu L$
at fixed $L$ and fixed five-dimensional couplings. These relations
apply as long as the light-state hierarchy and the perturbative
zero-mode--KK mixing conditions specified above remain valid.

%----------------------------------------------------------
\subsection{Brane-to-brane propagation and the radiative portal}
%----------------------------------------------------------

In this subsection, we describe the propagation of the bulk sterile
fermion between the visible and hidden branes. We then match the
light-state contribution onto the four-dimensional radiative portal,
determine its dependence on the zero-mode localization relative to a
flat profile, and finally discuss the additional contribution of the
massive KK tower.

The free right-handed second-order kernel in mixed Euclidean
momentum--position space has the spectral representation
\begin{equation}
G_R(p_E;0,L)
=
\sum_{n=0}^{\infty}
\frac{
f_R^{(n)}(0)f_R^{(n)}(L)
}{
p_E^2+m_n^2
},
\label{eq:Gsum_5D}
\end{equation}
where $p_E$ is the Euclidean momentum. For the strict orbifold, or
equivalently for the interval with the orbifold boundary conditions,
\begin{equation}
G_R(p_E;0,L)
=
\frac{
\omega
}{
p_E^2\sinh(\omega L)
},
\,\,\,
\omega
=
\sqrt{
p_E^2+M_5^2
}.
\label{eq:Gcsch_5D}
\end{equation}
This expression is the free kernel before the hidden-boundary
Majorana interaction is resummed. It is therefore written in terms of
the free KK masses $m_n$ and profiles $f_R^{(n)}$. After the boundary
mass is included, the poles and residues are instead those of the
physical Majorana eigenstates.

The explicit pole $1/p_E^2$ is the chiral zero-mode pole,
\begin{equation}
G_R(p_E;0,L)
=
\frac{
f_R^{(0)}(0)f_R^{(0)}(L)
}{
p_E^2
}
+
G_R^{\rm KK}(p_E;0,L).
\end{equation}
For $\omega L\gg1$,
\begin{equation}
G_R(p_E;0,L)
\simeq
\frac{2\omega}{p_E^2}
e^{-\omega L}.
\end{equation}
Since the portal loop contains one propagation from the visible brane
to the hidden brane and one propagation back, its genuinely nonlocal
large-momentum part contains an asymptotic damping proportional to
\begin{equation}
e^{-2\omega L}.
\end{equation}
This high-momentum damping renders the separated-boundary contribution
to the complete five-dimensional amplitude ultraviolet finite, after
possible local brane subdivergences have been renormalized.

At energies below the sterile-neutrino and KK thresholds, the
radiative effect is represented by
\begin{equation}
\mathcal L_{\rm EFT}
\supset
-\kappa_{\rm EFT}(\mu)
R^2\Phi^\dagger\Phi.
\end{equation}
In the absence of a fundamental tree-level portal, the matching
contribution to $\kappa_{\rm EFT}$ is generated by bulk-fermion loops.
In the one-state regime, the renormalized zero-momentum
$\overline{\rm MS}$ loop contribution is
\begin{equation}
\kappa_{\rm loop}^{(0)}(\mu)
=
-\frac{
|Y_N^{(00)}|^2
|Y_\nu^{(0)}|^2
}{
8\pi^2
}
\left[
\ln\left(
\frac{M_0^2}{\mu^2}
\right)
+2
\right].
\label{eq:kappa5D_mu}
\end{equation}
Matching at $\mu=M_0$ gives
\begin{equation}
\kappa_{5,\rm loop}^{(0)}(M_0)
=
-\frac{
|Y_N^{(00)}|^2
|Y_\nu^{(0)}|^2
}{
4\pi^2
}.
\label{eq:kappa5D_physical}
\end{equation}
Equivalently, in terms of the fundamental five-dimensional brane
coefficients,
\begin{equation}
\kappa_{5,\rm loop}^{(0)}(M_0)
=
-\frac{
|\widehat y_{5,N}|^2
|\widehat y_{5,\nu}|^2
}{
4\pi^2
}
|F_{R,0}(0)|^2
|F_{R,0}(L)|^4.
\label{eq:kappa5D_profiles}
\end{equation}
The two visible vertices provide $|F_{R,0}(0)|^2$, while the two
hidden vertices provide $|F_{R,0}(L)|^4$. These profile factors
already encode the effects of the brane separation. Therefore, no
additional exponential factor from the free brane-to-brane kernel
should be multiplied into a loop result already expressed in terms of
the physical four-dimensional couplings.

To display the dependence on $M_5L$, we now use the weak-mixing
approximation
\begin{equation}
F_{R,0}(y)\simeq f_R^{(0)}(y).
\end{equation}
We define dimensionless reference couplings by
\begin{equation}
y_\nu^{\rm ref}
=
\frac{\widehat y_{5,\nu}}{\sqrt L},
\qquad
y_N^{\rm ref}
=
\frac{\widehat y_{5,N}}{L},
\end{equation}
and the corresponding flat-profile four-dimensional threshold,
\begin{equation}
\kappa_{5,\rm loop}^{\rm flat}(M_0)
=
-\frac{
|y_N^{\rm ref}|^2
|y_\nu^{\rm ref}|^2
}{
4\pi^2
}.
\label{eq:kappa4D_flat_5D}
\end{equation}
This is the result obtained with the same fundamental brane
coefficients and a flat zero-mode profile
$f_{\rm flat}=1/\sqrt L$. Therefore, the ratio introduced below
compares localized and flat theories at fixed
$\widehat y_{5,\nu}$ and $\widehat y_{5,N}$. If the physical
four-dimensional couplings are instead held fixed, their profile
factors are already included and no additional sequestering factor
must be applied.

The localized result can be written as
\begin{equation}
\kappa_{5,\rm loop}^{(0)}(M_0)
=
\kappa_{\rm loop}^{\rm flat}(M_0)
\,
\mathcal F_0(M_5L),
\label{eq:kappa5D_kappa4D_general}
\end{equation}
where
\begin{equation}
\mathcal F_0(M_5L)
=
L^3
|f_R^{(0)}(0)|^2
|f_R^{(0)}(L)|^4.
\label{eq:F0_general_5D}
\end{equation}
For $M_5<0$, the zero mode is localized toward the hidden boundary,
which is the localization naturally associated with the sequestered
DM construction. In this case, defining
\begin{equation}
x=|M_5|L,
\end{equation}
the geometrical profile factor becomes
\begin{equation}
\mathcal F_0(x)
=
\left[
\frac{2x}{1-e^{-2x}}
\right]^3
e^{-2x},
\label{eq:F0_negative_5D}
\end{equation}
and hence
\begin{equation}
\kappa_{5,\rm loop}^{(0)}
=
\kappa_{\rm loop}^{\rm flat}
\left[
\frac{2|M_5|L}
{1-e^{-2|M_5|L}}
\right]^3
e^{-2|M_5|L}.
\label{eq:kappa5D_negative_final}
\end{equation}
In the strong-localization limit $x\gg1$,
Eq.~\eqref{eq:kappa5D_negative_final} becomes
\begin{equation}
\kappa_{5,\rm loop}^{(0)}
\simeq
\kappa_{\rm loop}^{\rm flat}
(2|M_5|L)^3
e^{-2|M_5|L}.
\label{eq:kappa5D_xx}
\end{equation}

In terms of the compactification scale
\begin{equation}
M_{\rm KK}\equiv\frac{\pi}{L},
\qquad
|M_5|L
=
\pi\frac{|M_5|}{M_{\rm KK}},
\label{eq:MKK_relation}
\end{equation}
the zero-mode contribution can be written as
\begin{equation}
\kappa_{5,\rm loop}^{(0)}
=
\kappa_{\rm loop}^{\rm flat}
\left[
\frac{
2\pi |M_5|/M_{\rm KK}
}{
1-\exp\!\left(-2\pi |M_5|/M_{\rm KK}\right)
}
\right]^3
\exp\!\left(
-2\pi\frac{|M_5|}{M_{\rm KK}}
\right).
\label{eq:kappa5D_negative_MKK}
\end{equation}

\begin{figure}
\centering
\includegraphics[width=0.95\columnwidth]
{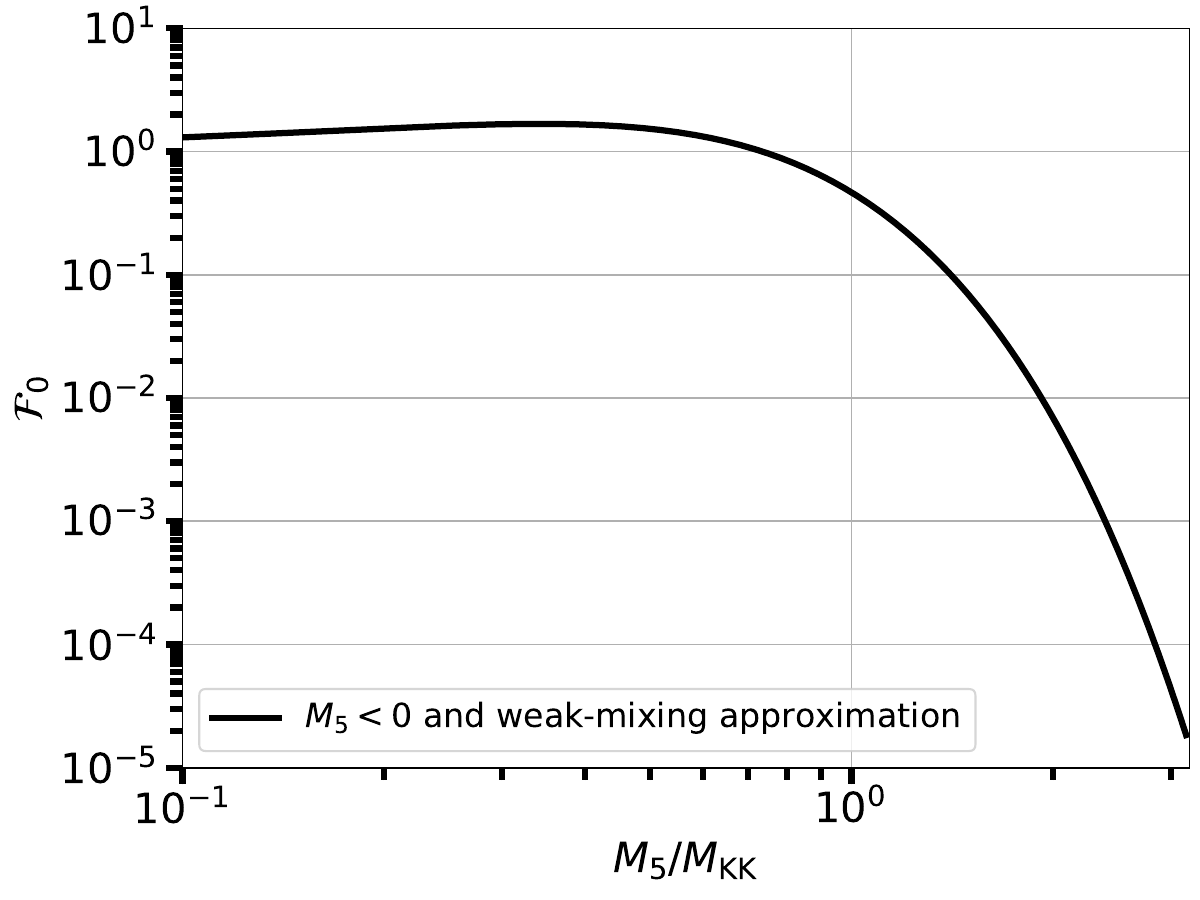}
\caption{Geometrical profile factor $\mathcal F_0$ as a function of
$|M_5|/M_{\rm KK}$ for the hidden-localized case $M_5<0$. The factor
approaches unity in the flat-profile limit, can be mildly enhanced for
intermediate localization, and becomes increasingly suppressed when
the bulk mass is comparable to or larger than the KK scale.}
\label{fig:sequestering}
\end{figure}

In Fig.~\ref{fig:sequestering}, we show the geometrical profile factor
$\mathcal F_0$ as a function of $|M_5|/M_{\rm KK}$ for the
hidden-localized case $M_5<0$. In the flat-profile limit,
$|M_5|/M_{\rm KK}\rightarrow0$, one has
$\mathcal F_0\rightarrow1$. For moderate localization, the enhancement
of the zero-mode profile at the hidden brane can overcompensate its
suppression at the visible brane. Consequently, $\mathcal F_0$
initially exceeds unity and reaches a maximum
$\mathcal F_0^{\rm max}\simeq1.68$ at
$|M_5|/M_{\rm KK}\simeq0.34$. For larger bulk masses, the exponential
suppression of the visible-brane overlap dominates and
$\mathcal F_0$ decreases rapidly. In particular, at $|M_5|=M_{\rm KK}$ the sequestering factor is $\mathcal F_0 \sim 0.5$ while for $|M_5|=2 M_{\rm KK}$ it becomes $\sim 7\times10^{-3}$.
Thus, when $|M_5|$ is comparable to or larger than the KK scale, the
localization of the sterile state provides an additional suppression
of the light-state contribution to the radiative portal.

The result above includes only the contribution of the light state in
the zero-mode-dominated regime. The full loop matching coefficient
also contains the threshold contribution of the massive KK tower,
\begin{equation}
\kappa_{\rm loop}(\mu)
=
\kappa_{\rm loop}^{(0)}(\mu)
+
\Delta\kappa_{\rm KK}(\mu).
\label{eq:kappa_full_KK_5D}
\end{equation}
The correction $\Delta\kappa_{\rm KK}$ includes pure-KK and mixed
light--KK contributions after the complete mass matrix has been
diagonalized and is not, in general, a universal multiplicative
factor multiplying the light-state result. The conditions
\begin{equation}
M_0\ll m_1,
\qquad
\epsilon_n\ll1,
\label{eq:cond}
\end{equation}
ensure that the lightest physical state is dominated by the free
zero mode. They do not, however, guarantee that the loop threshold from the remaining tower is negligible, since the boundary couplings
of the KK-dominated states are not controlled solely by
$\epsilon_n$. A finite KK truncation and the corresponding convergence
criteria are described in Appendix~\ref{app:KK_truncation}. The
single-state phenomenology assumes
$|\Delta\kappa_{\rm KK}|\ll|\kappa_{\rm loop}^{(0)}|$.

The numerical curves below retain only the light-state contribution.
The full KK threshold can change their normalization and may also modify the inferred numerical bounds. Its quantitative effect cannot
be determined without the complete regulated spectral sum; the zero-state curves should therefore be regarded as illustrative.
In any case the main conclusions of the paper are not affected by including the full contribution of the KK modes.

\subsection{Consequences of the five-dimensional sequestered model}
\label{sec:5D_consequences}

The five-dimensional construction has two distinct consequences.
First, the spatial separation of the visible and hidden sectors
explains the absence of a fundamental tree-level Higgs--dark-scalar
portal. Second, the localization of the bulk-neutrino wave functions
modifies the radiatively generated portal. In this subsection, we
summarize the main consistency requirements and phenomenological
implications of the compact fifth dimension.

For a transparent comparison with the four-dimensional theory, we
consider the regime in which one physical state, $\mathcal N_0$, is
much lighter than the first massive KK level and adopt the weak-mixing
approximation, for which the zero-mode localization factor is
$\mathcal F_0$ as defined in
Eq.~\eqref{eq:F0_negative_5D}. The complete matching coefficient also
contains mixed and pure-KK contributions from the massive tower.
These contributions may shift the normalization of the effective
portal, and hence the corresponding phenomenological bounds, by
factors of order $\mathcal{O}(1)$. In the following, we assume that they do not
alter the characteristic dependence on the heavy-neutrino mass and
localization parameters or the qualitative conclusions of the
zero-mode analysis.

The quantities $m_\nu^{\rm ref}$, $y_N^{\rm ref}$, and
$M_N^{\rm ref}$ used below denote the parameters of the flat-profile
four-dimensional reference theory constructed from the same
fundamental brane coefficients:
\begin{equation}
y_\nu^{\rm ref}
=
\frac{\widehat y_{5,\nu}}{\sqrt L},
\,\,\,
y_N^{\rm ref}
=
\frac{\widehat y_{5,N}}{L},
\,\,\,
M_N^{\rm ref}
=
\frac{y_N^{\rm ref}v_r}{\sqrt2}.
\end{equation}
These quantities should not be identified with the physical
couplings and masses of the localized light state. At fixed
fundamental brane coefficients and fixed $v_r$, the latter satisfy
\begin{equation}
Y_\nu^{(0)}
=
\sqrt L\,f_R^{(0)}(0)y_\nu^{\rm ref},
\,\,\,
Y_N^{(00)}
=
L|f_R^{(0)}(L)|^2y_N^{\rm ref},
\end{equation}
and
\begin{equation}
M_0
=
L|f_R^{(0)}(L)|^2M_N^{\rm ref},
\,\,\,
m_\nu^{5D}
=
e^{-2|M_5|L}m_\nu^{\rm ref}.
\end{equation}
With this prescription, the zero-mode contribution satisfies Eq.~\eqref{eq:kappa5D_xx}.

\subsubsection{Combined direct-detection and BBN constraints}

The main phenomenological effect of zero-mode localization is a
modification, and in the strongly localized regime a suppression, of
the physical Higgs--singlet mixing angle. Since
Eq.~\eqref{eq:eigenmix} implies
$\tan(2\alpha)\propto\kappa$ and
$\tan(2\alpha)\simeq2\sin\alpha$ for small mixing, the relation above
gives
\begin{equation}
\sin\alpha_{\rm phys}
\equiv
\sin\alpha_{5D}
\simeq
\mathcal F_0(|M_5|L)\,
\sin\alpha_{4D}^{\rm ref}
\left(
m_\nu^{\rm ref},
y_N^{\rm ref},
M_N^{\rm ref}
\right).
\label{eq:alpha_phys_def}
\end{equation}
Here $\alpha_{4D}^{\rm ref}$ is the mixing angle obtained from the
flat-profile reference parameters. If the physical localized
couplings $Y_\nu^{(0)}$ and $Y_N^{(00)}$ are instead used directly as
inputs, their wave-function dependence is already included and no
additional factor $\mathcal F_0$ should be applied.

\begin{figure}
\centering
\includegraphics[width=0.95\columnwidth]{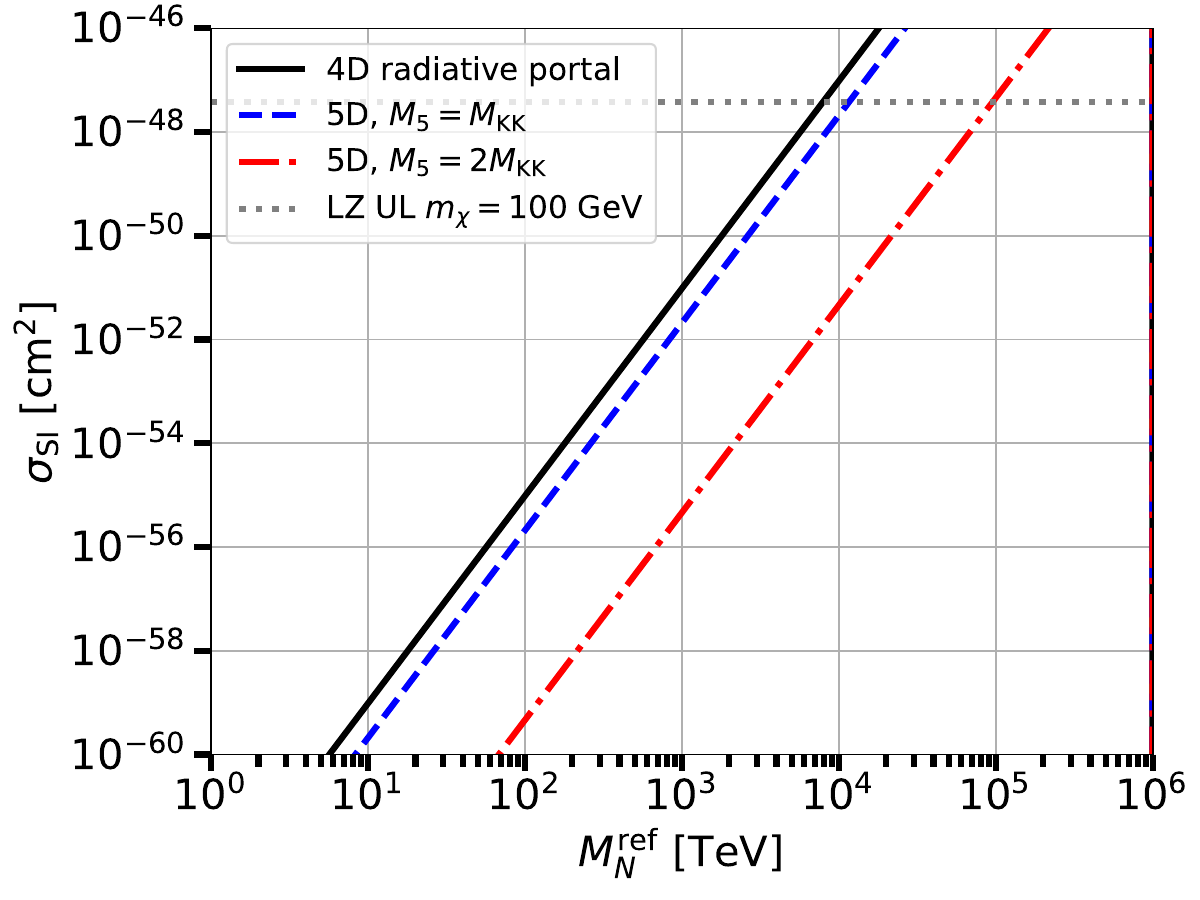}
\caption{Spin-independent direct-detection cross section as a function of the heavy-neutrino mass $M_N^{\rm{ref}}$ for $m_\chi=100,{\rm GeV}$, $y_p=0.4$, $m^{\rm{ref}}_\nu=0.05,{\rm eV}$, $y_N^{\rm ref}=1$, $m_H=125,{\rm GeV}$ and $m_{H_p}=10,{\rm GeV}$. The solid black line shows the four-dimensional radiative-portal prediction, while the dashed blue and dash-dotted red lines include the five-dimensional brane-to-brane suppression for $M_5=M_{\rm KK}$ and $M_5=2M_{\rm KK}$, respectively. The dashed horizontal line indicates the current LZ upper limit on $\sigma_{\rm SI}$ at $m_\chi=100 \,{\rm GeV}$.}
\label{fig:directdetection}
\end{figure}

Direct detection and BBN constrain the same physical quantity, but in opposite directions. The spin-independent direct-detection cross section can be written as
\begin{equation}
\sigma_{\rm SI}=C_{\rm SI}\sin^2\alpha_{\rm phys},
\label{eq:sigma_CSI_consequence}
\end{equation}
where $C_{\rm SI}$ depends on $m_\chi$, $m_{H_p}$, $y_p$, the nucleon scalar matrix element and the scalar propagators as given in Eq.~\eqref{eq:sigmaSI}. Therefore the LZ upper limit on $\sigma_{\rm SI}$ gives an upper bound on the physical mixing $\sin\alpha_{\rm phys}$.
For the benchmark $m_\chi=100\,{\rm GeV}$, $m_{H_p}=10\,{\rm GeV}$ and $y_p$ fixed by the secluded relic density ($y_p\sim 0.4$), the latest LZ limit implies an upper limit $\sin\alpha_{\rm phys}\lesssim 5.5\times10^{-5}$~\cite{Aalbers:2024LZ}. 
%This bound is weak for the multi-TeV benchmarks considered here, where the loop-induced mixing is much smaller than $2\times10^{-5}$, but it can become relevant for much larger heavy-neutrino masses because $\sin\alpha_{\rm phys}$ grows approximately as $M_N^2$.
For the benchmark used in Fig.~\ref{fig:directdetection}, we fix \(y_p=0.4\), consistently with the approximate secluded relic-density estimate in Eq.~\eqref{eq:yp_relic_scaling_short}. With this choice, the four-dimensional prediction scales as \(\sigma_{\rm SI}\propto M_N^4\). Including the sequestering factor gives \(\sigma_{5,\rm SI}=\mathcal{F}_0^2\sigma_{SI}\), so the corresponding upper bound on \(M_N\) is shifted by \(\mathcal{F}_0^{-1/2}\). The numerical bounds quoted below are obtained by applying this scaling to the LZ limit used in the figure.

Fig.~\ref{fig:directdetection} shows the scaling of the spin-independent direct-detection cross section as a function of the heavy-neutrino mass for the benchmark parameter values ($m_\chi=100~{\rm GeV}$, $m_{H_p}=10~{\rm GeV}$, $m_H=125~{\rm GeV}$, $m_\nu=0.05~{\rm eV}$ and $y_p\sim 0.4$). In the four-dimensional radiative-portal case, the cross section grows as $M_N^4$, reflecting the scaling of the loop-induced Higgs--singlet mixing in the aligned seesaw limit. In this case, the upper limit for the heavy neutrino mass is of the order of $M_N\lesssim 7$ PeV. The five-dimensional curves illustrate the effect of exponential sequestering: taking $M_5=M_{\rm KK}$ it produces a minimal effect, while for $M_5=2M_{\rm KK}$ it reduces significantly the physical portal and shifts the prediction for the nuclear cross section downward by several orders of magnitude and, as a consequence, increases significantly the upper limit for the heavy neutrini mass. 
In particular, we note that assuming $M_5=2M_{KK}$ shifts the upper limit for the heavy neutrino mass to $M_N\lesssim 100$ PeV.
For all of the cases shown, the bounds on the nuclear cross section could become relevant only for heavy-neutrino masses above tens of the PeV, lighter mediators, or weaker sequestering.

The BBN requirement gives instead a lower bound for $\sin\alpha_{\rm phys}$. If the mostly hidden scalar $H_p$ decays into SM particles only through Higgs--singlet mixing, its visible width is proportional to $\sin^2\alpha_{\rm phys}$ as in Eq.~\eqref{eq:Hp_width}
The corresponding $H_p$ lifetime is
\begin{equation}
\tau_{H_p}\simeq \frac{\hbar}{\sin^2\alpha_{\rm phys}\,\Gamma_H^{\rm SM}(m_{H_p})}.
\label{eq:tau_Hp_BBN_consequence}
\end{equation}
Requiring $\tau_{H_p}\lesssim \tau_{\rm BBN}$ gives a lower limit for $\sin^2\alpha_{\rm phys}$.
For $m_{H_p}=10\,{\rm GeV}$ and $\tau_{\rm BBN}=1\,{\rm s}$, using the fermionic decay width of $H_p$ into the kinematically open SM states gives approximately a lower limit for the physical mixing angle of $\sin\alpha_{\rm phys}\gtrsim 9\times10^{-11}$.
The precise BBN constraint depends on the $H_p$ abundance, its visible branching fractions and the hadronic/electromagnetic energy release, but lifetimes of order $0.1$--$1\,{\rm s}$ or longer are often constrained by light-element abundances~\cite{Kawasaki:2004qu,Jedamzik:2006xz,Pospelov:2010cw}.

The combined viability condition is therefore
\begin{equation}
\sin{\alpha_{\rm BBN}}\lesssim \mathcal{F}_0(M_5L)\sin\alpha_{4D}(M_N)\lesssim \sin{\alpha_{\rm DD}},
\label{eq:combined_BBN_DD_window}
\end{equation}
where $\alpha_{\rm BBN}$ and $\alpha_{\rm DD}$ are the mixing angle lower and upper limits coming from BBN and DD constraints.
In the aligned radiative-neutrino-portal limit, the four-dimensional mixing scales as
\begin{equation}
\sin\alpha_{4D}^{\rm ref}
\simeq
\frac{
y_N^{\rm ref}
m_\nu^{\rm ref}
\left(M_N^{\rm ref}\right)^2
}{
\sqrt2\pi^2v_h
\left|m_{H_p}^2-m_H^2\right|
}.
\end{equation}
Thus increasing $M_N$ increases the underlying radiative mixing and can compensate for a smaller sequestering factor. This is the key point: the brane suppression weakens direct detection, but it also lengthens the $H_p$ lifetime. Therefore too strong a suppression can be incompatible with BBN unless the heavy-neutrino scale is increased or additional decay channels for $H_p$ are present.

\begin{figure}
\centering
\includegraphics[width=0.99\columnwidth]{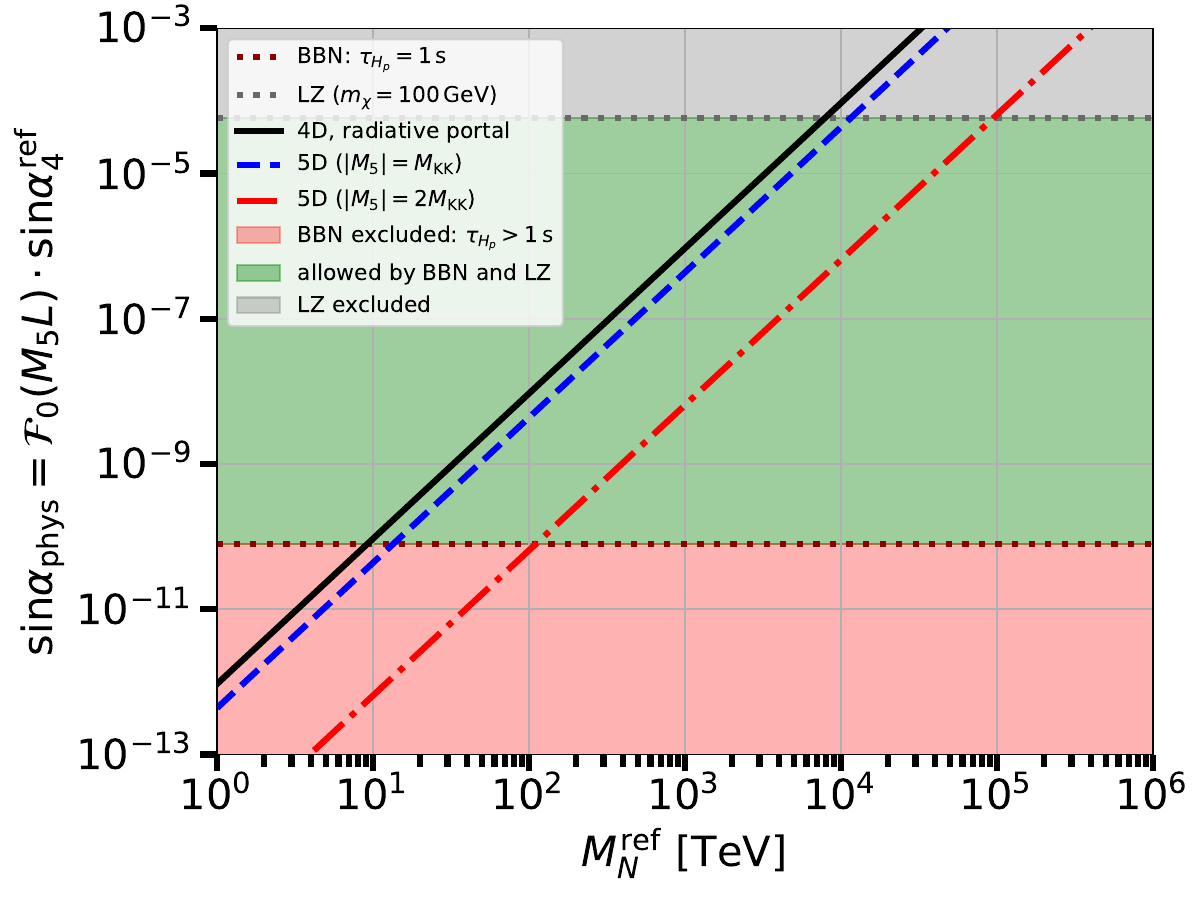}
\caption{Combined direct-detection and BBN constraints in the plane $(M_N,\sin\alpha_{\rm phys})$ for $m_\chi=100\,{\rm GeV}$ and $m_{H_p}=10\,{\rm GeV}$. The dark-sector coupling $y_p$ is fixed by the secluded relic abundance.
The red region has $\tau_{H_p}>1\,\mathrm{s}$ and fails the indicative pre-BBN decay criterion adopted here, while the grey region is excluded by the LZ spin-independent bound at $m_\chi=100\,{\rm GeV}$. The green band is allowed by both constraints. The solid black curve shows the four-dimensional prediction $F=1$, while the blue and red curves show the five-dimensional predictions for $M_5=M_{KK}$ and $M_5=2M_{KK}$, respectively. Stronger sequestering shifts the prediction to smaller physical mixing and therefore requires larger $M_N$ to satisfy the BBN lifetime constraint.}
\label{fig:DD_BBN_alpha_MN}
\end{figure}

Fig.~\ref{fig:DD_BBN_alpha_MN} illustrates the interplay between direct detection and the $H_p$ lifetime for the benchmark $m_{H_p}=10\,{\rm GeV}$ and $m_\chi=100\,{\rm GeV}$, with $y_p$ fixed by the secluded relic abundance. The red shaded region is excluded because the mediator is too long lived, $\tau_{H_p}>1\,{\rm s}$, while the grey shaded region is excluded by the LZ spin-independent bound. 
The red region indicates the benchmark lifetime criterion $\tau_{H_p}>1\,{\rm s}$. It should not be interpreted as a full BBN exclusion, which depends on the $H_p$ abundance and on its hadronic and electromagnetic energy release.
The green region satisfies both constraints. The solid black curve shows the four-dimensional prediction with $F=1$, whereas the dashed blue and dash-dotted red curves show two five-dimensional examples with $M_5=M_{\rm KK}$ and $M_5=2M_{\rm KK}$, respectively. The combined direct-detection and BBN constraints select the range of $M_N$ for which a given prediction curve lies inside the green region. Increasing the amount of sequestering suppresses the physical mixing at fixed $M_N$, shifting the prediction curves downward. As a result, larger values of $M_N$ are required to generate a sufficiently large $\sin\alpha_{\rm phys}$ for $H_p$ to decay before BBN while remaining below the direct-detection bound. Thus stronger sequestering shifts the viable region toward higher heavy-neutrino masses. This lower bound on $M_N$ can be relaxed if additional hidden-sector decay channels of $H_p$ are present.
As an example, if $M_5=2M_{\rm KK}$ then $M_N$ is constrained to be in the range:
\begin{equation}
100\,{\rm TeV}\lesssim M_0\lesssim100\,{\rm PeV}.,
\label{eq:MN_viable_5D}
\end{equation}
which is an order of magnitude higher in mass with respect to the one obtained in Eq.~\eqref{eq:MN_viable_rough} for the 4D model.

\subsubsection{Effect of five-dimensional sequestering on hidden-sector freeze-out}

The localization of the bulk-neutrino wave functions can modify the
rate at which the visible and hidden sectors exchange energy and may therefore shift the temperature at which the heavy-neutrino states become ineffective as thermal messengers and the two sectors decouple. This affects the detailed thermal history, but not the dominant annihilation process $\chi\bar\chi\to H_pH_p$, whose rate is controlled by $y_p$. Provided that the two sectors reach thermal
equilibrium before decoupling and retain approximately the same temperature until DM freeze-out, the value of $y_p$ required to
reproduce the observed relic abundance is therefore expected to remain essentially unchanged. An earlier decoupling could instead
generate a hidden-to-visible temperature ratio different from unity and induce an order-one shift in the preferred value of $y_p$; a
precise determination would require solving the coupled Boltzmann equations in the localized five-dimensional theory.
An indicative estimate of the effect of localization on thermal
contact can be obtained from the decay parameter of the light
sterile state,
\begin{equation}
K_0
\equiv
\left.
\frac{\Gamma(N_0\leftrightarrow L\Phi)}{H}
\right|_{T=M_0}
\simeq
\frac{\widetilde m_0}{m_\ast},
\qquad
m_\ast\simeq1.1\times10^{-3}\,{\rm eV}.
\end{equation}
At fixed fundamental five-dimensional brane coefficients, the localized zero-mode profiles give
\begin{equation}
\widetilde m_0
=
e^{-2|M_5|L}\widetilde m_{\rm flat}.
\end{equation}
For $\widetilde m_{\rm flat}\simeq0.05\,{\rm eV}$, this gives $K_0\simeq45e^{-2|M_5|L}$. Thermal contact through the light state is therefore expected to remain efficient for $|M_5|/M_{\rm KK}\lesssim0.4$, to become marginal around $|M_5|/M_{\rm KK}\simeq0.4$--$0.6$, and to be strongly suppressed for larger localization. These estimates assume that the massive KK states are not thermally populated. If the physical four-dimensional
seesaw parameters are instead held fixed, the profile suppression is already included in their definition and $K_0$ does not acquire the additional factor $e^{-2|M_5|L}$.
Provided that equilibrium is established, the resulting variation of the hidden-to-visible temperature ratio is expected to have only a mild effect on the relic-density coupling because $y_p\propto\zeta^{1/4}$. Even a factor-of-two change in $\zeta=T'/T$ modifies the preferred value of $y_p$ by less than about
$20\%$. For stronger localization, when the light state fails to thermalize the two sectors, $\zeta$ becomes dependent on the reheating history and the standard thermal-relic estimate must be replaced by a coupled Boltzmann analysis.

\subsubsection{Size of the compact dimension and other experimental limits}

The fifth dimension should be compactified, or at least effectively bounded, so that the model has a well-defined four-dimensional low-energy limit. The separated-brane construction requires a finite distance between the visible and hidden sectors, and the low-energy theory should contain a discrete KK tower rather than a continuum. In the minimal realization considered here, the compactification is an orbifold interval $S^1/\mathbb{Z}_2$, with endpoints $y=0$ and $y=L$. The geometrical KK spacing is
\begin{equation}
\MKK\equiv \frac{\pi}{L},
\qquad
L\simeq 6.2\times10^{-20}\,{\rm m}
\left(\frac{10\,{\rm TeV}}{\MKK}\right).
\label{eq:L_numeric_consequences}
\end{equation}
Thus, a compactification scale in the multi-TeV range corresponds to a microscopic length, far below the distances directly probed by macroscopic tests of gravity.

The upper limit on $L$ is model dependent. If gravity propagates in a large extra dimension, short-distance tests of Newton's law constrain new gravitational-strength interactions at distances of order tens of microns. The most relevant torsion-balance results find no deviation from the inverse-square law down to separations of order $50\,\mu{\rm m}$, excluding gravitational-strength Yukawa interactions with ranges above about $40$--$50\,\mu{\rm m}$~\cite{Lee:2020zjt,Tan:2020vpf}. These bounds, together with recent global updates of short-range inverse-square-law constraints~\cite{Murata:2026isl}, are much weaker than the particle-physics sizes in Eq.~\eqref{eq:L_numeric_consequences}. If SM gauge fields or charged fermions propagated in the bulk, collider and electroweak constraints on their KK excitations would instead typically require $\MKK$ in the TeV range or above, with the precise bound depending on the spectrum and boundary terms~\cite{Appelquist:2000nn,Deutschmann:2017bth,ParticleDataGroup:2024cfk}.

In the present setup, the SM fields are localized on the visible brane and the relevant bulk field is sterile-neutrino-like. The most direct particle-physics constraints therefore arise from active--sterile mixing with the neutrino KK tower, which can affect oscillations, cosmology and laboratory observables~\cite{Grossman:1999ra,Dienes:1998sb,Davoudiasl:2002fq,Cao:2003rm}. A simple way to remain in the decoupled regime considered in this work is to keep the first massive KK-dominated states heavy, for example at the multi-TeV scale or above, and to keep their active--sterile mixing small.

Before the hidden-brane Majorana interaction is included, the free massive KK tower satisfies
\begin{equation}
m_n^2=M_5^2+n^2\MKK^2,
\qquad n\geq 1.
\label{eq:KKtower_consequences}
\end{equation}
The chiral zero mode is a separate massless solution at this stage. Once $R$ acquires its vacuum expectation value, the boundary-localized Majorana interaction gives a mass to the zero mode and also shifts and mixes the massive KK levels. The physical masses must therefore be obtained by diagonalizing the complete Dirac-plus-Majorana mass matrix.

In the weak-mixing regime used in the phenomenological analysis, the physical state $\mathcal N_0$ is mostly composed of the free zero mode and is well separated from the first massive KK-dominated state,
\begin{equation}
M_0\ll m_1=\sqrt{M_5^2+\MKK^2}.
\label{eq:zero_KK_hierarchy_consequences}
\end{equation}
Whenever results are presented as functions of $M_0$ at fixed $|M_5|/\MKK$, the corresponding curves should therefore be understood as families of five-dimensional models in which $\MKK$ is chosen sufficiently above each value of $M_0$ to maintain the hierarchy in Eq.~\eqref{eq:zero_KK_hierarchy_consequences}. In particular, the reference value $\MKK=10\,{\rm TeV}$ used in Eq.~\eqref{eq:L_numeric_consequences} to illustrate the compactification length is not assumed to remain fixed throughout scans extending to much larger values of $M_0$.

The phenomenological impact of the physical sterile states depends on both their masses and their active--sterile mixing with the SM neutrinos. Denoting the physical Majorana mass eigenstates by $\mathcal N_a$, their mixing is schematically
\begin{equation}
\theta_{\alpha a}
\simeq
\frac{m_{D,\alpha a}}{M_a},
\qquad
m_{D,\alpha a}
=
\frac{v_h}{\sqrt{2}}
\left(\lambda_\nu\right)_{\alpha a},
\label{eq:active_sterile_mixing_KK}
\end{equation}
where $(\lambda_\nu)_{\alpha a}$ is the effective coupling of the physical state $\mathcal N_a$ to the visible brane. Before diagonalization, the corresponding coupling of a free KK mode is proportional to its visible-brane wave-function overlap,
\begin{equation}
\left(Y_\nu^{(n)}\right)_{\alpha I}
=
\left(\widehat Y_{5,\nu}\right)_{\alpha I}
f_{R,I}^{(n)}(0).
\end{equation}

In the general KK theory, the seesaw relation constrains a coherent sum over physical states,
\begin{equation}
\left(m_\nu\right)_{\alpha\beta}
\simeq
-\sum_a
\frac{m_{D,\alpha a}m_{D,\beta a}}{M_a}.
\label{eq:seesaw_sum_KK_consequences}
\end{equation}
Therefore, the estimate
$|\theta_{\alpha a}|^2\sim m_\nu/M_a$
applies only in the canonical seesaw regime in which an individual state gives an $\mathcal O(m_\nu)$ contribution without significant cancellations or approximate lepton-number symmetries. More generally, laboratory limits constrain cumulative combinations such as
\begin{equation}
\left(\Theta\Theta^\dagger\right)_{\alpha\beta}
=
\sum_a
\theta_{\alpha a}\theta_{\beta a}^{*}.
\label{eq:cumulative_KK_mixing}
\end{equation}

If the first massive KK-dominated state is well above the weak scale and the model lies in the canonical aligned-seesaw regime, the KK tower is effectively decoupled from direct laboratory searches and enters mainly through the threshold corrections and brane-to-brane propagation effects discussed above. If instead one or more sterile states are light, they must satisfy the corresponding sterile-neutrino constraints. Modes near the eV scale are strongly constrained by oscillation data and cosmology; sterile states in the keV range are constrained by structure formation and X-ray searches if sufficiently long lived; MeV--GeV modes are probed by meson decays, beam-dump experiments, beta-decay spectra and supernova cooling; and GeV--TeV modes are constrained by prompt and displaced heavy-neutral-lepton searches at colliders~\cite{Hagstotz:2020ukm,Abdullahi:2022jlv,Carenza:2023old,ParticleDataGroup:2024cfk}.

Complementary constraints on heavier modes arise from deviations from PMNS unitarity, electroweak precision and lepton-universality observables, and charged-lepton-flavour-violating processes. Since the physical sterile states are Majorana or pseudo-Dirac fermions, neutrinoless double-beta decay can also constrain their electron-flavour mixing. These limits are strongly suppressed in the canonical aligned-seesaw regime considered here, but can become relevant in the presence of enhanced mixing, flavour misalignment or cancellations in the light-neutrino mass matrix~\cite{Abdullahi:2022jlv,ParticleDataGroup:2024cfk}.

A further model-dependent issue is the cosmological production of the massive KK tower. If the reheating temperature exceeds the mass of the first massive KK-dominated state, several sterile modes may be thermally populated, and their decays and entropy injection must be included in the thermal history. In the regime considered here, one may consistently assume a hierarchy of the schematic form
\begin{equation}
M_0\lesssim T_{\rm RH}\ll m_1,
\label{eq:reheating_KK_hierarchy}
\end{equation}
so that the mostly zero-mode state can participate in the early thermal history while the massive KK tower is not appreciably populated. Alternatively, the effects of the thermally produced KK states must be treated explicitly. In the phenomenological analysis performed in this paper, we focus on the regime in which the sterile KK tower is sufficiently heavy and weakly mixed that these additional constraints are avoided.

If the five-dimensional metric is dynamical, there is also a modulus associated with fluctuations of the inter-brane separation. In four-dimensional language, this degree of freedom is the radion. A complete gravitational model should stabilize $L$, for example through a Goldberger--Wise-type mechanism~\cite{Goldberger:1999uk}. In the present phenomenological EFT, we take $L$ to be fixed, but a UV completion should ensure that the radion is sufficiently heavy or sufficiently weakly coupled to evade fifth-force, cosmological and collider constraints.

The minimal construction does not automatically predict an axion. An axion-like degree of freedom can arise if additional bulk gauge fields are introduced, because the fifth component $A_5$ behaves as a four-dimensional scalar. Its zero mode may be interpreted as a Wilson-line field and can be protected by higher-dimensional gauge invariance, as in the Hosotani mechanism and gauge-Higgs unification~\cite{Hosotani:1983xw,Hosotani:1983vn}. Therefore, an axion or axion-like particle is a possible extension, but it is not a mandatory consequence of the sequestered neutrino-portal setup.

The construction is also compatible with string-inspired intuition. String compactifications often contain branes, localized sectors and bulk fields, so the EFT picture used here is close in spirit to brane-world model building. However, no specific string embedding is assumed. A full string realization could provide a microscopic origin for the branes, the compactification and the stabilization of $L$, but the main sequestering argument only requires higher-dimensional locality.

\subsubsection{Gravity and the dark-matter interpretation}
\label{sec:gravity}

The sequestered construction suppresses local non-gravitational contact interactions between fields localized on different branes, but it does not remove their gravitational interactions. The visible and hidden sectors are separated along the compact coordinate $y$, while they share the same four-dimensional spacetime coordinates $x^\mu$. If gravity propagates in the five-dimensional bulk, both branes couple to the same higher-dimensional metric~\cite{ArkaniHamed:1998rs,Randall:1999ee,Randall:1999vf,Maartens:2010ar}. Hidden-brane matter can therefore contribute to the ordinary gravitational phenomena attributed to dark matter in the four-dimensional effective theory.

We assume that the compactification is stabilized and that the bulk cosmological constant, brane tensions and stabilization sector support the approximately flat background adopted above. Their detailed dynamics are not specified in the present phenomenological EFT.

Denoting the fundamental five-dimensional gravitational scale by $M_\ast$, in order to distinguish it from the sterile-neutrino bulk mass, the gravitational action is
\begin{equation}
S_{\rm grav}
=
\frac{M_\ast^3}{2}
\int \dd^4x\int_0^L\dd y\,
\sqrt{-g_5}\,\mathcal{R}_5 .
\label{eq:Sgrav5D_consequences}
\end{equation}
The complete action contains, in addition, the matter sectors localized on the two branes,
\begin{equation}
S
=
S_{\rm grav}
+
S_{\rm vis}
\left[
g_{\mu\nu}(x,0),\psi_{\rm vis}
\right]
+
S_{\rm hid}
\left[
g_{\mu\nu}(x,L),\psi_{\rm hid}
\right]
+\cdots .
\label{eq:full_gravity_brane_action}
\end{equation}
The visible and hidden actions are therefore functionals of the metric induced on their respective branes. Varying the action with respect to the five-dimensional metric gives
\begin{equation}
M_\ast^3 G_{5,AB}
=
T_{AB}^{\rm bulk}
+
T_{AB}^{\rm brane}.
\label{eq:Einstein5D_consequences}
\end{equation}
Schematically, the brane contribution contains sources localized at the two endpoints of the interval,
\begin{equation}
T_{AB}^{\rm brane}
\simeq
T_{\mu\nu}^{\rm vis}(x)
\delta_A^\mu\delta_B^\nu\delta(y)
+
T_{\mu\nu}^{\rm hid}(x)
\delta_A^\mu\delta_B^\nu\delta(y-L)
+\cdots .
\label{eq:TAB_brane_consequences}
\end{equation}
A fully covariant expression also contains the appropriate ratios of the induced and bulk metric determinants. These factors do not change the physical conclusion that both visible- and hidden-brane matter source the same five-dimensional gravitational field.

At distances much larger than the compactification length, $r\gg L$, gravity is dominated by the four-dimensional graviton zero mode. In the flat compactification considered here, its wave function is independent of $y$, while the massive gravitational KK modes give corrections suppressed at distances larger than their inverse masses. Keeping only the zero mode, the metric may be written schematically as
\begin{equation}
g_{\mu\nu}(x,y)
=
g_{\mu\nu}^{(0)}(x)
+
\text{massive gravitational KK modes}.
\label{eq:graviton_zero_mode_metric}
\end{equation}
The five-dimensional Einstein--Hilbert action then reduces to
\begin{equation}
S_{\rm grav}
\supset
\frac{M_\ast^3L}{2}
\int \dd^4x\,
\sqrt{-g_4}\,\mathcal{R}_4 .
\label{eq:dimensional_reduction_gravity}
\end{equation}
Comparing this expression with the four-dimensional Einstein--Hilbert action,
\begin{equation}
S_{\rm EH}^{(4)}
=
\frac{M_{\rm Pl}^2}{2}
\int \dd^4x\,
\sqrt{-g_4}\,\mathcal{R}_4 ,
\label{eq:EH4D_standard}
\end{equation}
where $M_{\rm Pl}$ denotes the reduced Planck mass, gives
\begin{equation}
M_{\rm Pl}^2
=
M_\ast^3L.
\label{eq:Mpl_5D_relation}
\end{equation}
This relation follows directly in the interval convention $0\leq y\leq L$ used here. A corresponding factor of two would appear if the action were instead integrated over the full covering circle.

Because the graviton zero mode couples to the stress-energy tensors localized on both branes, the long-distance four-dimensional Einstein equation contains both the visible and hidden contributions,
\begin{equation}
M_{\rm Pl}^2G_{\mu\nu}^{(4)}
\simeq
T_{\mu\nu}^{\rm vis}
+
T_{\mu\nu}^{\rm hid},
\label{eq:Einstein_hiddenbrane}
\end{equation}
where corrections from massive gravitational KK modes, radion dynamics and possible bulk sources have been omitted. In the Newtonian limit, this relation becomes
\begin{equation}
\nabla^2\Phi_N
=
4\pi G_N
\left(
\rho_{\rm vis}
+
\rho_{\rm hid}
\right).
\label{eq:Poisson_hiddenbrane}
\end{equation}

The separation of the two branes therefore does not make hidden-sector matter gravitationally invisible. A stable, non-relativistic hidden-brane population with the appropriate cosmological abundance contributes to the same long-distance gravitational potential as visible matter and can account for galaxy dynamics, gravitational lensing, cluster dynamics, structure formation and the cosmological expansion. What is suppressed by sequestering is instead the direct non-gravitational communication between the two sectors, such as the local Higgs--dark-scalar contact interaction. The universal gravitational interaction remains present because it is mediated by the bulk metric and, at long distances, by its four-dimensional zero mode.

%==========================================================
\section{Discussion and conclusions}
\label{sec:conclusion}
%==========================================================

The absence of a confirmed DM signal in direct detection, indirect searches and collider experiments has placed strong pressure on the simplest WIMP scenarios. In particular, models in which the same electroweak-scale interaction controls both thermal freeze-out and present-day laboratory signals are now highly constrained. This motivates scenarios in which the relic abundance remains thermal, but the connection to the SM is naturally weak. Secluded DM provides a simple realization of this idea: the relic abundance is set mainly by annihilations into lighter hidden-sector states, while the portal coupling connecting the hidden sector to the SM can be much smaller than a standard WIMP coupling.

In previous work~\cite{DiMauro:2025nmsdm}, a radiative neutrino-portal realization of secluded DM was introduced. The key idea is that the smallness of the Higgs--dark-scalar mixing is not imposed by hand, but is connected to the smallness of active-neutrino masses. Heavy Majorana neutrinos generate light neutrino masses through the seesaw mechanism and, at the same time, radiatively induce the Higgs--dark-scalar portal. In the aligned seesaw limit, the loop-induced portal is proportional to the light-neutrino masses and is therefore naturally suppressed. This explains why direct-detection signals can be extremely small even if the DM abundance has a thermal origin. However, the purely four-dimensional construction does not by itself explain why the tree-level Higgs--dark-scalar portal should vanish at the ultraviolet scale.

The main purpose of this work has been to provide a geometrical origin for this boundary condition. We have shown that a five-dimensional sequestered construction naturally realizes
\begin{equation}
\kappa(\Lambda_{\rm UV})=0 .
\end{equation}
The SM fields are localized on a visible brane, while the DM particle and the hidden scalar are localized on a hidden brane. Since the two sectors live at different points in the compact dimension, a local operator involving both the SM Higgs and the hidden scalar is forbidden by five-dimensional locality. The sterile neutrinos propagate in the bulk and provide the only non-gravitational communication between the two branes. The Higgs--dark-scalar portal is therefore absent at tree level, but is regenerated radiatively by the same heavy-neutrino sector responsible for the smallness of neutrino masses.

The compact dimension has several important consequences. First, it gives a geometrical explanation for the absence of the local Higgs portal at the cutoff scale. Second, before the hidden-brane Majorana interaction is included, it predicts a tower of massive sterile-neutrino KK excitations with free masses
\begin{equation}
m_n^2=M_5^2+n^2\MKK^2,
\qquad n\geq1.
\end{equation}
After hidden-sector symmetry breaking, the boundary Majorana interaction shifts and mixes these states, and the physical masses are obtained by diagonalizing the complete Dirac-plus-Majorana mass matrix. If the massive KK-dominated states are heavy, they decouple from the low-energy phenomenology except through threshold corrections and brane-to-brane propagation effects.

Third, in the light-state-dominated and weak-KK-mixing regime considered in the phenomenological analysis, the compact dimension can further suppress the radiatively generated Higgs--singlet portal through the geometrical profile factor. For the hidden-localized branch, defining $x=|M_5|L$, the strong-localization limit is
\begin{equation}
\mathcal{F}_0(x)\simeq (2x)^3\exp(-2x),
\qquad x\gg1.
\end{equation}
As a result, Higgs mixing, direct detection and other Higgs-portal observables can be much weaker than in the corresponding four-dimensional radiative model. Outside the light-state and weak-mixing regime, however, the physical KK profiles and the complete KK spectral sum must be retained.

The model is nevertheless not unconstrained. For the benchmarks considered in this work, direct detection and BBN provide the two most important phenomenological requirements. Direct detection gives an upper bound on the physical Higgs--singlet mixing, or equivalently on the combination of parameters controlling the loop-induced portal. For representative parameters, such as $m_\chi=100\,{\rm GeV}$, $m_{H_p}=10\,{\rm GeV}$, $m_\nu=0.05\,{\rm eV}$ and order-one hidden-sector Yukawa couplings, the spin-independent direct-detection cross section can approach present experimental sensitivities only for very large heavy-neutrino masses, typically in the multi-PeV range in the four-dimensional radiative model. In the five-dimensional sequestered realization, the same rate can be further suppressed by the geometrical profile factor, pushing the heavy-neutrino masses required to produce detectable direct-detection signals to still higher values.

The second and often more restrictive requirement comes from the lifetime of the hidden scalar $H_p$. If no additional hidden-sector decay channels are present, $H_p$ decays to SM particles through its Higgs component. The same small mixing that suppresses direct detection therefore also suppresses the decay rate of $H_p$. Requiring the mediator to decay before BBN implies a lower bound on the physical mixing and, in the neutrino-aligned scenario, a lower bound on the heavy-neutrino scale that is typically in the tens-of-TeV range in the four-dimensional model for the benchmark considered here. Increasing the amount of sequestering suppresses the physical mixing at fixed heavy-neutrino mass and therefore shifts the viable region toward larger values of $M_N$. This lower bound can be relaxed if $H_p$ has additional cosmologically safe hidden-sector decay channels.

The resulting phenomenology is therefore different from that of standard prompt Higgs-portal searches. The relic abundance is controlled mainly by secluded annihilations inside the hidden sector, while the most relevant observable constraints are set by the small residual connection to the SM. Direct detection constrains the portal from above, whereas BBN constrains it from below through the mediator lifetime. In the benchmarks considered in this work, the direct-detection rate is naturally suppressed, especially in the strongly localized five-dimensional regime, while the mediator lifetime provides the most important consistency condition.

Finally, the sequestered geometry does not make DM gravitationally invisible. Gravity propagates in the bulk, and the graviton zero mode couples to the stress-energy localized on both branes. Hidden-brane DM therefore contributes to the same long-distance four-dimensional gravitational potential as ordinary matter and, provided that it has the required abundance and is sufficiently cold, can account for the gravitational evidence for DM. What is suppressed by the brane separation is not gravity, but the direct particle-physics communication between the visible and hidden sectors.

In summary, the model provides a higher-dimensional realization of secluded DM in which the weakness of laboratory signals is not imposed by hand. The smallness of active-neutrino masses is generated through the seesaw mechanism, while the same heavy-neutrino sector radiatively generates the Higgs--dark-scalar portal. The tree-level Higgs portal is absent because of five-dimensional locality, and the nonzero low-energy portal is suppressed by the loop factor, the light-neutrino masses and, in the strongly localized light-state regime, the geometrical profile factor. The model can therefore accommodate a thermal secluded relic abundance while evading conventional direct-detection searches. Its viability is mainly controlled by the interplay between the direct-detection upper bound on the portal and the BBN lower bound required for the hidden mediator to decay sufficiently early.

\acknowledgments
M.D.M. acknowledges support from the research grant {\sl TAsP (Theoretical Astroparticle Physics)} funded by Istituto Nazionale di Fisica Nucleare (INFN).

\paragraph*{Personal note.}
On a personal note, this paper belongs to a small series of works indirectly inspired by my wife Chiara, who has repeatedly suggested that dark matter probably does not exist. Her skepticism motivated me to think more carefully about secluded dark sectors \cite{DiMauro:2025jsb,DiMauro:2025uxt}, where dark matter can be thermal and gravitationally present while remaining almost invisible to laboratory searches. I tried to explain the physical meaning of the present sequestered construction to her, but the idea of placing the dark sector on a different brane seems to have exhausted the remaining chances of convincing her. I can only hope that the referee will be more indulgent.

%==========================================================

\FloatBarrier
\appendix

%==========================================================
\section{Odd scalar operators and the independent dark-matter mass}
\label{app:odd_scalar_operators}
%==========================================================

In the simplified Lagrangian used in the main text, the real singlet
scalar $R$ couples both to the sterile neutrinos and to the Dirac
DM fermion,
\begin{equation}
\mathcal L
\supset
-\frac{1}{2}Y_N R\,\overline{N^c}N
-y_p R\,\overline{\chi}\chi .
\label{eq:odd_app_yukawas}
\end{equation}
Since $\chi$ is vector-like, its bare Dirac mass is also allowed,
\begin{equation}
\mathcal L
\supset
-m_{\chi,0}\overline{\chi}\chi .
\label{eq:odd_app_bare_dm_mass}
\end{equation}
After hidden-sector symmetry breaking, with
\begin{equation}
R=\frac{v_r+\rho}{\sqrt{2}},
\end{equation}
the physical dark-matter mass is
\begin{equation}
m_\chi
=
m_{\chi,0}
+
\frac{y_pv_r}{\sqrt{2}}.
\label{eq:odd_app_dm_mass}
\end{equation}
The bare mass therefore allows $m_\chi$ to be treated as a
renormalized input independently of the sterile-neutrino mass scale.
We note, however, that if $y_pv_r/\sqrt{2}\gg m_\chi$, maintaining a
weak-scale physical DM mass requires a cancellation against
$m_{\chi,0}$. This is a naturalness issue, but it does not prevent
$m_\chi$ and $M_N$ from being treated as independent phenomenological
parameters.

There are two possible ways of organizing the scalar sector. In the
first, no exact symmetry under $R\to-R$ is imposed and the effects of
the operators odd in $R$ are included in the effective theory. In the
second, such operators are forbidden by a $Z_4$ symmetry, which is
softly broken by the vector-like DM mass.

%----------------------------------------------------------
\subsection{General real-singlet theory without an exact $R$ parity}
\label{app:odd_scalar_general}
%----------------------------------------------------------

In the absence of an exact symmetry acting on $R$, the most general
renormalizable scalar potential contains, in addition to the even
operators displayed in the main text,
\begin{equation}
V_{\rm odd}(\Phi,R)
=
t_R R
+
\frac{\mu_3}{3!}R^3
+
\mu_{R\Phi}R\Phi^\dagger\Phi .
\label{eq:odd_app_general_potential}
\end{equation}
The tadpole $t_R$ and the cubic coupling $\mu_3$ are purely
hidden-sector parameters. They modify the vacuum conditions, the
relation between the Lagrangian parameters and the physical scalar
mass, and the scalar self-interactions. They can be included in the
renormalized hidden-sector potential and traded, together with the
even parameters, for $v_r$, $m_{H_p}$ and the relevant scalar
self-couplings. They do not by themselves communicate directly with
the visible sector.

The dimension-three interaction
\begin{equation}
V\supset \mu_{R\Phi}R\Phi^\dagger\Phi
\label{eq:odd_app_linear_portal}
\end{equation}
instead provides an additional Higgs--singlet portal. In the
five-dimensional realization, both
$R\Phi^\dagger\Phi$ and $R^2\Phi^\dagger\Phi$ involve fields
localized on opposite branes. Five-dimensional locality therefore
forbids both operators as fundamental local interactions at the UV
scale, and the corresponding boundary conditions may be written as
\begin{equation}
\kappa(\Lambda_{\rm UV})=0,
\qquad
\mu_{R\Phi}(\Lambda_{\rm UV})=0.
\label{eq:odd_app_uv_conditions}
\end{equation}
The two interactions can nevertheless be generated radiatively by
fields that propagate between the visible and hidden boundaries.

With the normalization used in the main text, the quartic and linear
portals contribute to the off-diagonal scalar mass as
\begin{equation}
\left(\mathcal M^2\right)_{h\rho}
=
v_h\left(
\kappa v_r+\frac{\mu_{R\Phi}}{\sqrt{2}}
\right).
\label{eq:odd_app_offdiag_mass}
\end{equation}
It is therefore convenient to define
\begin{equation}
\kappa_{\rm eff}
\equiv
\kappa+
\frac{\mu_{R\Phi}}{\sqrt{2}v_r},
\label{eq:odd_app_keff}
\end{equation}
so that
\begin{equation}
\tan(2\alpha)
=
\frac{2\kappa_{\rm eff}v_hv_r}
     {m_{H_p}^2-m_H^2}.
\label{eq:odd_app_mixing}
\end{equation}
Consequently, observables controlled only by the physical
Higgs--singlet mixing angle, including the leading direct-detection
cross section, the production of $H_p$, and its visible decay width,
constrain $\kappa_{\rm eff}$ rather than $\kappa$ and
$\mu_{R\Phi}$ separately.

In the absence of a symmetry forbidding the linear portal, a complete
matching calculation should include both its radiative coefficient
and the box-induced coefficient of $R^2\Phi^\dagger\Phi$. Schematically,
the heavy-neutrino sector can generate
\begin{equation}
\mu_{R\Phi}^{\rm loop}
\sim
\frac{1}{16\pi^2}
\operatorname{Tr}
\left[
Y_N M_N Y_\nu^\dagger Y_\nu
\right],
\label{eq:odd_app_mu_loop_scaling}
\end{equation}
up to numerical factors, flavour contractions and
renormalization-scheme-dependent finite terms. In the aligned seesaw
limit this contribution has the parametric behavior
\begin{equation}
\mu_{R\Phi}^{\rm loop}
\sim
\frac{y_Nm_\nu M_N^2}{8\pi^2v_h^2}.
\label{eq:odd_app_mu_aligned}
\end{equation}
After division by $v_r$, its contribution to
$\kappa_{\rm eff}$ can be of the same parametric order as the
box-induced quartic portal. Moreover, in the five-dimensional theory,
the linear and quartic portals need not carry the same geometrical
profile factor because they involve, respectively, one and two
hidden-brane scalar vertices.

The phenomenological formulae in the main text may therefore be
interpreted directly in terms of $\kappa_{\rm eff}$. The explicit box
calculation determines the quartic contribution to this combination.
If the theory is formulated without a symmetry forbidding
$R\Phi^\dagger\Phi$, a complete ultraviolet prediction for
$\kappa_{\rm eff}$ additionally requires the matching calculation for
$\mu_{R\Phi}$. Alternatively, the results obtained from the box
diagram may be regarded as the limit in which the radiative linear
portal is subleading or has been absorbed into the phenomenological
definition of $\kappa_{\rm eff}$.

The two operators are indistinguishable in observables that depend
only on the scalar mixing angle. They can, however, be distinguished
by observables sensitive to the scalar potential beyond quadratic
order, since $R\Phi^\dagger\Phi$ and $R^2\Phi^\dagger\Phi$ generate
different trilinear and quartic scalar interactions after symmetry
breaking.

%----------------------------------------------------------
\subsection{Softly broken $Z_4$ realization}
\label{app:odd_scalar_z4}
%----------------------------------------------------------

A second possibility is to forbid operators containing an odd number
of $R$ fields by means of a discrete $Z_4$ symmetry. One possible
assignment, in additive notation modulo four, is
\begin{equation}
\begin{array}{c|cccccc}
 & R & \Phi & L_\alpha & e_{R\alpha} & N_R
 & (\chi_L,\chi_R)\\
\hline
q_{Z_4}
 & 2 & 0 & 1 & 1 & 1 & (1,3)
\end{array}.
\label{eq:odd_app_z4_charges}
\end{equation}
Equivalently, under the generator of $Z_4$,
\begin{equation}
\begin{aligned}
R&\to-R,
&
N_R&\to iN_R,
\\
L_\alpha&\to iL_\alpha,
&
e_{R\alpha}&\to i e_{R\alpha},
\\
\chi_L&\to i\chi_L,
&
\chi_R&\to-i\chi_R,
\\
\Phi&\to\Phi .
\end{aligned}
\label{eq:odd_app_z4_transformations}
\end{equation}
The ordinary charged-lepton Yukawa interaction and the visible
neutrino Yukawa interaction remain invariant. The symmetry also
allows
\begin{equation}
R\,\overline{N^c}N,
\qquad
R\,\overline{\chi}\chi,
\qquad
R^2\Phi^\dagger\Phi,
\label{eq:odd_app_z4_allowed}
\end{equation}
while forbidding
\begin{equation}
R,
\qquad
R^3,
\qquad
R\Phi^\dagger\Phi,
\qquad
\overline{N^c}N,
\qquad
\overline{\chi}\chi.
\label{eq:odd_app_z4_forbidden}
\end{equation}
In the five-dimensional construction, the two chiral components of
the bulk Dirac field may be assigned the same $Z_4$ charge, so that
the bulk Dirac mass remains invariant.

An independent vector-like DM mass,
\begin{equation}
-m_{\chi,0}\overline{\chi}\chi,
\end{equation}
softly breaks $Z_4$. It permits the physical DM mass to be treated as
an independent renormalized parameter according to
Eq.~\eqref{eq:odd_app_dm_mass}. The Yukawa interaction and the bare
mass both preserve the continuous dark fermion number of $\chi$, so
the stability of the Dirac DM particle does not rely on the exact
$Z_4$ symmetry.

Because the soft-breaking spurion is confined to the $\chi$ sector,
it does not generate $R\Phi^\dagger\Phi$ through the leading
one-loop heavy-neutrino diagram. The field $\chi$ has no direct
coupling to the Higgs doublet or to the sterile neutrinos. A linear
Higgs portal therefore requires both an insertion of the
soft-breaking mass and additional interactions communicating between
the $\chi$ and visible sectors. It is higher order and proportional
to the soft-breaking spurion and to at least one already small
visible--hidden portal coupling. It may consequently be neglected
consistently at the order at which the one-loop box contribution to
$R^2\Phi^\dagger\Phi$ is retained.

The soft breaking does generate hidden-sector operators odd in $R$
through loops of $\chi$, including a scalar tadpole and a cubic
interaction. Schematically,
\begin{equation}
\delta t_R
\sim
\frac{y_pm_{\chi,0}^3}{16\pi^2},
\qquad
\delta\mu_3
\sim
\frac{y_p^3m_{\chi,0}}{16\pi^2},
\label{eq:odd_app_soft_operators}
\end{equation}
up to numerical factors and logarithmic terms. These contributions
are included through the corresponding hidden-sector counterterms.
The tadpole is absorbed into the renormalized vacuum condition,
whereas the cubic term modifies scalar self-interactions but does not
directly generate a leading visible-sector portal.

It is worth emphasizing that the symmetry forbids operators odd in
the unshifted field $R$, not all interactions odd in the physical
fluctuation $\rho$. Once $R$ acquires a vacuum expectation value,
the invariant operators $R^4$ and $R^2\Phi^\dagger\Phi$ generate,
respectively, cubic interactions in $\rho$ and the trilinear
interaction $\rho\Phi^\dagger\Phi$. The latter is precisely the
interaction responsible for Higgs--singlet mixing after the
radiative quartic portal is generated.

The softly broken $Z_4$ realization therefore preserves the
box-induced quartic portal as the leading visible-sector connection.
It also avoids an exact spontaneously broken discrete symmetry and
the associated stable domain-wall problem. Its remaining limitation
is that, whenever
\begin{equation}
\frac{y_pv_r}{\sqrt{2}}\gg m_\chi,
\end{equation}
a weak-scale physical DM mass requires a cancellation between the
bare and VEV-induced contributions in Eq.~\eqref{eq:odd_app_dm_mass}.

%----------------------------------------------------------
\subsection{Interpretation adopted in the phenomenological analysis}
\label{app:odd_scalar_interpretation}
%----------------------------------------------------------

The two possibilities lead to the same low-energy phenomenology when
all visible-sector observables are expressed in terms of the physical
mixing angle $\alpha$. In the general real-singlet theory, the
quantity constrained by direct detection and by the lifetime of
$H_p$ is $\kappa_{\rm eff}$ in
Eq.~\eqref{eq:odd_app_keff}. In the softly broken $Z_4$
realization, the linear portal is absent at leading order and
\begin{equation}
\kappa_{\rm eff}
\simeq
\kappa
\end{equation}
to the accuracy of the one-loop calculation.

The numerical analysis in the main text can therefore be interpreted
in either of the following ways:
\begin{enumerate}
\item as a phenomenological analysis of the total effective portal
      $\kappa_{\rm eff}$, without resolving separately the quartic
      and linear contributions; or
\item as the leading prediction of the softly broken $Z_4$
      realization, in which the box-induced coefficient of
      $R^2\Phi^\dagger\Phi$ provides the dominant visible portal.
\end{enumerate}
In both cases, $m_\chi$ is taken to be the renormalized physical DM
mass and is treated independently of the heavy-neutrino scale.

\section{One-loop matching calculation for the radiative scalar portal}
\label{app:loop_matching_complete}

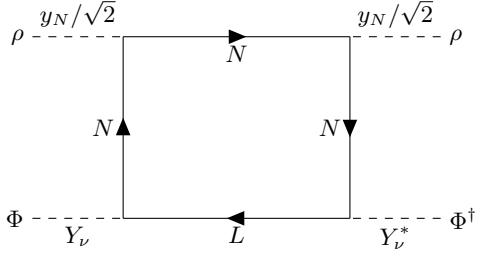
\begin{figure}[!t]
\centering
\begin{tikzpicture}[baseline={(current bounding box.center)}]
\begin{feynman}
  \vertex (A) at (0,  1.2);
  \vertex (B) at (3.0, 1.2);
  \vertex (C) at (3.0,-1.2);
  \vertex (D) at (0, -1.2);
  \vertex[left=1.2cm of A] (PhiL) {$\rho$};
  \vertex[right=1.2cm of B] (PhiR) {$\rho$};
  \vertex[right=1.2cm of C] (HbotR) {$\Phi^\dagger$};
  \vertex[left=1.2cm of D] (HbotL) {$\Phi$};
  \diagram*{
    (A) -- [fermion, edge label'=$N$] (B)
        -- [fermion, edge label'=$N$] (C)
        -- [fermion, edge label=$L$]   (D)
        -- [fermion, edge label=$N$]   (A),

    (PhiL)  -- [scalar, edge label=$y_N/\sqrt{2}$] (A),
    (PhiR)  -- [scalar, edge label'=$y_N/\sqrt{2}$] (B),
    (HbotR) -- [scalar, edge label=$Y_\nu^\ast$] (C),
    (HbotL) -- [scalar, edge label'=$Y_\nu$] (D),
  };
\end{feynman}
\end{tikzpicture}
\caption{One-loop box diagram generating the radiative Higgs--dark portal after hidden-sector symmetry breaking. The two external singlet fluctuations $\rho$ couple to the heavy Majorana neutrino through $y_N/\sqrt{2}$, while the two external Higgs fields couple through $Y_\nu$ and the lepton doublet. At momenta below the heavy-neutrino mass, the amplitude is local and matches onto the $\rho^2\Phi^\dagger\Phi$ component of $R^2\Phi^\dagger\Phi$. In the five-dimensional realization the same topology is present, with the heavy-neutrino propagators replaced by brane-to-brane bulk propagators or, equivalently, by a KK-tower sum.}
\label{fig:boxapp}
\end{figure}

In this appendix we provide the explicit steps leading to the loop-induced Higgs--dark-scalar portal discussed in Sec.~\ref{sec:radiativeloop} and shown in Fig.~\ref{fig:boxapp}. Before hidden-sector symmetry breaking, the sterile fields are right-handed gauge-singlet Weyl fermions $N_{RI}$. After $R$ acquires a vacuum expectation value, we denote the corresponding four-component Majorana mass eigenstates by $N_I=N_I^c$. The calculation below is performed after hidden-sector symmetry breaking, but in the unbroken electroweak phase.

In the baseline scenario with vanishing independent bare Majorana mass,
\begin{equation}
M_N=\frac{v_r}{\sqrt{2}}Y_N,
\qquad
Y_N=Y_N^T.
\label{eq:app_complete_mass_relation}
\end{equation}
A Takagi transformation diagonalizes both matrices,
\begin{eqnarray}
U_N^T M_N U_N
=
\operatorname{diag}(M_1,M_2,M_3),
\\
U_N^T Y_N U_N
=
\operatorname{diag}(y_{N,1},y_{N,2},y_{N,3}),
\label{eq:app_complete_takagi}
\end{eqnarray}
with $M_I>0$, $y_{N,I}>0$, and $M_I=y_{N,I}v_r/\sqrt{2}$. The visible Yukawa matrix is correspondingly replaced by $Y_\nu U_N$. For clarity, we first perform the matching for one heavy Majorana neutrino with mass $M_N$, visible Yukawa coupling $Y_\nu$, and real positive hidden Yukawa coupling $y_N$.

Writing
\begin{equation}
R=\frac{v_r+\rho}{\sqrt{2}},
\end{equation}
the interactions relevant for the matching are
\begin{align}
\mathcal{L}
\supset {}&
-Y_\nu\overline{L}\widetilde{\Phi}P_R N
-\frac{1}{2}M_N\overline{N}N
-\frac{y_N}{2\sqrt{2}}\rho\overline{N}N
+\mathrm{h.c.}-V,
\nonumber\\
V
\supset {}&
\frac{\kappa}{2}\rho^2\Phi^\dagger\Phi.
\label{eq:app_complete_interactions}
\end{align}
Here the Hermitian conjugate applies to the visible Yukawa interaction. If an independent bare Majorana mass is present and is not aligned with $Y_N$, the singlet coupling need not be diagonal in the physical mass basis. The simple mode-by-mode expressions below therefore apply to the baseline limit in Eq.~\eqref{eq:app_complete_mass_relation}, or more generally when the two matrices are simultaneously diagonalizable.

The matching is performed at vanishing external momenta. We neglect the masses of the SM lepton doublets and keep only the leading local term in the expansion in $p_i^2/M_N^2$, where $p_i$ are the external momenta. Derivative operators generated at higher order are not relevant for the renormalization of the local portal.

The effective low-energy portal operator is normalized as
\begin{equation}
\mathcal{L}_{\rm EFT}
\supset
-\frac{\kappa}{2}\rho^2\Phi^\dagger\Phi .
\label{eq:app_complete_EFT_operator}
\end{equation}
With this convention, the local EFT amplitude for two external real scalars $\rho$ and two external Higgs fields $\Phi,\Phi^\dagger$ is
\begin{equation}
\mathrm{i}\mathcal{M}_{\rm EFT}
=
-\mathrm{i}\kappa .
\label{eq:app_complete_EFT_amplitude}
\end{equation}
In the regime $p_i^2\ll M_N^2$, the loop amplitude admits the local expansion
\begin{equation}
\mathrm{i}\mathcal{M}_{\rm loop}
=
\mathrm{i}\mathcal{M}_{\rm loop}(0)
+
\mathcal{O}\left(\frac{p_i^2}{M_N^2}\right).
\label{eq:app_complete_local_expansion}
\end{equation}
The first term has the same field structure as the operator in Eq.~\eqref{eq:app_complete_EFT_operator} and can therefore be matched onto the Wilson coefficient $\kappa$.

The relevant propagators are
\begin{align}
S_N(q)&=\frac{\mathrm{i}(\slashed q+M_N)}{q^2-M_N^2},
&
S_L(q)&=\frac{\mathrm{i}\slashed q}{q^2}.
\label{eq:app_complete_props}
\end{align}
The lepton doublet can be treated as massless because the matching is performed in the unbroken electroweak phase. Electroweak-symmetry-breaking corrections are suppressed by $m_\ell^2/M_N^2$ or $m_\nu^2/M_N^2$.

The original sterile field is chiral, but the propagating field in this calculation is the physical Majorana state $N=N^c$. The singlet interaction
\begin{equation}
-\frac{y_N}{2\sqrt{2}}\rho\overline{N}N
\label{eq:app_complete_RNN_majorana}
\end{equation}
is a scalar Majorana mass-type interaction. After the coupling $y_N/\sqrt{2}$ has been factored out, each $\rho NN$ insertion acts as the identity in Dirac space. The chiral projectors occur only at the visible Yukawa vertices.

After extracting the scalar denominators from the three heavy-neutrino propagators and the massless lepton propagator, one representative fermion flow gives
\begin{equation}
\ii\mathcal{M}_{\rm loop}^{\rm bare}
=
-|y_N|^2|Y_\nu|^2
\int\frac{\mathrm{d}^d q}{(2\pi)^d}
\frac{\mathcal{N}(q)}
{q^2(q^2-M_N^2)^3}.
\label{eq:app_complete_bare_general}
\end{equation}
The two identical external $\rho$ fields give two equivalent contractions. Together with the vertex coupling $y_N/\sqrt{2}$, these contractions yield the overall factor $|y_N|^2$. The remaining minus sign is the standard closed-fermion-loop sign. After summing the equivalent Majorana fermion flows, the result is independent of the chosen flow assignment.

The numerator is
\begin{equation}
\mathcal{N}(q)
=
\operatorname{Tr}\left[
(\slashed q+M_N)(\slashed q+M_N)
P_L\slashed q\,P_R
(\slashed q+M_N)
\right].
\label{eq:app_complete_numerator}
\end{equation}
The two factors $(\slashed q+M_N)$ before the lepton propagator arise because the two $\rho NN$ insertions are scalar insertions on the heavy-neutrino line. By contrast, $P_L$ and $P_R$ appear at the visible Yukawa vertices. They are defined as
\begin{equation}
P_L=\frac{1-\gamma^5}{2},
\qquad
P_R=\frac{1+\gamma^5}{2}.
\label{eq:app_complete_projectors}
\end{equation}
Using
\begin{equation}
P_L\slashed q=\slashed q P_R,
\,\,\,
P_R\slashed q=\slashed q P_L,
\,\,
P_LP_R=0,
\,\,\,
P_RP_R=P_R,
\label{eq:app_complete_chiral_identities}
\end{equation}
one obtains
\begin{align}
\mathcal{N}(q)
&=
\operatorname{Tr}\left[
(\slashed q+M_N)^2
P_L\slashed q\,P_R
(\slashed q+M_N)
\right]
\nonumber\\
&=
\operatorname{Tr}\left[
(\slashed q+M_N)^2
\slashed q P_R
(\slashed q+M_N)
\right].
\label{eq:app_complete_N_step1}
\end{align}
Expanding the square gives
\begin{align}
\mathcal{N}(q)
&=
\operatorname{Tr}\left[
\left(q^2+2M_N\slashed q+M_N^2\right)
\slashed q P_R
(\slashed q+M_N)
\right]
\nonumber\\
&=
(q^2+M_N^2)
\operatorname{Tr}\left[
\slashed q P_R\slashed q
\right]
+
2M_N^2q^2
\operatorname{Tr}\left[P_R\right].
\label{eq:app_complete_N_step2}
\end{align}
The traces containing a single $\slashed q$ vanish. Moreover,
\begin{equation}
\operatorname{Tr}\left[
\slashed q P_R\slashed q
\right]
=
2q^2,
\qquad
\operatorname{Tr}\left[P_R\right]=2.
\label{eq:app_complete_traces}
\end{equation}
Therefore
\begin{equation}
\mathcal{N}(q)
=
2q^2(q^2+3M_N^2).
\label{eq:app_complete_N_result}
\end{equation}
The term proportional to $q^4$ generates the logarithmically divergent integral that renormalizes the local operator $R^2\Phi^\dagger\Phi$.

Substituting Eq.~\eqref{eq:app_complete_N_result} into Eq.~\eqref{eq:app_complete_bare_general}, the factor of $q^2$ cancels the massless lepton denominator and gives
\begin{align}
\mathrm{i}\mathcal{M}_{\rm loop}^{\rm bare}
&=
-2|y_N|^2|Y_\nu|^2
\int\frac{\mathrm{d}^d q}{(2\pi)^d}
\frac{q^2+3M_N^2}{(q^2-M_N^2)^3}
\nonumber\\
&=
-2|y_N|^2|Y_\nu|^2
\left[
I_2(M_N)+4M_N^2I_3(M_N)
\right].
\label{eq:app_complete_bare_I2I3}
\end{align}
We evaluate the loop integrals in dimensional regularization with $d=4-2\epsilon$ and use the $\overline{\rm MS}$ subtraction convention. The scale $\mu$ is the 't Hooft renormalization scale. We define
\begin{align}
I_2(M_N)
&=
\mu^{2\epsilon}
\int\frac{\mathrm{d}^d q}{(2\pi)^d}
\frac{1}{(q^2-M_N^2)^2},
\nonumber\\
I_3(M_N)
&=
\mu^{2\epsilon}
\int\frac{\mathrm{d}^d q}{(2\pi)^d}
\frac{1}{(q^2-M_N^2)^3}.
\label{eq:app_complete_I2I3_defs}
\end{align}
In obtaining Eq.~\eqref{eq:app_complete_bare_I2I3}, we used
\begin{equation}
\frac{q^2+3M_N^2}{(q^2-M_N^2)^3}
=
\frac{1}{(q^2-M_N^2)^2}
+
\frac{4M_N^2}{(q^2-M_N^2)^3}.
\label{eq:app_complete_decomposition}
\end{equation}
This decomposition makes the ultraviolet structure explicit: $I_2$ is logarithmically divergent, whereas $I_3$ is finite.

We define
\begin{equation}
\frac{1}{\overline{\epsilon}}
=
\frac{1}{\epsilon}
-\gamma_E
+\ln(4\pi),
\label{eq:app_complete_epsbar}
\end{equation}
where $\gamma_E$ is the Euler--Mascheroni constant. The relevant integrals are
\begin{align}
I_2(M_N)
&=
\frac{\mathrm{i}}{16\pi^2}
\left[
\frac{1}{\overline{\epsilon}}
-
\ln\left(\frac{M_N^2}{\mu^2}\right)
\right],
\nonumber\\
I_3(M_N)
&=
-\frac{\mathrm{i}}{32\pi^2M_N^2}
+
\mathcal{O}(\epsilon).
\label{eq:app_complete_integrals}
\end{align}
Therefore
\begin{equation}
\mathrm{i}\mathcal{M}_{\rm loop}^{\rm bare}
=
-\frac{\mathrm{i}|y_N|^2|Y_\nu|^2}{8\pi^2}
\left[
\frac{1}{\overline{\epsilon}}
-
\ln\left(\frac{M_N^2}{\mu^2}\right)
-
2
\right].
\label{eq:app_complete_bare_result}
\end{equation}

Although the renormalized tree-level portal is set to zero as a boundary condition in the radiative scenario, the bare loop amplitude contains a UV-divergent local contribution proportional to $R^2\Phi^\dagger\Phi$. Since this operator is allowed by the symmetries, it must be included as a counterterm. After hidden-sector symmetry breaking,
\begin{equation}
\mathcal{L}_{\rm ct}
\supset
-\frac{\delta\kappa}{2}\rho^2\Phi^\dagger\Phi,
\label{eq:app_complete_ct_lagrangian}
\end{equation}
whose contribution to the four-point amplitude is
\begin{equation}
\mathrm{i}\mathcal{M}_{\rm ct}
=
-\mathrm{i}\delta\kappa.
\label{eq:app_complete_ct_amplitude}
\end{equation}
In the $\overline{\mathrm{MS}}$ scheme, the divergent part is removed by
\begin{equation}
\delta\kappa
=
-\frac{|y_N|^2|Y_\nu|^2}{8\pi^2}
\frac{1}{\overline{\epsilon}}.
\label{eq:app_complete_delta_kappa}
\end{equation}
The renormalized zero-momentum amplitude is then
\begin{align}
\mathrm{i}\mathcal{M}_{\rm loop}^{\rm ren}(0)
\!=
\mathrm{i}\mathcal{M}_{\rm loop}^{\rm bare}
+
\mathrm{i}\mathcal{M}_{\rm ct}\!=
\!\frac{\mathrm{i}|y_N|^2|Y_\nu|^2}{8\pi^2}
\!\left[
\ln\left(\frac{M_N^2}{\mu^2}\right)
+
2
\right].
\label{eq:app_complete_ren_amplitude}
\end{align}
Matching to Eq.~\eqref{eq:app_complete_EFT_amplitude},
\begin{equation}
-\mathrm{i}\kappa_{\rm loop}(\mu)
=
\mathrm{i}\mathcal{M}_{\rm loop}^{\rm ren}(0),
\label{eq:app_complete_matching_condition}
\end{equation}
gives
\begin{equation}
\kappa_{\rm loop}(\mu)
=
-\frac{|y_N|^2|Y_\nu|^2}{8\pi^2}
\left[
\ln\left(\frac{M_N^2}{\mu^2}\right)
+
2
\right].
\label{eq:app_complete_kappa_mu}
\end{equation}
At the matching scale $\mu=M_N$,
\begin{equation}
\kappa_{\rm loop}(M_N)
=
-\frac{|y_N|^2|Y_\nu|^2}{4\pi^2}.
\label{eq:app_complete_kappa_threshold}
\end{equation}

The logarithmic term in Eq.~\eqref{eq:app_complete_kappa_mu} fixes the additive contribution to the running of the portal coupling in the full theory. The finite constant is convention and scheme dependent. In the $\overline{\rm MS}$ scheme with the zero-momentum matching convention used here, the finite threshold at $\mu=M_N$ is given by Eq.~\eqref{eq:app_complete_kappa_threshold}. Physical observables are independent of this convention once the renormalized portal coupling is specified at a given scale.

Since the physical amplitude is independent of the renormalization scale, the explicit scale dependence of $\kappa_{\rm loop}(\mu)$ is compensated by the running of the renormalized local coupling. From Eq.~\eqref{eq:app_complete_kappa_mu},
\begin{equation}
\frac{\mathrm{d}\kappa_{\rm loop}}{\mathrm{d}\ln\mu}
=
\frac{|y_N|^2|Y_\nu|^2}{4\pi^2}.
\label{eq:app_complete_dkloop}
\end{equation}
Therefore the additive contribution to the beta function is
\begin{equation}
\beta_\kappa^{\rm add}
=
\frac{\mathrm{d}\kappa}{\mathrm{d}\ln\mu}
=
-\frac{|y_N|^2|Y_\nu|^2}{4\pi^2},
\label{eq:app_complete_beta_single}
\end{equation}
or equivalently
\begin{equation}
16\pi^2\beta_\kappa^{\rm add}
=
-4|y_N|^2|Y_\nu|^2.
\label{eq:app_complete_beta_single_16pi}
\end{equation}
Below the heavy-neutrino threshold, the field $N$ is removed from the low-energy theory. Its effect remains in the threshold value of the Wilson coefficient, while the subsequent running is multiplicative,
\begin{equation}
\frac{\mathrm{d}\kappa_{\rm EFT}}{\mathrm{d}\ln\mu}
=
\gamma_\kappa\kappa_{\rm EFT}.
\label{eq:app_complete_low_energy_running}
\end{equation}

We finally rewrite the result in terms of seesaw parameters. In the single-generation aligned limit,
\begin{equation}
m_\nu\simeq\frac{m_D^2}{M_N},
\qquad
m_D=\frac{v_hY_\nu}{\sqrt{2}},
\end{equation}
and therefore
\begin{equation}
|Y_\nu|^2
=
\frac{2m_\nu M_N}{v_h^2}.
\label{eq:app_complete_Ynu_seesaw}
\end{equation}
Substituting this expression into Eq.~\eqref{eq:app_complete_kappa_threshold} gives
\begin{equation}
\kappa_{\rm loop}(M_N)
=
-\frac{y_N^2M_Nm_\nu}{2\pi^2v_h^2}.
\label{eq:app_complete_kappa_seesaw_single}
\end{equation}

For three heavy neutrinos, in the basis in which the heavy-neutrino mass matrix and the hidden Yukawa matrix are diagonal, the threshold correction is
\begin{equation}
\kappa_{\rm loop}
=
-\sum_I
\frac{y_{N,I}^2\left(Y_\nu^\dagger Y_\nu\right)_{II}}{4\pi^2},
\label{eq:app_complete_kappa_three_general}
\end{equation}
up to scheme-dependent finite terms if the matching is performed away from $\mu=M_I$.

Using the Casas--Ibarra parameterization,
\begin{equation}
m_D
=
\mathrm{i}\,U^*_\nu\sqrt{\widehat{m}_\nu}\,O\,\sqrt{\widehat{M}_N},
\label{eq:app_complete_CI}
\end{equation}
one obtains
\begin{equation}
\left(Y_\nu^\dagger Y_\nu\right)_{II}
=
\frac{2M_I}{v_h^2}
\sum_i m_{\nu,i}|O_{iI}|^2.
\label{eq:app_complete_YdagY_CI}
\end{equation}
In the aligned limit $O=\mathbf{1}$,
\begin{equation}
\left(Y_\nu^\dagger Y_\nu\right)_{II}
=
\frac{2M_I m_{\nu,I}}{v_h^2},
\label{eq:app_complete_YdagY_aligned}
\end{equation}
and therefore
\begin{equation}
\kappa_{\rm loop}
=
-\sum_I
\frac{y_{N,I}^2M_I m_{\nu,I}}{2\pi^2v_h^2}.
\label{eq:app_complete_kappa_three_aligned}
\end{equation}
This expression displays explicitly the loop suppression, the dependence on the hidden-sector Yukawa coupling, and the proportionality to the light-neutrino masses in the aligned seesaw limit.

% =====================================================================
% Globally reviewed five-dimensional appendices
% Conservative revision: all equations and calculations from the supplied
% draft are retained.  The sections are reordered and signposted so that
% each has a distinct role: free spectrum, 5D-to-4D matching, propagation,
% and the one-loop amplitude.
% =====================================================================

\section{Kaluza--Klein masses from the five-dimensional Dirac equation}
\label{app:5D_Dirac_KK}

In this appendix, we establish the notation used in Sec.~\ref{sec:5D},
present the five-dimensional model action and the boundary conditions
for the sterile fermion $\mathcal N$, and derive its free
Kaluza--Klein (KK) spectrum. The derivation follows the standard
treatment of fermions in flat extra dimensions; see, e.g.,
Refs.~\cite{Dienes:1998sb,Grossman:1999ra,ArkaniHamed:1999dc}.

For clarity, we suppress the sterile-generation index $I$ throughout the derivation. It can be restored, for example, by replacing $\mathcal{N} \rightarrow \mathcal{N}_I$, $M_5\rightarrow M_{5,I}$ and $m_n\rightarrow m_{I,n}$ when the bulk mass matrix is diagonal in generation space.

%----------------------------------------------------------
\subsection{Five-dimensional Dirac equation, orbifold parity, and boundary conditions}
\label{sec:5Ddd}
%----------------------------------------------------------

We consider a five-dimensional Dirac fermion
$\mathcal N(x,y)$ propagating in flat spacetime. As our default geometrical setup, the fifth coordinate $y$ is compactified on an $S^1/\mathbb Z_2$ orbifold. Starting from a covering circle of circumference $2L$ and quotienting by the reflection $y\rightarrow-y$, the physical fundamental domain is
\begin{equation}
0\leq y\leq L.
\end{equation}
The two orbifold fixed points become the boundaries at $y=0$ and $y=L$, while the $\mathbb Z_2$ transformation properties of the bulk fields determine the corresponding boundary conditions.

We will also comment on the results obtained when the extra dimension is instead treated as a genuine interval. By an interval geometry we mean that the five-dimensional spacetime is defined directly on the line segment $0\leq y\leq L$, without introducing a covering circle or identifying points under a $\mathbb Z_2$ reflection. In this case, the boundary conditions at $y=0$ and $y=L$ are imposed directly as part of the definition of the theory and are not derived from an
orbifold parity assignment. Although the orbifold and interval descriptions share the same coordinate domain, they need not impose the same restrictions on bulk operators, in particular on the form of the fermion bulk mass.

The minimal bulk action on the interval is
\begin{equation}
S_{\rm bulk}
=
\int\dd^4x\int_0^L\dd y\,
\overline{\mathcal N}(x,y)
\left(
\ii\Gamma^A\partial_A-M_5
\right)
\mathcal N(x,y),
\label{eq:app_bulk_action}
\end{equation}
where
\begin{equation}
A=0,1,2,3,5,
\qquad
\partial_5\equiv\partial_y,
\end{equation}
and we adopt the five-dimensional gamma-matrix convention
\begin{equation}
\Gamma^A
=
\left(
\gamma^\mu,\ii\gamma^5
\right).
\label{eq:app_gamma_convention}
\end{equation}
For the metric
\begin{equation}
\eta_{AB}
=
{\rm diag}(+,-,-,-,-),
\end{equation}
the five-dimensional gamma matrices satisfy the Clifford-algebra
relation
\begin{equation}
\left\{
\Gamma^A,\Gamma^B
\right\}
=
2\eta^{AB}.
\end{equation}
The parameter $M_{5}$ is a five-dimensional bulk Dirac mass. It
couples the two four-dimensional chiral components in the bulk and
controls the localization of the chiral zero mode, the KK spectrum,
and propagation along the fifth dimension.

More general five-dimensional effective actions may include gauge-covariant derivatives and additional orbifold-compatible brane operators~\cite{Hebecker2002}, position-dependent or orbifold-odd fermion masses that modify the localization of the chiral zero modes~\cite{GrossmanNeubert2000}, brane-localized kinetic or mass terms~\cite{Hebecker2002,delAguilaEtAl2004}, and higher-derivative or other higher-dimensional operators required by the effective-field-theory expansion and renormalization~\cite{GhilenceaLee2006,delAguilaEtAl2006}. Nevertheless, Eq.~\eqref{eq:app_bulk_action} captures the minimal structure needed for the sequestered construction considered here.

Consistency with the $S^1/\mathbb Z_2$ orbifold geometry requires the bulk fermion to be assigned a definite transformation law under the
reflection $y\to -y$. We choose
\begin{equation}
\mathcal N(x,-y)
=
\gamma^5\mathcal N(x,y),
\label{eq:app_orbifold_parity}
\end{equation}
which makes the right-handed component even and the left-handed component odd under the orbifold reflection.
To show explicitly the consequences of this parity assignment, we decompose the five-dimensional Dirac spinor into four-dimensional components of definite chirality,
\begin{equation}
\mathcal N
=
\mathcal N_L+\mathcal N_R,
\qquad
\mathcal N_{L,R}
=
P_{L,R}\mathcal N,
\label{eq:app_chiral_decomp}
\end{equation}
where
\begin{equation}
P_{L,R}
=
\frac{1\mp\gamma^5}{2},
\,\,\,
\gamma^5\mathcal N_L
=
-\mathcal N_L,
\,\,\,
\gamma^5\mathcal N_R
=
+\mathcal N_R.
\end{equation}
Equation~\eqref{eq:app_orbifold_parity} then implies
\begin{equation}
\mathcal N_R(x,-y) = +\mathcal N_R(x,y),
\,\,\,
\mathcal N_L(x,-y) = -\mathcal N_L(x,y).
\label{eq:app_chiral_parities}
\end{equation}
The right-handed component is therefore even under the orbifold reflection, whereas the left-handed component is odd.

An orbifold-odd field vanishes at both fixed points. At $y=0$, the reflection leaves the point unchanged, so odd parity implies $\mathcal N_L(x,0)=-\mathcal N_L(x,0)$. At the second fixed point, the periodicity of the covering circle identifies $-L$ with $L$; combining this identification with odd parity similarly gives $\mathcal N_L(x,L)=-\mathcal N_L(x,L)$.
Therefore, the odd component vanishes at the two fixed points:
\begin{equation}
\mathcal N_L(x,0) = \mathcal N_L(x,L) = 0.
\label{eq:app_NL_boundary_conditions}
\end{equation}
The right-handed component may instead be nonzero at both boundaries.
This is the parity choice used in the main text: it leaves a
right-handed chiral zero mode and allows the visible- and
hidden-boundary interactions to involve the same brane-accessible
component $\mathcal N_R$.

Within the interior of the fundamental interval, the bulk mass parameter $M_5$ can be written as a constant. Given the parity assignment in Eq.~\eqref{eq:app_orbifold_parity}, the fermion bilinear is odd under the orbifold reflection $y\rightarrow-y$:
\begin{equation}
\overline{\mathcal N}(x,-y)\mathcal N(x,-y)
=
-\overline{\mathcal N}(x,y)\mathcal N(x,y).
\end{equation}
Consequently, the coefficient of a diagonal Dirac mass term must also be odd on the covering circle so that the full mass term is invariant under the orbifold reflection. Therefore, in a strict covering-space
description of the $S^1/\mathbb Z_2$ orbifold, the theory is represented by an orbifold-odd kink mass,
\begin{equation}
M_5(y)
=
M_5\epsilon(y),
\end{equation}
where $\epsilon(y)$ is the sign function. Inside the fundamental
interval, $\epsilon(y)=+1$, while the discontinuities of the covering-space mass at the fixed points are encoded in the orbifold boundary conditions. The constant interval mass and the orbifold-odd covering-space mass are therefore two equivalent descriptions of the same free bulk problem.

Varying Eq.~\eqref{eq:app_bulk_action} with respect to
$\overline{\mathcal N}$ gives the five-dimensional Dirac equation
\begin{equation}
\left(
\ii\Gamma^A\partial_A-M_5
\right)
\mathcal N
=
0.
\label{eq:app_5D_dirac}
\end{equation}
Using the convention in Eq.~\eqref{eq:app_gamma_convention}, this can
be written as
\begin{equation}
\left(
\ii\gamma^\mu\partial_\mu
-
\gamma^5\partial_y
-
M_5
\right)
\mathcal N(x,y)
=
0.
\label{eq:app_5D_dirac_explicit}
\end{equation}

Projecting Eq.~\eqref{eq:app_5D_dirac_explicit} onto the two
four-dimensional chiralities gives
\begin{align}
\ii\slashed{\partial}_4\mathcal N_R
+
\left(
\partial_y-M_5
\right)
\mathcal N_L
&=
0,
\label{eq:app_coupled_R}
\\
\ii\slashed{\partial}_4\mathcal N_L
-
\left(
\partial_y+M_5
\right)
\mathcal N_R
&=
0,
\label{eq:app_coupled_L}
\end{align}
where $\slashed{\partial}_4 \equiv \gamma^\mu\partial_\mu$.
It is useful to introduce the first-order operators
\begin{equation}
Q_y
=
\partial_y+M_5,
\qquad
Q_y^\dagger
=
-\partial_y+M_5.
\label{eq:app_Q_operators}
\end{equation}
Equations~\eqref{eq:app_coupled_R} and
\eqref{eq:app_coupled_L} can then be written in the compact form
\begin{equation}
\ii\slashed{\partial}_4\mathcal N_R
=
Q_y^\dagger\mathcal N_L,
\qquad
\ii\slashed{\partial}_4\mathcal N_L
=
Q_y\mathcal N_R.
\label{eq:app_coupled_Q_form}
\end{equation}
The derivative along the fifth dimension therefore connects the two
four-dimensional chiralities in the same way as an ordinary Dirac
mass term does in four dimensions.

The opposite intrinsic parity choice,
\begin{equation}
\mathcal N(x,-y)
=
-\gamma^5\mathcal N(x,y),
\end{equation}
would instead make $\mathcal N_L$ even and $\mathcal N_R$ odd.  It
would retain a left-handed zero mode, while the right-handed component
would vanish at the fixed points.  The boundary interactions would
then have to be reformulated.  The choice of orbifold parity is
therefore not merely a relabeling once the brane-localized
interactions have been specified.

%----------------------------------------------------------
\subsection{KK decomposition and first-order profile equations}
%----------------------------------------------------------

Before including the hidden-boundary Majorana interaction, we begin
with a general KK decomposition in which both four-dimensional
chiralities are formally allowed to contain a zero mode:
\begin{align}
\mathcal N_R(x,y)
&=
\sum_{n=0}^{\infty}
N_R^{(n)}(x)f_R^{(n)}(y),
\label{eq:app_KK_expansion_R}
\\
\mathcal N_L(x,y)
&=
\sum_{n=0}^{\infty}
N_L^{(n)}(x)f_L^{(n)}(y).
\label{eq:app_KK_expansion_L}
\end{align}
At this stage, the presence or absence of either chiral zero mode has
not yet been determined. It will follow from the profile equations
together with the orbifold boundary conditions.

For each KK level, the four-dimensional fields are taken to satisfy
\begin{equation}
\ii\slashed{\partial}_4N_L^{(n)}
=
m_nN_R^{(n)},
\,\,\,
\ii\slashed{\partial}_4N_R^{(n)}
=
m_nN_L^{(n)},
\,\,\,
n\geq0,
\label{eq:app_4D_dirac_equations}
\end{equation}
where the zero-mode level corresponds to $m_0=0$. Consequently, $N_L^{(0)}$ and $N_R^{(0)}$ would describe independent massless Weyl fermions if both associated profiles were compatible with the boundary conditions.

Substituting
Eqs.~\eqref{eq:app_KK_expansion_R}--\eqref{eq:app_4D_dirac_equations}
into Eq.~\eqref{eq:app_coupled_Q_form} gives the first-order profile
equations
\begin{align}
Q_y f_R^{(n)}(y)
&=
m_n f_L^{(n)}(y),
\label{eq:app_profile_R}
\\
Q_y^\dagger f_L^{(n)}(y)
&=
m_n f_R^{(n)}(y),
\label{eq:app_profile_L}
\end{align}
or, equivalently,
\begin{align}
\left(
\partial_y+M_5
\right)
f_R^{(n)}
&=
m_nf_L^{(n)},
\\
\left(
-\partial_y+M_5
\right)
f_L^{(n)}
&=
m_nf_R^{(n)}.
\label{eq:app_profile_explicit}
\end{align}

The odd orbifold parity of $\mathcal N_L$ imposes Dirichlet boundary
conditions on all its profiles:
\begin{equation}
f_L^{(n)}(0)
=
f_L^{(n)}(L)
=
0.
\label{eq:app_fL_Dirichlet}
\end{equation}
For the zero mode, $m_0=0$, and
Eq.~\eqref{eq:app_profile_L} reduces to
\begin{equation}
\left(
-\partial_y+M_5
\right)
f_L^{(0)}(y)
=
0.
\end{equation}
Its general solution is
\begin{equation}
f_L^{(0)}(y)
=
C_L e^{M_5y}.
\end{equation}
The boundary condition $f_L^{(0)}(0)=0$ requires $C_L=0$, and hence
\begin{equation}
f_L^{(0)}(y)
=
0.
\label{eq:app_fL_zero_vanishes}
\end{equation}
The left-handed zero mode is therefore projected out by the orbifold
boundary conditions.

By contrast, the right-handed zero-mode equation is
\begin{equation}
\left(
\partial_y+M_5
\right)
f_R^{(0)}(y)
=
0,
\end{equation}
which admits the nonvanishing solution
\begin{equation}
f_R^{(0)}(y)
\propto
e^{-M_5y}.
\end{equation}
The orbifold-compatible KK decomposition can therefore be written as
\begin{align}
\mathcal N_R(x,y)
&=
N_R^{(0)}(x)f_R^{(0)}(y)
+
\sum_{n=1}^{\infty}
N_R^{(n)}(x)f_R^{(n)}(y),
\\
\mathcal N_L(x,y)
&=
\sum_{n=1}^{\infty}
N_L^{(n)}(x)f_L^{(n)}(y),
\label{eq:KKdec}
\end{align}
showing explicitly that only the right-handed component contains a physical zero mode.

For each massive level, $n\geq1$, the combination
\begin{equation}
N^{(n)}
=
N_L^{(n)}+N_R^{(n)}
\end{equation}
is a four-dimensional Dirac fermion of mass $m_n$. The quantities
$m_n$ are the free KK Dirac masses. They should be distinguished from
the physical Majorana masses $M_a$, which arise only after the
hidden-boundary interaction is included and $R$ acquires a vacuum
expectation value. The relation between the two spectra is discussed
at the end of this appendix.

Finally, using Eq.~\eqref{eq:app_profile_R} together with the
Dirichlet conditions in Eq.~\eqref{eq:app_fL_Dirichlet}, the
right-handed profiles satisfy the corresponding Robin boundary
conditions
\begin{equation}
\left.
\left(
\partial_y+M_5
\right)
f_R^{(n)}(y)
\right|_{y=0,L}
=
0.
\label{eq:app_fR_Robin}
\end{equation}
For $M_5\neq0$, the even component therefore does not obey an ordinary
Neumann condition. The Robin condition contains the information about
the bulk mass and is the interval counterpart of the kink-mass
boundary behavior on the covering orbifold.

%----------------------------------------------------------
\subsection{The chiral zero mode}
%----------------------------------------------------------

In this subsection, we focus on the $n=0$ KK level. We solve the
corresponding first-order profile equations, verify that the surviving
four-dimensional state is massless before hidden-boundary symmetry
breaking, and show explicitly how the orbifold boundary conditions
remove the left-handed zero mode. We then normalize the remaining
right-handed profile and discuss how its localization depends on the
bulk mass parameter $M_5$.

We first determine whether the boundary-value problem admits a
four-dimensional massless state.  A KK zero mode is defined by
\begin{equation}
m_0=0.
\end{equation}
The term ``zero mode'' refers to the vanishing four-dimensional mass
eigenvalue and does not imply that the corresponding profile is
constant along the extra dimension.

For $m_0=0$, the first-order equations decouple:
\begin{equation}
Q_yf_R^{(0)}=0,
\qquad
Q_y^\dagger f_L^{(0)}=0.
\label{eq:app_zero_mode_equations}
\end{equation}
Their local solutions are
\begin{equation}
f_R^{(0)}(y)
\propto
e^{-M_5y},
\qquad
f_L^{(0)}(y)
\propto
e^{+M_5y}.
\end{equation}
The left-handed solution is excluded by the odd orbifold parity and the
Dirichlet conditions at the two fixed points.  The right-handed
solution is even and normalizable, and therefore survives:
\begin{equation}
f_R^{(0)}(y)
=
\left[\frac{2M_5}{1-e^{-2M_5L}}\right]^{1/2} e^{-M_5y},
\label{eq:app_zero_profile_unormalized}
\end{equation}
In the case of $M_5=0$, $f_R^{(0)}(y)$ becomes
\begin{equation}
\lim_{M_5\rightarrow0}
f_R^{(0)}(y)
=
\frac{1}{\sqrt L}.
\end{equation}

Therefore, the zero mode is a four-dimensional right-handed chiral fermion with
\begin{equation}
f_R^{(0)}(y)\neq0,
\qquad
f_L^{(0)}(y)=0.
\end{equation}
It is massless at this stage because a four-dimensional Dirac mass
term necessarily couples left- and right-handed components:
\begin{equation}
\mathcal L_{\rm mass}^{(0)}
=
-m_0\left(
\overline{N_L^{(0)}}N_R^{(0)}
+
\overline{N_R^{(0)}}N_L^{(0)}
\right).
\label{eq:zero_mode_Dirac_mass}
\end{equation}
Since $f_L^{(0)}(y)=0$, the corresponding left-handed four-dimensional zero mode is absent, and the mass term cannot be formed.
The zero mode remains massless before hidden-boundary symmetry breaking even though the five-dimensional action contains the bulk parameter $M_5$. 

The parameter $M_5$ controls the localization of the
massless chiral state rather than its four-dimensional mass. From
Eq.~\eqref{eq:app_zero_profile_unormalized}, the profile decreases along
the extra dimension for $M_5>0$, and the zero mode is localized toward
the visible boundary at $y=0$. Each occurrence of a right-handed
zero-mode field in a brane-localized interaction is weighted by the
value of its profile at the position of that boundary. Consequently,
the coupling to operators at $y=0$ is comparatively unsuppressed,
whereas the coupling to operators at $y=L$ is exponentially
suppressed.

Conversely, for $M_5<0$, the profile increases toward $y=L$, and the
zero mode is localized near the hidden boundary. Interactions on the
hidden boundary are then unsuppressed, while the overlap with the
visible boundary is exponentially suppressed. For $M_5=0$, the
zero-mode profile is flat and no geometrical localization occurs.

For the sequestered construction considered here, the most natural
benchmark is $M_5<0$. The sterile zero mode is then localized toward
the hidden boundary, where its Majorana interaction with $R$ is
generated, while its coupling to the visible-sector fields at $y=0$
is naturally suppressed by the small wave-function overlap. This
provides a geometrical origin for a small visible-sector neutrino
Yukawa coupling without requiring an equally small fundamental
five-dimensional coefficient.

%----------------------------------------------------------
\subsection{Massive KK spectrum}
%----------------------------------------------------------

For $n\geq1$, the two first-order equations can be combined into
second-order equations.  Acting with $Q_y^\dagger$ on
Eq.~\eqref{eq:app_profile_R} and with $Q_y$ on
Eq.~\eqref{eq:app_profile_L} gives
\begin{align}
Q_y^\dagger Q_y f_R^{(n)}
&=
m_n^2 f_R^{(n)},
\label{eq:app_second_order_R_Q}
\\
Q_y Q_y^\dagger f_L^{(n)}
&=
m_n^2 f_L^{(n)}.
\label{eq:app_second_order_L_Q}
\end{align}
For a constant interval mass,
\begin{equation}
Q_y^\dagger Q_y
=
Q_yQ_y^\dagger
=
-\partial_y^2+M_5^2
\end{equation}
in the interior of the interval ($y\in[0,L]$). Therefore,
\begin{align}
\left(
-\partial_y^2+M_5^2
\right)
f_R^{(n)}
&=
m_n^2f_R^{(n)},
\label{eq:app_second_order_R}
\\
\left(
-\partial_y^2+M_5^2
\right)
f_L^{(n)}
&=
m_n^2f_L^{(n)}.
\label{eq:app_second_order_L}
\end{align}
In the covering-space description, the bulk mass is instead written as $M_5\epsilon(y)$. When the first-order operators are squared, derivatives of $\epsilon(y)$ generate delta-function contributions localized at the orbifold fixed points. In the interval formulation used here, these localized contributions are not written explicitly: their effect is equivalently implemented through the Dirichlet condition in Eq.~\eqref{eq:app_fL_Dirichlet} and the Robin condition in Eq.~\eqref{eq:app_fR_Robin}.

Defining
\begin{equation}
k_n^2
\equiv
m_n^2-M_5^2,
\end{equation}
the left-handed Dirichlet conditions in Eq.~\eqref{eq:app_fL_Dirichlet} require
\begin{equation}
f_L^{(n)}(y)
=
\sqrt{\frac{2}{L}}
\sin(k_ny),
\label{eq:moden}
\end{equation}
up to an unphysical overall phase.  The condition at $y=L$ quantizes
the momentum along the compact direction:
\begin{equation}
k_n
=
\frac{n\pi}{L},
\qquad
n=1,2,\ldots.
\label{eq:app_kn_quantization}
\end{equation}
The corresponding right-handed profile follows from
Eq.~\eqref{eq:app_profile_L}:
\begin{equation}
f_R^{(n)}(y)
=
\sqrt{\frac{2}{L}}\,
\frac{
M_5\sin(k_ny)-k_n\cos(k_ny)
}{
m_n
}.
\label{eq:app_massive_fR_profile}
\end{equation}
The profiles in
Eqs.~\eqref{eq:moden} and
\eqref{eq:app_massive_fR_profile} satisfy the first-order equations,
the Dirichlet condition for $f_L^{(n)}$, and the Robin condition for
$f_R^{(n)}$.  Their overall signs may be changed by rephasing the
four-dimensional KK fields.

The free massive spectrum is therefore
\begin{equation}
m_n^2
=
M_5^2
+
\frac{n^2\pi^2}{L^2},
\qquad
n\geq1.
\label{eq:app_mn_spectrum}
\end{equation}
Introducing
\begin{equation}
M_{\rm KK}
\equiv
\frac{\pi}{L},
\label{eq:app_MKK_definition}
\end{equation}
one may write
\begin{equation}
m_n^2
=
M_5^2+n^2M_{\rm KK}^2,
\qquad
n\geq1.
\label{eq:app_KK_spectrum}
\end{equation}

Equation~\eqref{eq:app_KK_spectrum} is exact for the minimal free
interval problem considered here.  It applies only to the massive
trigonometric tower with $n\geq1$.  The chiral zero mode is a separate
solution of the first-order equations and must not be obtained by
setting $n=0$ in Eq.~\eqref{eq:app_KK_spectrum}; doing so would
incorrectly give $m_0=|M_5|$.  The correct zero-mode mass is
$m_0=0$, with the nontrivial exponential profile given in
Eq.~\eqref{eq:app_zero_profile_unormalized}.

The physical interpretation is straightforward.  The quantity
$n\pi/L$ is the quantized momentum along the compact dimension.  From
the four-dimensional point of view, this momentum contributes to the
invariant mass of the KK excitation.  Each nonzero level contains both
four-dimensional chiralities and forms a Dirac fermion with mass
$m_n$.  Inserting the normalized profiles into the free action gives
schematically
\begin{equation}
S_{\rm bulk}
=
\int\dd^4x
\left[
\overline{N_R^{(0)}}\,
\ii\slashed{\partial}_4N_R^{(0)}
+
\sum_{n=1}^{\infty}
\overline{N^{(n)}}
\left(
\ii\slashed{\partial}_4-m_n
\right)
N^{(n)}
\right].
\label{eq:app_reduced_free_action}
\end{equation}
Thus the orbifold theory contains one massless right-handed Weyl mode
and a tower of massive four-dimensional Dirac fermions.

%----------------------------------------------------------
\subsection{Relation to the boundary Majorana mass and physical spectrum}
%----------------------------------------------------------

The spectrum derived above is the free spectrum obtained before the
hidden scalar acquires a vacuum expectation value. After hidden
symmetry breaking,
\begin{equation}
R(x)
=
\frac{1}{\sqrt{2}}
\left[
v_r+\rho(x)
\right],
\end{equation}
the hidden-sector action then contains the boundary-localized mass term
\begin{equation}
S_{\rm hid}
\supset
-
\frac{v_r}{2\sqrt{2}}
\int\dd^{4}x\,
\left(
\widehat{Y}_{5,N}
\right)_{IJ}
\overline{\mathcal N_{IR}^{c}(x,L)}
\mathcal N_{JR}(x,L)
+
{\rm h.c.}
\label{eq:hidden_boundary_mass_after_vev}
\end{equation}

Substituting the KK decomposition in Eq.~\eqref{eq:KKdec} into
Eq.~\eqref{eq:hidden_boundary_mass_after_vev} gives
\begin{equation}
\mathcal{L}_{\rm KK}^{\rm Majorana}
=
-
\frac{1}{2}
\sum_{I,J}
\sum_{m,n=0}^{\infty}
\left(
\mathcal{M}_{N}^{mn}
\right)_{IJ}
\overline{
\left(
N_{IR}^{(m)}
\right)^{c}
}
N_{JR}^{(n)}
+
{\rm h.c.},
\label{eq:KK_Majorana_Lagrangian}
\end{equation}
where
\begin{equation}
\left(
\mathcal M_N^{mn}
\right)_{IJ}
=
\frac{v_r}{\sqrt2}
\left(
\widehat Y_{5,N}
\right)_{IJ}
f_{R,I}^{(m)}(L)
f_{R,J}^{(n)}(L).
\label{eq:app_boundary_Majorana_matrix}
\end{equation}
This is the Majorana mass matrix written in the basis of the free KK
eigenstates. The indices $I,J$ label the sterile generations, whereas
$m,n$ label the free KK levels. Since the matrix elements depend on
the boundary values of two KK profiles, the interaction is generally
non-diagonal in both generation space and KK space.

The complete four-dimensional mass Lagrangian contains both the free
KK Dirac masses and the boundary-induced Majorana entries:
\begin{align}
\mathcal L_{\rm mass}^{\rm 4D}
={}&
-
\sum_{I}
\sum_{n=1}^{\infty}
m_{I,n}
\left[
\overline{N_{IL}^{(n)}}N_{IR}^{(n)}
+
{\rm h.c.}
\right]
\nonumber\\
&-
\frac12
\sum_{I,J}
\sum_{m,n=0}^{\infty}
\left(
\mathcal M_N^{mn}
\right)_{IJ}
\overline{
\left(
N_{IR}^{(m)}
\right)^c
}
N_{JR}^{(n)}
+
{\rm h.c.}
\label{eq:app_full_mass_lagrangian}
\end{align}
The boundary term gives a Majorana mass to the chiral zero mode,
splits the massive Dirac KK levels into Majorana states, and generally
mixes different KK levels and sterile generations.

The physical masses therefore cannot be obtained by diagonalizing
only the zero-mode entry $\mathcal M_N^{00}$, or the Majorana block
$\mathcal M_N^{mn}$ in isolation. The complete mass matrix must also
include the free Dirac KK masses. In a left-handed Weyl basis of the
schematic form
\begin{equation}
\Xi^T
=
\left(
N_{R}^{(0)c},
N_{R}^{(1)c},
N_{R}^{(2)c},
\ldots,
N_L^{(1)},
N_L^{(2)},
\ldots
\right),
\end{equation}
the sterile-sector mass Lagrangian can be written as
\begin{equation}
\mathcal L_{\rm mass}^{\rm sterile}
=
-\frac{1}{2}
\Xi^T C\,
\mathcal M_{\rm full}
\Xi
+
{\rm h.c.},
\end{equation}
where the complete matrix has the schematic block structure
\begin{equation}
\mathcal M_{\rm full}
=
\begin{pmatrix}
\mathcal M_N
&
\mathcal D_{\rm KK}
\\[2mm]
\mathcal D_{\rm KK}^{T}
&
0
\end{pmatrix}.
\label{eq:5D4D_full_mass_matrix}
\end{equation}
Here $\mathcal M_N$ is the boundary-induced Majorana matrix defined in
Eq.~\eqref{eq:app_boundary_Majorana_matrix}. Its entries connect pairs
of right-handed free KK modes and are generally non-diagonal in both
generation and KK space. By contrast, $\mathcal D_{\rm KK}$ contains
the free Dirac masses that pair $N_{IR}^{(n)}$ with
$N_{IL}^{(n)}$ for each massive level $n\geq1$. In the free KK basis,
and assuming a generation-diagonal bulk problem, its entries are
\begin{equation}
\left(\mathcal D_{\rm KK}\right)_{(I,m)(J,n)}
=
\delta_{IJ}\delta_{mn}\,m_{I,n},
\qquad
m,n\geq1.
\end{equation}
The row associated with $N_{IR}^{(0)c}$ vanishes because the orbifold
projection removes the left-handed zero mode, so there is no
$N_{IL}^{(0)}$ with which the right-handed zero mode could form a
Dirac mass.

The matrix $\mathcal M_{\rm full}$ is complex and symmetric and is
therefore diagonalized by a Takagi transformation:
\begin{equation}
U_N^T
\mathcal M_{\rm full}
U_N
=
\widehat M_N
=
{\rm diag}
\left(
M_0,M_1,M_2,\ldots
\right),
\,\,\,
M_a\geq0.
\label{eq:app_full_Takagi}
\end{equation}
The quantities $M_a$ are the physical nonnegative Majorana masses.
They should not be identified with either the bulk parameters
$M_{5,I}$ or the free Dirac KK masses $m_{I,n}$ derived in
Eq.~\eqref{eq:app_mn_spectrum}.

The same transformation determines the wave functions of the
physical Majorana eigenstates. Their right-handed profiles are
\begin{equation}
F_{R,Ia}(y)
=
\sum_n
f_{R,I}^{(n)}(y)
\left(
U_N^R
\right)_{(I,n)a},
\label{eq:5D4D_physical_profiles}
\end{equation}
where $U_N^R$ denotes the part of the Takagi transformation acting on
the free right-handed KK fields. The original bulk field may then be
written as
\begin{equation}
\mathcal N_{IR}(x,y)
=
\sum_a
F_{R,Ia}(y)
P_RN_a(x).
\label{eq:physicalNR}
\end{equation}
Thus, the lower-case profiles $f_{R,I}^{(n)}$ describe the free KK
basis, whereas the capital profiles $F_{R,Ia}$ describe the physical
Majorana eigenstates obtained after diagonalizing the complete
Dirac-plus-Majorana mass matrix.

In the weak-mixing regime, the lightest physical state is dominated
by the free chiral zero mode. More precisely, this regime is realized
when the boundary-induced Majorana entries connecting the zero mode
to the excited KK states are small compared with the corresponding
mass separations. For one sterile generation, this requires
\begin{equation}
M_0^{(0)}
=
\mathcal M_N^{00}
\ll
m_1,
\,\,
\epsilon_n
\equiv
\frac{\left|\mathcal M_N^{0n}\right|}
{\left|m_n-M_0^{(0)}\right|}
\ll1,
\,\, n\geq1.
\label{eq:weak_KK_mixing_conditions}
\end{equation}
In the phenomenologically relevant limit
$M_0^{(0)}\ll m_n$, one has
$\epsilon_n\simeq|\mathcal M_N^{0n}|/m_n$. The admixture of the
$n$th excited KK state in the lightest physical eigenstate is then of
order $\epsilon_n$, while corrections to its mass and normalization
start at higher order in these ratios. Consequently, the lightest
state has a physical profile satisfying
\begin{equation}
F_{R,0}(y)
=
f_R^{(0)}(y)
+
\mathcal O(\epsilon_n),
\end{equation}
and its mass is approximately
\begin{equation}
M_0
\simeq
\frac{v_r}{\sqrt2}
\widehat y_{5,N}
\left[
f_R^{(0)}(L)
\right]^2.
\end{equation}
Outside this regime, the lightest eigenstate can contain a substantial
admixture of excited KK modes, and neither its mass nor its boundary
couplings can be inferred from the free zero-mode profile alone.
The complete Dirac-plus-Majorana mass matrix must then be
diagonalized.

The physical interpretation can therefore be summarized as follows.
The bulk parameter $M_5$ controls the localization of the chiral zero
mode and contributes to the free masses of the massive Dirac KK
tower, but it does not by itself generate a four-dimensional mass for
the chiral zero mode. The zero-mode Majorana mass, the splitting of
the massive Dirac levels, and the mixing among KK levels and sterile
generations instead arise from the hidden-boundary interaction after
$R$ acquires its vacuum expectation value.

%----------------------------------------------------------
\subsection{Brane interactions and dimensions of the couplings}
%----------------------------------------------------------

It is useful to distinguish the five-dimensional brane coefficients
$\widehat Y_{5,\nu}$ and $\widehat Y_{5,N}$ from the dimensionless
four-dimensional Yukawa couplings obtained after compactification.
With the orbifold parity assignment used in this work, the relevant
brane interactions involving the fields $\mathcal N_I$ are
\begin{widetext}
\begin{equation}
\begin{aligned}
S_{\rm vis}
&\supset
-\int\dd^4x\int_0^L\dd y\,
\delta(y)
\left[
\left(
\widehat Y_{5,\nu}
\right)_{\alpha I}
\overline{L_\alpha}(x)
\widetilde\Phi(x)
\mathcal N_{IR}(x,y)
+
{\rm h.c.}
\right],
\\
S_{\rm hid}
&\supset
-\int\dd^4x\int_0^L\dd y\,
\delta(y-L)
\left[
\frac{1}{2}
\left(
\widehat Y_{5,N}
\right)_{IJ}
R(x)
\overline{
\mathcal N_{IR}^{\,c}
}(x,y)
\mathcal N_{JR}(x,y)
+
{\rm h.c.}
\right].
\end{aligned}
\label{eq:5D4D_brane_interactions}
\end{equation}
\end{widetext}
We now insert the KK expansion in Eq.~\eqref{eq:KKdec} into the
visible-boundary interaction.
The delta function evaluates the bulk field at $y=0$:
\begin{widetext}
\begin{align}
S_{\rm vis}
\supset
-\int\dd^4x
\left[
\left(
\widehat Y_{5,\nu}
\right)_{\alpha I}
\overline{L_\alpha}
\widetilde\Phi
\left(
N_{IR}^{(0)}
f_{R,I}^{(0)}(0)
+
\sum_{n=1}^{\infty}
N_{IR}^{(n)}
f_{R,I}^{(n)}(0)
\right)
+
{\rm h.c.}
\right].
\label{eq:5D4D_visible_KK_inserted}
\end{align}
\end{widetext}
The effective visible coupling of the $n$th free KK mode is therefore
\begin{equation}
\left(
Y_{4,\nu}^{(n)}
\right)_{\alpha I}
=
\left(
\widehat Y_{5,\nu}
\right)_{\alpha I}
f_{R,I}^{(n)}(0).
\label{eq:5D4D_Ynu_KK}
\end{equation}
For the free zero mode,
\begin{equation}
\left(
Y_{4,\nu}^{(0)}
\right)_{\alpha I}
=
\left(
\widehat Y_{5,\nu}
\right)_{\alpha I}
f_{R,I}^{(0)}(0).
\label{eq:5D4D_Ynu_zero}
\end{equation}

The hidden-boundary interaction evaluates the profiles at $y=L$:
\begin{widetext}
\begin{align}
S_{\rm hid}
\supset
-\frac{1}{2}
\int\dd^4x\,
R
\sum_{I,J}
\sum_{m,n=0}^{\infty}
\left[
\left(
\widehat Y_{5,N}
\right)_{IJ}
f_{R,I}^{(m)}(L)
f_{R,J}^{(n)}(L)
\overline{
\left(
N_{IR}^{(m)}
\right)^c
}
N_{JR}^{(n)}
+
{\rm h.c.}
\right].
\label{eq:5D4D_hidden_KK_inserted}
\end{align}
\end{widetext}
The hidden coupling matrix in the free KK basis is
\begin{equation}
\left(
Y_{4,N}^{mn}
\right)_{IJ}
=
\left(
\widehat Y_{5,N}
\right)_{IJ}
f_{R,I}^{(m)}(L)
f_{R,J}^{(n)}(L).
\label{eq:5D4D_YN_KK}
\end{equation}
For the zero-mode block,
\begin{equation}
\left(
Y_{4,N}^{00}
\right)_{IJ}
=
\left(
\widehat Y_{5,N}
\right)_{IJ}
f_{R,I}^{(0)}(L)
f_{R,J}^{(0)}(L).
\label{eq:5D4D_YN_zero}
\end{equation}
The matrix in Eq.~\eqref{eq:5D4D_YN_KK} is generally non-diagonal in
the free KK labels.  The hidden boundary therefore mixes the chiral
zero modes with the massive tower and also mixes different massive KK
levels.

In the approximately flat-profile limit, $|M_5|L\ll1$,
\begin{equation}
f_{\rm flat}(y)
=
\frac{1}{\sqrt L},
\end{equation}
the expressions above reduce to
\begin{equation}
Y_{4,\nu}
=
\frac{
\widehat Y_{5,\nu}
}{
\sqrt L
},
\qquad
Y_{4,N}
=
\frac{
\widehat Y_{5,N}
}{
L
}.
\end{equation}
These are the dimensionless reference couplings used when comparing
the localized theory with a flat-profile four-dimensional limit.

\section{Brane-to-brane propagation and exponential sequestering}
\label{sec:brane_to_brane_propagation}

In this section we analyze the propagation of the bulk sterile fermion $\mathcal{N}_I$ between the
visible boundary at $y=0$ and the hidden boundary at $y=L$, using the
operators, boundary conditions, and free profiles derived in
Appendix~\ref{app:5D_Dirac_KK}.  Five-dimensional locality forbids a
fundamental contact operator containing both $\Phi(x)$ and $R(x)$ because
these fields are localized at different points of the compact dimension.
The two sectors can nevertheless communicate radiatively through the bulk
fermions $\mathcal N_I$, and the nonlocal part of the loop is controlled by
propagation from one boundary to the other and back.
For simplicity, we suppress the sterile-generation index in the
derivation of the free kernel. 

Two related but distinct effects must be separated. First, the exact
five-dimensional propagator is exponentially damped at Euclidean
momenta for which the propagation length is shorter than the boundary
separation. This damping makes the genuinely nonlocal part of the
five-dimensional amplitude ultraviolet soft. Second, below the first
KK threshold, the portal can be described in terms of the light chiral
mode. Its effective four-dimensional couplings are controlled by the
values of its wave function at the two boundaries. The
momentum-dependent propagator suppression and the low-energy
zero-mode overlap are not identical quantities, although they have
the same geometrical origin.

%It can be restored by replacing $M_5\rightarrow M_{5,I}$ and by promoting the boundary couplings to matrices in sterile-flavor space.

%----------------------------------------------------------
\subsection{Right-handed second-order Green kernel}
%----------------------------------------------------------

To compute amplitudes involving interactions localized on different
boundaries, we need the propagator of the bulk fermion between two
points in the fifth dimension $y$. In particular, a visible-boundary
vertex probes the field at $y=0$, whereas a hidden-boundary vertex
probes it at $y=L$. The quantity relevant for communication between
the two sectors is therefore the bulk propagator evaluated between
these two positions.

Four-dimensional translational invariance allows us to Fourier
transform the ordinary spacetime coordinates $x^\mu$, while retaining
the explicit dependence on the fifth-dimensional coordinates.
\begin{equation}
\mathcal{N}(p_E,y)
=
\int\dd^4x_E\,
e^{-\ii p_E\cdot x_E}
\mathcal{N}(x_E,y),
\end{equation}
We therefore work in mixed momentum--position space,
\begin{equation}
\mathcal S(p_E;y,y'),
\end{equation}
where $p_E$ is the Euclidean four-momentum and $y,y'$ denote the
positions of the two fermionic insertions along the extra dimension.
For example, $\mathcal S(p_E;0,L)$ describes propagation from one
boundary to the other.

Directly inverting the first-order five-dimensional Dirac operator is
possible, but it is more convenient to express its chiral components
in terms of Green kernels for the corresponding second-order
operators. The resulting scalar kernels contain the full dependence
on the KK spectrum and on the positions $y$ and $y'$, while the
spinorial structure is restored by acting with the appropriate
four-dimensional momentum and fifth-dimensional differential
operators.

The orbifold parity used in the main text makes $\mathcal N_R$ even
and $\mathcal N_L$ odd. The free profiles obey
Eqs.~\eqref{eq:app_profile_R} and \eqref{eq:app_profile_L}. The scalar
kernel associated with the right-handed second-order operator is
defined in mixed Euclidean momentum--position space by
\begin{equation}
\left[
p_E^2+Q_y^\dagger Q_y
\right]
G_R(p_E;y,y')
=
\delta(y-y').
\label{eq:prop_app_GR_definition}
\end{equation}
Thus, $G_R$ is the inverse of the quadratic operator governing the
right-handed profiles.
For a constant interval mass,
\begin{equation}
Q_y^\dagger Q_y
=
-\partial_y^2+M_5^2
\end{equation}
in the interior of the interval. It is useful to define
\begin{equation}
\omega
\equiv
\sqrt{p_E^2+M_5^2}.
\label{eq:prop_app_omega}
\end{equation}
Equation~\eqref{eq:prop_app_GR_definition} then becomes
\begin{equation}
\left[
-\partial_y^2+\omega^2
\right]
G_R(p_E;y,y')
=
\delta(y-y').
\label{eq:prop_app_green_equation}
\end{equation}

The differential equation alone does not uniquely determine the
kernel: the boundary conditions must also be specified. These follow
from the orbifold parity assignment. Since the odd left-handed
profiles vanish at both fixed points,
\begin{equation}
f_L^{(n)}(0)
=
f_L^{(n)}(L)
=
0,
\end{equation}
the first-order relation between the chiral profiles implies the
Robin conditions
\begin{equation}
\left.
Q_yG_R(p_E;y,y')
\right|_{y=0,L}
=
0
\label{eq:prop_app_GR_Robin}
\end{equation}
with respect to the first coordinate. The corresponding condition
holds with respect to the second coordinate. For $M_5\neq0$, these
are not ordinary Neumann conditions because the operator $Q_y$
contains the bulk mass.

The kernel $G_R$ is scalar in four-dimensional spinor space and is
therefore not, by itself, the complete fermion propagator. It contains
the KK poles, the boundary wave-function factors, and the dependence
on the separation in the fifth dimension. The spinorial numerator is
restored when the appropriate chiral block of the fermion propagator
is constructed. For the free lepton-number-conserving right-handed
contraction,
\begin{equation}
\mathcal S_{RR}(p_E;y,y')
=
-\ii\slashed p_EP_L\,
G_R(p_E;y,y').
\label{eq:prop_app_SRR_from_GR}
\end{equation}
Other chiral blocks contain $Q_y G_R$ or the corresponding operator
acting on $y'$. Therefore, solving for $G_R$ determines the essential
fifth-dimensional part of the free bulk propagator, while the full
fermionic structure follows from the first-order Dirac equations.

After the hidden-boundary Majorana interaction is included,
lepton-number-conserving and lepton-number-violating contractions must
instead be collected into the full Nambu--Gorkov propagator. The free
kernel derived here nevertheless remains the basic building block:
the effects of the boundary Majorana mass can be incorporated by
resumming insertions localized at $y=L$.

%----------------------------------------------------------
\subsection{Spectral representation and exact orbifold kernel}
%----------------------------------------------------------

The right-handed Green kernel can be represented either as a sum over
the free KK eigenstates or, equivalently, in a closed form obtained by
solving the differential equation directly. The spectral
representation derived from the KK decomposition in
Appendix~\ref{app:5D_Dirac_KK} is
\begin{equation}
G_R(p_E;y,y')
=
\frac{
f_R^{(0)}(y)f_R^{(0)}(y')
}{
p_E^2
}
+
\sum_{n=1}^{\infty}
\frac{
f_R^{(n)}(y)f_R^{(n)}(y')
}{
p_E^2+m_n^2
}.
\label{eq:prop_app_GR_spectral}
\end{equation}
The first term is the contribution of the massless chiral zero mode,
whereas the remaining terms describe the massive KK tower. In
particular, the zero-mode denominator is $p_E^2$, rather than
$p_E^2+M_5^2$, because the bulk parameter $M_5$ determines the
localization of the zero mode but does not generate its
four-dimensional mass.

Using the normalized zero-mode profile in
Eq.~\eqref{eq:app_zero_profile_unormalized} and the massive profiles in
Eqs.~\eqref{eq:app_massive_fR_profile} and \eqref{eq:moden}, the
opposite-boundary kernel becomes
\begin{equation}
G_R(p_E;0,L)
=
\frac{
M_5
}{
p_E^2\sinh(M_5L)
}
+
\frac{2}{L}
\sum_{n=1}^{\infty}
\frac{
(-1)^n k_n^2
}{
m_n^2
\left(
p_E^2+m_n^2
\right)
}.
\label{eq:prop_app_GR_spectral_explicit}
\end{equation}
This expression displays separately the massless chiral pole and the
contribution of the massive KK states.

The same spectral sum can be resummed by solving
Eq.~\eqref{eq:prop_app_green_equation} directly. Away from the source
point $y=y'$, the kernel satisfies the homogeneous equation
\begin{equation}
\left(
-\partial_y^2+\omega^2
\right)G_R=0.
\end{equation}
To impose the two Robin boundary conditions, we introduce a solution
$u(y)$ satisfying the condition at $y=0$ and a solution $v(y)$
satisfying the condition at $y=L$. A convenient normalization is
\begin{align}
u(y)
&=
\cosh(\omega y)
-
\frac{M_5}{\omega}
\sinh(\omega y),
\\
v(y)
&=
\cosh\!\left[\omega(L-y)\right]
+
\frac{M_5}{\omega}
\sinh\!\left[\omega(L-y)\right],
\end{align}
for which
\begin{equation}
u(0)=v(L)=1.
\end{equation}
Their Wronskian,
\begin{equation}
W[u,v]
\equiv
u\,\partial_yv
-
v\,\partial_yu,
\end{equation}
is independent of $y$ and evaluates to
\begin{equation}
W[u,v]
=
-\frac{p_E^2}{\omega}
\sinh(\omega L).
\end{equation}

The Green kernel is constructed by using $u$ on the side of the
source closer to the left boundary and $v$ on the side closer to the
right boundary. Continuity at $y=y'$ and the derivative jump implied
by Eq.~\eqref{eq:prop_app_green_equation} then give
\begin{equation}
G_R(p_E;y,y')
=
\frac{
\omega
}{
p_E^2\sinh(\omega L)
}
u(y_<)\,v(y_>),
\label{eq:prop_app_GR_general}
\end{equation}
where
\begin{equation}
y_<
=
\min(y,y'),
\qquad
y_>
=
\max(y,y').
\end{equation}
At opposite boundaries, $u(0)=v(L)=1$, and the exact kernel reduces to
\begin{equation}
G_R(p_E;0,L)
=
\frac{
\omega
}{
p_E^2\sinh(\omega L)
}.
\label{eq:prop_app_GR_exact}
\end{equation}
Equation~\eqref{eq:prop_app_GR_exact} is therefore the closed-form
resummation of the zero-mode and KK contributions in
Eq.~\eqref{eq:prop_app_GR_spectral_explicit}. It is the exact
right-handed second-order kernel for the strict orbifold, or
equivalently for an interval endowed with the orbifold Robin boundary
conditions.

The explicit factor $1/p_E^2$ represents the pole of the massless
chiral zero mode. Indeed, the exact kernel can be separated as
\begin{equation}
G_R(p_E;0,L)
=
\frac{
f_R^{(0)}(0)f_R^{(0)}(L)
}{
p_E^2
}
+
G_R^{\rm KK}(p_E;0,L),
\label{eq:prop_app_zero_KK_split}
\end{equation}
where $G_R^{\rm KK}$ is regular at $p_E^2=0$. The residue of the pole
obtained from Eq.~\eqref{eq:prop_app_GR_exact} is
\begin{equation}
\lim_{p_E^2\rightarrow0}
p_E^2G_R(p_E;0,L)
=
\frac{
|M_5|
}{
\sinh(|M_5|L)
}
=
f_R^{(0)}(0)f_R^{(0)}(L),
\end{equation}
in agreement with the spectral representation.

%----------------------------------------------------------
\subsection{Momentum regimes and brane-to-brane suppression}
%----------------------------------------------------------

The behavior of the exact cross-boundary kernel depends on the
dimensionless quantity $\omega L=\sqrt{p_E^2+M_5^2} L$.
This quantity compares the separation between the two boundaries with
the characteristic Euclidean propagation length $1/\omega$. Two
distinct regimes can be identified.

For
\begin{equation}
\omega L\ll1,
\end{equation}
the propagation length is much larger than the size of the extra
dimension. Expanding the exact kernel gives
\begin{equation}
G_R(p_E;0,L)
=
\frac{1}{p_E^2L}
\left[
1
-
\frac{
(\omega L)^2
}{
6
}
+
\mathcal O\!\left(
(\omega L)^4
\right)
\right].
\label{eq:prop_app_GR_small}
\end{equation}
The leading contribution is
\begin{equation}
G_R^{(0)}(p_E;0,L)
\simeq
\frac{1}{p_E^2L},
\end{equation}
which corresponds to the propagation of an approximately flat
massless chiral zero mode. In this regime, the bulk field probes the
entire compact dimension coherently, and the finite separation between
the two boundaries does not produce an exponential suppression.
Radiative communication between the visible and hidden sectors is
therefore controlled mainly by the ordinary four-dimensional
zero-mode pole.

The opposite behavior occurs when
\begin{equation}
\omega L\gg1.
\end{equation}
In this regime, the Euclidean propagation length $1/\omega$ is much
shorter than the distance between the two boundaries. The exact
cross-boundary kernel then behaves as
\begin{equation}
G_R(p_E;0,L)
\simeq
\frac{
2\omega
}{
p_E^2
}
e^{-\omega L}.
\label{eq:prop_app_GR_large}
\end{equation}
A single propagation from one boundary to the other is therefore
exponentially damped. Physically, modes with characteristic inverse
propagation length $\omega$ cannot efficiently communicate across an
interval of length $L$ when $\omega L$ is large.

The portal loop contains one sterile propagation from the visible
boundary to the hidden boundary and a second propagation in the
opposite direction. Consequently, the product of the two
cross-boundary kernels behaves as
\begin{equation}
G_R(p_E;0,L)G_R(p_E;L,0)
\simeq
\frac{
4\omega^2
}{
p_E^4
}
e^{-2\omega L}.
\label{eq:prop_app_cross_product_large}
\end{equation}
At large Euclidean momentum,
\begin{equation}
p_EL\gg1,
\end{equation}
one has $\omega\simeq p_E$, and hence
\begin{equation}
G_R(p_E;0,L)G_R(p_E;L,0)
\propto
e^{-2p_EL}.
\end{equation}
The contribution of large loop momenta to a genuinely
separated-boundary amplitude is therefore exponentially suppressed.
After the ultraviolet divergences associated with local
brane-supported subdiagrams have been renormalized, the remaining
nonlocal contribution connecting the two different boundaries is
ultraviolet finite.

The two regimes thus have different physical consequences. When
$\omega L\ll1$, the bulk fermion effectively behaves as a
four-dimensional light state extending across the compact dimension,
and no exponential separation effect is present. When
$\omega L\gg1$, propagation across the interval is exponentially
suppressed, and the two boundaries become increasingly sequestered
from one another.

%The high-momentum damping depends on $M_5$ only through
%\begin{equation}
%\omega
%=
%\sqrt{p_E^2+M_5^2},
%\end{equation}
%and is therefore insensitive to the sign of $M_5$. The sign of $M_5$
%instead becomes important in the low-energy theory, because it
%determines toward which boundary the chiral zero mode is localized.
%For the sequestered configuration considered here, $M_5<0$ localizes
%the light state toward the hidden boundary and suppresses its overlap
%with the visible boundary.

%----------------------------------------------------------
\subsection{Low-energy zero-mode propagation}
%----------------------------------------------------------

For Euclidean momenta well below the first KK mass,
\begin{equation}
p_E^2\ll m_1^2,
\end{equation}
the excited KK contributions are suppressed relative to the zero-mode
contribution. Indeed,
for every mode with $n\geq1$,
\begin{equation}
\frac{1}{p_E^2+m_n^2}
\simeq
\frac{1}{m_n^2},
\qquad
p_E^2\ll m_n^2,
\end{equation}
so that the corresponding contribution remains finite as
$p_E^2\rightarrow0$. By contrast, the free zero-mode contribution is
\begin{equation}
G_R^{(0)}(p_E;0,L)
=
\frac{
f_R^{(0)}(0)f_R^{(0)}(L)
}{
p_E^2
},
\label{eq:prop_app_zero_mode_low_energy}
\end{equation}
which increases as the momentum decreases. Therefore, sufficiently
below the first KK threshold, propagation between the two boundaries
is dominated by the massless chiral zero mode, while the excited KK
tower gives only subleading corrections suppressed by the heavy
masses $m_n$.

After the hidden-boundary Majorana interaction is included, the
surviving chiral state acquires the physical mass $M_0$. In the
weak-mixing regime, where the lightest physical state remains
predominantly the free zero mode, the corresponding pole is
schematically shifted according to
\begin{equation}
\frac{1}{p_E^2}
\longrightarrow
\frac{1}{p_E^2+M_0^2},
\end{equation}
with additional corrections arising from its mixing with the excited
KK states. Thus, at energies below the KK scale, the five-dimensional
propagation is effectively described by the exchange of the lightest
four-dimensional Majorana state.

For the sequestered model considered here, the physically relevant
choice is $M_5=-\mu$ with $\mu>0$, for which the zero-mode wave function is localized toward the hidden
boundary at $y=L$. Here, localization does not mean that the particle
occupies a fixed position in the fifth dimension. Rather, it means
that the probability density associated with the zero-mode profile,
$\left|f_R^{(0)}(y)\right|^2$, is largest near $y=L$ and exponentially
smaller near $y=0$.

Using the zero-mode profile derived previously, the values at the two
boundaries satisfy
\begin{equation}
\frac{
f_R^{(0)}(0)
}{
f_R^{(0)}(L)
}
=
e^{-\mu L}.
\label{eq:prop_app_hidden_boundary_ratio}
\end{equation}
Consequently, the zero mode has an unsuppressed overlap with operators
localized on the hidden boundary, whereas its overlap with
visible-boundary operators is exponentially suppressed.

More explicitly, each sterile zero-mode field appearing in a
brane-localized interaction contributes one factor of its wave
function evaluated at that boundary. Equation~\eqref{eq:prop_app_hidden_boundary_ratio}
therefore shows that each visible-boundary sterile field carries an
additional factor $e^{-\mu L}$ relative to a hidden-boundary sterile
field. The visible Yukawa interaction contains one sterile field,
whereas the hidden Majorana interaction contains two and is
proportional to $\left[f_R^{(0)}(L)\right]^2$. The latter is therefore
not exponentially suppressed in the hidden-localized configuration.

The residue of the cross-boundary zero-mode propagator is
\begin{equation}
f_R^{(0)}(0)f_R^{(0)}(L)
=
\frac{
\mu
}{
\sinh(\mu L)
},
\label{eq:prop_app_zero_boundary_product}
\end{equation}
which contains one visible-boundary overlap and one hidden-boundary
overlap. In the limit $\mu L\gg1$,
\begin{equation}
f_R^{(0)}(0)f_R^{(0)}(L)
\simeq
2\mu e^{-\mu L}.
\end{equation}
The propagation of the light state from one boundary to the other is
therefore exponentially suppressed. This is the geometrical origin of
sequestering in the present construction: the sterile state couples
strongly to the hidden boundary, where it is localized, but
communicates only weakly with the visible boundary because its wave
function there is exponentially small.

\section{One-loop derivation of the five-dimensional radiative portal}
\label{app:5D_loop_derivation}

We now derive the one-loop contribution to the operator
\begin{equation}
\mathcal L_{\rm EFT}
\supset
-\kappa_{5,\rm loop}
R^2\Phi^\dagger\Phi
\label{eq:app_portal_operator}
\end{equation}
generated by the bulk sterile fermions.  We first discuss the physical
Majorana eigenstates obtained after the hidden scalar acquires a vacuum
expectation value.  We then introduce the boundary Feynman rules and
write the complete loop amplitude explicitly in mixed
momentum--position space.  Finally, we rewrite the result in the
physical mass basis and consider one five-dimensional sterile
generation and the perturbative single-state limit.

%The mixed five-dimensional representation and the physical four-dimensional spectral representation are two equivalent ways of organizing the same calculation. The former makes the propagation between the two boundaries explicit, while the latter is convenient for matching to the four-dimensional theory.  They must not be added as independent contributions.

%----------------------------------------------------------
\subsection{Majorana mass, physical eigenstates, and physical profiles}
\label{app:loop_physical_states}
%----------------------------------------------------------

Once the Majorana interaction is included, sterile-fermion number is
not conserved and the physical mass eigenstates are Majorana
fermions, for which the particle and antiparticle are not independent
degrees of freedom. Consequently, in addition to the ordinary
contraction
$\langle\mathcal N\overline{\mathcal N}\rangle$, one must also include
the anomalous contractions involving $\mathcal N$ and
$\mathcal N^c$. It is therefore convenient to organize the field and
its charge conjugate into the Nambu--Gorkov multiplet
\begin{equation}
\Psi_N
=
\begin{pmatrix}
\mathcal N\\
\mathcal N^c
\end{pmatrix}.
\end{equation}
Here the charge-conjugate field is defined by
\begin{equation}
\mathcal N^c
=
C\overline{\mathcal N}^{\,T},
\qquad
C^{-1}\gamma^\mu C
=
-\left(\gamma^\mu\right)^T,
\end{equation}
where $C$ is the charge-conjugation matrix. This doubling does not
introduce additional physical degrees of freedom; it provides a
compact representation of the propagators and mass insertions in the
presence of Majorana interactions.
The propagator of the Nambu--Gorkov multiplet is
\begin{equation}
\mathbb S_N(p_E;y,y')
=
\left\langle
\Psi_N(p_E,y)
\overline{\Psi_N}(-p_E,y')
\right\rangle.
\label{eq:app_NG_propagator}
\end{equation}
It contains both ordinary and anomalous contractions:
\begin{equation}
\mathbb S_N
=
\begin{pmatrix}
\langle\mathcal N\overline{\mathcal N}\rangle
&
\langle\mathcal N\overline{\mathcal N^c}\rangle
\\[2mm]
\langle\mathcal N^c\overline{\mathcal N}\rangle
&
\langle\mathcal N^c\overline{\mathcal N^c}\rangle
\end{pmatrix}_{\rm NG}.
\end{equation}

The physical profiles determine the residues of the propagator poles.
Because the physical states are Majorana fermions, two different
right-handed contractions must be considered.

The first is the ordinary, lepton-number-conserving contraction,
\begin{equation}
\mathcal S_{RR,IJ}^{\rm LNC}(p_E;y,y')
\equiv
\int\dd^4x\,
e^{-\ii p_E\cdot(x-x')}
\left\langle
T\,
\mathcal N_{IR}(x,y)
\overline{\mathcal N}_{JR}(x',y')
\right\rangle.
\label{eq:app_Majorana_LNC_definition}
\end{equation}
Using the expansion in physical Majorana eigenstates, it becomes
\begin{equation}
\mathcal S_{RR,IJ}^{\rm LNC}(p_E;y,y')
=
\sum_a
F_{R,Ia}(y)
F_{R,Ja}^*(y')
\frac{
-\ii\slashed p_E P_L
}{
p_E^2+M_a^2
}.
\label{eq:app_Majorana_LNC_block}
\end{equation}
This contraction connects a sterile field with its Dirac conjugate
and therefore preserves the assigned sterile-fermion number. 

The Majorana mass also allows the anomalous contraction
\begin{equation}
\mathcal S_{RR,IJ}^{\rm LNV}(p_E;y,y')
\equiv
\int\dd^4x\,
e^{-\ii p_E\cdot(x-x')}
\left\langle
T\,
\mathcal N_{IR}(x,y)
\mathcal N_{JR}^{T}(x',y')
\right\rangle.
\label{eq:app_Majorana_LNV_definition}
\end{equation}
This contraction connects two fields carrying the same sterile
fermion number and therefore violates that number by two units. It is
nonzero only because the physical states have Majorana masses:
\begin{equation}
\mathcal S_{RR,IJ}^{\rm LNV}(p_E;y,y')
=
\sum_a
F_{R,Ia}(y)
F_{R,Ja}(y')
\frac{
M_a P_R C
}{
p_E^2+M_a^2
}.
\label{eq:app_Majorana_LNV_block}
\end{equation}
The distinction between the two propagator blocks is therefore
physical. The LNC block describes propagation without a Majorana-mass
conversion and contains a momentum numerator, whereas the LNV block
describes propagation in which the fermion is converted into its
charge conjugate and contains a Majorana-mass numerator. In the
Dirac limit $M_a\rightarrow0$, the anomalous contraction vanishes.

%----------------------------------------------------------
\subsection{Boundary Feynman rules and loop topology}
\label{app:loop_rules_topology}
%----------------------------------------------------------

Since the visible-boundary interaction Lagrangian is written as
\begin{equation}
\mathcal L_{\rm vis}
\supset
-\delta(y)
\left(
\widehat Y_{5,\nu}
\right)_{\alpha I}
\overline{L_\alpha}
\widetilde\Phi\,
\mathcal N_{IR}
+
{\rm h.c.}
\label{eq:app_visible_interaction}
\end{equation}
it is convenient to define the transition operators
\begin{align}
\left(
\Gamma_{\widetilde\Phi}^{LN}
\right)_{\alpha I}
&=
\left(
\widehat Y_{5,\nu}
\right)_{\alpha I}
P_R,
\label{eq:app_Gamma_Phi_LN}
\\
\left(
\Gamma_{\widetilde\Phi^\dagger}^{NL}
\right)_{I\alpha}
&=
\left(
\widehat Y_{5,\nu}^\dagger
\right)_{I\alpha}
P_L.
\label{eq:app_Gamma_Phdagger_NL}
\end{align}

In terms of these operators, the visible-boundary interaction can be
written in the compact form
\begin{equation}
\mathcal L_{\rm vis}
\supset
-\delta(y)
\left[
\overline{L_\alpha}\,
\widetilde\Phi\,
\left(
\Gamma_{\widetilde\Phi}^{LN}
\right)_{\alpha I}
\mathcal N_I
+
\overline{\mathcal N}_I
\left(
\Gamma_{\widetilde\Phi^\dagger}^{NL}
\right)_{I\alpha}
\widetilde\Phi^\dagger
L_\alpha
\right].
\label{eq:app_visible_interaction_Gamma}
\end{equation}

The first maps sterile-flavor space into lepton-flavor space, while
the second maps lepton-flavor space back into sterile-flavor space.
The corresponding Minkowski-space rules are
\begin{align}
\widetilde\Phi\,\mathcal N_I\longrightarrow L_\alpha
&:
\qquad
-\ii
\left(
\Gamma_{\widetilde\Phi}^{LN}
\right)_{\alpha I},
\\
\widetilde\Phi^\dagger L_\alpha\longrightarrow\mathcal N_I
&:
\qquad
-\ii
\left(
\Gamma_{\widetilde\Phi^\dagger}^{NL}
\right)_{I\alpha}.
\end{align}
The two visible vertices enter the loop in the ordered combination
\begin{equation}
\Gamma_{\widetilde\Phi^\dagger}^{NL}
S_L(p_E)
\Gamma_{\widetilde\Phi}^{LN},
\label{eq:app_visible_ordered_product}
\end{equation}
which maps sterile-flavor space back into itself.

In Nambu--Gorkov space, the hidden interaction is
\begin{equation}
\mathcal L_{\rm hid}
\supset
-\frac{1}{2}
\delta(y-L)
R\,
\overline{\Psi_N}
\bm{\Gamma}_{RNN}
\Psi_N,
\end{equation}
with
\begin{equation}
\bm{\Gamma}_{RNN}
=
\begin{pmatrix}
0
&
\widehat Y_{5,N}^\dagger P_L
\\[2mm]
\widehat Y_{5,N}P_R
&
0
\end{pmatrix}_{\rm NG}.
\label{eq:app_Gamma_RNN_NG}
\end{equation}
The corresponding Minkowski-space rule for one external unshifted
field $R$ is
\begin{equation}
R\,\Psi_N\,\Psi_N:
\qquad
-\ii\bm{\Gamma}_{RNN}.
\end{equation}
Since $R$ is real, the same Nambu--Gorkov matrix is inserted at both
hidden vertices.  One does not replace the second vertex by
$\bm{\Gamma}_{RNN}^\dagger$, because
Eq.~\eqref{eq:app_Gamma_RNN_NG} already contains the interaction and
its Hermitian conjugate.

If the external scalar is instead the physical fluctuation $\rho$, the
vertex matrix is
\begin{equation}
\bm{\Gamma}_{\rho NN}
=
\frac{1}{\sqrt{2}}
\bm{\Gamma}_{RNN}.
\end{equation}
Two external-$\rho$ vertices therefore carry an overall factor $1/2$
relative to two unshifted-$R$ vertices.  The EFT matching
normalization changes by the same factor:
\begin{equation}
\Gamma_{RR\Phi\Phi^\dagger}^{\rm EFT}(0)
=
-2\kappa_{5,\rm loop},
\qquad
\Gamma_{\rho\rho\Phi\Phi^\dagger}^{\rm EFT}(0)
=
-\kappa_{5,\rm loop}.
\label{eq:app_R_r_matching}
\end{equation}
Both choices consequently give the same Wilson coefficient when the
normalizations are used consistently.

At vanishing external momenta, the loop contains two hidden vertices,
two visible vertices, three sterile propagators, and one lepton
propagator.  The ordered vertex--propagator chain is
\begin{align}
&
\bm{\Gamma}_{RNN}
\mathbb S_N(p_E;L,L)
\bm{\Gamma}_{RNN}
\mathbb S_N(p_E;L,0)
\nonumber\\
&\hspace{15mm}\times
\Gamma_{\widetilde\Phi^\dagger}^{NL}
S_L(p_E)
\Gamma_{\widetilde\Phi}^{LN}
\mathbb S_N(p_E;0,L).
\label{eq:app_ordered_loop_chain}
\end{align}
The propagator $\mathbb S_N(L,L)$ joins the two hidden vertices.  The
propagators $\mathbb S_N(L,0)$ and $\mathbb S_N(0,L)$ cross the
interval, while the lepton propagator joins the two visible vertices
at $y=0$.

%----------------------------------------------------------
\subsection{General one-loop amplitude}
\label{app:loop_general_amplitude}
%----------------------------------------------------------

The complete mixed-representation one-loop amplitude, with all
external fields stripped off, is
\begin{widetext}
\begin{align}
\Gamma^{(1)}_{RR\Phi\Phi^\dagger}(0)
=
-\mathcal N_{\rm comb}
\int
\frac{\dd^4p_E}{(2\pi)^4}
\operatorname{Tr}_{\rm spin,flav,NG}
\Big[
\bm{\Gamma}_{RNN}
\mathbb S_N(p_E;L,L)
\bm{\Gamma}_{RNN}
\mathbb S_N(p_E;L,0)
\Gamma_{\widetilde\Phi^\dagger}^{NL}
S_L(p_E)
\Gamma_{\widetilde\Phi}^{LN}
\mathbb S_N(p_E;0,L)
\Big].
\label{eq:app_full_5D_amplitude}
\end{align}
\end{widetext}
Matching this one-loop vertex to
Eq.~\eqref{eq:app_R_r_matching} gives
\begin{equation}
\kappa_{5,\rm loop}
=
-\frac{1}{2}
\Gamma^{(1)}_{RR\Phi\Phi^\dagger}(0).
\label{eq:app_loop_to_kappa_matching}
\end{equation}
for two external unshifted $R$ fields.

The explicit minus sign is the closed-fermion-loop sign.  The trace is
over spinor, sterile-flavor, lepton-flavor, and Nambu--Gorkov indices.
The factor $\mathcal N_{\rm comb}$ denotes the normalization associated
with the chosen Majorana or Nambu--Gorkov convention.  It must be fixed
consistently and must not be supplemented by an independent sum over
Majorana-flow orientations if those orientations are already included
in the propagator trace. In our case its value is $\mathcal N_{\rm comb}=1/2$.

The exact spectral representation in the physical basis is obtained by
inserting the expansion of the bulk field in terms of the Majorana
mass eigenstates into each sterile propagator as in Eq.~\eqref{eq:physicalNR}.
When this expansion is evaluated at a boundary, the fundamental
five-dimensional Yukawa couplings $\widehat Y_{5,\nu}$ and
$\widehat Y_{5,N}$ are multiplied by the corresponding boundary
values of the physical profiles. These profile factors therefore arise
from projecting the bulk fields onto the physical Majorana
eigenstates; they do not represent additional interactions.

It is convenient to define the resulting visible- and hidden-boundary
couplings as
\begin{align}
\left(
\lambda_\nu
\right)_{\alpha a}
&=
\sum_I
\left(
\widehat Y_{5,\nu}
\right)_{\alpha I}
F_{R,Ia}(0),
\label{eq:app_physical_visible_couplings}
\\
\left(
g_N
\right)_{ab}
&=
\sum_{I,J}
F_{R,Ia}(L)
\left(
\widehat Y_{5,N}
\right)_{IJ}
F_{R,Jb}(L).
\label{eq:app_physical_hidden_couplings}
\end{align}
Thus, $\lambda_\nu$ is the coupling of the physical Majorana state
$N_a$ to the visible boundary at $y=0$, while $g_N$ is the coupling
of the physical states $N_a$ and $N_b$ to the hidden scalar at
$y=L$. The matrix $g_N$ is symmetric when
$\widehat Y_{5,N}$ is symmetric.

In terms of these couplings, the amplitude can be organized as
\begin{equation}
\Gamma^{(1)}_{RR\Phi\Phi^\dagger}(0)
=
-\mathcal N_{\rm comb}
\sum_{\alpha,a,b,c}
\left(
g_N
\right)_{ab}
\left(
g_N^\dagger
\right)_{ca}
\left(
\lambda_\nu^\dagger
\right)_{b\alpha}
\left(
\lambda_\nu
\right)_{\alpha c}
\mathcal I_{abc}.
\label{eq:app_general_mass_eigenstate_amplitude}
\end{equation}
Here $\mathcal I_{abc}$ is the renormalized four-dimensional loop
kernel containing three sterile propagators with masses
$M_a,M_b,M_c$ and one lepton propagator.  Its generic form is
\begin{equation}
\mathcal I_{abc}
=
\int
\frac{\dd^4p_E}{(2\pi)^4}
\frac{
\mathcal P_{abc}
\left(
p_E;M_a,M_b,M_c
\right)
}{
p_E^2
\left(
p_E^2+M_a^2
\right)
\left(
p_E^2+M_b^2
\right)
\left(
p_E^2+M_c^2
\right)
}.
\label{eq:app_Iabc}
\end{equation}
The numerator $\mathcal P_{abc}$ is fixed by the chiral projectors and
by the LNC and LNV components of the sterile propagators.  Different
Majorana-flow conventions can rearrange transposes and complex
conjugations, but the complete amplitude is convention independent.
Equation~\eqref{eq:app_general_mass_eigenstate_amplitude} is the exact
result for an arbitrary number of sterile generations and for the
complete physical KK tower.

The fifth-dimensional dependence is already contained in the physical
masses and in the profiles entering
Eqs.~\eqref{eq:app_physical_visible_couplings} and
\eqref{eq:app_physical_hidden_couplings}.  Therefore, no additional
wave-function overlap or sequestering factor should be multiplied onto
Eq.~\eqref{eq:app_general_mass_eigenstate_amplitude}.

%----------------------------------------------------------
\subsection{One five-dimensional sterile generation}
\label{app:loop_one_generation}
%----------------------------------------------------------

To obtain an expression that is sufficiently simple for a transparent
analytical and numerical evaluation, we first restrict the discussion
to a single five-dimensional sterile generation. This assumption
removes the additional flavor structure associated with several bulk
fermions and isolates the effects of the KK tower and of the
boundary-induced Majorana mixing. It should nevertheless be stressed
that this is not a single-particle approximation.

Indeed, even for one five-dimensional sterile generation,
compactification and the boundary Majorana interaction produce an
infinite set of physical Majorana states. Thus, one five-dimensional
generation must not be confused with the single-state approximation
considered below.

Suppressing the original sterile-generation index, the physical
couplings become
\begin{equation}
\lambda_a
=
\widehat y_{5,\nu}
F_{R,a}(0),
\qquad
g_{ab}
=
\widehat y_{5,N}
F_{R,a}(L)
F_{R,b}(L).
\label{eq:app_one_generation_couplings}
\end{equation}
The exact amplitude is then
\begin{equation}
\Gamma^{(1)}_{RR\Phi\Phi^\dagger}(0)
=
-\mathcal N_{\rm comb}
\sum_{a,b,c}
g_{ab}
g_{ca}^{*}
\lambda_b^{*}
\lambda_c
\mathcal I_{abc}.
\label{eq:app_one_generation_amplitude}
\end{equation}
Substituting the boundary couplings gives
\begin{align}
\Gamma^{(1)}_{RR\Phi\Phi^\dagger}(0)
={}&
-\mathcal N_{\rm comb}
\left|
\widehat y_{5,N}
\right|^2
\left|
\widehat y_{5,\nu}
\right|^2
\sum_{a,b,c}
\left|
F_{R,a}(L)
\right|^2
\nonumber\\
&\times
F_{R,b}(L)
F_{R,b}^{*}(0)
F_{R,c}^{*}(L)
F_{R,c}(0)
\mathcal I_{abc}.
\label{eq:app_one_generation_profile_amplitude}
\end{align}
This expression is exact for one five-dimensional sterile field. It
includes the light-state contribution, the pure-KK terms, and all
mixed light--KK terms. The explicit prefactor displays the two
visible-boundary and two hidden-boundary couplings. The physical masses $M_a$, the
profiles $F_{R,a}$, and the loop functions also depend implicitly on
$v_r\widehat y_{5,N}$ through the diagonalization of the complete mass
matrix.

%----------------------------------------------------------
\subsection{Perturbative single-state limit}
\label{app:loop_single_state}
%----------------------------------------------------------

Although the one-generation result derived above still contains the
complete infinite tower of physical Majorana states, it is useful to
identify a controlled limit in which the amplitude can be approximated
by the contribution of a single light state. This limit provides a
simple analytical expression that can be evaluated directly and
compared with the corresponding four-dimensional theory. It is valid
only when the lightest state is well separated from the KK tower and
its mixing with the excited modes is small.

We therefore suppose that one physical Majorana state $N_0$ is much
lighter than the first massive KK level,
\begin{equation}
M_0
\ll
m_1,
\qquad
m_1
=
\sqrt{
M_5^2+\frac{\pi^2}{L^2}
}.
\label{eq:app_light_state_hierarchy}
\end{equation}
In addition, we assume that the boundary-induced mixing between the
free zero mode and each massive KK level is perturbative:
\begin{equation}
\epsilon_n
\equiv
\frac{
\left|
\mathcal M_N^{0n}
\right|
}{
\left|m_n-M_0\right|
}
\ll1,
\qquad
n\geq1.
\label{eq:app_small_mixing}
\end{equation}
Under these conditions, $N_0$ is predominantly the free chiral zero
mode, while the admixture of the $n$th excited KK state is of order
$\epsilon_n$. The effects of the remaining tower can then be treated
as perturbative corrections to the single-state result.

Since $M_0\ll m_n$, one may estimate
\begin{equation}
\epsilon_n
\simeq
\frac{
\left|
\mathcal M_N^{0n}
\right|
}{
m_n
}.
\end{equation}
The physical light-state profile can be expanded as
\begin{equation}
F_{R,0}(y)
=
U_{00}^R
f_R^{(0)}(y)
+
\sum_{n=1}^{\infty}
U_{n0}^R
f_R^{(n)}(y),
\label{eq:app_light_profile_expansion}
\end{equation}
where $\epsilon\equiv\max_{n\geq1}\epsilon_n$. The mixing
coefficients satisfy
\begin{equation}
U_{00}^R
=
1+\mathcal O(\epsilon^2),
\qquad
U_{n0}^R
=
\mathcal O(\epsilon_n).
\end{equation}
Therefore,
\begin{equation}
F_{R,0}(y)
=
f_R^{(0)}(y)
+
\mathcal O(\epsilon_n).
\label{eq:app_F_approx_f}
\end{equation}

For brevity, we denote the zero-mode couplings
$Y_{4,\nu}^{(0)}$ and $Y_{4,N}^{00}$ introduced previously by
$Y_\nu^{(0)}$ and $Y_N^{(00)}$, respectively. The physical couplings
of the light state are
\begin{align}
Y_\nu^{(0)}
&=
\widehat y_{5,\nu}
F_{R,0}(0),
\label{eq:app_Ynu_zero}
\\
Y_N^{(00)}
&=
\widehat y_{5,N}
\left[
F_{R,0}(L)
\right]^2,
\label{eq:app_YN_zero}
\end{align}
and its Majorana mass is approximately
\begin{equation}
M_0
\simeq
\frac{v_r}{\sqrt{2}}
Y_N^{(00)}.
\label{eq:app_M0}
\end{equation}

In this limit, the physical spectral sum is dominated by
\begin{equation}
a=b=c=0.
\end{equation}
The loop therefore reduces to the ordinary four-dimensional
one-Majorana-state portal diagram evaluated with the physical
couplings $Y_\nu^{(0)}$ and $Y_N^{(00)}$.  In the zero-momentum
$\overline{\rm MS}$ matching convention used in the main text, the
renormalized result is
\begin{equation}
\kappa_{5,\rm loop}^{(0)}(\mu)
=
-\frac{
\left|
Y_N^{(00)}
\right|^2
\left|
Y_\nu^{(0)}
\right|^2
}{
8\pi^2
}
\left[
\ln\left(
\frac{M_0^2}{\mu^2}
\right)
+2
\right].
\label{eq:app_single_state_threshold}
\end{equation}
At the natural matching scale $\mu=M_0$,
\begin{equation}
\kappa_{5,\rm loop}^{(0)}(M_0)
=
-\frac{
\left|
Y_N^{(00)}
\right|^2
\left|
Y_\nu^{(0)}
\right|^2
}{
4\pi^2
}.
\label{eq:app_single_state_matching_scale}
\end{equation}
This is precisely the ordinary four-dimensional Majorana-neutrino
threshold written in terms of the physical light-state parameters.
The five-dimensional origin enters through $M_0$ and through the
boundary-profile factors contained in the physical couplings.

Expressed in terms of the fundamental brane coefficients,
\begin{equation}
\kappa_{5,\rm loop}^{(0)}(M_0)
=
-\frac{
\left|
\widehat y_{5,N}
\right|^2
\left|
\widehat y_{5,\nu}
\right|^2
}{
4\pi^2
}
\left|
F_{R,0}(0)
\right|^2
\left|
F_{R,0}(L)
\right|^4.
\label{eq:app_single_state_profiles}
\end{equation}
The two visible vertices contribute
$|F_{R,0}(0)|^2$, while the two hidden vertices contribute
$|F_{R,0}(L)|^4$.

%----------------------------------------------------------
\subsection{Matching to the flat-profile four-dimensional reference result}
\label{app:loop_4D_matching}
%----------------------------------------------------------

In the weak-mixing approximation of
Eq.~\eqref{eq:app_F_approx_f}, the lightest physical state is
dominated by the free zero mode, so that
\begin{equation}
F_{R,0}(y)
\simeq
f_R^{(0)}(y),
\end{equation}
where the normalized right-handed zero-mode profile is given in
Eq.~\eqref{eq:app_zero_profile_unormalized}.

At the matching scale $\mu=M_0$, the single-state contribution to the
portal is therefore
\begin{equation}
\kappa_{5,\rm loop}^{(0)}(M_0)
=
-\frac{
\left|Y_N^{(00)}\right|^2
\left|Y_\nu^{(0)}\right|^2
}{
4\pi^2
}.
\label{eq:app_kappa_physical_couplings}
\end{equation}
Its dependence on the zero-mode profile is proportional to
\begin{equation}
\left|
f_R^{(0)}(0)
\right|^2
\left|
f_R^{(0)}(L)
\right|^4.
\end{equation}
Using the normalized profile, this product is
\begin{equation}
\left|
f_R^{(0)}(0)
\right|^2
\left|
f_R^{(0)}(L)
\right|^4
=
\left[
\frac{
2M_5
}{
1-e^{-2M_5L}
}
\right]^3
e^{-4M_5L}.
\label{eq:app_profile_factor_signed}
\end{equation}
Although written in terms of the signed parameter $M_5$, the
right-hand side is positive for either sign of $M_5$.

The five-dimensional brane coefficients have mass dimensions
\begin{equation}
\left[
\widehat y_{5,\nu}
\right]
=
-\frac12,
\qquad
\left[
\widehat y_{5,N}
\right]
=
-1.
\end{equation}
It is therefore convenient to define the dimensionless reference
parameters
\begin{equation}
y_\nu^{\rm ref}
=
\frac{
\widehat y_{5,\nu}
}{
\sqrt L
},
\qquad
y_N^{\rm ref}
=
\frac{
\widehat y_{5,N}
}{
L
}.
\label{eq:app_reference_couplings}
\end{equation}
These quantities provide a dimensionless parametrization of the
fundamental five-dimensional coefficients. They are not, in general,
equal to the physical four-dimensional couplings. Indeed,
Eqs.~\eqref{eq:app_Ynu_zero} and
\eqref{eq:app_YN_zero} can be written as
\begin{align}
Y_\nu^{(0)}
&=
y_\nu^{\rm ref}
\left[
\sqrt L\,f_R^{(0)}(0)
\right],
\label{eq:app_Ynu_ref_general}
\\
Y_N^{(00)}
&=
y_N^{\rm ref}
\left[
L\left(f_R^{(0)}(L)\right)^2
\right].
\label{eq:app_YN_ref_general}
\end{align}
Only in the flat-profile limit,
\begin{equation}
|M_5|L\ll1,
\qquad
f_R^{(0)}(y)
\simeq
\frac{1}{\sqrt L},
\end{equation}
do the overlap factors approach unity and the physical couplings
reduce to
\begin{equation}
Y_\nu^{(0)}
\simeq
y_\nu^{\rm ref},
\qquad
Y_N^{(00)}
\simeq
y_N^{\rm ref}.
\end{equation}

The corresponding flat-profile four-dimensional reference result is
therefore
\begin{equation}
\kappa_{4,\rm loop}(M_0)
=
-\frac{
\left|
y_N^{\rm ref}
\right|^2
\left|
y_\nu^{\rm ref}
\right|^2
}{
4\pi^2
}.
\label{eq:app_kappa_flat}
\end{equation}
At fixed fundamental five-dimensional brane coefficients, the
localized zero-mode result can be factorized as
\begin{equation}
\kappa_{5,\rm loop}^{(0)}(M_0)
=
\kappa_{4,\rm loop}(M_0)
\mathcal F_0(M_5L),
\label{eq:app_factorized_kappa}
\end{equation}
where
\begin{equation}
\mathcal F_0(M_5L)
=
L^3
\left|
f_R^{(0)}(0)
\right|^2
\left|
f_R^{(0)}(L)
\right|^4.
\label{eq:app_F0_definition}
\end{equation}

For the sequestered configuration considered here, the zero mode is
localized toward the hidden boundary and $M_5<0$.
Defining $x=|M_5|L$, the exact overlap factor becomes
\begin{equation}
\mathcal F_0(x)
=
\left[
\frac{
2x
}{
1-e^{-2x}
}
\right]^3
e^{-2x}.
\label{eq:app_F0_negative}
\end{equation}
The factor $e^{-2x}$ originates from the two visible-boundary
couplings in the squared amplitude. The hidden-boundary coupling is
not exponentially suppressed because the zero mode is localized near
$y=L$, although its normalization produces the polynomial prefactor
in Eq.~\eqref{eq:app_F0_negative}.

In the strongly sequestered regime, $x\gg1$, one obtains
\begin{equation}
\mathcal F_0(x)
\simeq
(2x)^3e^{-2x},
\end{equation}
and hence
\begin{equation}
\kappa_{5,\rm loop}^{(0)}(M_0)
\simeq
\kappa_{4,\rm loop}(M_0)
(2|M_5|L)^3
e^{-2|M_5|L}.
\label{eq:app_kappa_strong_sequestering}
\end{equation}
The portal is therefore exponentially suppressed in this regime. By
contrast, for weak localization, $x\ll1$,
\begin{equation}
\mathcal F_0(x)
\longrightarrow
1,
\end{equation}
so that
\begin{equation}
\kappa_{5,\rm loop}^{(0)}(M_0)
\simeq
\kappa_{4,\rm loop}(M_0).
\end{equation}
This is the flat-profile limit in which the zero mode has comparable
overlap with the two boundaries.

It is essential to specify which quantities are kept fixed in this
comparison. The factor $\mathcal F_0$ compares the localized and flat
profiles while holding the fundamental five-dimensional brane
coefficients $\widehat y_{5,\nu}$ and $\widehat y_{5,N}$ fixed. If,
instead, the physical four-dimensional couplings $Y_\nu^{(0)}$ and
$Y_N^{(00)}$ are held fixed, the threshold is already given by
Eq.~\eqref{eq:app_kappa_physical_couplings}, or equivalently by
Eq.~\eqref{eq:app_single_state_matching_scale}, and no additional
overlap factor should be inserted.

Using this definition, the localization parameter can be written as
\begin{equation}
x
=
|M_5|L
=
\pi
\frac{
|M_5|
}{
M_{\rm KK}
}.
\label{eq:prop_app_x_MKK}
\end{equation}
For the hidden-localized branch, the single-state result therefore
takes the form
\begin{equation}
\begin{aligned}
\kappa_{5,\rm loop}^{(0)}(M_0)
={}&
\kappa_{4,\rm loop}(M_0)
\left[
\frac{
2\pi |M_5|/M_{\rm KK}
}{
1-\exp\!\left(
-2\pi |M_5|/M_{\rm KK}
\right)
}
\right]^3
\\
&\times
\exp\!\left(
-2\pi
\frac{
|M_5|
}{
M_{\rm KK}
}
\right).
\end{aligned}
\label{eq:prop_app_kappa_MKK}
\end{equation}
This form makes explicit that the strength of the geometrical
suppression is controlled by the ratio of the localization scale
$|M_5|$ to the compactification scale $M_{\rm KK}$.

%----------------------------------------------------------
\subsection{Beyond the single-state approximation}
\label{app:loop_beyond_single_state}
%----------------------------------------------------------

The single-state formulas are reliable only when
\begin{equation}
M_0\ll m_1,
\qquad
\epsilon_n\ll1.
\end{equation}
If the boundary-induced mixing is not perturbative, the physical
profiles $F_{R,Ia}$ must be retained and can differ substantially from
the free profiles.  The complete coefficient may be organized as
\begin{equation}
\kappa_{5,\rm loop}(\mu)
=
\kappa_{5,\rm loop}^{(0)}(\mu)
+
\Delta\kappa_{\rm KK}(\mu),
\label{eq:app_full_KK_split}
\end{equation}
Here $\Delta\kappa_{\rm KK}$ contains the terms in the physical
spectral sum with at least one KK-dominated state.  It includes both
pure-KK and mixed light--KK contributions and is not, in general, a
universal multiplicative factor multiplying the zero-mode result.

For several sterile generations, the Yukawa matrices remain ordered in
Eq.~\eqref{eq:app_general_mass_eigenstate_amplitude}.  Only when the
relevant coupling matrices are simultaneously diagonal, or when the
loop functions become degenerate in the required subspace, can the
result be reduced to a simple trace or to a sum of independent
single-state thresholds.

The two cross-boundary propagators control the nonlocal ultraviolet
behavior.  At $p_EL\gg1$, each propagator is exponentially damped, so
their product contains $e^{-2p_EL}$.  After local brane subdivergences
are renormalized, the genuinely nonlocal visible-to-hidden amplitude
is ultraviolet finite.  The low-energy four-dimensional EFT still
contains the usual logarithmic running of
$\kappa_{\rm loop}(\mu)$; the finite difference between the complete
five-dimensional result and the light-state EFT is the KK matching
threshold.

Finally, the Nambu--Gorkov calculation and an explicit
Majorana-fermion-flow calculation are equivalent descriptions of the
same loop.  They must not be combined or double counted.

%----------------------------------------------------------
\section{From the five-dimensional sequestered theory to the
four-dimensional effective model}
\label{sec:5D_to_4D_matching}
%----------------------------------------------------------

The purpose of this section is to summarize how the five-dimensional
sequestered construction reduces to the four-dimensional model used
for the phenomenological analysis. The bulk equations, KK profiles,
boundary couplings, and complete mass diagonalization have been
derived in the preceding sections and will not be repeated here.

It is useful to distinguish two conceptually different steps. First,
compactification rewrites the five-dimensional theory as a
four-dimensional theory containing an infinite KK tower. Second, at
energies below the first KK threshold, the heavy tower can be
integrated out, leaving a finite four-dimensional effective theory
containing only the light physical Majorana states. This low-energy
reduction is valid independently of whether the surviving zero-mode
profiles are flat or localized. The flat-profile limit
$|M_5|L\ll1$ provides a particular four-dimensional reference case
in which the geometrical overlap factors approach unity.

%----------------------------------------------------------
\subsection{Dimensional reduction of the action}
%----------------------------------------------------------

The five-dimensional action is the sum of the bulk sterile-fermion
action and the visible- and hidden-boundary actions,
\begin{equation}
S_{5}
=
S_{\rm bulk}
+
S_{\rm vis}
+
S_{\rm hid}.
\label{eq:5D4D_total_action_summary}
\end{equation}
The explicit form of these terms was given in
Eqs.~\eqref{eq:app_bulk_action} and
\eqref{eq:5D4D_brane_interactions}. The visible and hidden fields are
already four-dimensional because they are confined to the boundaries.
The nontrivial dimensional reduction therefore concerns the bulk
sterile fermions $\mathcal N_I(x,y)$.

Substituting the KK decomposition in Eq.~\eqref{eq:KKdec} into the
bulk action and integrating over the fifth coordinate first produces
an infinite four-dimensional theory containing the chiral zero modes
and the massive Dirac KK tower. After the hidden scalar develops its
vacuum expectation value, the boundary Majorana interaction mixes
these free KK states. The complete Dirac-plus-Majorana mass matrix is
then diagonalized through the Takagi transformation in
Eq.~\eqref{eq:app_full_Takagi}.

In terms of the resulting physical Majorana states, the right-handed
bulk field is
\begin{equation}
\mathcal N_{IR}(x,y)
=
\sum_a
F_{R,Ia}(y)P_RN_a(x),
\label{eq:5D4D_physical_expansion_summary}
\end{equation}
where the profiles $F_{R,Ia}(y)$ were defined in
Eq.~\eqref{eq:5D4D_physical_profiles}. The complete dimensionally
reduced theory can therefore be written schematically as
\begin{align}
S_{\rm KK}^{(4)}
=
\int\dd^4x
\Bigg\{&
\mathcal L_{\rm vis}
+
\mathcal L_{\rm hid}
+
\frac12
\sum_a
\overline{N_a}
\left(
\ii\slashed{\partial}-M_a
\right)
N_a
\nonumber\\
&-
\left[
\left(\lambda_\nu\right)_{\alpha a}
\overline{L_\alpha}
\widetilde\Phi\,P_RN_a
+
{\rm h.c.}
\right]
\nonumber\\
&-
\frac12 R\,
\left(g_N\right)_{ab}
\overline{N_a}P_RN_b
+
{\rm h.c.}
\Bigg\}.
\label{eq:5D4D_full_reduced_action}
\end{align}
Equation~\eqref{eq:5D4D_full_reduced_action} is already a
four-dimensional action, but it still contains the infinite set of
physical states generated by the KK tower.

The effective couplings appearing in this action are
\begin{align}
\left(
\lambda_\nu
\right)_{\alpha a}
&=
\sum_I
\left(
\widehat Y_{5,\nu}
\right)_{\alpha I}
F_{R,Ia}(0),
\label{eq:5D4D_lambda_summary}
\\
\left(
g_N
\right)_{ab}
&=
\sum_{I,J}
F_{R,Ia}(L)
\left(
\widehat Y_{5,N}
\right)_{IJ}
F_{R,Jb}(L).
\label{eq:5D4D_gN_summary}
\end{align}
These are not new fundamental parameters. They are the original
five-dimensional boundary couplings projected onto the physical
Majorana eigenstates. The factors $F_{R,Ia}(0)$ and
$F_{R,Ia}(L)$ arise because the corresponding interactions probe the
bulk wave functions at the visible and hidden boundaries,
respectively.

%----------------------------------------------------------
\subsection{Low-energy four-dimensional limit}
%----------------------------------------------------------

A finite four-dimensional effective theory is obtained when the
characteristic energy of the process is much smaller than the first
KK mass:
\begin{equation}
E
\ll
m_1,
\qquad
m_1
=
\sqrt{
M_5^2+\frac{\pi^2}{L^2}
}.
\label{eq:5D4D_low_energy_condition}
\end{equation}
The KK-dominated states can then be integrated out. Their effects are
encoded in operators suppressed by inverse powers of the KK scale and
in finite threshold corrections to the parameters of the
four-dimensional theory.

If the boundary-induced mixing is also perturbative, as specified in
Eq.~\eqref{eq:weak_KK_mixing_conditions}, the lightest physical state
associated with each five-dimensional sterile generation is dominated
by the corresponding free chiral zero mode:
\begin{equation}
F_{R,I0}(y)
\simeq
f_{R,I}^{(0)}(y).
\label{eq:5D4D_light_profile_limit}
\end{equation}
Restricting Eqs.~\eqref{eq:5D4D_lambda_summary} and
\eqref{eq:5D4D_gN_summary} to these light states gives
\begin{align}
\left(
Y_\nu^{(0)}
\right)_{\alpha I}
&\simeq
\left(
\widehat Y_{5,\nu}
\right)_{\alpha I}
f_{R,I}^{(0)}(0),
\label{eq:5D4D_low_energy_Ynu}
\\
\left(
Y_N^{(00)}
\right)_{IJ}
&\simeq
f_{R,I}^{(0)}(L)
\left(
\widehat Y_{5,N}
\right)_{IJ}
f_{R,J}^{(0)}(L).
\label{eq:5D4D_low_energy_YN}
\end{align}
The resulting low-energy Lagrangian has the same form as the
four-dimensional sterile-neutrino model:
\begin{align}
\mathcal L_{\rm EFT}^{(4)}
\supset{}&
\frac12
\overline{N_I}
\left(
\ii\slashed{\partial}-M_I
\right)
N_I
\nonumber\\
&-
\left[
\left(
Y_\nu^{(0)}
\right)_{\alpha I}
\overline{L_\alpha}
\widetilde\Phi\,P_RN_I
+
{\rm h.c.}
\right]
\nonumber\\
&-
\frac12 R
\left(
Y_N^{(00)}
\right)_{IJ}
\overline{N_I}P_RN_J
+
{\rm h.c.}
+
\mathcal O\!\left(\frac{E^2}{m_1^2}\right).
\label{eq:5D4D_low_energy_lagrangian}
\end{align}
Here the index $I$ labels the light physical states retained in the
effective theory. Thus, the four-dimensional Yukawa matrices are not
independent of the extra-dimensional construction: they contain the
boundary values of the normalized bulk profiles.

%----------------------------------------------------------
\subsection{Relation between the five- and four-dimensional masses}
%----------------------------------------------------------

Before hidden symmetry breaking, the five-dimensional theory contains
a massless right-handed chiral zero mode and a tower of free Dirac
states with masses
\begin{equation}
m_{I,n}^2
=
M_{5,I}^2
+
\frac{n^2\pi^2}{L^2},
\qquad
n\geq1,
\end{equation}
as derived in Eq.~\eqref{eq:app_mn_spectrum}. The bulk parameters
$M_{5,I}$ therefore determine the localization of the zero modes and
contribute to the masses of the excited KK levels, but they do not
directly give four-dimensional masses to the chiral zero modes.

After
\begin{equation}
R
=
\frac{1}{\sqrt2}
\left(
v_r+\rho
\right),
\end{equation}
the hidden-boundary interaction generates the Majorana matrix in
Eq.~\eqref{eq:app_boundary_Majorana_matrix}. The exact physical masses
$M_a$ are the singular values obtained by diagonalizing the complete
Dirac-plus-Majorana matrix, and must not be identified with either
$M_{5,I}$ or $m_{I,n}$.

In the weak-mixing limit, the mass matrix of the light states reduces
approximately to the zero-mode block,
\begin{equation}
\left(
M_N^{\rm light}
\right)_{IJ}
\simeq
\frac{v_r}{\sqrt2}
\left(
Y_N^{(00)}
\right)_{IJ}
=
\frac{v_r}{\sqrt2}
f_{R,I}^{(0)}(L)
\left(
\widehat Y_{5,N}
\right)_{IJ}
f_{R,J}^{(0)}(L).
\label{eq:5D4D_light_mass_matrix}
\end{equation}
Diagonalizing this finite matrix gives the Majorana masses $M_I$
appearing in Eq.~\eqref{eq:5D4D_low_energy_lagrangian}. Corrections
from the KK tower are suppressed by the mixing parameters
$\epsilon_n$ and by the separation between the light and excited
levels.

The correspondence between the different masses is therefore
\begin{equation}
\begin{aligned}
M_{5,I}
&:\quad
\text{bulk localization parameter},
\\
m_{I,n}
&:\quad
\text{free Dirac KK masses},
\\
M_a
&:\quad
\text{exact physical Majorana masses},
\\
M_I
&:\quad
\text{light Majorana masses retained in the 4D EFT}.
\end{aligned}
\label{eq:5D4D_mass_dictionary}
\end{equation}

%----------------------------------------------------------
\subsection{Sequestering and the effective four-dimensional couplings}
%----------------------------------------------------------

For the sequestered branch relevant to the present model,
\begin{equation}
M_5=-\mu,
\qquad
\mu>0,
\end{equation}
the light sterile state is localized toward the hidden boundary at
$y=L$. Its boundary values satisfy
\begin{equation}
\frac{
f_R^{(0)}(0)
}{
f_R^{(0)}(L)
}
=
e^{-\mu L}.
\label{eq:5D4D_sequestered_ratio}
\end{equation}
Equation~\eqref{eq:5D4D_low_energy_Ynu} then shows that the effective
visible Yukawa coupling is exponentially suppressed. By contrast,
Eq.~\eqref{eq:5D4D_low_energy_YN} contains the profile evaluated at
the boundary toward which the state is localized, and the hidden
Majorana interaction is therefore not exponentially suppressed.

The low-energy theory has the same operator content as the ordinary
four-dimensional model, but the small visible-sector coupling now has
a geometrical origin:
\begin{equation}
\widehat Y_{5,\nu}
\quad\longrightarrow\quad
Y_\nu^{(0)}
=
\widehat Y_{5,\nu}f_R^{(0)}(0),
\qquad
Y_\nu^{(0)}
\propto
e^{-\mu L}.
\label{eq:5D4D_sequestered_Ynu}
\end{equation}
Thus, sequestering does not remove the visible interaction from the
four-dimensional theory. Instead, it determines its effective
strength through the exponentially small tail of the bulk wave
function.

%----------------------------------------------------------
\subsection{Reduction of the radiative portal}
%----------------------------------------------------------

In the five-dimensional theory, a local tree-level operator
$R^2\Phi^\dagger\Phi$ is absent because $R$ and $\Phi$ are localized
on different boundaries. The corresponding ultraviolet boundary
condition is therefore
\begin{equation}
\kappa_{\rm tree}(\Lambda_{\rm UV})
=
0.
\end{equation}
After compactification, the physical Majorana states couple to both
boundaries through $\lambda_\nu$ and $g_N$. Their exchange consequently
generates the portal radiatively.

The exact five-dimensional result contains the complete physical
Majorana spectrum and can be separated as
\begin{equation}
\kappa_{5,\rm loop}
=
\kappa_{5,\rm loop}^{(0)}
+
\Delta\kappa_{\rm KK},
\label{eq:5D4D_portal_split_summary}
\end{equation}
where $\kappa_{5,\rm loop}^{(0)}$ denotes the contribution of the
light state or light states, while $\Delta\kappa_{\rm KK}$ contains
the pure-KK and mixed light--KK terms.

At energies below the KK threshold and in the weak-mixing regime,
$\Delta\kappa_{\rm KK}$ is absorbed into the matching corrections of
the four-dimensional effective theory. For one light Majorana state,
the remaining threshold has exactly the usual four-dimensional form,
\begin{equation}
\kappa_{5,\rm loop}^{(0)}(\mu)
=
-\frac{
\left|Y_N^{(00)}\right|^2
\left|Y_\nu^{(0)}\right|^2
}{
8\pi^2
}
\left[
\ln\left(\frac{M_0^2}{\mu^2}\right)
+2
\right].
\label{eq:5D4D_portal_low_energy}
\end{equation}
At the matching scale $\mu=M_0$,
\begin{equation}
\kappa_{5,\rm loop}^{(0)}(M_0)
=
-\frac{
\left|Y_N^{(00)}\right|^2
\left|Y_\nu^{(0)}\right|^2
}{
4\pi^2
}.
\label{eq:5D4D_portal_threshold}
\end{equation}
Once the physical four-dimensional couplings are used, this is
identical to the portal obtained directly in the four-dimensional
model. No additional wave-function factor should be multiplied by
Eq.~\eqref{eq:5D4D_portal_threshold}, because the profile dependence
is already contained in $Y_\nu^{(0)}$ and $Y_N^{(00)}$.

A separate overlap factor is useful only when comparing the localized
and flat theories while keeping the fundamental five-dimensional
brane coefficients fixed. Introducing
\begin{equation}
x
=
|M_5|L,
\end{equation}
one then has, for the hidden-localized branch,
\begin{equation}
\kappa_{5,\rm loop}^{(0)}
=
\kappa_{4,\rm loop}^{\rm ref}
\mathcal F_0(x),
\qquad
\mathcal F_0(x)
=
\left[
\frac{2x}{1-e^{-2x}}
\right]^3
e^{-2x},
\label{eq:5D4D_portal_overlap_summary}
\end{equation}
as derived previously. In the flat-profile limit,
\begin{equation}
x\longrightarrow0,
\qquad
\mathcal F_0(x)\longrightarrow1,
\end{equation}
and the five-dimensional zero-mode result approaches the flat
four-dimensional reference result. In the strongly sequestered
regime, $x\gg1$,
\begin{equation}
\mathcal F_0(x)
\simeq
(2x)^3e^{-2x},
\end{equation}
so the portal is exponentially suppressed when expressed in terms of
the fundamental five-dimensional coefficients.

The relation between the two descriptions can therefore be summarized
as
\begin{equation}
\begin{aligned}
\text{5D theory}
&\xrightarrow{\text{KK reduction}}
\text{infinite 4D KK theory}
\\
&\xrightarrow[\epsilon_n\ll1]{E\ll m_1}
\text{low-energy 4D Majorana theory}
\\
&\xrightarrow{|M_5|L\to0}
\text{flat-profile 4D reference model}.
\end{aligned}
\label{eq:5D4D_reduction_summary}
\end{equation}
The first two steps define the general four-dimensional effective
description. The final step is an additional limit in which the
geometrical localization disappears and the effective couplings
reduce to their flat-profile reference values.

%==========================================================
\section{Finite KK truncation of the radiative portal}
\label{app:KK_truncation}
%==========================================================

In this appendix we give a finite-dimensional prescription for
estimating the contribution of the massive KK tower to the radiative
portal. We restrict the discussion to one five-dimensional sterile
fermion and retain the chiral zero mode together with the first
$N_{\rm KK}$ massive Dirac levels. A finite truncation is useful as a
diagnostic, although the physical result requires a prescription
consistent with the cutoff and locality of the five-dimensional EFT.

After the hidden scalar develops a vacuum expectation value, the
boundary Majorana entries in the free KK basis are
\begin{equation}
\mu_{mn}
=
\frac{v_r}{\sqrt{2}}\,
\widehat y_{5,N}
 f_R^{(m)}(L)f_R^{(n)}(L),
\,\,\,
m,n=0,\ldots,N_{\rm KK}.
\label{eq:KKtr_mu_mn}
\end{equation}
In the left-handed Weyl basis
\begin{equation}
\Xi^T
=
\left(
(N_R^{(0)})^c,\ldots,(N_R^{(N_{\rm KK})})^c,
N_L^{(1)},\ldots,N_L^{(N_{\rm KK})}
\right),
\end{equation}
the truncated symmetric mass matrix is
\begin{equation}
\mathcal M^{(N_{\rm KK})}
=
\begin{pmatrix}
\mu & D\\
D^T & 0
\end{pmatrix},
\label{eq:KKtr_mass_matrix}
\end{equation}
where
\begin{equation}
D_{0n}=0,
\qquad
D_{mn}=m_n\delta_{mn},
\qquad
m,n\geq1.
\label{eq:KKtr_D_matrix}
\end{equation}
A Takagi transformation gives
\begin{equation}
U^T\mathcal M^{(N_{\rm KK})}U
=
\operatorname{diag}(M_0,\ldots,M_{2N_{\rm KK}}),
\,\,\,
M_a\geq0.
\label{eq:KKtr_Takagi}
\end{equation}
If $U_R$ denotes the upper $(N_{\rm KK}+1)$ rows of $U$, the
right-handed profile of the physical state $N_a$ is
\begin{equation}
F_{R,a}^{(N_{\rm KK})}(y)
=
\sum_{n=0}^{N_{\rm KK}}
 f_R^{(n)}(y)(U_R)_{na}.
\label{eq:KKtr_physical_profile}
\end{equation}
The visible- and hidden-boundary couplings are consequently
\begin{equation}
\lambda_a
=
\widehat y_{5,\nu}F_{R,a}^{(N_{\rm KK})}(0),
\,\,\,
g_{ab}
=
\widehat y_{5,N}
F_{R,a}^{(N_{\rm KK})}(L)
F_{R,b}^{(N_{\rm KK})}(L).
\label{eq:KKtr_couplings}
\end{equation}
The complete truncated coefficient can then be written schematically
as
\begin{equation}
\kappa_{\rm loop}^{(N_{\rm KK})}(\mu)
=
\sum_{a,b,c=0}^{2N_{\rm KK}}
 g_{ab}g_{ca}^{*}
 \lambda_b^{*}\lambda_c
 \mathcal I_{abc}(\mu),
\label{eq:KKtr_kappa_full}
\end{equation}
where $\mathcal I_{abc}$ is the renormalized four-dimensional loop
kernel for three physical Majorana masses $M_a,M_b,M_c$ and one
massless lepton propagator. Equation~\eqref{eq:KKtr_kappa_full}
contains the light-state term, pure-KK thresholds, and mixed
light--KK and KK--KK contributions. The finite KK correction is
defined by
\begin{equation}
\Delta\kappa_{\rm KK}^{(N_{\rm KK})}(\mu)
=
\kappa_{\rm loop}^{(N_{\rm KK})}(\mu)
-
\kappa_{\rm loop}^{(0)}(\mu).
\label{eq:KKtr_delta_kappa}
\end{equation}

The first nontrivial truncation retains one massive Dirac level. In
the basis $((N_R^{(0)})^c,(N_R^{(1)})^c,N_L^{(1)})$, the mass matrix
is
\begin{equation}
\mathcal M^{(1)}
=
\begin{pmatrix}
\mu_{00} & \mu_{01} & 0\\
\mu_{01} & \mu_{11} & m_1\\
0 & m_1 & 0
\end{pmatrix}.
\label{eq:KKtr_first_mass_matrix}
\end{equation}
For one sterile generation the boundary matrix factorizes,
\begin{equation}
\mu_{01}=\mu_{00}r_1,
\qquad
\mu_{11}=\mu_{00}r_1^2,
\qquad
r_1=\frac{f_R^{(1)}(L)}{f_R^{(0)}(L)}.
\label{eq:KKtr_first_entries}
\end{equation}
When
\begin{equation}
\epsilon_1
\equiv
\frac{|\mu_{01}|}{m_1}
\simeq
\frac{M_0}{m_1}|r_1|
\ll1,
\label{eq:KKtr_epsilon1}
\end{equation}
the lightest state is zero-mode dominated. This condition does not,
however, imply that the threshold generated by the massive level is
small.

%----------------------------------------------------------
\subsection{Pure first-level estimate}
\label{app:KK_first_level_estimate}
%----------------------------------------------------------

A simple analytic diagnostic is obtained by neglecting the
boundary-induced mixing in the massive sector and retaining only the
pure contribution of one free KK level. For the $n$-th level, the
field-dependent Weyl mass matrix may be written as
\begin{equation}
\mathcal M_n(R,\phi)
=
\begin{pmatrix}
0 & \lambda_n\phi & 0\\
\lambda_n\phi & g_{nn}R & m_n\\
0 & m_n & 0
\end{pmatrix},
\label{eq:KKtr_single_level_field_matrix}
\end{equation}
where $\phi$ denotes a fixed neutral component of the Higgs doublet,
\begin{equation}
\lambda_n=\widehat y_{5,\nu}f_R^{(n)}(0),
\qquad
g_{nn}=\widehat y_{5,N}[f_R^{(n)}(L)]^2.
\label{eq:KKtr_single_level_couplings}
\end{equation}
The two nonzero signed eigenvalues of
Eq.~\eqref{eq:KKtr_single_level_field_matrix} are
\begin{equation}
\mathcal M_{n,\pm}
=
\frac{1}{2}
\left[
g_{nn}R
\pm
\sqrt{g_{nn}^2R^2+4(m_n^2+|\lambda_n\phi|^2)}
\right].
\label{eq:KKtr_single_level_eigenvalues}
\end{equation}
Expanding the one-loop Coleman--Weinberg potential to order
$R^2|\phi|^2$ gives
\begin{equation}
\Delta\kappa_n^{\rm pure}(\mu)
=
-
\frac{|g_{nn}|^2|\lambda_n|^2}{16\pi^2}
\left[
1+2\ln\left(\frac{m_n^2}{\mu^2}\right)
\right].
\label{eq:KKtr_single_level_threshold}
\end{equation}
At the natural matching scale $\mu=m_n$,
\begin{equation}
\Delta\kappa_n^{\rm pure}(m_n)
=
-
\frac{|g_{nn}|^2|\lambda_n|^2}{16\pi^2}.
\label{eq:KKtr_single_level_threshold_scale}
\end{equation}
The light-state threshold at $\mu=M_0$ is
\begin{equation}
\kappa_{\rm loop}^{(0)}(M_0)
=
-
\frac{|g_{00}|^2|\lambda_0|^2}{4\pi^2}.
\label{eq:KKtr_zero_threshold_compare}
\end{equation}
The pure $n$-th-level diagnostic ratio is therefore
\begin{equation}
\mathcal R_n^{\rm pure}
\equiv
\left|
\frac{\Delta\kappa_n^{\rm pure}(m_n)}
{\kappa_{\rm loop}^{(0)}(M_0)}
\right|
=
\frac{1}{4}
\frac{
|f_R^{(n)}(0)|^2|f_R^{(n)}(L)|^4
}{
|f_R^{(0)}(0)|^2|f_R^{(0)}(L)|^4
}.
\label{eq:KKtr_Rn_general_profiles}
\end{equation}

For the hidden-localized branch, define
\begin{equation}
x=|M_5|L,
\qquad
M_{\rm KK}=\frac{\pi}{L}.
\end{equation}
The free profiles at the two boundaries satisfy
\begin{align}
|f_R^{(0)}(0)|^2
&=
\frac{2x}{L(e^{2x}-1)}, \,\,\,
|f_R^{(0)}(L)|^2
=
\frac{2x}{L(1-e^{-2x})},
\nonumber\\
|f_R^{(n)}(0)|^2
&=
|f_R^{(n)}(L)|^2
=
\frac{2}{L}
\frac{n^2\pi^2}{x^2+n^2\pi^2}.
\label{eq:KKtr_boundary_profiles_n}
\end{align}
Consequently,
\begin{equation}
\mathcal R_n^{\rm pure}(x)
=
\frac{
 n^6\pi^6(1-e^{-2x})^3e^{2x}
}{
4x^3(x^2+n^2\pi^2)^3
}
\label{eq:KKtr_Rn_closed}
\end{equation}
For the first two levels,
\begin{align}
\mathcal R_1^{\rm pure}(x)
&=
\frac{
\pi^6(1-e^{-2x})^3e^{2x}
}{
4x^3(x^2+\pi^2)^3
},
\nonumber\\
\mathcal R_2^{\rm pure}(x)
&=
\frac{
16\pi^6(1-e^{-2x})^3e^{2x}
}{
x^3(x^2+4\pi^2)^3
}.
\label{eq:KKtr_R1_R2}
\end{align}
For $|M_5|/M_{\rm KK}=1$, one finds
$\mathcal R_1^{\rm pure}\simeq0.54$ and
$\mathcal R_2^{\rm pure}\simeq2.20$, whereas for
$|M_5|/M_{\rm KK}=2$ the corresponding values are approximately
$2.31$ and $36.1$. Thus even an individual massive level need not be
smaller than the light-state contribution. These numbers are only
diagnostics: they omit mixed terms and compare thresholds evaluated
at different natural matching scales.

%----------------------------------------------------------
\subsection{KK cutoff and interpretation of the truncated sum}
\label{app:KK_cutoff}
%----------------------------------------------------------

The compact continuum theory possesses a formally infinite KK
spectrum. The five-dimensional description is, however, an effective
field theory valid only below a cutoff $\Lambda_5$. Only modes
satisfying
\begin{equation}
m_n^2=M_5^2+n^2M_{\rm KK}^2<\Lambda_5^2
\label{eq:KKtr_cutoff_condition}
\end{equation}
belong to the controlled EFT. The maximal mode number is therefore
\begin{equation}
N_{\rm KK}^{\rm max}
=
\left\lfloor
\frac{\sqrt{\Lambda_5^2-M_5^2}}{M_{\rm KK}}
\right\rfloor
\simeq
\frac{\Lambda_5}{M_{\rm KK}},
\label{eq:KK_max_cutoff}
\end{equation}
where the last expression applies for
$\Lambda_5\gg|M_5|$. The cutoff is set by the lowest among the scale
of the ultraviolet completion, the perturbative strong-coupling
scale of the brane interactions, the inverse brane thickness, and
the fundamental gravitational scale. At the order-of-magnitude
level, perturbativity of the brane couplings requires
\begin{equation}
\frac{|\widehat y_{5,\nu}|^2\Lambda_5}{16\pi^2}\lesssim1,
\qquad
\frac{|\widehat y_{5,N}|^2\Lambda_5^2}{16\pi^2}\lesssim1,
\label{eq:KKtr_NDA_cutoff}
\end{equation}
up to convention-dependent NDA factors.

The pure diagonal contribution does not decrease at large mode
number. Indeed,
\begin{equation}
\lim_{n\to\infty}\mathcal R_n^{\rm pure}(x)
=
\frac{(1-e^{-2x})^3e^{2x}}{4x^3}.
\label{eq:KKtr_Rn_asymptotic}
\end{equation}
Consequently, the formal sum
\begin{equation}
\sum_{n=1}^{N_{\rm KK}}\Delta\kappa_n^{\rm pure}
\label{eq:KKtr_diagonal_sum}
\end{equation}
does not converge when $N_{\rm KK}\to\infty$. This diagonal-only sum
is not the complete five-dimensional matching coefficient and should
not be identified with $\Delta\kappa_{\rm KK}$. In particular, it
removes the relative phases of the boundary wave functions and omits
the mixed terms in Eq.~\eqref{eq:KKtr_kappa_full}.

The importance of these phases can be illustrated by the
opposite-boundary propagator of a massless scalar on a flat interval,
\begin{equation}
\frac{1}{L}
\left[
\frac{1}{p_E^2}
+
2\sum_{n=1}^{\infty}
\frac{(-1)^n}
{p_E^2+n^2\pi^2/L^2}
\right]
=
\frac{1}{p_E\sinh(p_EL)}.
\label{eq:KK_opposite_boundary_identity}
\end{equation}
Although a sum of the absolute values of the individual terms would
not converge, the signed KK sum reconstructs a propagator that is
exponentially suppressed at large Euclidean momentum,
\begin{equation}
G(p_E;0,L)
\simeq
\frac{2}{p_E}e^{-p_EL},
\qquad
p_EL\gg1.
\label{eq:KKtr_nonlocal_asymptotic}
\end{equation}
The fermionic propagator relevant to the present model has a
different numerator and depends on $M_5$, but it exhibits the same
nonlocal exponential damping. Since the portal loop contains two
propagations between the separated boundaries, its genuinely
nonlocal ultraviolet contribution is suppressed as
$e^{-2p_EL}$. After local bulk and single-brane subdivergences are
renormalized, the brane-to-brane contribution is therefore finite.
This behavior is obscured when only the positive diagonal pieces of
the KK expansion are retained; analogous locality-preserving tower
calculations are discussed, for example, in
Refs.~\cite{Rattazzi:2003rj,Gregoire:2004nn,Falkowski:2005zv}.

A consistent finite truncation must consequently retain all physical
states below a common cutoff, diagonalize the complete
Dirac-plus-Majorana mass matrix, and include both diagonal and mixed
terms at a common renormalization scale. A useful numerical test is
\begin{equation}
\mathcal R_{N_{\rm KK}}
=
\left|
\frac{
\Delta\kappa_{\rm KK}^{(N_{\rm KK})}
}{
\kappa_{\rm loop}^{(0)}
}
\right|,
\,\,\,
\mathcal C_{N_{\rm KK}}
=
\left|
\frac{
\kappa_{\rm loop}^{(N_{\rm KK})}
-
\kappa_{\rm loop}^{(N_{\rm KK}-1)}
}{
\kappa_{\rm loop}^{(N_{\rm KK})}
}
\right|.
\label{eq:KKtr_diagnostics}
\end{equation}
The single-state approximation requires
$\mathcal R_{N_{\rm KK}}\ll1$ together with stability under increasing
$N_{\rm KK}$ while remaining below
$N_{\rm KK}^{\rm max}$. The $n=1$ and $n=2$ estimates above show only
that individual massive levels are not parametrically negligible;
they are not a substitute for the complete locality-preserving tower
sum.

\section{Gravitational coupling of the visible and hidden branes}
\label{app:gravity_derivation}

In this appendix, we provide an explicit derivation of the gravitational statements used in Sec.~\ref{sec:gravity}. We consider the flat five-dimensional interval introduced in Sec.~\ref{sec:5D}, with coordinates
\begin{equation}
X^A=(x^\mu,y),
\qquad
A=0,1,2,3,5,
\qquad
y\in[0,L].
\label{eq:app_gravity_coordinates}
\end{equation}
The visible and hidden branes are located at
\begin{equation}
y_{\rm vis}=0,
\qquad
y_{\rm hid}=L.
\label{eq:app_gravity_brane_positions}
\end{equation}
Gravity is assumed to propagate in the full five-dimensional bulk, while the visible- and hidden-sector matter fields are localized on the corresponding boundaries of the interval. This is the standard brane-world structure considered in Refs.~\cite{ArkaniHamed:1998rs,Randall:1999ee,Randall:1999vf,Maartens:2010ar}.

Throughout this appendix, $M_\ast$ denotes the reduced five-dimensional gravitational scale. We assume that the size of the compact dimension is stabilized and that the background geometry can consistently be approximated by the direct product of four-dimensional spacetime and a flat interval. Possible effects associated with a dynamical radion, nonzero brane tensions, a bulk cosmological constant, and gravitational backreaction are discussed at the end of the appendix. In deriving the Newtonian and cosmological limits, we focus on macroscopic sources localized on the two branes. Any non-negligible stress-energy carried by light bulk modes can be included as an additional four-dimensional source after dimensional reduction.

\subsection{Five-dimensional action and induced metrics}

The gravitational action on the interval is
\begin{equation}
S_{\rm grav}
=
\frac{M_\ast^3}{2}
\int_{\mathcal{M}_5}
\dd^5X\,
\sqrt{|g_5|}\,
\mathcal{R}_5
+
S_{\rm GHY},
\label{eq:app_gravity_action}
\end{equation}
where
\begin{equation}
\dd^5X\equiv \dd^4x\,\dd y,
\qquad
g_5\equiv \det g_{AB},
\label{eq:app_gravity_measure}
\end{equation}
and $S_{\rm GHY}$ denotes the Gibbons--Hawking--York boundary term required for a well-defined metric variation on a manifold with boundaries~\cite{York:1972sj,Gibbons:1976ue}. Explicitly,
\begin{equation}
S_{\rm GHY}
=
M_\ast^3
\sum_{i={\rm vis,hid}}
\epsilon_i
\int_{y=y_i}
\dd^4x\,
\sqrt{|\gamma_i|}\,
K_i,
\label{eq:app_gravity_GHY}
\end{equation}
where $\gamma_{i,\mu\nu}$ is the metric induced on the $i$-th brane, $K_i$ is the trace of its extrinsic curvature, and $\epsilon_i$ specifies the orientation of the outward-pointing normal. The boundary term cancels the normal-derivative terms arising from the variation of the Einstein--Hilbert action and does not modify the bulk Einstein equations.

The embedding of the $i$-th brane is denoted by
\begin{equation}
X_i^A(x)
=
\left(
x^\mu,y_i
\right).
\label{eq:app_gravity_embedding}
\end{equation}
The corresponding tangent vectors are
\begin{equation}
e^A_{i,\mu}
\equiv
\frac{\partial X_i^A}{\partial x^\mu}
=
\delta^A_\mu,
\label{eq:app_gravity_tangent_vectors}
\end{equation}
and the induced metric is therefore
\begin{equation}
\gamma_{i,\mu\nu}(x)
=
g_{AB}\left(X_i(x)\right)e^A_{i,\mu}e^B_{i,\nu}
=
g_{\mu\nu}(x,y_i).
\label{eq:app_gravity_induced_metric}
\end{equation}
The visible- and hidden-brane actions can then be written covariantly as
\begin{equation}
S_i
=
\int_{y=y_i}
\dd^4x\,
\sqrt{|\gamma_i|}\,
\mathcal{L}_i
\left(
\gamma_{i,\mu\nu},\psi_i
\right),
\qquad
i={\rm vis,hid}.
\label{eq:app_gravity_brane_actions}
\end{equation}
The complete action relevant for the gravitational discussion is
\begin{equation}
S
=
S_{\rm grav}
+
S_{\rm bulk}
+
S_{\rm vis}
+
S_{\rm hid},
\label{eq:app_gravity_total_action}
\end{equation}
where $S_{\rm bulk}$ includes the bulk sterile-neutrino sector and any additional fields required to stabilize the compact dimension. Its stress-energy is retained in the five-dimensional field equations below.

\subsection{Covariant stress-energy tensors localized on the branes}

The five-dimensional bulk stress-energy tensor is defined by
\begin{equation}
T_{AB}^{\rm bulk}
\equiv
-\frac{2}{\sqrt{|g_5|}}
\frac{\delta S_{\rm bulk}}{\delta g^{AB}}.
\label{eq:app_gravity_bulk_stress_definition}
\end{equation}
Similarly, the four-dimensional stress-energy tensor localized on the $i$-th brane is
\begin{equation}
T_{i}^{\mu\nu}
\equiv
\frac{2}{\sqrt{|\gamma_i|}}
\frac{\delta S_i}
{\delta\gamma_{i,\mu\nu}}.
\label{eq:app_gravity_brane_stress_definition}
\end{equation}
The variation of the brane action is therefore
\begin{equation}
\delta S_i
=
\frac{1}{2}
\int_{y=y_i}
\dd^4x\,
\sqrt{|\gamma_i|}\,
T_i^{\mu\nu}
\delta\gamma_{i,\mu\nu}.
\label{eq:app_gravity_brane_variation_1}
\end{equation}
Using
\begin{equation}
\delta\gamma_{i,\mu\nu}
=
e^A_{i,\mu}e^B_{i,\nu}
\delta g_{AB}(x,y_i),
\label{eq:app_gravity_induced_metric_variation}
\end{equation}
and inserting
\begin{equation}
1
=
\int_0^L\dd y\,
\delta(y-y_i),
\label{eq:app_gravity_delta_identity}
\end{equation}
Eq.~\eqref{eq:app_gravity_brane_variation_1} becomes
\begin{align}
\delta S_i
&=
\frac{1}{2}
\int\dd^4x\int_0^L\dd y\,
\sqrt{|\gamma_i|}
\delta(y-y_i)
T_i^{\mu\nu}
e^A_{i,\mu}e^B_{i,\nu}
\delta g_{AB}(x,y)
\\
&=
\frac{1}{2}
\int\dd^5X\,
\sqrt{|g_5|}\,
T_{i,{\rm loc}}^{AB}(X)
\delta g_{AB}(X),
\label{eq:app_gravity_brane_variation_2}
\end{align}
where the corresponding five-dimensional localized source is
\begin{equation}
T_{i,{\rm loc}}^{AB}(X)
=
\frac{\sqrt{|\gamma_i|}}{\sqrt{|g_5|}}
\,
T_i^{\mu\nu}(x)
e^A_{i,\mu}e^B_{i,\nu}
\delta(y-y_i).
\label{eq:app_gravity_covariant_localized_stress}
\end{equation}
The total brane-localized contribution is consequently
\begin{equation}
T_{\rm brane}^{AB}
=
T_{{\rm vis},{\rm loc}}^{AB}
+
T_{{\rm hid},{\rm loc}}^{AB}.
\label{eq:app_gravity_total_brane_stress}
\end{equation}

In Gaussian-normal coordinates,
\begin{equation}
\dd s_5^2
=
g_{\mu\nu}(x,y)\dd x^\mu\dd x^\nu
-b^2(x,y)\dd y^2,
\label{eq:app_gravity_gaussian_normal_metric}
\end{equation}
one has
\begin{equation}
\sqrt{|g_5|}
=
b\sqrt{|\gamma_i|}
\label{eq:app_gravity_determinant_relation}
\end{equation}
at the position of the brane. Equation~\eqref{eq:app_gravity_covariant_localized_stress} then reduces to
\begin{equation}
T_{i,{\rm loc}}^{AB}
=
\frac{\delta(y-y_i)}{b(x,y)}
T_i^{\mu\nu}
\delta^A_\mu\delta^B_\nu.
\label{eq:app_gravity_localized_stress_gaussian}
\end{equation}
For the flat background used in the paper, $b=1$, and hence
\begin{equation}
T_{\rm brane}^{AB}
=
T_{\rm vis}^{\mu\nu}(x)\delta^A_\mu\delta^B_\nu\delta(y)
+
T_{\rm hid}^{\mu\nu}(x)\delta^A_\mu\delta^B_\nu\delta(y-L).
\label{eq:app_gravity_brane_stress_flat}
\end{equation}
This is the covariant origin of the delta-function expression used in
the main text. In the Einstein equations below, the covariant tensor is
obtained as
$T_{AB}^{\rm brane}=g_{AC}g_{BD}T_{\rm brane}^{CD}$.

\subsection{Five-dimensional Einstein equations}

The variation of the gravitational action with respect to the inverse metric gives
\begin{equation}
\delta S_{\rm grav}
=
\frac{M_\ast^3}{2}
\int\dd^5X\,
\sqrt{|g_5|}\,
G_{5,AB}\,
\delta g^{AB},
\label{eq:app_gravity_EH_variation}
\end{equation}
after including the boundary term in Eq.~\eqref{eq:app_gravity_GHY}. The five-dimensional Einstein tensor is
\begin{equation}
G_{5,AB}
=
R_{5,AB}
-\frac{1}{2}g_{AB}\mathcal{R}_5.
\label{eq:app_gravity_Einstein_tensor}
\end{equation}
The variation of the matter actions is
\begin{equation}
\delta S_{\rm matter}
=
-\frac{1}{2}
\int\dd^5X\,
\sqrt{|g_5|}
\left(
T_{AB}^{\rm bulk}
+
T_{AB}^{\rm brane}
\right)
\delta g^{AB}.
\label{eq:app_gravity_matter_variation}
\end{equation}
The condition
\begin{equation}
\delta S=0
\label{eq:app_gravity_stationarity}
\end{equation}
therefore gives the five-dimensional Einstein equation
\begin{equation}
M_\ast^3G_{5,AB}
=
T_{AB}^{\rm bulk}
+
T_{AB}^{\rm brane}.
\label{eq:app_gravity_5D_Einstein_equation}
\end{equation}
Substituting Eq.~\eqref{eq:app_gravity_brane_stress_flat}, the components parallel to the branes satisfy
\begin{equation}
M_\ast^3G_{5,\mu\nu}
=
T_{\mu\nu}^{\rm bulk}
+
T_{\mu\nu}^{\rm vis}(x)\delta(y)
+
T_{\mu\nu}^{\rm hid}(x)\delta(y-L).
\label{eq:app_gravity_5D_Einstein_parallel}
\end{equation}
Thus the visible- and hidden-brane stress-energy tensors enter the same five-dimensional Einstein equation. Their localization at different values of $y$ does not remove either source from the gravitational field equations.

More generally, one may include a bulk cosmological constant $\Lambda_5$ and brane tensions $\lambda_i$ through
\begin{equation}
S_{\Lambda,\lambda}
=
-M_\ast^3
\int\dd^5X\,
\sqrt{|g_5|}\,
\Lambda_5
-
\sum_i
\int_{y=y_i}
\dd^4x\,
\sqrt{|\gamma_i|}\,
\lambda_i.
\label{eq:app_gravity_cosmological_tension_action}
\end{equation}
The Einstein equation would then become
\begin{widetext}
\begin{equation}
M_\ast^3\left(G_{5,AB}+\Lambda_5g_{AB}\right)
=
T_{AB}^{\rm bulk}
+
T_{AB}^{\rm brane}
-
\sum_i\lambda_i
\frac{\sqrt{|\gamma_i|}}{\sqrt{|g_5|}}
\gamma_{i,\mu\nu}e^\mu_{i,A}e^\nu_{i,B}\delta(y-y_i).
\label{eq:app_gravity_Einstein_with_tensions}
\end{equation}
\end{widetext}
Here $e^\mu_{i,A}$ denotes the dual tangent basis on the $i$-th brane. The flat-background approximation adopted in this work corresponds either to vanishing $\Lambda_5$ and $\lambda_i$ or to a completion in which their contributions are appropriately cancelled or tuned. These background terms do not alter the conclusion that matter on both branes sources the common bulk metric.

\subsection{Dimensional reduction and the four-dimensional Planck scale}

We next derive the relation between the five- and four-dimensional
gravitational scales. At energies below the first gravitational KK
threshold, given in Eq.~\eqref{eq:app_gravity_first_KK_scale}, the
metric is dominated by its $y$-independent zero mode, and the
background can be approximated by the direct-product ansatz
\begin{equation}
\dd s_5^2
=
g_{\mu\nu}^{(4)}(x)
\dd x^\mu\dd x^\nu
-\dd y^2.
\label{eq:app_gravity_product_metric}
\end{equation}
The four-dimensional metric is independent of $y$, and all mixed metric components vanish:
\begin{equation}
\partial_y g_{\mu\nu}^{(4)}=0,
\qquad
g_{\mu5}=0,
\qquad
g_{55}=-1.
\label{eq:app_gravity_product_conditions}
\end{equation}
For this direct-product geometry, the only nonzero Christoffel symbols are the ordinary four-dimensional ones,
\begin{equation}
\Gamma^\rho_{\mu\nu}[g_5]
=
\Gamma^\rho_{\mu\nu}[g_4],
\label{eq:app_gravity_christoffel_reduction}
\end{equation}
while
\begin{equation}
\Gamma^5_{AB}=0,
\qquad
\Gamma^\mu_{A5}=0.
\label{eq:app_gravity_mixed_christoffel}
\end{equation}
It follows directly that
\begin{equation}
R_{5,\mu\nu}
=
R_{4,\mu\nu},
\qquad
R_{5,\mu5}=0,
\qquad
R_{5,55}=0,
\label{eq:app_gravity_Ricci_components}
\end{equation}
and therefore
\begin{equation}
\mathcal{R}_5
=
\mathcal{R}_4.
\label{eq:app_gravity_Ricci_reduction}
\end{equation}
The determinant also factorizes:
\begin{equation}
\sqrt{|g_5|}
=
\sqrt{|g_4|}.
\label{eq:app_gravity_determinant_reduction}
\end{equation}
For the unwarped direct-product metric, the extrinsic curvature of the
constant-$y$ boundaries vanishes, so the Gibbons--Hawking--York term
does not contribute to the background reduction. Substituting
Eqs.~\eqref{eq:app_gravity_Ricci_reduction} and
\eqref{eq:app_gravity_determinant_reduction} into
Eq.~\eqref{eq:app_gravity_action} gives
\begin{equation}
S_{\rm grav}
=
\frac{M_\ast^3}{2}
\int\dd^4x\int_0^L\dd y\,\sqrt{|g_4|}\,\mathcal{R}_4
=
\frac{M_\ast^3L}{2}
\int\dd^4x\,\sqrt{|g_4|}\,\mathcal{R}_4.
\label{eq:app_gravity_reduced_action}
\end{equation}
The canonically normalized four-dimensional Einstein--Hilbert action is
\begin{equation}
S_{\rm EH}^{(4)}
=
\frac{M_{\rm Pl}^2}{2}
\int\dd^4x\,
\sqrt{|g_4|}\,
\mathcal{R}_4.
\label{eq:app_gravity_4D_EH}
\end{equation}
Comparison with Eq.~\eqref{eq:app_gravity_reduced_action} yields the standard volume-suppression relation~\cite{ArkaniHamed:1998rs,Maartens:2010ar}
\begin{equation}
M_{\rm Pl}^2
=
M_\ast^3L.
\label{eq:app_gravity_Planck_relation}
\end{equation}
Here $M_{\rm Pl}$ is the reduced four-dimensional Planck mass,
\begin{equation}
M_{\rm Pl}
=
\frac{1}{\sqrt{8\pi G_N}}.
\label{eq:app_gravity_reduced_Planck_definition}
\end{equation}
If the action is instead integrated over the full covering circle of length $2L$, the corresponding relation is
\begin{equation}
M_{\rm Pl}^2
=
2M_\ast^3L.
\label{eq:app_gravity_Planck_covering_circle}
\end{equation}
Equation~\eqref{eq:app_gravity_Planck_relation} is the appropriate convention when the physical integration domain is the orbifold interval $y\in[0,L]$, as in the present paper.

\subsection{Kaluza--Klein decomposition of the graviton}

The universal coupling of the graviton zero mode to both branes can be seen directly by expanding the five-dimensional metric around the flat background:
\begin{equation}
g_{AB}(x,y)
=
\eta_{AB}
+
\frac{2}{M_\ast^{3/2}}
h_{AB}(x,y).
\label{eq:app_gravity_linearized_metric}
\end{equation}
The normalization in Eq.~\eqref{eq:app_gravity_linearized_metric} is chosen so that the five-dimensional graviton field has a canonical kinetic term. Under the orbifold parity, the tensor components have the assignments
\begin{equation}
h_{\mu\nu}(x,-y)
=
+h_{\mu\nu}(x,y),
\label{eq:app_gravity_tensor_parity}
\end{equation}
\begin{equation}
h_{\mu5}(x,-y)
=
-h_{\mu5}(x,y),
\label{eq:app_gravity_vector_parity}
\end{equation}
and
\begin{equation}
h_{55}(x,-y)
=
+h_{55}(x,y).
\label{eq:app_gravity_scalar_parity}
\end{equation}
Consequently, $h_{\mu\nu}$ can be nonzero on both branes and contains a massless four-dimensional tensor zero mode. The odd field $h_{\mu5}$ vanishes at the fixed points and has no zero mode. The even scalar component $h_{55}$ contains the radion degree of freedom, which is assumed to be stabilized.

Focusing on the transverse-traceless tensor fluctuations, we write
\begin{equation}
h_{\mu\nu}(x,y)
=
\sum_{n=0}^{\infty}
h_{\mu\nu}^{(n)}(x)
f_n^{\rm grav}(y).
\label{eq:app_gravity_KK_expansion}
\end{equation}
For a flat interval, the mode functions satisfy
\begin{equation}
-\frac{\dd^2f_n^{\rm grav}}{\dd y^2}
=
\left(m_n^{\rm grav}\right)^2f_n^{\rm grav}
\label{eq:app_gravity_mode_equation}
\end{equation}
with Neumann boundary conditions
\begin{equation}
\left.
\frac{\dd f_n^{\rm grav}}{\dd y}
\right|_{y=0}
=
\left.
\frac{\dd f_n^{\rm grav}}{\dd y}
\right|_{y=L}
=
0.
\label{eq:app_gravity_Neumann_conditions}
\end{equation}
They are normalized according to
\begin{equation}
\int_0^L\dd y\,
f_n^{\rm grav}(y)f_m^{\rm grav}(y)
=
\delta_{nm}.
\label{eq:app_gravity_mode_orthonormality}
\end{equation}
The normalized solutions are
\begin{equation}
f_0^{\rm grav}(y)
=
\frac{1}{\sqrt{L}},
\qquad
m_0^{\rm grav}=0,
\label{eq:app_gravity_zero_mode}
\end{equation}
and
\begin{equation}
f_n^{\rm grav}(y)
=
\sqrt{\frac{2}{L}}
\cos\left(
\frac{n\pi y}{L}
\right),
\qquad
m_n^{\rm grav}
=
\frac{n\pi}{L},
\qquad
n\geq1.
\label{eq:app_gravity_massive_modes}
\end{equation}
The first gravitational KK scale is consequently
\begin{equation}
M_{\rm KK}^{\rm grav}
=
\frac{\pi}{L}.
\label{eq:app_gravity_first_KK_scale}
\end{equation}
In the minimal flat construction,
$M_{\rm KK}^{\rm grav}$ coincides with the geometrical
compactification scale $M_{\rm KK}$ defined in
Eq.~\eqref{eq:app_MKK_definition}. Bulk masses and boundary terms can
modify the fermionic spectrum without changing the gravitational
tensor eigenfunctions~\cite{ArkaniHamed:1998rs,Maartens:2010ar}.

The zero-mode wave function is constant along the fifth dimension:
\begin{equation}
f_0^{\rm grav}(0)
=
f_0^{\rm grav}(L)
=
\frac{1}{\sqrt{L}}.
\label{eq:app_gravity_zero_mode_brane_values}
\end{equation}
The massive-mode wave functions evaluated on the two branes are instead
\begin{equation}
f_n^{\rm grav}(0)
=
\sqrt{\frac{2}{L}},
\label{eq:app_gravity_massive_visible_value}
\end{equation}
and
\begin{equation}
f_n^{\rm grav}(L)
=
(-1)^n
\sqrt{\frac{2}{L}}.
\label{eq:app_gravity_massive_hidden_value}
\end{equation}
The factor $(-1)^n$ affects the relative coupling of a massive KK mode to the two branes, but it is absent for the zero mode.

\subsection{Coupling of the graviton modes to brane matter}

The coupling of the metric perturbation to matter follows directly from the variation of the brane action. At linear order,
\begin{equation}
\delta\gamma_{i,\mu\nu}
=
\frac{2}{M_\ast^{3/2}}
h_{\mu\nu}(x,y_i).
\label{eq:app_gravity_linear_induced_variation}
\end{equation}
Substituting this relation into Eq.~\eqref{eq:app_gravity_brane_variation_1} gives
\begin{equation}
S_{{\rm int},i}
=
-\frac{1}{M_\ast^{3/2}}
\int\dd^4x\,
h_{\mu\nu}(x,y_i)
T_i^{\mu\nu}(x).
\label{eq:app_gravity_single_brane_interaction}
\end{equation}
Using the KK decomposition in Eq.~\eqref{eq:app_gravity_KK_expansion},
\begin{equation}
S_{{\rm int},i}
=
-\frac{1}{M_\ast^{3/2}}
\sum_{n=0}^{\infty}
f_n^{\rm grav}(y_i)
\int\dd^4x\,
h_{\mu\nu}^{(n)}
T_i^{\mu\nu}.
\label{eq:app_gravity_single_brane_KK_interaction}
\end{equation}
Summing over the visible and hidden branes gives
\begin{equation}
S_{\rm int}
=
-\frac{1}{M_\ast^{3/2}}
\sum_{n=0}^{\infty}
\int\dd^4x\,
h_{\mu\nu}^{(n)}
\left[
f_n^{\rm grav}(0)T_{\rm vis}^{\mu\nu}
+
f_n^{\rm grav}(L)T_{\rm hid}^{\mu\nu}
\right].
\label{eq:app_gravity_total_KK_interaction}
\end{equation}
For the zero mode, Eqs.~\eqref{eq:app_gravity_zero_mode_brane_values} and \eqref{eq:app_gravity_Planck_relation} imply
\begin{align}
S_{\rm int}^{(0)}
&=
-\frac{1}{M_\ast^{3/2}\sqrt{L}}
\int\dd^4x\,
h_{\mu\nu}^{(0)}
\left(
T_{\rm vis}^{\mu\nu}
+
T_{\rm hid}^{\mu\nu}
\right)
\nonumber\\
&=
-\frac{1}{M_{\rm Pl}}
\int\dd^4x\,
h_{\mu\nu}^{(0)}
\left(
T_{\rm vis}^{\mu\nu}
+
T_{\rm hid}^{\mu\nu}
\right).
\label{eq:app_gravity_zero_mode_interaction}
\end{align}
This equation gives the explicit result underlying the discussion in the main text: the four-dimensional massless graviton couples with exactly the same gravitational strength to the visible- and hidden-brane stress-energy tensors.

For the massive modes, Eqs.~\eqref{eq:app_gravity_massive_visible_value} and \eqref{eq:app_gravity_massive_hidden_value} give
\begin{equation}
S_{\rm int}^{\rm massive}
=
-\frac{\sqrt{2}}{M_{\rm Pl}}
\sum_{n=1}^{\infty}
\int\dd^4x\,
h_{\mu\nu}^{(n)}
\left[
T_{\rm vis}^{\mu\nu}
+
(-1)^nT_{\rm hid}^{\mu\nu}
\right].
\label{eq:app_gravity_massive_mode_interaction}
\end{equation}
The relative sign of the massive-mode coupling reflects the values of the cosine wave functions at opposite ends of the interval. These modes are irrelevant for ordinary long-distance gravity when
\begin{equation}
r
\gg
\frac{1}{M_{\rm KK}^{\rm grav}}
=
\frac{L}{\pi},
\label{eq:app_gravity_long_distance_condition}
\end{equation}
because their exchange is Yukawa suppressed.

\subsection{Low-energy four-dimensional effective action}

At energies
\begin{equation}
E
\ll
M_{\rm KK}^{\rm grav},
\label{eq:app_gravity_low_energy_condition}
\end{equation}
the massive gravitational KK modes can be integrated out. Assuming that the radion is stabilized, the leading low-energy effective action is
\begin{widetext}
\begin{equation}
S_{\rm eff}^{(4)}
=
\int\dd^4x\,\sqrt{|g_4|}
\left[
\frac{M_{\rm Pl}^2}{2}\mathcal{R}_4
+
\mathcal{L}_{\rm vis}
+
\mathcal{L}_{\rm hid}
+
\mathcal{L}_{\rm bulk,eff}^{(4)}
\right]
+
\Delta S_{\rm KK}.
\label{eq:app_gravity_low_energy_action}
\end{equation}
\end{widetext}
Here $\mathcal{L}_{\rm bulk,eff}^{(4)}$ contains any light bulk zero
modes, while $\Delta S_{\rm KK}$ collects higher-derivative operators
generated by the massive gravitational tower and suppressed by powers
of $E/M_{\rm KK}^{\rm grav}$.

Varying Eq.~\eqref{eq:app_gravity_low_energy_action} with respect to $g_4^{\mu\nu}$ gives
\begin{equation}
M_{\rm Pl}^2G_{\mu\nu}^{(4)}
=
T_{\mu\nu}^{\rm vis}
+
T_{\mu\nu}^{\rm hid}
+
T_{\mu\nu}^{\rm bulk,eff}
+
\Delta T_{\mu\nu}^{\rm KK}.
\label{eq:app_gravity_effective_Einstein_equation}
\end{equation}
Here $\Delta T_{\mu\nu}^{\rm KK}$ denotes the corresponding
higher-derivative corrections and is suppressed by
$E^2/(M_{\rm KK}^{\rm grav})^2$. Thus, all light
four-dimensional degrees of freedom source the long-distance
gravitational field. In the remainder of this appendix
we set $T_{\mu\nu}^{\rm bulk,eff}=0$ in order to isolate the
gravitational response to visible- and hidden-brane matter.

The four-dimensional Bianchi identity,
\begin{equation}
\nabla^\mu
G_{\mu\nu}^{(4)}
=
0,
\label{eq:app_gravity_Bianchi_identity}
\end{equation}
implies
\begin{equation}
\nabla^\mu
\left(
T_{\mu\nu}^{\rm vis}
+
T_{\mu\nu}^{\rm hid}
\right)
=
0.
\label{eq:app_gravity_total_conservation}
\end{equation}
If non-gravitational energy transfer between the two branes and the
bulk is negligible, the visible- and hidden-sector stress-energy
tensors are separately conserved to leading order:
\begin{equation}
\nabla^\mu T_{\mu\nu}^{\rm vis}
=
0,
\qquad
\nabla^\mu T_{\mu\nu}^{\rm hid}
=
0.
\label{eq:app_gravity_separate_conservation}
\end{equation}

\subsection{Newtonian limit}

To obtain the Newtonian limit, we write the weak-field four-dimensional metric as
\begin{equation}
\dd s_4^2
=
\left(
1+2\Phi_N
\right)\dd t^2
-
\left(
1-2\Psi_N
\right)
\delta_{ij}\dd x^i\dd x^j.
\label{eq:app_gravity_Newtonian_metric}
\end{equation}
For nonrelativistic matter with negligible anisotropic stress,
\begin{equation}
T_{00}
\simeq
\rho,
\qquad
|T_{0i}|
\ll
T_{00},
\qquad
|T_{ij}|
\ll
T_{00},
\label{eq:app_gravity_nonrelativistic_stress}
\end{equation}
and
\begin{equation}
\Phi_N
=
\Psi_N.
\label{eq:app_gravity_equal_potentials}
\end{equation}
At linear order, the $00$ component of the Einstein tensor is
\begin{equation}
G_{00}^{(4)}
\simeq
2\nabla^2\Phi_N.
\label{eq:app_gravity_linear_G00}
\end{equation}
Equation~\eqref{eq:app_gravity_effective_Einstein_equation} therefore gives
\begin{equation}
2M_{\rm Pl}^2
\nabla^2\Phi_N
=
\rho_{\rm vis}
+
\rho_{\rm hid}.
\label{eq:app_gravity_Poisson_intermediate}
\end{equation}
Using
\begin{equation}
G_N
=
\frac{1}{8\pi M_{\rm Pl}^2},
\label{eq:app_gravity_Newton_constant}
\end{equation}
one obtains
\begin{equation}
\nabla^2\Phi_N
=
4\pi G_N
\left(
\rho_{\rm vis}
+
\rho_{\rm hid}
\right).
\label{eq:app_gravity_Poisson_equation}
\end{equation}
For a visible point mass $m_{\rm vis}$ and a hidden-brane point mass $m_{\rm hid}$ located at four-dimensional positions $\mathbf{x}_{\rm vis}$ and $\mathbf{x}_{\rm hid}$,
\begin{equation}
\rho_{\rm vis}
=
m_{\rm vis}
\delta^{(3)}
\left(
\mathbf{x}-\mathbf{x}_{\rm vis}
\right),
\label{eq:app_gravity_visible_point_source}
\end{equation}
\begin{equation}
\rho_{\rm hid}
=
m_{\rm hid}
\delta^{(3)}
\left(
\mathbf{x}-\mathbf{x}_{\rm hid}
\right),
\label{eq:app_gravity_hidden_point_source}
\end{equation}
the long-distance solution is
\begin{equation}
\Phi_N(\mathbf{x})
=
-\frac{G_Nm_{\rm vis}}
{|\mathbf{x}-\mathbf{x}_{\rm vis}|}
-\frac{G_Nm_{\rm hid}}
{|\mathbf{x}-\mathbf{x}_{\rm hid}|}.
\label{eq:app_gravity_Newtonian_solution}
\end{equation}
The hidden-brane source therefore produces the same inverse-distance gravitational potential as an ordinary visible-sector source.

\subsection{Static bulk Green function and KK corrections}

The approach to the four-dimensional limit can be displayed explicitly
through the static Green function. The scalar kernel below captures the
spatial and KK-mass dependence common to the tensor propagator; the
polarization structure is discussed separately. It is determined by
\begin{equation}
\left[
-\nabla_{\mathbf{r}}^2
-\partial_y^2
\right]
\mathcal{G}
\left(
\mathbf{r};y,y'
\right)
=
\delta^{(3)}(\mathbf{r})
\delta(y-y'),
\label{eq:app_gravity_static_Green_equation}
\end{equation}
with Neumann boundary conditions at $y=0$ and $y=L$. Expanding in the normalized eigenfunctions gives
\begin{equation}
\mathcal{G}
\left(
r;y,y'
\right)
=
\sum_{n=0}^{\infty}
f_n^{\rm grav}(y)f_n^{\rm grav}(y')
\frac{\exp\!\left(-m_n^{\rm grav}r\right)}{4\pi r},
\label{eq:app_gravity_static_Green_KK}
\end{equation}
where
\begin{equation}
r
=
|\mathbf{r}|.
\label{eq:app_gravity_radial_distance}
\end{equation}

For two sources localized on the visible brane,
\begin{equation}
\mathcal{G}(r;0,0)
=
\frac{1}{4\pi Lr}
\left[
1
+
2\sum_{n=1}^{\infty}
\exp\left(
-\frac{n\pi r}{L}
\right)
\right].
\label{eq:app_gravity_Green_same_sum}
\end{equation}
Defining
\begin{equation}
q
\equiv
\exp\left(
-\frac{\pi r}{L}
\right),
\label{eq:app_gravity_q_definition}
\end{equation}
the geometric series gives
\begin{equation}
1+2\sum_{n=1}^{\infty}q^n
=
\frac{1+q}{1-q}
=
\coth\left(
\frac{\pi r}{2L}
\right).
\label{eq:app_gravity_same_series}
\end{equation}
Hence
\begin{equation}
\mathcal{G}(r;0,0)
=
\frac{1}{4\pi Lr}
\coth\left(
\frac{\pi r}{2L}
\right).
\label{eq:app_gravity_Green_same_exact}
\end{equation}

For one source on each brane, the alternating factor in Eq.~\eqref{eq:app_gravity_massive_hidden_value} gives
\begin{equation}
\mathcal{G}(r;0,L)
=
\frac{1}{4\pi Lr}
\left[
1
+
2\sum_{n=1}^{\infty}
(-1)^n
\exp\left(
-\frac{n\pi r}{L}
\right)
\right].
\label{eq:app_gravity_Green_opposite_sum}
\end{equation}
The alternating series satisfies
\begin{equation}
1
+
2\sum_{n=1}^{\infty}
(-1)^nq^n
=
\frac{1-q}{1+q}
=
\tanh\left(
\frac{\pi r}{2L}
\right),
\label{eq:app_gravity_opposite_series}
\end{equation}
so that
\begin{equation}
\mathcal{G}(r;0,L)
=
\frac{1}{4\pi Lr}
\tanh\left(
\frac{\pi r}{2L}
\right).
\label{eq:app_gravity_Green_opposite_exact}
\end{equation}

At distances much larger than the compactification length,
\begin{equation}
r\gg L,
\label{eq:app_gravity_large_r}
\end{equation}
the two Green functions become
\begin{equation}
\mathcal{G}(r;0,0)
=
\frac{1}{4\pi Lr}
\left[
1
+
2\exp\left(
-\frac{\pi r}{L}
\right)
+
\mathcal{O}
\left(
\exp\left(
-\frac{2\pi r}{L}
\right)
\right)
\right],
\label{eq:app_gravity_Green_same_long}
\end{equation}
and
\begin{equation}
\mathcal{G}(r;0,L)
=
\frac{1}{4\pi Lr}
\left[
1
-
2\exp\left(
-\frac{\pi r}{L}
\right)
+
\mathcal{O}
\left(
\exp\left(
-\frac{2\pi r}{L}
\right)
\right)
\right].
\label{eq:app_gravity_Green_opposite_long}
\end{equation}
Their leading terms are identical:
\begin{equation}
\mathcal{G}(r;0,0)
\simeq
\mathcal{G}(r;0,L)
\simeq
\frac{1}{4\pi Lr}.
\label{eq:app_gravity_Green_universal_long}
\end{equation}
This is the Green-function manifestation of the universal zero-mode coupling. Differences between same-brane and opposite-brane gravity are exponentially suppressed by the lightest KK mass,
\begin{equation}
\exp\left(
-\frac{\pi r}{L}
\right)
=
\exp\left(
-M_{\rm KK}^{\rm grav}r
\right).
\label{eq:app_gravity_KK_Yukawa_suppression}
\end{equation}

The exact tensor gravitational potential includes polarization-dependent coefficients and possible radion exchange in addition to the scalar Green function derived above. These factors affect the numerical coefficients multiplying the massive-mode corrections, but not the universal massless contribution. Consequently, after radion stabilization, the potential between any two nonrelativistic sources satisfies
\begin{equation}
V_{ij}(r)
=
-\frac{G_Nm_im_j}{r}
\left[
1
+
\mathcal{O}
\left(
\exp\left(
-M_{\rm KK}^{\rm grav}r
\right)
\right)
\right],
\label{eq:app_gravity_general_long_potential}
\end{equation}
independently of whether the sources are localized on the same brane or on opposite branes.

At distances much shorter than the compactification length, the same-brane Green function behaves as
\begin{equation}
\mathcal{G}(r;0,0)
\simeq
\frac{1}{2\pi^2r^2},
\qquad
r\ll L,
\label{eq:app_gravity_same_short_distance}
\end{equation}
which displays the inverse-square potential characteristic of four spatial dimensions, including the image contribution associated with a source on an orbifold boundary. In contrast, $\mathcal{G}(r;0,L)$ remains finite as $r\rightarrow0$ because the two sources remain separated by the nonzero proper distance $L$ in the fifth dimension:
\begin{equation}
\mathcal{G}(r;0,L)
\longrightarrow
\frac{1}{8L^2},
\qquad
r\rightarrow0.
\label{eq:app_gravity_opposite_short_distance}
\end{equation}
These short-distance expressions are not relevant for galactic or cosmological gravity, for which $r\gg L$ by many orders of magnitude.

\subsection{Cosmological equations}

The same conclusion follows in a homogeneous and isotropic
cosmological background. Once the compactification radius is stabilized,
the ordinary four-dimensional Friedmann equations are recovered at
energies below the gravitational KK and radion scales
\cite{Csaki:1999mp,Maartens:2010ar}. The four-dimensional metric is
\begin{equation}
\dd s_4^2
=
\dd t^2
-
a^2(t)
\delta_{ij}\dd x^i\dd x^j.
\label{eq:app_gravity_FRW_metric}
\end{equation}
The total stress-energy tensor is
\begin{equation}
T^\mu{}_\nu
=
\operatorname{diag}
\left(
\rho_{\rm tot},
-p_{\rm tot},
-p_{\rm tot},
-p_{\rm tot}
\right),
\label{eq:app_gravity_total_perfect_fluid}
\end{equation}
with
\begin{equation}
\rho_{\rm tot}
=
\rho_{\rm vis}
+
\rho_{\rm hid},
\qquad
p_{\rm tot}
=
p_{\rm vis}
+
p_{\rm hid}.
\label{eq:app_gravity_total_density_pressure}
\end{equation}
The $00$ component of Eq.~\eqref{eq:app_gravity_effective_Einstein_equation} gives
\begin{equation}
3M_{\rm Pl}^2H^2
=
\rho_{\rm vis}
+
\rho_{\rm hid},
\label{eq:app_gravity_Friedmann_equation}
\end{equation}
where
\begin{equation}
H
\equiv
\frac{\dot a}{a}.
\label{eq:app_gravity_Hubble_definition}
\end{equation}
The spatial components give
\begin{equation}
-2M_{\rm Pl}^2\dot H
=
\rho_{\rm vis}
+
p_{\rm vis}
+
\rho_{\rm hid}
+
p_{\rm hid}.
\label{eq:app_gravity_Raychaudhuri_equation}
\end{equation}
If the two sectors do not exchange energy appreciably, their continuity equations are
\begin{equation}
\dot\rho_{\rm vis}
+
3H
\left(
\rho_{\rm vis}
+
p_{\rm vis}
\right)
=
0,
\label{eq:app_gravity_visible_continuity}
\end{equation}
and
\begin{equation}
\dot\rho_{\rm hid}
+
3H
\left(
\rho_{\rm hid}
+
p_{\rm hid}
\right)
=
0.
\label{eq:app_gravity_hidden_continuity}
\end{equation}
For nonrelativistic hidden-sector dark matter,
\begin{equation}
p_{\rm hid}
\ll
\rho_{\rm hid},
\label{eq:app_gravity_pressureless_hidden}
\end{equation}
and therefore
\begin{equation}
\rho_{\rm hid}
\propto
a^{-3}.
\label{eq:app_gravity_hidden_matter_scaling}
\end{equation}
Hidden-brane matter therefore contributes to the background expansion
as ordinary cold dark matter at leading order in the low-energy
four-dimensional expansion.

At the level of scalar perturbations and on scales well inside the Hubble radius, the gravitational potential obeys
\begin{equation}
-\frac{k^2}{a^2}\Phi_N
=
\frac{1}{2M_{\rm Pl}^2}
\left(
\delta\rho_{\rm vis}
+
\delta\rho_{\rm hid}
\right),
\label{eq:app_gravity_cosmological_Poisson}
\end{equation}
up to corrections suppressed by the gravitational KK and radion masses. The hidden-sector density perturbations therefore contribute to gravitational clustering and lensing through the same four-dimensional metric perturbations as visible matter.

\subsection{Radion stabilization and range of validity}

The metric component $g_{55}$ contains a scalar modulus associated with fluctuations of the physical brane separation. Before stabilization, this radion can mediate an additional long-range force coupled to the trace of the brane stress-energy tensors. A consistent phenomenological construction must therefore fix the size
of the compact dimension and give the radion a sufficiently large mass,
as in standard stabilization mechanisms for compactified brane-world
models~\cite{Goldberger:1999uk,Csaki:1999mp,Maartens:2010ar}.

Independently of the details of the stabilization sector, its low-energy contribution can be parameterized schematically as
\begin{widetext}
\begin{equation}
S_{\rm radion}^{(4)}
=
\int\dd^4x\,\sqrt{|g_4|}
\left[
\frac{1}{2}\partial_\mu\varphi\partial^\mu\varphi
-
\frac{1}{2}m_\varphi^2\varphi^2
+
\frac{\varphi}{\Lambda_\varphi}
\left(T_{\rm vis}+c_{\rm hid}T_{\rm hid}\right)
\right].
\label{eq:app_gravity_radion_effective_action}
\end{equation}
\end{widetext}
where
\begin{equation}
T_i
\equiv
T_i{}^\mu{}_{\mu},
\label{eq:app_gravity_stress_trace}
\end{equation}
and the scale $\Lambda_\varphi$ and relative coefficient $c_{\rm hid}$ depend on the stabilization mechanism and the radion wave function. For
\begin{equation}
m_\varphi r
\gg
1,
\label{eq:app_gravity_heavy_radion_condition}
\end{equation}
radion exchange is Yukawa suppressed and does not affect the long-distance equations derived above. The main text implicitly assumes this stabilized regime.

The four-dimensional description used in Sec.~\ref{sec:gravity} is therefore valid under the conditions
\begin{equation}
E
\ll
M_{\rm KK}^{\rm grav},
\qquad
E
\ll
m_\varphi,
\label{eq:app_gravity_validity_energy}
\end{equation}
or, equivalently for static gravitational phenomena,
\begin{equation}
r
\gg
\frac{1}{M_{\rm KK}^{\rm grav}},
\qquad
r
\gg
\frac{1}{m_\varphi}.
\label{eq:app_gravity_validity_distance}
\end{equation}
In this regime, the only relevant gravitational degree of freedom from
the metric sector is the ordinary four-dimensional massless graviton,
whose coupling is given by
Eq.~\eqref{eq:app_gravity_zero_mode_interaction}. The visible- and hidden-brane stress-energy tensors therefore enter with equal strength in the long-distance Einstein equations. Geometrical sequestering suppresses local non-gravitational contact operators, but it does not suppress the universal gravitational interaction carried by the graviton zero mode.

\bibliographystyle{apsrev4-2}
\bibliography{main}

\begin{thebibliography}{86}%
\makeatletter
\providecommand \@ifxundefined [1]{%
 \@ifx{#1\undefined}
}%
\providecommand \@ifnum [1]{%
 \ifnum #1\expandafter \@firstoftwo
 \else \expandafter \@secondoftwo
 \fi
}%
\providecommand \@ifx [1]{%
 \ifx #1\expandafter \@firstoftwo
 \else \expandafter \@secondoftwo
 \fi
}%
\providecommand \natexlab [1]{#1}%
\providecommand \enquote  [1]{``#1''}%
\providecommand \bibnamefont  [1]{#1}%
\providecommand \bibfnamefont [1]{#1}%
\providecommand \citenamefont [1]{#1}%
\providecommand \href@noop [0]{\@secondoftwo}%
\providecommand \href [0]{\begingroup \@sanitize@url \@href}%
\providecommand \@href[1]{\@@startlink{#1}\@@href}%
\providecommand \@@href[1]{\endgroup#1\@@endlink}%
\providecommand \@sanitize@url [0]{\catcode `\\12\catcode `\$12\catcode
  `\&12\catcode `\#12\catcode `\^12\catcode `\_12\catcode `\%12\relax}%
\providecommand \@@startlink[1]{}%
\providecommand \@@endlink[0]{}%
\providecommand \url  [0]{\begingroup\@sanitize@url \@url }%
\providecommand \@url [1]{\endgroup\@href {#1}{\urlprefix }}%
\providecommand \urlprefix  [0]{URL }%
\providecommand \Eprint [0]{\href }%
\providecommand \doibase [0]{https://doi.org/}%
\providecommand \selectlanguage [0]{\@gobble}%
\providecommand \bibinfo  [0]{\@secondoftwo}%
\providecommand \bibfield  [0]{\@secondoftwo}%
\providecommand \translation [1]{[#1]}%
\providecommand \BibitemOpen [0]{}%
\providecommand \bibitemStop [0]{}%
\providecommand \bibitemNoStop [0]{.\EOS\space}%
\providecommand \EOS [0]{\spacefactor3000\relax}%
\providecommand \BibitemShut  [1]{\csname bibitem#1\endcsname}%
\let\auto@bib@innerbib\@empty
%</preamble>
\bibitem [{\citenamefont {Zwicky}(1933)}]{Zwicky:1933gu}%
  \BibitemOpen
  \bibfield  {author} {\bibinfo {author} {\bibfnamefont {F.}~\bibnamefont
  {Zwicky}},\ }\href@noop {} {\bibfield  {journal} {\bibinfo  {journal} {Helv.
  Phys. Acta}\ }\textbf {\bibinfo {volume} {6}},\ \bibinfo {pages} {110}
  (\bibinfo {year} {1933})}\BibitemShut {NoStop}%
\bibitem [{\citenamefont {Rubin}\ and\ \citenamefont
  {Ford}(1970)}]{Rubin:1970zza}%
  \BibitemOpen
  \bibfield  {author} {\bibinfo {author} {\bibfnamefont {V.~C.}\ \bibnamefont
  {Rubin}}\ and\ \bibinfo {author} {\bibfnamefont {J.}~\bibnamefont {Ford},
  \bibfnamefont {W.~Kent}},\ }\href {https://doi.org/10.1086/150317} {\bibfield
   {journal} {\bibinfo  {journal} {Astrophys. J.}\ }\textbf {\bibinfo {volume}
  {159}},\ \bibinfo {pages} {379} (\bibinfo {year} {1970})}\BibitemShut
  {NoStop}%
\bibitem [{\citenamefont {Clowe}\ \emph {et~al.}(2006)\citenamefont {Clowe},
  \citenamefont {Bradac}, \citenamefont {Gonzalez}, \citenamefont {Markevitch},
  \citenamefont {Randall}, \citenamefont {Jones},\ and\ \citenamefont
  {Zaritsky}}]{Clowe:2006eq}%
  \BibitemOpen
  \bibfield  {author} {\bibinfo {author} {\bibfnamefont {D.}~\bibnamefont
  {Clowe}}, \bibinfo {author} {\bibfnamefont {M.}~\bibnamefont {Bradac}},
  \bibinfo {author} {\bibfnamefont {A.~H.}\ \bibnamefont {Gonzalez}}, \bibinfo
  {author} {\bibfnamefont {M.}~\bibnamefont {Markevitch}}, \bibinfo {author}
  {\bibfnamefont {S.~W.}\ \bibnamefont {Randall}}, \bibinfo {author}
  {\bibfnamefont {C.}~\bibnamefont {Jones}},\ and\ \bibinfo {author}
  {\bibfnamefont {D.}~\bibnamefont {Zaritsky}},\ }\href
  {https://doi.org/10.1086/508162} {\bibfield  {journal} {\bibinfo  {journal}
  {Astrophys. J. Lett.}\ }\textbf {\bibinfo {volume} {648}},\ \bibinfo {pages}
  {L109} (\bibinfo {year} {2006})},\ \Eprint
  {https://arxiv.org/abs/astro-ph/0608407} {arXiv:astro-ph/0608407}
  \BibitemShut {NoStop}%
\bibitem [{\citenamefont {Aghanim}\ \emph {et~al.}(2020)\citenamefont {Aghanim}
  \emph {et~al.}}]{Planck:2018vyg}%
  \BibitemOpen
  \bibfield  {author} {\bibinfo {author} {\bibfnamefont {N.}~\bibnamefont
  {Aghanim}} \emph {et~al.} (\bibinfo {collaboration} {Planck}),\ }\href
  {https://doi.org/10.1051/0004-6361/201833910} {\bibfield  {journal} {\bibinfo
   {journal} {Astron. Astrophys.}\ }\textbf {\bibinfo {volume} {641}},\
  \bibinfo {pages} {A6} (\bibinfo {year} {2020})},\ \Eprint
  {https://arxiv.org/abs/1807.06209} {arXiv:1807.06209 [astro-ph.CO]}
  \BibitemShut {NoStop}%
\bibitem [{\citenamefont {Jungman}\ \emph {et~al.}(1996)\citenamefont
  {Jungman}, \citenamefont {Kamionkowski},\ and\ \citenamefont
  {Griest}}]{Jungman:1995df}%
  \BibitemOpen
  \bibfield  {author} {\bibinfo {author} {\bibfnamefont {G.}~\bibnamefont
  {Jungman}}, \bibinfo {author} {\bibfnamefont {M.}~\bibnamefont
  {Kamionkowski}},\ and\ \bibinfo {author} {\bibfnamefont {K.}~\bibnamefont
  {Griest}},\ }\href {https://doi.org/10.1016/0370-1573(95)00058-5} {\bibfield
  {journal} {\bibinfo  {journal} {Phys. Rept.}\ }\textbf {\bibinfo {volume}
  {267}},\ \bibinfo {pages} {195} (\bibinfo {year} {1996})},\ \Eprint
  {https://arxiv.org/abs/hep-ph/9506380} {arXiv:hep-ph/9506380} \BibitemShut
  {NoStop}%
\bibitem [{\citenamefont {Bertone}\ \emph {et~al.}(2005)\citenamefont
  {Bertone}, \citenamefont {Hooper},\ and\ \citenamefont
  {Silk}}]{Bertone:2004pz}%
  \BibitemOpen
  \bibfield  {author} {\bibinfo {author} {\bibfnamefont {G.}~\bibnamefont
  {Bertone}}, \bibinfo {author} {\bibfnamefont {D.}~\bibnamefont {Hooper}},\
  and\ \bibinfo {author} {\bibfnamefont {J.}~\bibnamefont {Silk}},\ }\href
  {https://doi.org/10.1016/j.physrep.2004.08.031} {\bibfield  {journal}
  {\bibinfo  {journal} {Phys. Rept.}\ }\textbf {\bibinfo {volume} {405}},\
  \bibinfo {pages} {279} (\bibinfo {year} {2005})},\ \Eprint
  {https://arxiv.org/abs/hep-ph/0404175} {arXiv:hep-ph/0404175} \BibitemShut
  {NoStop}%
\bibitem [{\citenamefont {Cirelli}\ \emph {et~al.}(2024)\citenamefont
  {Cirelli}, \citenamefont {Strumia},\ and\ \citenamefont
  {Zupan}}]{Cirelli:2024ssz}%
  \BibitemOpen
  \bibfield  {author} {\bibinfo {author} {\bibfnamefont {M.}~\bibnamefont
  {Cirelli}}, \bibinfo {author} {\bibfnamefont {A.}~\bibnamefont {Strumia}},\
  and\ \bibinfo {author} {\bibfnamefont {J.}~\bibnamefont {Zupan}},\
  }\href@noop {} {\  (\bibinfo {year} {2024})},\ \Eprint
  {https://arxiv.org/abs/2406.01705} {arXiv:2406.01705 [hep-ph]} \BibitemShut
  {NoStop}%
\bibitem [{\citenamefont {Schumann}(2019)}]{Schumann:2019eaa}%
  \BibitemOpen
  \bibfield  {author} {\bibinfo {author} {\bibfnamefont {M.}~\bibnamefont
  {Schumann}},\ }\href {https://doi.org/10.1088/1361-6471/ab2ea5} {\bibfield
  {journal} {\bibinfo  {journal} {J. Phys. G}\ }\textbf {\bibinfo {volume}
  {46}},\ \bibinfo {pages} {103003} (\bibinfo {year} {2019})},\ \Eprint
  {https://arxiv.org/abs/1903.03026} {arXiv:1903.03026 [astro-ph.CO]}
  \BibitemShut {NoStop}%
\bibitem [{\citenamefont {Boveia}\ and\ \citenamefont
  {Doglioni}(2018)}]{Boveia:2018yeb}%
  \BibitemOpen
  \bibfield  {author} {\bibinfo {author} {\bibfnamefont {A.}~\bibnamefont
  {Boveia}}\ and\ \bibinfo {author} {\bibfnamefont {C.}~\bibnamefont
  {Doglioni}},\ }\href {https://doi.org/10.1146/annurev-nucl-101917-021008}
  {\bibfield  {journal} {\bibinfo  {journal} {Ann. Rev. Nucl. Part. Sci.}\
  }\textbf {\bibinfo {volume} {68}},\ \bibinfo {pages} {429} (\bibinfo {year}
  {2018})},\ \Eprint {https://arxiv.org/abs/1810.12238} {arXiv:1810.12238
  [hep-ex]} \BibitemShut {NoStop}%
\bibitem [{\citenamefont {Gaskins}(2016)}]{Gaskins:2016cha}%
  \BibitemOpen
  \bibfield  {author} {\bibinfo {author} {\bibfnamefont {J.~M.}\ \bibnamefont
  {Gaskins}},\ }\href {https://doi.org/10.1080/00107514.2016.1175160}
  {\bibfield  {journal} {\bibinfo  {journal} {Contemp. Phys.}\ }\textbf
  {\bibinfo {volume} {57}},\ \bibinfo {pages} {496} (\bibinfo {year} {2016})},\
  \Eprint {https://arxiv.org/abs/1604.00014} {arXiv:1604.00014 [astro-ph.HE]}
  \BibitemShut {NoStop}%
\bibitem [{\citenamefont {Aprile}\ \emph {et~al.}(2023)\citenamefont {Aprile}
  \emph {et~al.}}]{Aprile:2023XENONnT}%
  \BibitemOpen
  \bibfield  {author} {\bibinfo {author} {\bibfnamefont {E.}~\bibnamefont
  {Aprile}} \emph {et~al.} (\bibinfo {collaboration} {XENON}),\ }\href
  {https://doi.org/10.1103/PhysRevLett.131.041003} {\bibfield  {journal}
  {\bibinfo  {journal} {Phys. Rev. Lett.}\ }\textbf {\bibinfo {volume} {131}},\
  \bibinfo {pages} {041003} (\bibinfo {year} {2023})},\ \Eprint
  {https://arxiv.org/abs/2303.14729} {arXiv:2303.14729 [hep-ex]} \BibitemShut
  {NoStop}%
\bibitem [{\citenamefont {Aalbers}\ \emph {et~al.}(2024)\citenamefont {Aalbers}
  \emph {et~al.}}]{Aalbers:2024LZ}%
  \BibitemOpen
  \bibfield  {author} {\bibinfo {author} {\bibfnamefont {J.}~\bibnamefont
  {Aalbers}} \emph {et~al.} (\bibinfo {collaboration} {LUX-ZEPLIN}),\
  }\href@noop {} {\  (\bibinfo {year} {2024})},\ \Eprint
  {https://arxiv.org/abs/2410.17036} {arXiv:2410.17036 [hep-ex]} \BibitemShut
  {NoStop}%
\bibitem [{\citenamefont {Di~Mauro}\ \emph {et~al.}(2023)\citenamefont
  {Di~Mauro}, \citenamefont {Arina}, \citenamefont {Fornengo}, \citenamefont
  {Heisig},\ and\ \citenamefont {Massaro}}]{DiMauro:2023tho}%
  \BibitemOpen
  \bibfield  {author} {\bibinfo {author} {\bibfnamefont {M.}~\bibnamefont
  {Di~Mauro}}, \bibinfo {author} {\bibfnamefont {C.}~\bibnamefont {Arina}},
  \bibinfo {author} {\bibfnamefont {N.}~\bibnamefont {Fornengo}}, \bibinfo
  {author} {\bibfnamefont {J.}~\bibnamefont {Heisig}},\ and\ \bibinfo {author}
  {\bibfnamefont {D.}~\bibnamefont {Massaro}},\ }\href
  {https://doi.org/10.1103/PhysRevD.108.095008} {\bibfield  {journal} {\bibinfo
   {journal} {Phys. Rev. D}\ }\textbf {\bibinfo {volume} {108}},\ \bibinfo
  {pages} {095008} (\bibinfo {year} {2023})},\ \Eprint
  {https://arxiv.org/abs/2305.11937} {arXiv:2305.11937 [hep-ph]} \BibitemShut
  {NoStop}%
\bibitem [{\citenamefont {Arcadi}\ \emph {et~al.}(2024)\citenamefont {Arcadi},
  \citenamefont {Cabo-Almeida}, \citenamefont {Dutra}, \citenamefont {Ghosh},
  \citenamefont {Lindner}, \citenamefont {Mambrini}, \citenamefont {Neto},
  \citenamefont {Pierre}, \citenamefont {Profumo},\ and\ \citenamefont
  {Queiroz}}]{Arcadi:2024ukq}%
  \BibitemOpen
  \bibfield  {author} {\bibinfo {author} {\bibfnamefont {G.}~\bibnamefont
  {Arcadi}}, \bibinfo {author} {\bibfnamefont {D.}~\bibnamefont
  {Cabo-Almeida}}, \bibinfo {author} {\bibfnamefont {M.}~\bibnamefont {Dutra}},
  \bibinfo {author} {\bibfnamefont {P.}~\bibnamefont {Ghosh}}, \bibinfo
  {author} {\bibfnamefont {M.}~\bibnamefont {Lindner}}, \bibinfo {author}
  {\bibfnamefont {Y.}~\bibnamefont {Mambrini}}, \bibinfo {author}
  {\bibfnamefont {J.~P.}\ \bibnamefont {Neto}}, \bibinfo {author}
  {\bibfnamefont {M.}~\bibnamefont {Pierre}}, \bibinfo {author} {\bibfnamefont
  {S.}~\bibnamefont {Profumo}},\ and\ \bibinfo {author} {\bibfnamefont {F.~S.}\
  \bibnamefont {Queiroz}},\ }\href@noop {} {\  (\bibinfo {year} {2024})},\
  \Eprint {https://arxiv.org/abs/2403.15860} {arXiv:2403.15860 [hep-ph]}
  \BibitemShut {NoStop}%
\bibitem [{\citenamefont {Kong}\ and\ \citenamefont
  {Di~Mauro}(2026)}]{Kong:2025ccv}%
  \BibitemOpen
  \bibfield  {author} {\bibinfo {author} {\bibfnamefont {C.}~\bibnamefont
  {Kong}}\ and\ \bibinfo {author} {\bibfnamefont {M.}~\bibnamefont
  {Di~Mauro}},\ }\href {https://doi.org/10.1103/v9vr-9skd} {\bibfield
  {journal} {\bibinfo  {journal} {Phys. Rev. D}\ }\textbf {\bibinfo {volume}
  {113}},\ \bibinfo {pages} {043031} (\bibinfo {year} {2026})},\ \Eprint
  {https://arxiv.org/abs/2511.21808} {arXiv:2511.21808 [hep-ph]} \BibitemShut
  {NoStop}%
\bibitem [{\citenamefont {Di~Mauro}\ and\ \citenamefont
  {Xie}(2026)}]{DiMauro:2025jia}%
  \BibitemOpen
  \bibfield  {author} {\bibinfo {author} {\bibfnamefont {M.}~\bibnamefont
  {Di~Mauro}}\ and\ \bibinfo {author} {\bibfnamefont {B.}~\bibnamefont {Xie}},\
  }\href {https://doi.org/10.1103/pdd9-jkl6} {\bibfield  {journal} {\bibinfo
  {journal} {Phys. Rev. D}\ }\textbf {\bibinfo {volume} {113}},\ \bibinfo
  {pages} {015034} (\bibinfo {year} {2026})},\ \Eprint
  {https://arxiv.org/abs/2510.08677} {arXiv:2510.08677 [hep-ph]} \BibitemShut
  {NoStop}%
\bibitem [{\citenamefont {Koechler}\ and\ \citenamefont
  {Di~Mauro}(2025)}]{Koechler:2025ryv}%
  \BibitemOpen
  \bibfield  {author} {\bibinfo {author} {\bibfnamefont {J.}~\bibnamefont
  {Koechler}}\ and\ \bibinfo {author} {\bibfnamefont {M.}~\bibnamefont
  {Di~Mauro}},\ }\href {https://doi.org/10.1103/nw98-38vr} {\bibfield
  {journal} {\bibinfo  {journal} {Phys. Rev. D}\ }\textbf {\bibinfo {volume}
  {112}},\ \bibinfo {pages} {115016} (\bibinfo {year} {2025})},\ \Eprint
  {https://arxiv.org/abs/2508.02775} {arXiv:2508.02775 [hep-ph]} \BibitemShut
  {NoStop}%
\bibitem [{\citenamefont {Griest}\ and\ \citenamefont
  {Seckel}(1991)}]{Griest:1990kh}%
  \BibitemOpen
  \bibfield  {author} {\bibinfo {author} {\bibfnamefont {K.}~\bibnamefont
  {Griest}}\ and\ \bibinfo {author} {\bibfnamefont {D.}~\bibnamefont
  {Seckel}},\ }\href {https://doi.org/10.1103/PhysRevD.43.3191} {\bibfield
  {journal} {\bibinfo  {journal} {Phys. Rev. D}\ }\textbf {\bibinfo {volume}
  {43}},\ \bibinfo {pages} {3191} (\bibinfo {year} {1991})}\BibitemShut
  {NoStop}%
\bibitem [{\citenamefont {Arcadi}\ \emph {et~al.}(2018)\citenamefont {Arcadi},
  \citenamefont {Dutra}, \citenamefont {Ghosh}, \citenamefont {Lindner},
  \citenamefont {Mambrini}, \citenamefont {Pierre}, \citenamefont {Profumo},\
  and\ \citenamefont {Queiroz}}]{Arcadi:2017kky}%
  \BibitemOpen
  \bibfield  {author} {\bibinfo {author} {\bibfnamefont {G.}~\bibnamefont
  {Arcadi}}, \bibinfo {author} {\bibfnamefont {M.}~\bibnamefont {Dutra}},
  \bibinfo {author} {\bibfnamefont {P.}~\bibnamefont {Ghosh}}, \bibinfo
  {author} {\bibfnamefont {M.}~\bibnamefont {Lindner}}, \bibinfo {author}
  {\bibfnamefont {Y.}~\bibnamefont {Mambrini}}, \bibinfo {author}
  {\bibfnamefont {M.}~\bibnamefont {Pierre}}, \bibinfo {author} {\bibfnamefont
  {S.}~\bibnamefont {Profumo}},\ and\ \bibinfo {author} {\bibfnamefont {F.~S.}\
  \bibnamefont {Queiroz}},\ }\href
  {https://doi.org/10.1140/epjc/s10052-018-5662-y} {\bibfield  {journal}
  {\bibinfo  {journal} {Eur. Phys. J. C}\ }\textbf {\bibinfo {volume} {78}},\
  \bibinfo {pages} {203} (\bibinfo {year} {2018})},\ \Eprint
  {https://arxiv.org/abs/1703.07364} {arXiv:1703.07364 [hep-ph]} \BibitemShut
  {NoStop}%
\bibitem [{\citenamefont {Arcadi}\ \emph {et~al.}(2020)\citenamefont {Arcadi},
  \citenamefont {Djouadi},\ and\ \citenamefont {Raidal}}]{Arcadi:2019lka}%
  \BibitemOpen
  \bibfield  {author} {\bibinfo {author} {\bibfnamefont {G.}~\bibnamefont
  {Arcadi}}, \bibinfo {author} {\bibfnamefont {A.}~\bibnamefont {Djouadi}},\
  and\ \bibinfo {author} {\bibfnamefont {M.}~\bibnamefont {Raidal}},\ }\href
  {https://doi.org/10.1016/j.physrep.2019.11.003} {\bibfield  {journal}
  {\bibinfo  {journal} {Phys. Rept.}\ }\textbf {\bibinfo {volume} {842}},\
  \bibinfo {pages} {1} (\bibinfo {year} {2020})},\ \Eprint
  {https://arxiv.org/abs/1903.03616} {arXiv:1903.03616 [hep-ph]} \BibitemShut
  {NoStop}%
\bibitem [{\citenamefont {Pospelov}\ \emph {et~al.}(2008)\citenamefont
  {Pospelov}, \citenamefont {Ritz},\ and\ \citenamefont
  {Voloshin}}]{Pospelov:2007mp}%
  \BibitemOpen
  \bibfield  {author} {\bibinfo {author} {\bibfnamefont {M.}~\bibnamefont
  {Pospelov}}, \bibinfo {author} {\bibfnamefont {A.}~\bibnamefont {Ritz}},\
  and\ \bibinfo {author} {\bibfnamefont {M.~B.}\ \bibnamefont {Voloshin}},\
  }\href {https://doi.org/10.1016/j.physletb.2008.02.052} {\bibfield  {journal}
  {\bibinfo  {journal} {Phys. Lett. B}\ }\textbf {\bibinfo {volume} {662}},\
  \bibinfo {pages} {53} (\bibinfo {year} {2008})},\ \Eprint
  {https://arxiv.org/abs/0711.4866} {arXiv:0711.4866 [hep-ph]} \BibitemShut
  {NoStop}%
\bibitem [{\citenamefont {Pospelov}\ and\ \citenamefont
  {Ritz}(2009)}]{Pospelov:2008jd}%
  \BibitemOpen
  \bibfield  {author} {\bibinfo {author} {\bibfnamefont {M.}~\bibnamefont
  {Pospelov}}\ and\ \bibinfo {author} {\bibfnamefont {A.}~\bibnamefont
  {Ritz}},\ }\href {https://doi.org/10.1016/j.physletb.2008.12.012} {\bibfield
  {journal} {\bibinfo  {journal} {Phys. Lett. B}\ }\textbf {\bibinfo {volume}
  {671}},\ \bibinfo {pages} {391} (\bibinfo {year} {2009})},\ \Eprint
  {https://arxiv.org/abs/0810.1502} {arXiv:0810.1502 [hep-ph]} \BibitemShut
  {NoStop}%
\bibitem [{\citenamefont {Di~Mauro}\ and\ \citenamefont
  {Wang}(2026)}]{DiMauro:2025jsb}%
  \BibitemOpen
  \bibfield  {author} {\bibinfo {author} {\bibfnamefont {M.}~\bibnamefont
  {Di~Mauro}}\ and\ \bibinfo {author} {\bibfnamefont {Y.}~\bibnamefont
  {Wang}},\ }\href {https://doi.org/10.1103/44ls-ss78} {\bibfield  {journal}
  {\bibinfo  {journal} {Phys. Rev. D}\ }\textbf {\bibinfo {volume} {113}},\
  \bibinfo {pages} {075003} (\bibinfo {year} {2026})},\ \Eprint
  {https://arxiv.org/abs/2510.23771} {arXiv:2510.23771 [hep-ph]} \BibitemShut
  {NoStop}%
\bibitem [{\citenamefont {Cooke}\ \emph {et~al.}(2018)\citenamefont {Cooke},
  \citenamefont {Pettini},\ and\ \citenamefont {Steidel}}]{Cooke:2018qzw}%
  \BibitemOpen
  \bibfield  {author} {\bibinfo {author} {\bibfnamefont {R.~J.}\ \bibnamefont
  {Cooke}}, \bibinfo {author} {\bibfnamefont {M.}~\bibnamefont {Pettini}},\
  and\ \bibinfo {author} {\bibfnamefont {C.~C.}\ \bibnamefont {Steidel}},\
  }\href {https://doi.org/10.3847/1538-4357/aaab53} {\bibfield  {journal}
  {\bibinfo  {journal} {Astrophys. J.}\ }\textbf {\bibinfo {volume} {855}},\
  \bibinfo {pages} {102} (\bibinfo {year} {2018})},\ \Eprint
  {https://arxiv.org/abs/1710.11129} {arXiv:1710.11129 [astro-ph.CO]}
  \BibitemShut {NoStop}%
\bibitem [{\citenamefont {Kurichin}\ \emph {et~al.}(2021)\citenamefont
  {Kurichin}, \citenamefont {Kislitsyn}, \citenamefont {Klimenko},
  \citenamefont {Balashev},\ and\ \citenamefont {Ivanchik}}]{Kurichin:2021}%
  \BibitemOpen
  \bibfield  {author} {\bibinfo {author} {\bibfnamefont {O.~A.}\ \bibnamefont
  {Kurichin}}, \bibinfo {author} {\bibfnamefont {P.~A.}\ \bibnamefont
  {Kislitsyn}}, \bibinfo {author} {\bibfnamefont {V.~V.}\ \bibnamefont
  {Klimenko}}, \bibinfo {author} {\bibfnamefont {S.~A.}\ \bibnamefont
  {Balashev}},\ and\ \bibinfo {author} {\bibfnamefont {A.~V.}\ \bibnamefont
  {Ivanchik}},\ }\href {https://doi.org/10.1093/mnras/stab215} {\bibfield
  {journal} {\bibinfo  {journal} {Mon. Not. Roy. Astron. Soc.}\ }\textbf
  {\bibinfo {volume} {502}},\ \bibinfo {pages} {3045} (\bibinfo {year}
  {2021})},\ \Eprint {https://arxiv.org/abs/2101.09127} {arXiv:2101.09127
  [astro-ph.CO]} \BibitemShut {NoStop}%
\bibitem [{\citenamefont {Sbordone}\ \emph {et~al.}(2010)\citenamefont
  {Sbordone} \emph {et~al.}}]{Sbordone:2010}%
  \BibitemOpen
  \bibfield  {author} {\bibinfo {author} {\bibfnamefont {L.}~\bibnamefont
  {Sbordone}} \emph {et~al.},\ }\href
  {https://doi.org/10.1051/0004-6361/200913282} {\bibfield  {journal} {\bibinfo
   {journal} {Astron. Astrophys.}\ }\textbf {\bibinfo {volume} {522}},\
  \bibinfo {pages} {A26} (\bibinfo {year} {2010})},\ \Eprint
  {https://arxiv.org/abs/1003.4510} {arXiv:1003.4510 [astro-ph.GA]}
  \BibitemShut {NoStop}%
\bibitem [{\citenamefont {Aalbers}\ \emph {et~al.}(2016)\citenamefont {Aalbers}
  \emph {et~al.}}]{DARWIN:2016hyl}%
  \BibitemOpen
  \bibfield  {author} {\bibinfo {author} {\bibfnamefont {J.}~\bibnamefont
  {Aalbers}} \emph {et~al.} (\bibinfo {collaboration} {DARWIN}),\ }\href
  {https://doi.org/10.1088/1475-7516/2016/11/017} {\bibfield  {journal}
  {\bibinfo  {journal} {JCAP}\ }\textbf {\bibinfo {volume} {11}},\ \bibinfo
  {pages} {017}},\ \Eprint {https://arxiv.org/abs/1606.07001} {arXiv:1606.07001
  [astro-ph.IM]} \BibitemShut {NoStop}%
\bibitem [{\citenamefont {Di~Mauro}(2025{\natexlab{a}})}]{DiMauro:2025nmsdm}%
  \BibitemOpen
  \bibfield  {author} {\bibinfo {author} {\bibfnamefont {M.}~\bibnamefont
  {Di~Mauro}},\ }\href {https://doi.org/10.48550/arXiv.2511.19622} {\bibinfo
  {title} {{Two Puzzles, One Solution: Neutrino Mass and Secluded Dark
  Matter}}} (\bibinfo {year} {2025}{\natexlab{a}}),\ \Eprint
  {https://arxiv.org/abs/2511.19622} {arXiv:2511.19622 [hep-ph]} \BibitemShut
  {NoStop}%
\bibitem [{\citenamefont {Minkowski}(1977)}]{Minkowski:1977sc}%
  \BibitemOpen
  \bibfield  {author} {\bibinfo {author} {\bibfnamefont {P.}~\bibnamefont
  {Minkowski}},\ }\href {https://doi.org/10.1016/0370-2693(77)90435-X}
  {\bibfield  {journal} {\bibinfo  {journal} {Phys. Lett. B}\ }\textbf
  {\bibinfo {volume} {67}},\ \bibinfo {pages} {421} (\bibinfo {year}
  {1977})}\BibitemShut {NoStop}%
\bibitem [{\citenamefont {Yanagida}(1979)}]{Yanagida:1979as}%
  \BibitemOpen
  \bibfield  {author} {\bibinfo {author} {\bibfnamefont {T.}~\bibnamefont
  {Yanagida}},\ }in\ \href@noop {} {\emph {\bibinfo {booktitle} {{Proceedings
  of the Workshop on Unified Theory and Baryon Number in the Universe}}}}\
  (\bibinfo {year} {1979})\ pp.\ \bibinfo {pages} {95--99}\BibitemShut
  {NoStop}%
\bibitem [{\citenamefont {Gell-Mann}\ \emph {et~al.}(1979)\citenamefont
  {Gell-Mann}, \citenamefont {Ramond},\ and\ \citenamefont
  {Slansky}}]{GellMann:1979}%
  \BibitemOpen
  \bibfield  {author} {\bibinfo {author} {\bibfnamefont {M.}~\bibnamefont
  {Gell-Mann}}, \bibinfo {author} {\bibfnamefont {P.}~\bibnamefont {Ramond}},\
  and\ \bibinfo {author} {\bibfnamefont {R.}~\bibnamefont {Slansky}},\ }in\
  \href@noop {} {\emph {\bibinfo {booktitle} {{Supergravity}}}}\ (\bibinfo
  {year} {1979})\ pp.\ \bibinfo {pages} {315--321},\ \Eprint
  {https://arxiv.org/abs/1306.4669} {arXiv:1306.4669 [hep-th]} \BibitemShut
  {NoStop}%
\bibitem [{\citenamefont {Glashow}(1980)}]{Glashow:1979vf}%
  \BibitemOpen
  \bibfield  {author} {\bibinfo {author} {\bibfnamefont {S.~L.}\ \bibnamefont
  {Glashow}},\ }in\ \href@noop {} {\emph {\bibinfo {booktitle} {{Quarks and
  Leptons}}}}\ (\bibinfo {year} {1980})\ p.\ \bibinfo {pages} {687}\BibitemShut
  {NoStop}%
\bibitem [{\citenamefont {Mohapatra}\ and\ \citenamefont
  {Senjanovic}(1980)}]{Mohapatra:1979ia}%
  \BibitemOpen
  \bibfield  {author} {\bibinfo {author} {\bibfnamefont {R.~N.}\ \bibnamefont
  {Mohapatra}}\ and\ \bibinfo {author} {\bibfnamefont {G.}~\bibnamefont
  {Senjanovic}},\ }\href {https://doi.org/10.1103/PhysRevLett.44.912}
  {\bibfield  {journal} {\bibinfo  {journal} {Phys. Rev. Lett.}\ }\textbf
  {\bibinfo {volume} {44}},\ \bibinfo {pages} {912} (\bibinfo {year}
  {1980})}\BibitemShut {NoStop}%
\bibitem [{\citenamefont {Casas}\ and\ \citenamefont
  {Ibarra}(2001)}]{Casas:2001sr}%
  \BibitemOpen
  \bibfield  {author} {\bibinfo {author} {\bibfnamefont {J.~A.}\ \bibnamefont
  {Casas}}\ and\ \bibinfo {author} {\bibfnamefont {A.}~\bibnamefont {Ibarra}},\
  }\href {https://doi.org/10.1016/S0550-3213(01)00475-8} {\bibfield  {journal}
  {\bibinfo  {journal} {Nucl. Phys. B}\ }\textbf {\bibinfo {volume} {618}},\
  \bibinfo {pages} {171} (\bibinfo {year} {2001})},\ \Eprint
  {https://arxiv.org/abs/hep-ph/0103065} {arXiv:hep-ph/0103065} \BibitemShut
  {NoStop}%
\bibitem [{\citenamefont {Kaluza}(1921)}]{Kaluza:1921tu}%
  \BibitemOpen
  \bibfield  {author} {\bibinfo {author} {\bibfnamefont {T.}~\bibnamefont
  {Kaluza}},\ }\href@noop {} {\bibfield  {journal} {\bibinfo  {journal}
  {Sitzungsber. Preuss. Akad. Wiss. Berlin (Math. Phys.)}\ }\textbf {\bibinfo
  {volume} {1921}},\ \bibinfo {pages} {966} (\bibinfo {year}
  {1921})}\BibitemShut {NoStop}%
\bibitem [{\citenamefont {Klein}(1926)}]{Klein:1926tv}%
  \BibitemOpen
  \bibfield  {author} {\bibinfo {author} {\bibfnamefont {O.}~\bibnamefont
  {Klein}},\ }\href {https://doi.org/10.1007/BF01397481} {\bibfield  {journal}
  {\bibinfo  {journal} {Z. Phys.}\ }\textbf {\bibinfo {volume} {37}},\ \bibinfo
  {pages} {895} (\bibinfo {year} {1926})}\BibitemShut {NoStop}%
\bibitem [{\citenamefont {Arkani-Hamed}\ \emph {et~al.}(1998)\citenamefont
  {Arkani-Hamed}, \citenamefont {Dimopoulos},\ and\ \citenamefont
  {Dvali}}]{ArkaniHamed:1998rs}%
  \BibitemOpen
  \bibfield  {author} {\bibinfo {author} {\bibfnamefont {N.}~\bibnamefont
  {Arkani-Hamed}}, \bibinfo {author} {\bibfnamefont {S.}~\bibnamefont
  {Dimopoulos}},\ and\ \bibinfo {author} {\bibfnamefont {G.}~\bibnamefont
  {Dvali}},\ }\href {https://doi.org/10.1016/S0370-2693(98)00466-3} {\bibfield
  {journal} {\bibinfo  {journal} {Phys. Lett. B}\ }\textbf {\bibinfo {volume}
  {429}},\ \bibinfo {pages} {263} (\bibinfo {year} {1998})},\ \Eprint
  {https://arxiv.org/abs/hep-ph/9803315} {arXiv:hep-ph/9803315} \BibitemShut
  {NoStop}%
\bibitem [{\citenamefont {Randall}\ and\ \citenamefont
  {Sundrum}(1999{\natexlab{a}})}]{Randall:1999ee}%
  \BibitemOpen
  \bibfield  {author} {\bibinfo {author} {\bibfnamefont {L.}~\bibnamefont
  {Randall}}\ and\ \bibinfo {author} {\bibfnamefont {R.}~\bibnamefont
  {Sundrum}},\ }\href {https://doi.org/10.1103/PhysRevLett.83.3370} {\bibfield
  {journal} {\bibinfo  {journal} {Phys. Rev. Lett.}\ }\textbf {\bibinfo
  {volume} {83}},\ \bibinfo {pages} {3370} (\bibinfo {year}
  {1999}{\natexlab{a}})},\ \Eprint {https://arxiv.org/abs/hep-ph/9905221}
  {arXiv:hep-ph/9905221} \BibitemShut {NoStop}%
\bibitem [{\citenamefont {Randall}\ and\ \citenamefont
  {Sundrum}(1999{\natexlab{b}})}]{Randall:1999vf}%
  \BibitemOpen
  \bibfield  {author} {\bibinfo {author} {\bibfnamefont {L.}~\bibnamefont
  {Randall}}\ and\ \bibinfo {author} {\bibfnamefont {R.}~\bibnamefont
  {Sundrum}},\ }\href {https://doi.org/10.1103/PhysRevLett.83.4690} {\bibfield
  {journal} {\bibinfo  {journal} {Phys. Rev. Lett.}\ }\textbf {\bibinfo
  {volume} {83}},\ \bibinfo {pages} {4690} (\bibinfo {year}
  {1999}{\natexlab{b}})},\ \Eprint {https://arxiv.org/abs/hep-th/9906064}
  {arXiv:hep-th/9906064} \BibitemShut {NoStop}%
\bibitem [{\citenamefont {Randall}\ and\ \citenamefont
  {Sundrum}(1999{\natexlab{c}})}]{Randall:1998uk}%
  \BibitemOpen
  \bibfield  {author} {\bibinfo {author} {\bibfnamefont {L.}~\bibnamefont
  {Randall}}\ and\ \bibinfo {author} {\bibfnamefont {R.}~\bibnamefont
  {Sundrum}},\ }\href {https://doi.org/10.1016/S0550-3213(99)00359-4}
  {\bibfield  {journal} {\bibinfo  {journal} {Nucl. Phys. B}\ }\textbf
  {\bibinfo {volume} {557}},\ \bibinfo {pages} {79} (\bibinfo {year}
  {1999}{\natexlab{c}})},\ \Eprint {https://arxiv.org/abs/hep-th/9810155}
  {arXiv:hep-th/9810155} \BibitemShut {NoStop}%
\bibitem [{\citenamefont {Kaplan}\ \emph {et~al.}(2000)\citenamefont {Kaplan},
  \citenamefont {Kribs},\ and\ \citenamefont {Schmaltz}}]{Kaplan:1999ac}%
  \BibitemOpen
  \bibfield  {author} {\bibinfo {author} {\bibfnamefont {D.~E.}\ \bibnamefont
  {Kaplan}}, \bibinfo {author} {\bibfnamefont {G.~D.}\ \bibnamefont {Kribs}},\
  and\ \bibinfo {author} {\bibfnamefont {M.}~\bibnamefont {Schmaltz}},\ }\href
  {https://doi.org/10.1103/PhysRevD.62.035010} {\bibfield  {journal} {\bibinfo
  {journal} {Phys. Rev. D}\ }\textbf {\bibinfo {volume} {62}},\ \bibinfo
  {pages} {035010} (\bibinfo {year} {2000})},\ \Eprint
  {https://arxiv.org/abs/hep-ph/9911293} {arXiv:hep-ph/9911293} \BibitemShut
  {NoStop}%
\bibitem [{\citenamefont {Dienes}\ \emph {et~al.}(1999)\citenamefont {Dienes},
  \citenamefont {Dudas},\ and\ \citenamefont {Gherghetta}}]{Dienes:1998sb}%
  \BibitemOpen
  \bibfield  {author} {\bibinfo {author} {\bibfnamefont {K.~R.}\ \bibnamefont
  {Dienes}}, \bibinfo {author} {\bibfnamefont {E.}~\bibnamefont {Dudas}},\ and\
  \bibinfo {author} {\bibfnamefont {T.}~\bibnamefont {Gherghetta}},\ }\href
  {https://doi.org/10.1016/S0550-3213(99)00377-6} {\bibfield  {journal}
  {\bibinfo  {journal} {Nucl. Phys. B}\ }\textbf {\bibinfo {volume} {557}},\
  \bibinfo {pages} {25} (\bibinfo {year} {1999})},\ \Eprint
  {https://arxiv.org/abs/hep-ph/9811428} {arXiv:hep-ph/9811428} \BibitemShut
  {NoStop}%
\bibitem [{\citenamefont {Grossman}\ and\ \citenamefont
  {Neubert}(2000{\natexlab{a}})}]{Grossman:1999ra}%
  \BibitemOpen
  \bibfield  {author} {\bibinfo {author} {\bibfnamefont {Y.}~\bibnamefont
  {Grossman}}\ and\ \bibinfo {author} {\bibfnamefont {M.}~\bibnamefont
  {Neubert}},\ }\href {https://doi.org/10.1016/S0370-2693(00)00054-X}
  {\bibfield  {journal} {\bibinfo  {journal} {Phys. Lett. B}\ }\textbf
  {\bibinfo {volume} {474}},\ \bibinfo {pages} {361} (\bibinfo {year}
  {2000}{\natexlab{a}})},\ \Eprint {https://arxiv.org/abs/hep-ph/9912408}
  {arXiv:hep-ph/9912408} \BibitemShut {NoStop}%
\bibitem [{\citenamefont {Neubert}(2000)}]{Neubert:2000zb}%
  \BibitemOpen
  \bibfield  {author} {\bibinfo {author} {\bibfnamefont {M.}~\bibnamefont
  {Neubert}},\ }\href@noop {} {\  (\bibinfo {year} {2000})},\ \Eprint
  {https://arxiv.org/abs/hep-ph/0011063} {arXiv:hep-ph/0011063} \BibitemShut
  {NoStop}%
\bibitem [{\citenamefont {Ruegg}\ and\ \citenamefont
  {Ruiz-Altaba}(2004)}]{Ruegg:2003yd}%
  \BibitemOpen
  \bibfield  {author} {\bibinfo {author} {\bibfnamefont {H.}~\bibnamefont
  {Ruegg}}\ and\ \bibinfo {author} {\bibfnamefont {M.}~\bibnamefont
  {Ruiz-Altaba}},\ }\href {https://doi.org/10.1142/S0217751X04019755}
  {\bibfield  {journal} {\bibinfo  {journal} {Int. J. Mod. Phys. A}\ }\textbf
  {\bibinfo {volume} {19}},\ \bibinfo {pages} {3265} (\bibinfo {year}
  {2004})},\ \Eprint {https://arxiv.org/abs/hep-th/0304245}
  {arXiv:hep-th/0304245 [hep-th]} \BibitemShut {NoStop}%
\bibitem [{\citenamefont {Kors}\ and\ \citenamefont
  {Nath}(2004)}]{Kors:2004dx}%
  \BibitemOpen
  \bibfield  {author} {\bibinfo {author} {\bibfnamefont {B.}~\bibnamefont
  {Kors}}\ and\ \bibinfo {author} {\bibfnamefont {P.}~\bibnamefont {Nath}},\
  }\href {https://doi.org/10.1016/j.physletb.2004.02.051} {\bibfield  {journal}
  {\bibinfo  {journal} {Phys. Lett. B}\ }\textbf {\bibinfo {volume} {586}},\
  \bibinfo {pages} {366} (\bibinfo {year} {2004})},\ \Eprint
  {https://arxiv.org/abs/hep-ph/0402047} {arXiv:hep-ph/0402047 [hep-ph]}
  \BibitemShut {NoStop}%
\bibitem [{\citenamefont {Holdom}(1986)}]{Holdom:1985ag}%
  \BibitemOpen
  \bibfield  {author} {\bibinfo {author} {\bibfnamefont {B.}~\bibnamefont
  {Holdom}},\ }\href {https://doi.org/10.1016/0370-2693(86)91377-8} {\bibfield
  {journal} {\bibinfo  {journal} {Phys. Lett. B}\ }\textbf {\bibinfo {volume}
  {166}},\ \bibinfo {pages} {196} (\bibinfo {year} {1986})}\BibitemShut
  {NoStop}%
\bibitem [{\citenamefont {Essig}\ \emph {et~al.}(2013)\citenamefont {Essig}
  \emph {et~al.}}]{Essig:2013lka}%
  \BibitemOpen
  \bibfield  {author} {\bibinfo {author} {\bibfnamefont {R.}~\bibnamefont
  {Essig}} \emph {et~al.},\ }\href@noop {} {\  (\bibinfo {year} {2013})},\
  \Eprint {https://arxiv.org/abs/1311.0029} {arXiv:1311.0029 [hep-ph]}
  \BibitemShut {NoStop}%
\bibitem [{\citenamefont {Aalbers}\ \emph {et~al.}(2025)\citenamefont {Aalbers}
  \emph {et~al.}}]{LZ:2024zvo}%
  \BibitemOpen
  \bibfield  {author} {\bibinfo {author} {\bibfnamefont {J.}~\bibnamefont
  {Aalbers}} \emph {et~al.} (\bibinfo {collaboration} {LZ}),\ }\href
  {https://doi.org/10.1103/4dyc-z8zf} {\bibfield  {journal} {\bibinfo
  {journal} {Phys. Rev. Lett.}\ }\textbf {\bibinfo {volume} {135}},\ \bibinfo
  {pages} {011802} (\bibinfo {year} {2025})},\ \Eprint
  {https://arxiv.org/abs/2410.17036} {arXiv:2410.17036 [hep-ex]} \BibitemShut
  {NoStop}%
\bibitem [{\citenamefont {Cheung}\ \emph {et~al.}(2015)\citenamefont {Cheung},
  \citenamefont {Ko}, \citenamefont {Lee},\ and\ \citenamefont
  {Tseng}}]{Cheung:2015dta}%
  \BibitemOpen
  \bibfield  {author} {\bibinfo {author} {\bibfnamefont {K.}~\bibnamefont
  {Cheung}}, \bibinfo {author} {\bibfnamefont {P.}~\bibnamefont {Ko}}, \bibinfo
  {author} {\bibfnamefont {J.~S.}\ \bibnamefont {Lee}},\ and\ \bibinfo {author}
  {\bibfnamefont {P.-Y.}\ \bibnamefont {Tseng}},\ }\href
  {https://doi.org/10.1007/JHEP10(2015)057} {\bibfield  {journal} {\bibinfo
  {journal} {JHEP}\ }\textbf {\bibinfo {volume} {10}},\ \bibinfo {pages}
  {057}},\ \Eprint {https://arxiv.org/abs/1507.06158} {arXiv:1507.06158
  [hep-ph]} \BibitemShut {NoStop}%
\bibitem [{\citenamefont {Arina}\ \emph {et~al.}(2024)\citenamefont {Arina},
  \citenamefont {Di~Mauro}, \citenamefont {Fornengo}, \citenamefont {Heisig},
  \citenamefont {Jueid},\ and\ \citenamefont {de~Austri}}]{Arina:2023eic}%
  \BibitemOpen
  \bibfield  {author} {\bibinfo {author} {\bibfnamefont {C.}~\bibnamefont
  {Arina}}, \bibinfo {author} {\bibfnamefont {M.}~\bibnamefont {Di~Mauro}},
  \bibinfo {author} {\bibfnamefont {N.}~\bibnamefont {Fornengo}}, \bibinfo
  {author} {\bibfnamefont {J.}~\bibnamefont {Heisig}}, \bibinfo {author}
  {\bibfnamefont {A.}~\bibnamefont {Jueid}},\ and\ \bibinfo {author}
  {\bibfnamefont {R.~R.}\ \bibnamefont {de~Austri}},\ }\href
  {https://doi.org/10.1088/1475-7516/2024/03/035} {\bibfield  {journal}
  {\bibinfo  {journal} {JCAP}\ }\textbf {\bibinfo {volume} {03}},\ \bibinfo
  {pages} {035}},\ \Eprint {https://arxiv.org/abs/2312.01153} {arXiv:2312.01153
  [astro-ph.HE]} \BibitemShut {NoStop}%
\bibitem [{\citenamefont {Vissani}(1998)}]{Vissani:1997ys}%
  \BibitemOpen
  \bibfield  {author} {\bibinfo {author} {\bibfnamefont {F.}~\bibnamefont
  {Vissani}},\ }\href {https://doi.org/10.1103/PhysRevD.57.7027} {\bibfield
  {journal} {\bibinfo  {journal} {Phys. Rev. D}\ }\textbf {\bibinfo {volume}
  {57}},\ \bibinfo {pages} {7027} (\bibinfo {year} {1998})},\ \Eprint
  {https://arxiv.org/abs/hep-ph/9709409} {arXiv:hep-ph/9709409} \BibitemShut
  {NoStop}%
\bibitem [{\citenamefont {Clarke}\ \emph {et~al.}(2015)\citenamefont {Clarke},
  \citenamefont {Foot},\ and\ \citenamefont {Volkas}}]{Clarke:2015hta}%
  \BibitemOpen
  \bibfield  {author} {\bibinfo {author} {\bibfnamefont {J.~D.}\ \bibnamefont
  {Clarke}}, \bibinfo {author} {\bibfnamefont {R.}~\bibnamefont {Foot}},\ and\
  \bibinfo {author} {\bibfnamefont {R.~R.}\ \bibnamefont {Volkas}},\ }\href
  {https://doi.org/10.1103/PhysRevD.91.073009} {\bibfield  {journal} {\bibinfo
  {journal} {Phys. Rev. D}\ }\textbf {\bibinfo {volume} {91}},\ \bibinfo
  {pages} {073009} (\bibinfo {year} {2015})},\ \Eprint
  {https://arxiv.org/abs/1502.01352} {arXiv:1502.01352 [hep-ph]} \BibitemShut
  {NoStop}%
\bibitem [{\citenamefont {Appelquist}\ \emph {et~al.}(2001)\citenamefont
  {Appelquist}, \citenamefont {Cheng},\ and\ \citenamefont
  {Dobrescu}}]{Appelquist:2000nn}%
  \BibitemOpen
  \bibfield  {author} {\bibinfo {author} {\bibfnamefont {T.}~\bibnamefont
  {Appelquist}}, \bibinfo {author} {\bibfnamefont {H.-C.}\ \bibnamefont
  {Cheng}},\ and\ \bibinfo {author} {\bibfnamefont {B.~A.}\ \bibnamefont
  {Dobrescu}},\ }\href {https://doi.org/10.1103/PhysRevD.64.035002} {\bibfield
  {journal} {\bibinfo  {journal} {Phys. Rev. D}\ }\textbf {\bibinfo {volume}
  {64}},\ \bibinfo {pages} {035002} (\bibinfo {year} {2001})},\ \Eprint
  {https://arxiv.org/abs/hep-ph/0012100} {arXiv:hep-ph/0012100} \BibitemShut
  {NoStop}%
\bibitem [{\citenamefont {Csaki}(2004)}]{Csaki:2004ay}%
  \BibitemOpen
  \bibfield  {author} {\bibinfo {author} {\bibfnamefont {C.}~\bibnamefont
  {Csaki}},\ }\href@noop {} {\  (\bibinfo {year} {2004})},\ \Eprint
  {https://arxiv.org/abs/hep-ph/0404096} {arXiv:hep-ph/0404096 [hep-ph]}
  \BibitemShut {NoStop}%
\bibitem [{\citenamefont {Rubakov}(2001)}]{Rubakov:2001kp}%
  \BibitemOpen
  \bibfield  {author} {\bibinfo {author} {\bibfnamefont {V.~A.}\ \bibnamefont
  {Rubakov}},\ }\href {https://doi.org/10.1070/PU2001v044n09ABEH001000}
  {\bibfield  {journal} {\bibinfo  {journal} {Phys. Usp.}\ }\textbf {\bibinfo
  {volume} {44}},\ \bibinfo {pages} {871} (\bibinfo {year} {2001})},\ \Eprint
  {https://arxiv.org/abs/hep-ph/0104152} {arXiv:hep-ph/0104152} \BibitemShut
  {NoStop}%
\bibitem [{\citenamefont {Arkani-Hamed}\ and\ \citenamefont
  {Schmaltz}(2000)}]{ArkaniHamed:1999dc}%
  \BibitemOpen
  \bibfield  {author} {\bibinfo {author} {\bibfnamefont {N.}~\bibnamefont
  {Arkani-Hamed}}\ and\ \bibinfo {author} {\bibfnamefont {M.}~\bibnamefont
  {Schmaltz}},\ }\href {https://doi.org/10.1103/PhysRevD.61.033005} {\bibfield
  {journal} {\bibinfo  {journal} {Phys. Rev. D}\ }\textbf {\bibinfo {volume}
  {61}},\ \bibinfo {pages} {033005} (\bibinfo {year} {2000})},\ \Eprint
  {https://arxiv.org/abs/hep-ph/9903417} {arXiv:hep-ph/9903417} \BibitemShut
  {NoStop}%
\bibitem [{\citenamefont {Kawasaki}\ \emph {et~al.}(2005)\citenamefont
  {Kawasaki}, \citenamefont {Kohri},\ and\ \citenamefont
  {Moroi}}]{Kawasaki:2004qu}%
  \BibitemOpen
  \bibfield  {author} {\bibinfo {author} {\bibfnamefont {M.}~\bibnamefont
  {Kawasaki}}, \bibinfo {author} {\bibfnamefont {K.}~\bibnamefont {Kohri}},\
  and\ \bibinfo {author} {\bibfnamefont {T.}~\bibnamefont {Moroi}},\ }\href
  {https://doi.org/10.1103/PhysRevD.71.083502} {\bibfield  {journal} {\bibinfo
  {journal} {Phys. Rev. D}\ }\textbf {\bibinfo {volume} {71}},\ \bibinfo
  {pages} {083502} (\bibinfo {year} {2005})},\ \Eprint
  {https://arxiv.org/abs/astro-ph/0408426} {arXiv:astro-ph/0408426}
  \BibitemShut {NoStop}%
\bibitem [{\citenamefont {Jedamzik}(2006)}]{Jedamzik:2006xz}%
  \BibitemOpen
  \bibfield  {author} {\bibinfo {author} {\bibfnamefont {K.}~\bibnamefont
  {Jedamzik}},\ }\href {https://doi.org/10.1103/PhysRevD.74.103509} {\bibfield
  {journal} {\bibinfo  {journal} {Phys. Rev. D}\ }\textbf {\bibinfo {volume}
  {74}},\ \bibinfo {pages} {103509} (\bibinfo {year} {2006})},\ \Eprint
  {https://arxiv.org/abs/hep-ph/0604251} {arXiv:hep-ph/0604251} \BibitemShut
  {NoStop}%
\bibitem [{\citenamefont {Pospelov}\ and\ \citenamefont
  {Pradler}(2010)}]{Pospelov:2010cw}%
  \BibitemOpen
  \bibfield  {author} {\bibinfo {author} {\bibfnamefont {M.}~\bibnamefont
  {Pospelov}}\ and\ \bibinfo {author} {\bibfnamefont {J.}~\bibnamefont
  {Pradler}},\ }\href {https://doi.org/10.1146/annurev.nucl.012809.104521}
  {\bibfield  {journal} {\bibinfo  {journal} {Ann. Rev. Nucl. Part. Sci.}\
  }\textbf {\bibinfo {volume} {60}},\ \bibinfo {pages} {539} (\bibinfo {year}
  {2010})},\ \Eprint {https://arxiv.org/abs/1011.1054} {arXiv:1011.1054
  [hep-ph]} \BibitemShut {NoStop}%
\bibitem [{\citenamefont {Lee}\ \emph {et~al.}(2020)\citenamefont {Lee},
  \citenamefont {Adelberger}, \citenamefont {Cook}, \citenamefont {Fleischer},\
  and\ \citenamefont {Heckel}}]{Lee:2020zjt}%
  \BibitemOpen
  \bibfield  {author} {\bibinfo {author} {\bibfnamefont {J.~G.}\ \bibnamefont
  {Lee}}, \bibinfo {author} {\bibfnamefont {E.~G.}\ \bibnamefont {Adelberger}},
  \bibinfo {author} {\bibfnamefont {T.~S.}\ \bibnamefont {Cook}}, \bibinfo
  {author} {\bibfnamefont {S.~M.}\ \bibnamefont {Fleischer}},\ and\ \bibinfo
  {author} {\bibfnamefont {B.~R.}\ \bibnamefont {Heckel}},\ }\href
  {https://doi.org/10.1103/PhysRevLett.124.101101} {\bibfield  {journal}
  {\bibinfo  {journal} {Phys. Rev. Lett.}\ }\textbf {\bibinfo {volume} {124}},\
  \bibinfo {pages} {101101} (\bibinfo {year} {2020})},\ \Eprint
  {https://arxiv.org/abs/2002.11761} {arXiv:2002.11761 [hep-ex]} \BibitemShut
  {NoStop}%
\bibitem [{\citenamefont {Tan}\ \emph {et~al.}(2020)\citenamefont {Tan} \emph
  {et~al.}}]{Tan:2020vpf}%
  \BibitemOpen
  \bibfield  {author} {\bibinfo {author} {\bibfnamefont {W.-H.}\ \bibnamefont
  {Tan}} \emph {et~al.},\ }\href
  {https://doi.org/10.1103/PhysRevLett.124.051301} {\bibfield  {journal}
  {\bibinfo  {journal} {Phys. Rev. Lett.}\ }\textbf {\bibinfo {volume} {124}},\
  \bibinfo {pages} {051301} (\bibinfo {year} {2020})}\BibitemShut {NoStop}%
\bibitem [{\citenamefont {Murata}\ \emph {et~al.}(2026)\citenamefont {Murata},
  \citenamefont {Fujiie},\ and\ \citenamefont {Suzuki}}]{Murata:2026isl}%
  \BibitemOpen
  \bibfield  {author} {\bibinfo {author} {\bibfnamefont {J.}~\bibnamefont
  {Murata}}, \bibinfo {author} {\bibfnamefont {T.}~\bibnamefont {Fujiie}},\
  and\ \bibinfo {author} {\bibfnamefont {S.}~\bibnamefont {Suzuki}},\
  }\href@noop {} {\  (\bibinfo {year} {2026})},\ \Eprint
  {https://arxiv.org/abs/2605.18212} {arXiv:2605.18212 [gr-qc]} \BibitemShut
  {NoStop}%
\bibitem [{\citenamefont {Deutschmann}\ \emph {et~al.}(2017)\citenamefont
  {Deutschmann}, \citenamefont {Flacke},\ and\ \citenamefont
  {Kim}}]{Deutschmann:2017bth}%
  \BibitemOpen
  \bibfield  {author} {\bibinfo {author} {\bibfnamefont {N.}~\bibnamefont
  {Deutschmann}}, \bibinfo {author} {\bibfnamefont {T.}~\bibnamefont
  {Flacke}},\ and\ \bibinfo {author} {\bibfnamefont {J.~S.}\ \bibnamefont
  {Kim}},\ }\href {https://doi.org/10.1016/j.physletb.2017.06.004} {\bibfield
  {journal} {\bibinfo  {journal} {Phys. Lett. B}\ }\textbf {\bibinfo {volume}
  {771}},\ \bibinfo {pages} {515} (\bibinfo {year} {2017})},\ \Eprint
  {https://arxiv.org/abs/1702.00410} {arXiv:1702.00410 [hep-ph]} \BibitemShut
  {NoStop}%
\bibitem [{\citenamefont {Navas}\ \emph {et~al.}(2024)\citenamefont {Navas}
  \emph {et~al.}}]{ParticleDataGroup:2024cfk}%
  \BibitemOpen
  \bibfield  {author} {\bibinfo {author} {\bibfnamefont {S.}~\bibnamefont
  {Navas}} \emph {et~al.} (\bibinfo {collaboration} {Particle Data Group}),\
  }\href {https://doi.org/10.1103/PhysRevD.110.030001} {\bibfield  {journal}
  {\bibinfo  {journal} {Phys. Rev. D}\ }\textbf {\bibinfo {volume} {110}},\
  \bibinfo {pages} {030001} (\bibinfo {year} {2024})}\BibitemShut {NoStop}%
\bibitem [{\citenamefont {Davoudiasl}\ \emph {et~al.}(2002)\citenamefont
  {Davoudiasl}, \citenamefont {Langacker},\ and\ \citenamefont
  {Perelstein}}]{Davoudiasl:2002fq}%
  \BibitemOpen
  \bibfield  {author} {\bibinfo {author} {\bibfnamefont {H.}~\bibnamefont
  {Davoudiasl}}, \bibinfo {author} {\bibfnamefont {P.}~\bibnamefont
  {Langacker}},\ and\ \bibinfo {author} {\bibfnamefont {M.}~\bibnamefont
  {Perelstein}},\ }\href {https://doi.org/10.1103/PhysRevD.65.105015}
  {\bibfield  {journal} {\bibinfo  {journal} {Phys. Rev. D}\ }\textbf {\bibinfo
  {volume} {65}},\ \bibinfo {pages} {105015} (\bibinfo {year} {2002})},\
  \Eprint {https://arxiv.org/abs/hep-ph/0201128} {arXiv:hep-ph/0201128}
  \BibitemShut {NoStop}%
\bibitem [{\citenamefont {Cao}\ \emph {et~al.}(2004)\citenamefont {Cao},
  \citenamefont {Gopalakrishna},\ and\ \citenamefont {Yuan}}]{Cao:2003rm}%
  \BibitemOpen
  \bibfield  {author} {\bibinfo {author} {\bibfnamefont {Q.-H.}\ \bibnamefont
  {Cao}}, \bibinfo {author} {\bibfnamefont {S.}~\bibnamefont {Gopalakrishna}},\
  and\ \bibinfo {author} {\bibfnamefont {C.-P.}\ \bibnamefont {Yuan}},\ }\href
  {https://doi.org/10.1103/PhysRevD.69.115003} {\bibfield  {journal} {\bibinfo
  {journal} {Phys. Rev. D}\ }\textbf {\bibinfo {volume} {69}},\ \bibinfo
  {pages} {115003} (\bibinfo {year} {2004})},\ \Eprint
  {https://arxiv.org/abs/hep-ph/0312339} {arXiv:hep-ph/0312339} \BibitemShut
  {NoStop}%
\bibitem [{\citenamefont {Hagstotz}\ \emph {et~al.}(2021)\citenamefont
  {Hagstotz}, \citenamefont {de~Salas}, \citenamefont {Gariazzo}, \citenamefont
  {Gerbino}, \citenamefont {Lattanzi}, \citenamefont {Vagnozzi}, \citenamefont
  {Freese},\ and\ \citenamefont {Pastor}}]{Hagstotz:2020ukm}%
  \BibitemOpen
  \bibfield  {author} {\bibinfo {author} {\bibfnamefont {S.}~\bibnamefont
  {Hagstotz}}, \bibinfo {author} {\bibfnamefont {P.~F.}\ \bibnamefont
  {de~Salas}}, \bibinfo {author} {\bibfnamefont {S.}~\bibnamefont {Gariazzo}},
  \bibinfo {author} {\bibfnamefont {M.}~\bibnamefont {Gerbino}}, \bibinfo
  {author} {\bibfnamefont {M.}~\bibnamefont {Lattanzi}}, \bibinfo {author}
  {\bibfnamefont {S.}~\bibnamefont {Vagnozzi}}, \bibinfo {author}
  {\bibfnamefont {K.}~\bibnamefont {Freese}},\ and\ \bibinfo {author}
  {\bibfnamefont {S.}~\bibnamefont {Pastor}},\ }\href
  {https://doi.org/10.1103/PhysRevD.104.123524} {\bibfield  {journal} {\bibinfo
   {journal} {Phys. Rev. D}\ }\textbf {\bibinfo {volume} {104}},\ \bibinfo
  {pages} {123524} (\bibinfo {year} {2021})},\ \Eprint
  {https://arxiv.org/abs/2003.02289} {arXiv:2003.02289 [astro-ph.CO]}
  \BibitemShut {NoStop}%
\bibitem [{\citenamefont {Abdullahi}\ \emph {et~al.}(2023)\citenamefont
  {Abdullahi} \emph {et~al.}}]{Abdullahi:2022jlv}%
  \BibitemOpen
  \bibfield  {author} {\bibinfo {author} {\bibfnamefont {A.~M.}\ \bibnamefont
  {Abdullahi}} \emph {et~al.},\ }\href
  {https://doi.org/10.1088/1361-6633/ac98f9} {\bibfield  {journal} {\bibinfo
  {journal} {Rept. Prog. Phys.}\ }\textbf {\bibinfo {volume} {86}},\ \bibinfo
  {pages} {014201} (\bibinfo {year} {2023})},\ \Eprint
  {https://arxiv.org/abs/2203.08039} {arXiv:2203.08039 [hep-ph]} \BibitemShut
  {NoStop}%
\bibitem [{\citenamefont {Carenza}\ \emph {et~al.}(2024)\citenamefont
  {Carenza}, \citenamefont {Lucente}, \citenamefont {Mastrototaro},
  \citenamefont {Mirizzi},\ and\ \citenamefont {Serpico}}]{Carenza:2023old}%
  \BibitemOpen
  \bibfield  {author} {\bibinfo {author} {\bibfnamefont {P.}~\bibnamefont
  {Carenza}}, \bibinfo {author} {\bibfnamefont {G.}~\bibnamefont {Lucente}},
  \bibinfo {author} {\bibfnamefont {L.}~\bibnamefont {Mastrototaro}}, \bibinfo
  {author} {\bibfnamefont {A.}~\bibnamefont {Mirizzi}},\ and\ \bibinfo {author}
  {\bibfnamefont {P.~D.}\ \bibnamefont {Serpico}},\ }\href
  {https://doi.org/10.1103/PhysRevD.109.063010} {\bibfield  {journal} {\bibinfo
   {journal} {Phys. Rev. D}\ }\textbf {\bibinfo {volume} {109}},\ \bibinfo
  {pages} {063010} (\bibinfo {year} {2024})},\ \Eprint
  {https://arxiv.org/abs/2311.00033} {arXiv:2311.00033 [hep-ph]} \BibitemShut
  {NoStop}%
\bibitem [{\citenamefont {Goldberger}\ and\ \citenamefont
  {Wise}(1999)}]{Goldberger:1999uk}%
  \BibitemOpen
  \bibfield  {author} {\bibinfo {author} {\bibfnamefont {W.~D.}\ \bibnamefont
  {Goldberger}}\ and\ \bibinfo {author} {\bibfnamefont {M.~B.}\ \bibnamefont
  {Wise}},\ }\href {https://doi.org/10.1103/PhysRevLett.83.4922} {\bibfield
  {journal} {\bibinfo  {journal} {Phys. Rev. Lett.}\ }\textbf {\bibinfo
  {volume} {83}},\ \bibinfo {pages} {4922} (\bibinfo {year} {1999})},\ \Eprint
  {https://arxiv.org/abs/hep-ph/9907447} {arXiv:hep-ph/9907447} \BibitemShut
  {NoStop}%
\bibitem [{\citenamefont {Hosotani}(1983{\natexlab{a}})}]{Hosotani:1983xw}%
  \BibitemOpen
  \bibfield  {author} {\bibinfo {author} {\bibfnamefont {Y.}~\bibnamefont
  {Hosotani}},\ }\href {https://doi.org/10.1016/0370-2693(83)90170-3}
  {\bibfield  {journal} {\bibinfo  {journal} {Phys. Lett. B}\ }\textbf
  {\bibinfo {volume} {126}},\ \bibinfo {pages} {309} (\bibinfo {year}
  {1983}{\natexlab{a}})}\BibitemShut {NoStop}%
\bibitem [{\citenamefont {Hosotani}(1983{\natexlab{b}})}]{Hosotani:1983vn}%
  \BibitemOpen
  \bibfield  {author} {\bibinfo {author} {\bibfnamefont {Y.}~\bibnamefont
  {Hosotani}},\ }\href {https://doi.org/10.1016/0370-2693(83)90140-3}
  {\bibfield  {journal} {\bibinfo  {journal} {Phys. Lett. B}\ }\textbf
  {\bibinfo {volume} {129}},\ \bibinfo {pages} {193} (\bibinfo {year}
  {1983}{\natexlab{b}})}\BibitemShut {NoStop}%
\bibitem [{\citenamefont {Maartens}\ and\ \citenamefont
  {Koyama}(2010)}]{Maartens:2010ar}%
  \BibitemOpen
  \bibfield  {author} {\bibinfo {author} {\bibfnamefont {R.}~\bibnamefont
  {Maartens}}\ and\ \bibinfo {author} {\bibfnamefont {K.}~\bibnamefont
  {Koyama}},\ }\href {https://doi.org/10.12942/lrr-2010-5} {\bibfield
  {journal} {\bibinfo  {journal} {Living Rev. Rel.}\ }\textbf {\bibinfo
  {volume} {13}},\ \bibinfo {pages} {5} (\bibinfo {year} {2010})},\ \Eprint
  {https://arxiv.org/abs/1004.3962} {arXiv:1004.3962 [hep-th]} \BibitemShut
  {NoStop}%
\bibitem [{\citenamefont {Di~Mauro}(2025{\natexlab{b}})}]{DiMauro:2025uxt}%
  \BibitemOpen
  \bibfield  {author} {\bibinfo {author} {\bibfnamefont {M.}~\bibnamefont
  {Di~Mauro}},\ }\href@noop {} {\  (\bibinfo {year} {2025}{\natexlab{b}})},\
  \Eprint {https://arxiv.org/abs/2511.19622} {arXiv:2511.19622 [hep-ph]}
  \BibitemShut {NoStop}%
\bibitem [{\citenamefont {Hebecker}(2002)}]{Hebecker2002}%
  \BibitemOpen
  \bibfield  {author} {\bibinfo {author} {\bibfnamefont {A.}~\bibnamefont
  {Hebecker}},\ }\href {https://doi.org/10.1016/S0550-3213(02)00253-5}
  {\bibfield  {journal} {\bibinfo  {journal} {Nucl. Phys. B}\ }\textbf
  {\bibinfo {volume} {632}},\ \bibinfo {pages} {101} (\bibinfo {year}
  {2002})},\ \Eprint {https://arxiv.org/abs/hep-ph/0112230}
  {arXiv:hep-ph/0112230 [hep-ph]} \BibitemShut {NoStop}%
\bibitem [{\citenamefont {Grossman}\ and\ \citenamefont
  {Neubert}(2000{\natexlab{b}})}]{GrossmanNeubert2000}%
  \BibitemOpen
  \bibfield  {author} {\bibinfo {author} {\bibfnamefont {Y.}~\bibnamefont
  {Grossman}}\ and\ \bibinfo {author} {\bibfnamefont {M.}~\bibnamefont
  {Neubert}},\ }\href {https://doi.org/10.1016/S0370-2693(00)00054-X}
  {\bibfield  {journal} {\bibinfo  {journal} {Phys. Lett. B}\ }\textbf
  {\bibinfo {volume} {474}},\ \bibinfo {pages} {361} (\bibinfo {year}
  {2000}{\natexlab{b}})},\ \Eprint {https://arxiv.org/abs/hep-ph/9912408}
  {arXiv:hep-ph/9912408 [hep-ph]} \BibitemShut {NoStop}%
\bibitem [{\citenamefont {del Aguila}\ \emph {et~al.}(2004)\citenamefont {del
  Aguila}, \citenamefont {Perez-Victoria},\ and\ \citenamefont
  {Santiago}}]{delAguilaEtAl2004}%
  \BibitemOpen
  \bibfield  {author} {\bibinfo {author} {\bibfnamefont {F.}~\bibnamefont {del
  Aguila}}, \bibinfo {author} {\bibfnamefont {M.}~\bibnamefont
  {Perez-Victoria}},\ and\ \bibinfo {author} {\bibfnamefont {J.}~\bibnamefont
  {Santiago}},\ }\href {https://doi.org/10.1140/epjcd/s2003-03-821-9}
  {\bibfield  {journal} {\bibinfo  {journal} {Eur. Phys. J. C}\ }\textbf
  {\bibinfo {volume} {33}},\ \bibinfo {pages} {S773} (\bibinfo {year}
  {2004})},\ \Eprint {https://arxiv.org/abs/hep-ph/0310352}
  {arXiv:hep-ph/0310352 [hep-ph]} \BibitemShut {NoStop}%
\bibitem [{\citenamefont {Ghilencea}\ and\ \citenamefont
  {Lee}(2006)}]{GhilenceaLee2006}%
  \BibitemOpen
  \bibfield  {author} {\bibinfo {author} {\bibfnamefont {D.~M.}\ \bibnamefont
  {Ghilencea}}\ and\ \bibinfo {author} {\bibfnamefont {H.~M.}\ \bibnamefont
  {Lee}},\ }\href {https://doi.org/10.1142/S0217732306020147} {\bibfield
  {journal} {\bibinfo  {journal} {Mod. Phys. Lett. A}\ }\textbf {\bibinfo
  {volume} {21}},\ \bibinfo {pages} {769} (\bibinfo {year} {2006})},\ \Eprint
  {https://arxiv.org/abs/hep-ph/0508209} {arXiv:hep-ph/0508209 [hep-ph]}
  \BibitemShut {NoStop}%
\bibitem [{\citenamefont {del Aguila}\ \emph {et~al.}(2006)\citenamefont {del
  Aguila}, \citenamefont {Perez-Victoria},\ and\ \citenamefont
  {Santiago}}]{delAguilaEtAl2006}%
  \BibitemOpen
  \bibfield  {author} {\bibinfo {author} {\bibfnamefont {F.}~\bibnamefont {del
  Aguila}}, \bibinfo {author} {\bibfnamefont {M.}~\bibnamefont
  {Perez-Victoria}},\ and\ \bibinfo {author} {\bibfnamefont {J.}~\bibnamefont
  {Santiago}},\ }\href {https://doi.org/10.1088/1126-6708/2006/10/056}
  {\bibfield  {journal} {\bibinfo  {journal} {JHEP}\ }\textbf {\bibinfo
  {volume} {10}},\ \bibinfo {pages} {056}},\ \Eprint
  {https://arxiv.org/abs/hep-ph/0601222} {arXiv:hep-ph/0601222 [hep-ph]}
  \BibitemShut {NoStop}%
\bibitem [{\citenamefont {Rattazzi}\ \emph {et~al.}(2003)\citenamefont
  {Rattazzi}, \citenamefont {Scrucca},\ and\ \citenamefont
  {Strumia}}]{Rattazzi:2003rj}%
  \BibitemOpen
  \bibfield  {author} {\bibinfo {author} {\bibfnamefont {R.}~\bibnamefont
  {Rattazzi}}, \bibinfo {author} {\bibfnamefont {C.~A.}\ \bibnamefont
  {Scrucca}},\ and\ \bibinfo {author} {\bibfnamefont {A.}~\bibnamefont
  {Strumia}},\ }\href {https://doi.org/10.1016/j.nuclphysb.2003.09.034}
  {\bibfield  {journal} {\bibinfo  {journal} {Nucl. Phys. B}\ }\textbf
  {\bibinfo {volume} {674}},\ \bibinfo {pages} {171} (\bibinfo {year}
  {2003})},\ \Eprint {https://arxiv.org/abs/hep-th/0305184}
  {arXiv:hep-th/0305184} \BibitemShut {NoStop}%
\bibitem [{\citenamefont {Gregoire}\ \emph {et~al.}(2005)\citenamefont
  {Gregoire}, \citenamefont {Rattazzi}, \citenamefont {Scrucca}, \citenamefont
  {Strumia},\ and\ \citenamefont {Trincherini}}]{Gregoire:2004nn}%
  \BibitemOpen
  \bibfield  {author} {\bibinfo {author} {\bibfnamefont {T.}~\bibnamefont
  {Gregoire}}, \bibinfo {author} {\bibfnamefont {R.}~\bibnamefont {Rattazzi}},
  \bibinfo {author} {\bibfnamefont {C.~A.}\ \bibnamefont {Scrucca}}, \bibinfo
  {author} {\bibfnamefont {A.}~\bibnamefont {Strumia}},\ and\ \bibinfo {author}
  {\bibfnamefont {E.}~\bibnamefont {Trincherini}},\ }\href
  {https://doi.org/10.1016/j.nuclphysb.2005.05.001} {\bibfield  {journal}
  {\bibinfo  {journal} {Nucl. Phys. B}\ }\textbf {\bibinfo {volume} {720}},\
  \bibinfo {pages} {3} (\bibinfo {year} {2005})},\ \Eprint
  {https://arxiv.org/abs/hep-th/0411216} {arXiv:hep-th/0411216} \BibitemShut
  {NoStop}%
\bibitem [{\citenamefont {Falkowski}(2005)}]{Falkowski:2005zv}%
  \BibitemOpen
  \bibfield  {author} {\bibinfo {author} {\bibfnamefont {A.}~\bibnamefont
  {Falkowski}},\ }\href {https://doi.org/10.1088/1126-6708/2005/05/073}
  {\bibfield  {journal} {\bibinfo  {journal} {JHEP}\ }\textbf {\bibinfo
  {volume} {05}},\ \bibinfo {pages} {073}},\ \Eprint
  {https://arxiv.org/abs/hep-th/0502072} {arXiv:hep-th/0502072} \BibitemShut
  {NoStop}%
\bibitem [{\citenamefont {York}(1972)}]{York:1972sj}%
  \BibitemOpen
  \bibfield  {author} {\bibinfo {author} {\bibfnamefont {J.~W.}\ \bibnamefont
  {York}},\ }\href {https://doi.org/10.1103/PhysRevLett.28.1082} {\bibfield
  {journal} {\bibinfo  {journal} {Phys. Rev. Lett.}\ }\textbf {\bibinfo
  {volume} {28}},\ \bibinfo {pages} {1082} (\bibinfo {year}
  {1972})}\BibitemShut {NoStop}%
\bibitem [{\citenamefont {Gibbons}\ and\ \citenamefont
  {Hawking}(1977)}]{Gibbons:1976ue}%
  \BibitemOpen
  \bibfield  {author} {\bibinfo {author} {\bibfnamefont {G.~W.}\ \bibnamefont
  {Gibbons}}\ and\ \bibinfo {author} {\bibfnamefont {S.~W.}\ \bibnamefont
  {Hawking}},\ }\href {https://doi.org/10.1103/PhysRevD.15.2752} {\bibfield
  {journal} {\bibinfo  {journal} {Phys. Rev. D}\ }\textbf {\bibinfo {volume}
  {15}},\ \bibinfo {pages} {2752} (\bibinfo {year} {1977})}\BibitemShut
  {NoStop}%
\bibitem [{\citenamefont {Csaki}\ \emph {et~al.}(2000)\citenamefont {Csaki},
  \citenamefont {Graesser}, \citenamefont {Randall},\ and\ \citenamefont
  {Terning}}]{Csaki:1999mp}%
  \BibitemOpen
  \bibfield  {author} {\bibinfo {author} {\bibfnamefont {C.}~\bibnamefont
  {Csaki}}, \bibinfo {author} {\bibfnamefont {M.}~\bibnamefont {Graesser}},
  \bibinfo {author} {\bibfnamefont {L.}~\bibnamefont {Randall}},\ and\ \bibinfo
  {author} {\bibfnamefont {J.}~\bibnamefont {Terning}},\ }\href
  {https://doi.org/10.1103/PhysRevD.62.045015} {\bibfield  {journal} {\bibinfo
  {journal} {Phys. Rev. D}\ }\textbf {\bibinfo {volume} {62}},\ \bibinfo
  {pages} {045015} (\bibinfo {year} {2000})},\ \Eprint
  {https://arxiv.org/abs/hep-ph/9911406} {arXiv:hep-ph/9911406} \BibitemShut
  {NoStop}%
\end{thebibliography}%

\end{document}